\documentclass[a4paper,11pt,twoside]{book}

\pdfoutput=1

\usepackage[dvips,a4paper,portrait,twoside,inner=40mm,outer=25mm,headheight=15mm,headsep=5mm,bottom=40mm,top=40mm]{geometry}
\usepackage{fancyhdr}
\usepackage{makeidx}
\usepackage{nomencl}
\usepackage{setspace}	
\usepackage{url}
\usepackage{latexsym}
\usepackage{amsfonts}
\usepackage{amssymb}
\usepackage{amsmath}
\usepackage{graphicx}
\usepackage{amsthm}
\usepackage{float}
\usepackage{times}
\usepackage{subfig}	
\usepackage{hyperref}	

\newcommand{\system}{\bar{s}=\bar{B}(\bar{s})}
\newcommand{\orderedsetof}[1]{\langle#1\rangle}		
\newcommand{\WDB}{\textrm{WDB}}
\newcommand{\FALSE}{\mbox{\bf false}}
\newcommand{\TRUE}{\mbox{\bf true}}
\newcommand{\setof}[1]{\{#1\}}				
\newcommand{\arrr}{\longrightarrow}
\newcommand{\Dec}{\mbox{\sf Dec}}
\newcommand{\derivablefrom}{\mathrel{:-}}
\newcommand{\bis}{\approx}
\newcommand{\SNames}{\textit{SNames}}
\newcommand{\LLL}{L}
\newcommand{\question}[2]{#1\stackrel{?}{\bis} #2}

\newcommand{\AND}{\mathrel{\&}}

\newcommand{\MID}{\mathrel{\rule[-0.25em]{0.1em}{1em}}}
\newcommand{\TC}{\mbox{\sf TC}}
\newcommand{\Rec}{\mbox{\sf Rec}}
\newcommand{\LOOP}{\circlearrowleft}

\newtheorem{note}{Note}
\theoremstyle{plain}
\newtheorem{ass:bnf}{Assertion}
\theoremstyle{definition}
\newtheorem{defn}{Definition}

\newtheorem{definition}{Definition}

\floatstyle{ruled}
\newfloat{xml-file}{thp}{lop}
\floatname{xml-file}{XML-WDB file}
\floatstyle{ruled}
\newfloat{xml-schema}{thp}{lop}
\floatname{xml-schema}{XML schema}

\makeindex

\begin{document}


\pagestyle{empty}		

\begin{titlepage}
  \begin{center}
    \includegraphics[width=0.80\textwidth]{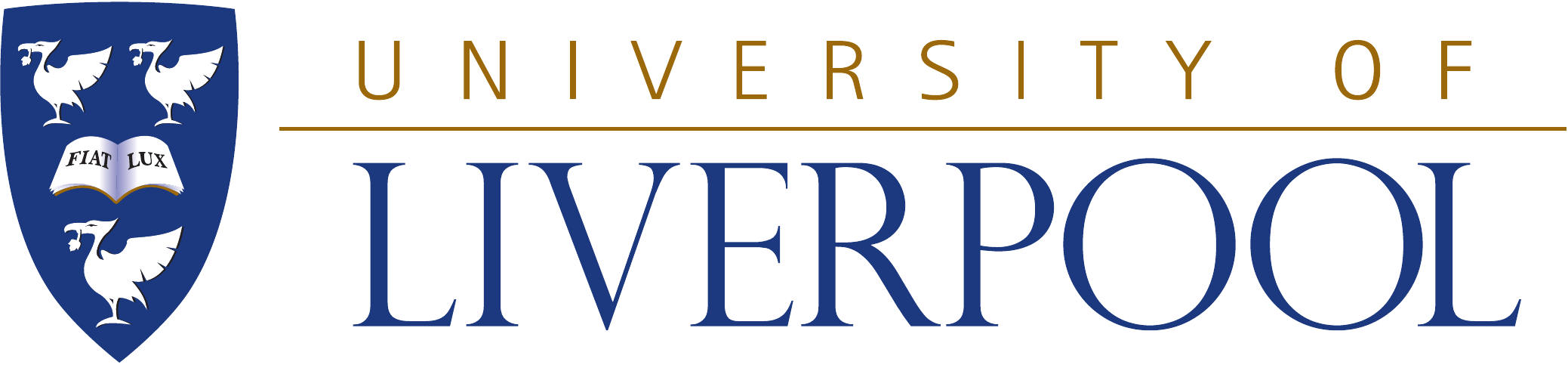}
    \\[2.5cm]
    {\huge \bf Hyperset Approach to Semi-structured Databases and the Experimental Implementation of the Query Language Delta}
    \\[1.75cm]
    {\Large Thesis submitted in accordance with the \\
    requirements of the University of Liverpool \\
    for the degree of Doctor in Philosophy by \\
    {\bf Richard Molyneux}
    \\[1.0cm]
    Thesis Supervisors: Dr. Vladimir Sazonov \\ Dr. Alexei Lisitsa
    \\[1.0cm]
    External examiner: Dr. Ulrich Berger \\ Internal examiner: Dr. Grant Malcolm
    \\[2.0cm]
    Department of Computer Science \\
    The University of Liverpool \\
    January, 2009
    } \\
  \end{center}
\end{titlepage}

\frontmatter
\pagestyle{plain}		

\chapter*{Abstract}

{\large
This thesis presents practical suggestions towards the implementation of the hyperset approach to semi-structured databases and the associated query language $\Delta$ (Delta). This work can be characterised as part of a top-down approach to semi-structured databases, from theory to practice.

\medskip

Over the last decade the rise of the World-Wide Web has lead to the suggestion for a shift from structured relational databases to semi-structured databases, which can query distributed and heterogeneous data having unfixed/non-rigid structure in contrast to ordinary relational databases. In principle, the World-Wide Web can be considered as a large distributed semi-structured database where arbitrary hyperlinking between Web pages can be interpreted as graph edges (inspiring the synonym
\linebreak
\mbox{`Web-like'} for `semi-structured' databases also called here WDB). In fact, most approaches to semi-structured databases are based on graphs, whereas the hyperset approach presented here represents such graphs as systems of set equations. This is more than just a style of notation, but rather a style of thought and the corresponding mathematical background leads to considerable differences with other approaches to semi-structured databases. The hyperset approach to such databases and to querying them has clear semantics based on the well established tradition of set theory and logic, and, in particular, on non-well-founded set theory because semi-structured data allow arbitrary graphs and hence cycles.

\medskip

The main original part of this work consisted in implementation of the hyperset $\Delta$-query language to semi-structured databases, including worked example queries.
In fact, the goal was to demonstrate the practical details of this approach and language.
The required development of an extended, practical version of the language based on the existing theoretical version, and the corresponding operational semantics.
Here we present detailed description of the most essential steps of the implementation.
Another crucial problem for this approach was to demonstrate how to deal in reality with the concept of the equality relation between (hyper)sets, which is computationally realised by the bisimulation relation. In fact, this expensive procedure, especially in the case of distributed  semi-structured data, required some additional theoretical considerations and practical suggestions for efficient implementation. To this end the ``local/global'' strategy for computing the bisimulation relation over distributed semi-structured data was developed and its efficiency was experimentally confirmed.

\medskip

Finally, the XML-WDB format for representing any distributed WDB as system of set equations was developed so that arbitrary XML elements can participate and, hence, queried by the $\Delta$-language.

\medskip

The query system with the syntax of the language and several example queries from this thesis is available online at

\begin{center}
\url{http://www.csc.liv.ac.uk/~molyneux/t/}
\end{center}

\bigskip

\noindent
{\sl Keywords:} Semi-structured, Web-like, distributed databases, hypersets, bisimulation, query language $\Delta$ (Delta)
}

\clearpage
\thispagestyle{empty}

\chapter*{Dedication}

{\large
This thesis is dedicated to my loving grandparents.
}

\clearpage
\thispagestyle{empty}

\chapter*{Acknowledgement}

{\large
The research presented in this thesis was undertaken at the Department of Computer Science under the supervision of Dr. Vladimir Sazonov and Dr. Alexei Lisitsa.

\medskip

This work was inspired by the research of my primary supervisor Dr. Vladimir Sazonov, his encouragement and dedication was invaluable in developing those ideas presented here. Additionally, I am grateful to the help and support given by Dr. Alexei Lisitsa and Prof. Michael Fisher. This work was made possible by the scholarship awarded to me by the Department of Computer Science.

\medskip

I wish to thank my parents 
whose love and support has been the foundation of all my achievements. Also, to my brothers and sister 
for their encouragement and support.
}

\clearpage
\thispagestyle{empty}


\setcounter{tocdepth}{2}
\tableofcontents

\clearpage
\thispagestyle{empty}

%


\mainmatter

\fancyhead[RO]{{\footnotesize{\nouppercase{\rightmark}}}\hspace{2em}\thepage}
\fancyhead[LE]{\thepage\hspace{2em}\footnotesize{\nouppercase{\leftmark}}}
\pagestyle{fancy}




\chapter{Introduction}

Before the emergence of the database culture in the late 1960's data processing involved the ad~hoc manipulation of data on tape or disk. The complexity of developing and managing such systems inspired new research into the principles of data organisation. Three models were suggested during the late 1960's and early 1970's: {i)} the hierarchical model  \cite{TL76}, {ii)} the network model \cite{TF76} proposed by the Data Base Task Group, and {iii)} Codd's relational model~\cite{C83}.

\medskip

The hierarchical and network models are closely related to the notion of \mbox{\emph{object-orientation}} as is argued in \cite{U88} and are, in fact, based on the idea of object identity, i.e.\ an object whose meaning is determined not only by records of values of its fields (or attributes) but also by a pointer or address of this object within files or memory. Note that, two objects are identical if they have the same address or pointer, whereas two objects are equivalent if they share the same fields. Links $T_1\rightarrow T_2$ denoting many-to-one relationships between record types constitute a graph in the case of the network model, and a forest (consisting of trees) in the case of the hierarchical model. Physically, each such graph or tree edge is represented by real relationships between OIDs of records of types $T_1$ and $T_2$. 

\medskip

On the other hand, the great success of Codd's relational model, which can be considered as a value-oriented approach, was based on taking the most fundamental concepts of logic and set theory as its foundation. Thus, any relation is a set of tuples, with each tuple also being represented%
\footnote{
under our interpretation
} 
by a set of a special kind (a set of attribute labelled values). In fact, this approach assumes an abstract view on data values where the concept of object identity is not needed. (Note that the concept of object identity may play a role in implementation but not in the abstract model itself.) The relational model was further extended by object-orientation during the early 1990's \cite{GMUW08}, thus again absorbing the idea of object identity and additionally allowing complex data values with possibly nested structure and the idea of abstract data type with encapsulated methods. 

\medskip

However, object-relational databases are still restricted by an imposed relational schema, that is they have a rigid structure. Note that complex, nested structures considered in this approach are somewhat related with the idea of semi-structured databases discussed in this thesis, but the latter approach does not assume in general a rigid structure. 
Moreover, the hyperset approach to semi-structured databases presented in this thesis is crucially based on the value-oriented rather than the object-oriented view

\section*{From relations to semi-structured or Web-like data}

From the second half of the 1990s a new idea of semi-structured databases emerged (see \cite{ABS00} as a general reference). In the age of the Internet and the World-Wide Web (WWW), allowing accessibility of remote and heterogeneous databases, the relational paradigm has become too narrow and restrictive. Indeed, the structure of the data over the WWW is typically \mbox{non-fixed} or \mbox{non-uniform}. The idea of graph representation of data was introduced with the interpretation of graph edges like hyperlinks on the Web. Due to this analogy such graph-like semi-structured databases can also be reasonably called Web-like databases (WDB) \cite{LS97}.

\medskip

An important example of the graph approach (in its pure form) is the system Lore \cite{MAGQW97lore} and the corresponding query language Lorel \cite{AQMWW97lorel}, which considers graph vertices as object identities (OIDs) with equality between vertices understood as essentially literal coincidence of OIDs irrespectively of their information content (presented by outgoing edges according to our hyperset approach). In fact, this is typical for most semi-structured database approaches  \cite{AQMWW97lorel,XQuery,CCDFPT99,CRF01,XSLT,CM90,CDQT02,XPath,FFKLS98,FFLS97,GMPQRSUVW97,GMW99,MAGQW97lore,PGMW95}, except in the case of the query language UnQL~\cite{BFS00} (as discussed briefly below).

\medskip

On the other hand, because of this idea of browsing by ``picturing'' the informational content (data value) of a graph vertex, considering such graphs merely as a binary (or ternary, if taking labels on edges into account) relation is not fully adequate in this context. Thus, we view the notion of semi-structured data as more than just a relation, that is more than just a graph where vertices are (uniquely presented by) object identities. In our hyperset theoretic approach, which is value-oriented, it becomes more appropriate to consider those
target vertices of outgoing edges from any given vertex $v$ as children or even as \emph{elements} of $v$ with $v$ understood as a \emph{set} of its elements. It is the latter view on graph vertices which makes it value oriented. In fact, similar terminology is used in Extensible Markup Language (XML), which is a widely adopted approach to semi-structured data. However, this is only a superficial similarity with the set theoretic approach. XML only allows to syntactically represent semi-structured data whereas treating such data as sets requires an additional level of abstraction (supported by an appropriate technique such as some set theoretic query language) which is more than just using the rudiments of set theoretic terminology.

\medskip

XML documents, in fact, represent ordered tree structured data rather than arbitrary graph structured data, however, using the attributes \verb+id+ and \verb+ref+ allows one to imitate in XML arbitrary graphs as well. Considering the ordering of data in XML documents as an essential feature is related mainly with numerous software implementations which are deliberately sensitive to the order of such data. But, XML documents can also be treated as unordered, as we do in this thesis. 
Note that XML plays only an auxiliary role in our approach as a particular way of representing semi-structured data (XML-WDB format). Our main terminology and abstract data model is based on (hyper)set theory.

\section*{The graph model and set theoretic model}

The interpretation of graph vertices as sets of their ``children'' leads us again to a set theoretic idea of representation of data, semi-structured data, a far going generalisation of the relational (value-oriented) approach. It is also worth noting that in the foundations of mathematics the previous century was marked by the triumph of the set theoretic approach for representing mathematical data as well as the style of mathematical language and reasoning. Mathematical logicians also developed generalised computability theory over abstract sets (of sets of sets, etc.) in the form of admissible set theory \cite{B75}. In computer science, the set theoretic programming language SETL \cite{SDDS86,S70} was created, quite naturally, for the case of finite sets only. Also some theoretical considerations on computability and query languages over hereditarily finite sets were done in \cite{DM86,DM92,LS99,S87,S93,S97,S06} with the perspective of a generalised set-theoretically presented databases -- in fact semi-structured -- even before the term ``semi-structured databases'' had arisen. Moreover, the set theoretic approach is closely related with a special version of the graph approach when graphs are considered up to bisimulation (see below).

\medskip

The first mathematical result relating both the set and graph approaches was Mostowski's Collapsing Lemma, allowing the interpretation of graph vertices as sets of sets corresponding to children of these vertices. This, however, worked properly only for well-founded graphs and sets (which in the finite case, especially interesting for database applications, means the absence of cycles).  But arbitrary graphs with cycles can also be ``collapsed'' into sets (interrelated by the membership relation) in the more general non-well-founded set theory also called hyperset theory \cite{A88,BM96}. Here, for example the set $\Omega=\{\Omega\}$ consisting of itself is quite natural and meaningful, and corresponds to the simplest graph cycle {\LARGE$\LOOP$}.

\medskip

These two trends, from abstract set theory to more concrete graph model of semi-structured data (which is closer to implementation), and vice versa were called in \cite{S06} top-down and bottom-up approaches. They meet most closely in the work on UnQL query language \cite{BFS00} which is devoted to a specific graph model approach to semi-structured data considered up to bisimulation. The latter concept is also the key one in the works \cite{LS97,LS99,S87,S93,S06} (serving as the theoretical background for this thesis) for interpreting graph vertices as a system of (hyper)sets belonging one to another according to the graph edges. Nevertheless, \cite{BFS00} is still rather a graph approach than hyperset one according to the special, however related to, but not a genuine set theoretical way in which \cite{BFS00} treats graphs (see Section~\ref{sec:comparisons_unql} and \cite{S06}). The main difference is that graphs considered in \cite{BFS00} have multiple ``input'' and ``output'' vertices, whereas graphs as considered in our hyperset approach have only one ``input'' corresponding to the set itself (and possibly one ``output'' corresponding to the empty set if it is contained in the transitive closure of this set). In fact, working with these ``inputs'' and ``outputs'' (used for appending one graph to another, etc.) is conceptually rather graph-theoretical than
\linebreak
set-theoretical.

\section*{Hyperset approach to semi-structured or Web-like databases}

As discussed above, the hyperset approach to semi-structured databases interprets graph structured data as abstract hypersets. Moreover, for the purposes of implementation, such graphs are represented as systems of set equations e.g. $\Omega=\setof{\Omega}$ for the graph {\LARGE$\LOOP$}. In fact, arbitrary finite graphs can be rewritten into systems of set equations and vice versa, where graph vertices (or object identities) represent set names. Moreover, elements of sets in these set equations should be labelled according to labelling of graph edges, and, in fact, these labels are the carriers of atomic information in the hyperset approach to semi-structured databases. Furthermore, graph structure or, respectively, set-element nesting organises such atomic data, just like relational tables in the relational or nested relational approaches. The notion of equality between sets can be represented in graph terms by the bisimulation relation on vertices or set names whose idea consists, roughly speaking, in (recursively) ignoring the order and repetition. Thus, any two graph vertices or set names denote the same set if they are bisimilar, that is contain the same (recursively, up to bisimulation) elements. In fact, the bisimulation relation is very important in our approach being a fundamental concept underlying hyperset theory.

\subsection*{Hyperset query language $\Delta$}

The associated $\Delta$-query language is based on set theory and predicate logic, being an extension of the basic or rudimentary operations \cite{G74,J72} -- the \emph{core} fragment of $\Delta$. The set theoretic operators of the $\Delta$-language, like in the relational calculus, have clear and well-understood semantics. In fact, the expressive power of $\Delta$ (the core fragment plus transitive closure, decoration and recursion set theoretic operations) was shown in \cite{S93} and \cite{LS99,S95} to capture all polynomial time computable operations over hereditarily finite sets and, respectively, hypersets. Also, another version of the language was shown in \cite{LS01,LS97logspace} to capture exactly all LogSpace computable operations over hereditarily finite sets (without cycles). Therefore, in principle, the $\Delta$-query language can be reasonably considered as computationally viable and worthy of implementation.

\bigskip

Some earlier preliminary work on the implementation of the $\Delta$-query language to WDB was done earlier by Yuri Serdyuk in \cite{S96}, as well as in some practical attempt towards a new implementation based on multiple distributed agents working cooperately over the Internet \cite{SS02} (taking into account the earlier theoretical work \cite{S01}). More recently the implementation work leading to this thesis was done in \cite{M04}. However, the latter implementation was insufficiently perfect. This antecedent work subsequently inspired the proposal for further research and the development of a sufficiently detailed implementation, that is, the point of the work done here. Note that some details of the implementation described here were published in \cite{MS07}.

\subsection*{Implementation of the hyperset approach}

The goal of this work was to demonstrate how the hyperset approach to semi-structured or Web-like databases could be implemented, with the aim of presenting this approach in a practical rather than theoretical context and making it accessible to a more practically oriented audience. In particular, the practical characteristic of this work assumes representation of hyperset data as files distributed over the World-Wide Web and the implementation of the hyperset query language $\Delta$ allowing queries over such distributed data. Importantly, the implemented language should preserve the original high level, declarative character%
\footnote{
Recall that, for example, Prolog initially intended to be a logical, declarative programming language, eventually has both declarative and imperative features. This mixture of ideologies was the result of making this language more efficient.
}
and retain its set theoretic style. Further, this approach should demonstrate the power of the set theoretic style of thought towards semi-structured databases. Note that the query system (which is implemented in Java) and the example queries described in this thesis can be found at

\begin{center}
\url{http://www.csc.liv.ac.uk/~molyneux/t/}
\end{center}

\subsection*{Efficiency issues}

Another goal consisted in the subsequent investigation of theoretical considerations arising from this experimental implementation, specifically the problem of efficient implementation of the equality or the bisimulation relation -- which crucially underlies this hyperset theoretic approach. Moreover, our proposed solution was restricted to making the bisimulation relation efficient only in context of distributed WDB which may require numerous and particularly expensive downloads of files from the World-Wide Web. However, this work does not consider the problem of efficiency in the non-distributed case, especially taking into account the previous works on efficient bisimulation algorithms that, on the other hand, do not consider distribution \cite{DPP04,F89}. Note that, many other aspects of efficiency of the implementation (such as indexing, hashing and other physical data organisation techniques \cite{U88}) as well as various other questions which should be resolved for creating a sufficiently realistic database management system were inevitably postponed here. In fact, the primary aim of this work was the correct and meaningful implementation of a non-trivial and user friendly version of the $\Delta$-language.

\section*{Organisation of the thesis}

Details of the implementation are rather technical, thus it makes sense to firstly explain the intuitive (or high level) meaning of the hyperset approach and demonstrate example queries of the implemented $\Delta$-query language. Secondly, technical details of the implementation appear towards the end of the thesis detailing the lower level aspects of our approach. Note that, the material presented in this thesis follows an intuitive perception of this approach towards semi-structured databases rather than a strict logical dependency. 

\medskip

\noindent
The thesis is organised into four parts:

\medskip

Part~\ref{part:theory}, ``Hyperset approach to querying Web-like databases'', gives an overview of the
\linebreak
implemented hyperset approach to semi-structured or Web-like databases and the associated query language $\Delta$, including worked example queries. The point of this part is to introduce this approach on an intuitive level before discussing the technical details of implementation.

\medskip

Part~\ref{part:local_global}, ``Local/global approach to optimise bisimulation and querying'', is concerned with the problem of efficient implementation of the  equality or bisimulation relation. Here two joint strategies were suggested for resolving this problem: i) implementation of an Internet service for resolving bisimulation questions, and ii) the computation of bisimulation approximations on fragments of distributed Web-like databases to aid the computation of global bisimulation. The viability of these suggestions as solutions is supported by empirical testing.

\medskip

Part~\ref{part:implementation}, ``Implementation issues'', presents the technical details of the implementation of the hyperset approach towards semi-structured or Web-like databases. We start by detailing query execution (which we feel is potentially more important for readers) followed with query parsing and contextual analysis, although query execution is, in fact, formally dependent on the latter syntactical considerations. Finally, XML representation of WDB systems of set equation has a quite isolated role in our approach and is presented at the end of this technical material, but this discussion is actually quite self-contained and can be read independently of the rest of this thesis.

\pagebreak

Part~\ref{part:evaluation}, ``Evaluation'', concludes with comparative analysis with other known approaches towards semi-structured databases, and finishes with some future prospects and closing
\linebreak
remarks.


%
%


\part{Hyperset approach to querying Web-like databases}\label{part:theory}


\chapter{Semi-structured or Web-like databases}\label{chap:ssd_wdb}

The term \emph{semi-structured data} denotes data which has a characteristically unfixed or non-rigid structure, thus semi-structured data is considered as ``schemaless'' or ``self-describing''%
\footnote{
The consideration of semi-structured data as ``self-describing'' is somewhat misleading as it might be wrongly thought to suggest clear semantic description of such data. In particular, when considering the graph representation of semi-structured data, labels have only an informal meaning dependant on subjective interpretation of language, e.g. the imprecise term ``location'' could have many interpretations -- address, map coordinates,
URI, anatomical, etc.
}%
having no complete structural description or schema \cite{ABS00}. However, typically semi-structured data is similar to structured data e.g. relational data (as described below) but without strictly imposed structure. More specifically our approach to semi-structured databases is based on (hyper)set theory \cite{A88,BM96}.

\section{Set theoretic view of structured and semi-structured data}\label{sec:sd_and_ssd}

\subsection{Structured relational data}\label{sec:relational_set_theoretic}

\begin{figure}[ht]
\centering
\includegraphics{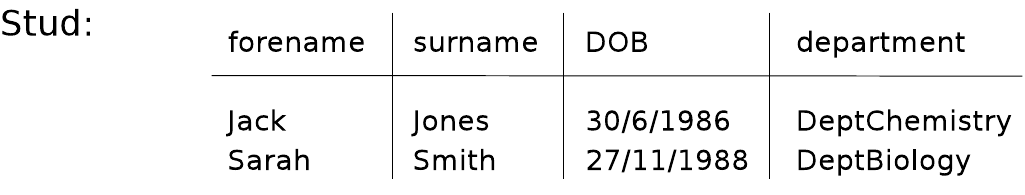}
\caption{Relational table of students.}
\label{fig:relational_data}
\end{figure}

Structured data has a fixed and rigid structure such as relational data \cite{CB02} described by relational schema  $R(A_1,A_2,...,A_n)$, where $R$ is \emph{relation} name and $A_i$ are \emph{attributes} (constrained by the domain $D_i$). In the relational model, relations are naturally represented as tables with attributes as named columns of a table. For example, the \verb+Stud+ relation shown in Figure~\ref{fig:relational_data} has the attributes \verb+forename+, \verb+surname+, \verb+DOB+ (date of birth) and \verb+department+.

\noindent
The relational approach is essentially based on set theory, as well as on logic. For example, the \verb+Stud+ relation (above) can be represented as set of student \emph{tuples} (rows or records),
\begin{small}
\begin{verbatim}
    Stud  = { st1, st2, ... }
\end{verbatim}
\end{small}

\noindent
or, better, as
\begin{small}
\begin{verbatim}
    Stud  = { student:st1, student:st2, ... }
\end{verbatim}
\end{small}

\noindent
where each student tuple is represented as a set of labelled atomic values, with labels being \emph{attribute names}, and \emph{attribute values} as atomic values (strings of symbols between quotation marks to distinguish them from set names and attribute names),
\begin{small}
\begin{verbatim}    
    st1  = { forename:"Jack", surname:"Jones",
             DOB:"30/6/1986", department:"DeptChemistry" }

    st2  = { forename:"Sarah", surname:"Smith",
             DOB:"27/11/1988", department:"DeptBiology" }.
\end{verbatim}
\end{small}

\noindent
Let us consider the relational database \verb+Univ+ as the following set of (labelled) relations,
\begin{small}
\begin{verbatim}
    Univ = { departments:Dept, students:Stud, lecturers:Lect, 
             modules:Mod, courses:Course, ... }.
\end{verbatim}
\end{small}

\noindent
The relations \verb+Dept+, \verb+Lect+, \verb+Mod+ and \verb+Course+ will not be further described, they are plausible example relations, like \verb+Stud+, that could belong to a University database. Here the labels (or attributes) \verb+departments+, \verb+students+, \verb+lecturers+, etc., give an informal description of what the sets \verb+Dept+, \verb+Stud+,  \verb+Lect+, etc., are about. These sets could be denoted differently, say as \verb+D+, \verb+S+, \verb+L+, etc. Thus, strictly speaking the denotation of sets does not necessarily carry informational content. Hence the important role of labels (attributes e.g. \verb+forename+) and atomic values (e.g. \verb+"Jack"+), which are the proper carriers of basic information.

\subsection{Relaxation of structural restrictions on relational data}

Relational data with the given schema $R(A_1,A_2,...,A_n)$ has a rigid structure with mandatory attributes $A_i$ for associated tuple components.
It is also known of the more general approaches to \emph{nested} relational databases \cite{PBGG89,RKS88,TF86} where attribute values could be relations. Say, in the above example we could reconsider \verb+DeptChemistry+ as a set (instead of an atomic value) by omitting the quotation marks around \verb+DeptChemistry+ and adding the corresponding set equation further detailing the chemistry department: 
\begin{small}
\begin{verbatim}    
    DeptChemistry  = { name:"Department_of_Chemistry",  
                       lecturers:ChemLect, 
                       modules:ChemMod,  
                       ... }.
\end{verbatim}
\end{small}

\medskip

\noindent
Moreover, we could relax the requirement on students tuples to have a value for each attribute \verb+forename+, \verb+surname+, \verb+age+ and \verb+department+. For example, the DOB of a student could be absent by some reason, but some other information could be present, such as 

\begin{small}
\begin{verbatim}
    email:"jones@liv.ac.uk"
\end{verbatim}
\end{small}

\noindent
or,

\begin{small}
\begin{verbatim}
    sex:"male".
\end{verbatim}
\end{small}

\noindent
Thus, relaxation of traditional structural restrictions on relational databases leads naturally to \mbox{semi-structured} databases, in fact, to the set theoretic 
approach where such data are considered as \emph{arbitrary} set of (labelled) sets of sets, etc., to any depth, represented by set equations like above.

\subsection{Semi-structured data}

For simplicity, we consider semi-structured data as systems of \emph{flat} set equations where a set equation consists of set name $s_i$ equated to a bracket expression $B_{i}(\bar{s})$ like those considered in the above example. In vector form this can be summarised as
\begin{align*}
\bar{s} = \bar{B}(\bar{s}).
\end{align*}

\noindent
Flat bracket expression $\{l_1:s_{i_1},\ldots,l_n:s_{i_n}\}$ is thought of as a set of labelled elements. In the flat (non-nested form) only set names $s_i$ from the list of all set names $\bar{s} = s_1, s_2, ..., s_n$, may participate as elements. Labels $l_j$ can be considered as analogous to attributes in the relational approach, however, element labelling is optional with the default label being the empty label $\Box$ (or \texttt{null}) which can be considered as invisible, such as the absence of labelling in the \verb+Stud+ set above. Formally our general approach does not consider atomic values such as \verb+"Jack"+, \verb+"Jones"+, etc., from the example above. However, any atomic value can be simulated as a set consisting of one labelled empty set \cite{LS97,S93,S06}, such as 
\begin{small}
\begin{verbatim}
    "Jack" = {'Jack':{}}.
\end{verbatim}
\end{small}

\noindent
Strictly speaking, we should use single quotation marks for labels (often omitted for simplicity) and double quotation marks for atomic values. Of course, we can still use the denotation for atomic data like \verb+"Jack"+, but it should be understood as above.

\subsection{Syntactical and conceptual set nesting}

In the case where nesting is allowed (like the participation of \verb+{}+ in the above definition of atomic values, and also in more complicated cases) any set name $s_i$ can be substituted with the corresponding nested bracket expression $B_i$, and vice versa. For example, the \verb+Stud+ set equation could be rewritten with the nested right-hand side (and adding the \verb+student+ attribute) as follows,
\begin{small}
\begin{verbatim}    
    Stud = {
            student:{ forename:"Jack", surname:"Jones",
                      DOB:"30/6/1986", department:"DeptChemistry" },

            student:{ forename:"Sarah", surname:"Smith",
                      DOB:"27/11/1988", department:"DeptBiology" }
           }.
\end{verbatim}
\end{small}

\noindent
Here the nesting of data inside the \verb+Stud+ set equation proves useful in avoiding the introduction of new set names, and thus eliminating \verb+st1+ and \verb+st2+. Moveover, this demonstrates that set names in set equations play an auxiliary role, and can even be readily renamed in an analogous way to renaming variables in any ordinary algebraic equations. Thus the real information of such semi-structured data is carried by labels and set/element nesting. More generally, we could allow (and, in fact, will consider later) arbitrary nesting in the right-hand sides of set equations $\bar{s}=\bar{B}(\bar{s})$. This can be evidently ``unnested'' or ``flattened'' by introducing new (fresh) set names and appropriate set equations. So, our restriction for non-nested systems of set equations (i.e.\ with non-nested right-hand sides) is not essential, but can simplify some considerations.

\medskip

In fact, the notion of non-nested or flat system of set equations is only syntactical and, conceptually, flat systems of set equations allow arbitrary nesting with the participation of set names (corresponding to set equation) as elements

\section{Hyperset theoretic view of semi-structured data}

In the above approach to semi-structured data via systems of set equations $\bar{s}=\bar{B}(\bar{s})$ there was, in fact, no restriction on the 
form of these equations. Thus allowing not only arbitrarily nested, but also cycling data like in the simplest example of a set consisting of itself
\begin{align*}
\Omega=\{\Omega\}.
\end{align*}

\noindent
Mathematically, such kind of sets are considered as non-traditional, although they have already been deeply investigated in \emph{hyperset theory}, as represented in the books \cite{A88,BM96}. From the point of view of semi-structured data there is nothing strange in such sets. Imagine that we have a relational table where some cells can represent other relational tables, etc. Such nesting can be implemented so that ``clicking'' on such a cell leads to the corresponding nested relational table shown instead of the original table. There is no technical or conceptual problem to have such a situation that after several such ``clicks'' we will arrive back to the original table we started ``clicking'' with -- like in the World-Wide Web by successive ``clicking'' we can possibly return to the Web page we started with. Moreover, from the informational or database point of view this can be quite meaningful.

\medskip

For example, let us consider the University database where formally the student set \verb+st1+ has the chemistry department set \verb+DeptChemistry+ as the member, and (possibly many) students are members of the \verb+ChemStud+ set of enrolled chemistry students, as described by mutually recursive set definitions,
\begin{small}
\begin{verbatim}    
    st1  = { forename:"Jack", surname:"Jones",
             DOB:"30/6/1986", department:DeptChemistry }

    DeptChemistry = { ..., enrolled:ChemStud, ... }

    ChemStud = { student:st1, ... }
\end{verbatim}
\end{small}%

\noindent
with \verb+ChemStud+ a subset of the set \verb+Stud+ of all university students. Any set (name) $s_i$ can be defined by referring to other set (names) as elements, etc., so that eventually we could possibly come to the original set $s_i$ -- thus, arbitrary cycling is allowed. 

\medskip

There is more to say about the hyperset approach to semi-structured data on the conceptual level, in particular, on the concept of equality between sets (possibly denoted by different set names) but we will postpone this discussion to Section~\ref{sec:preliminary_bisimulation}. On the current very preliminary level of consideration sets are thought simply as syntactical bracket expressions, or as represented by formal systems of set equations. In fact, we need an abstract concept of hypersets amongst which we could find a (unique) solution to any given system of set equations.

\section{Graph or Web-like view}\label{sec:graph_or_web-like_view}

\subsection{Graph representation of systems of set equations}

Representation of semi-structured databases by systems of set equations presents a clear and mathematically well-understood%
\footnote{
taking into account Section~\ref{sec:hypersets_abstractly}
}
conceptual view of semi-structured data as (hyper)sets. But it also makes sense to consider visualisation of systems of set equations by the equivalent representation as (finite) labelled directed graphs. In fact, it is important for all considerations of this work that any given system of set equations can be considered as a labelled directed graph.

\pagebreak

\begin{figure}[!ht]
\centering
\hspace{-4em}			
\includegraphics{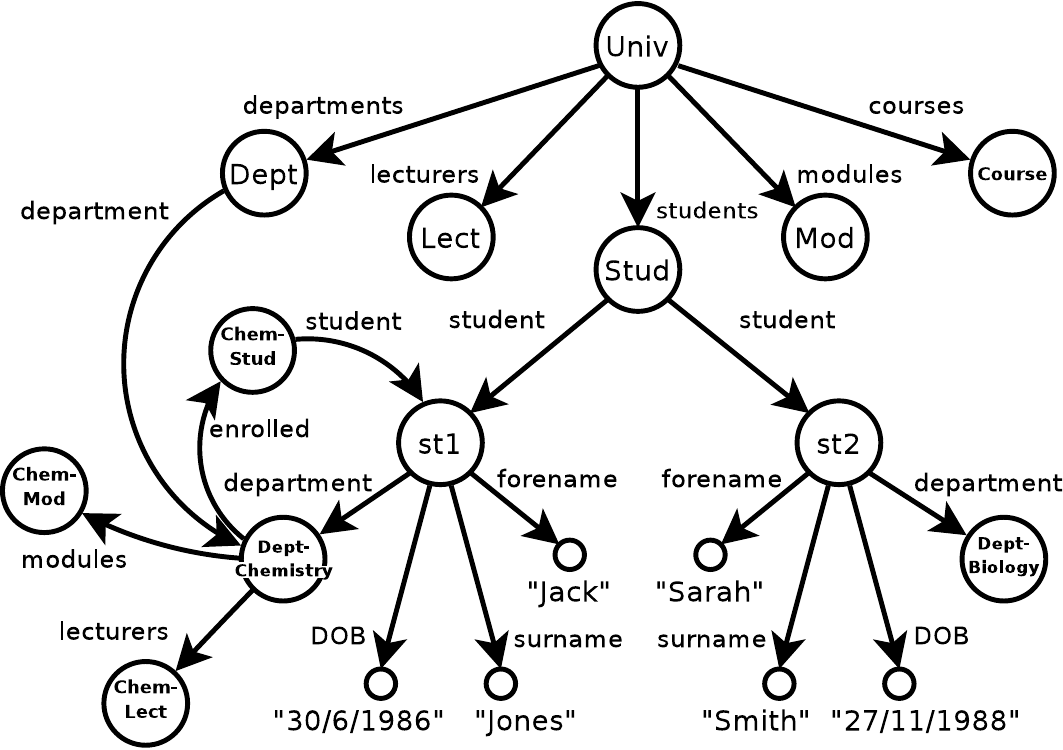}
\caption{Semi-structured database {\tt Univ} represented as directed graph.}
\label{fig:WDB-university}
\end{figure}

\noindent
In fact, most approaches to semi-structured databases typically consider them as labelled directed graphs, that is, semi-structured data is modelled as (finite) directed graph $G = \orderedsetof{N,E}$ with $L$-labelled edges, where $L$ is an infinite set of possible labels ($l_1, l_2,\ldots$, etc., and the empty label $\Box$), $N$ is a finite set of nodes ($s_1, s_2,\ldots$, etc.), and $E$ is a finite set of edges with each edge $s_i\stackrel{l_k}{\rightarrow}s_j$ being formally an ordered triple of the form $\orderedsetof{s_i, s_j, l_k}$. For example, the University database considered in Section~\ref{sec:sd_and_ssd} has the corresponding representation by directed graph shown in Figure~\ref{fig:WDB-university}.

\medskip

\begin{figure}[!ht]
\centering
\includegraphics{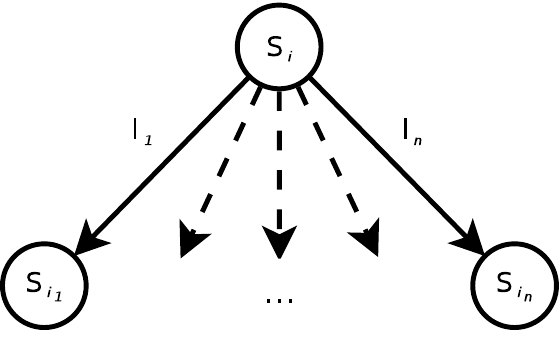}
\caption{Forking of labelled edges generated by the set equation $s_i=\{l_1\!:\!s_{i_1},\ldots,l_n\!:\!s_{i_n}\}$.}
\label{fig:graph_sse_forking}
\end{figure}

The membership of labelled element $label\!:\!s_2$ to the set $s_1$ ($label\!:\!s_2 \in s_1$) corresponds to the labelled edge $s_1 \stackrel{label}{\longrightarrow} s_2$ (and vice versa), where set names $s_i$ serve as (the unique names of) graph nodes. In general, each set equation $s_i=\{l_1\!:\!s_{i_1},\ldots,l_n\!:\!s_{i_n}\}$ from the system generates a fork of labelled edges
$
s_i \stackrel{l_1}{\longrightarrow} s_{i_1},
\ldots,
s_i \stackrel{l_n}{\longrightarrow} s_{i_n}
$
outgoing from $s_i$, as depicted in Figure~\ref{fig:graph_sse_forking}. All those forks generated from every set equation give the corresponding representation as graph. Vice versa, any graph with labelled edges is evidently visualising a system of set equations, with one equation for each node so that each node is thought as a (hyper)set. Thus, graphs and (formal) systems of set equations are essentially equivalent concepts.

\subsection{Graphs or systems of set equations as Web-like databases}

\begin{figure}[!ht]
\centering
\includegraphics[width=8.5cm]{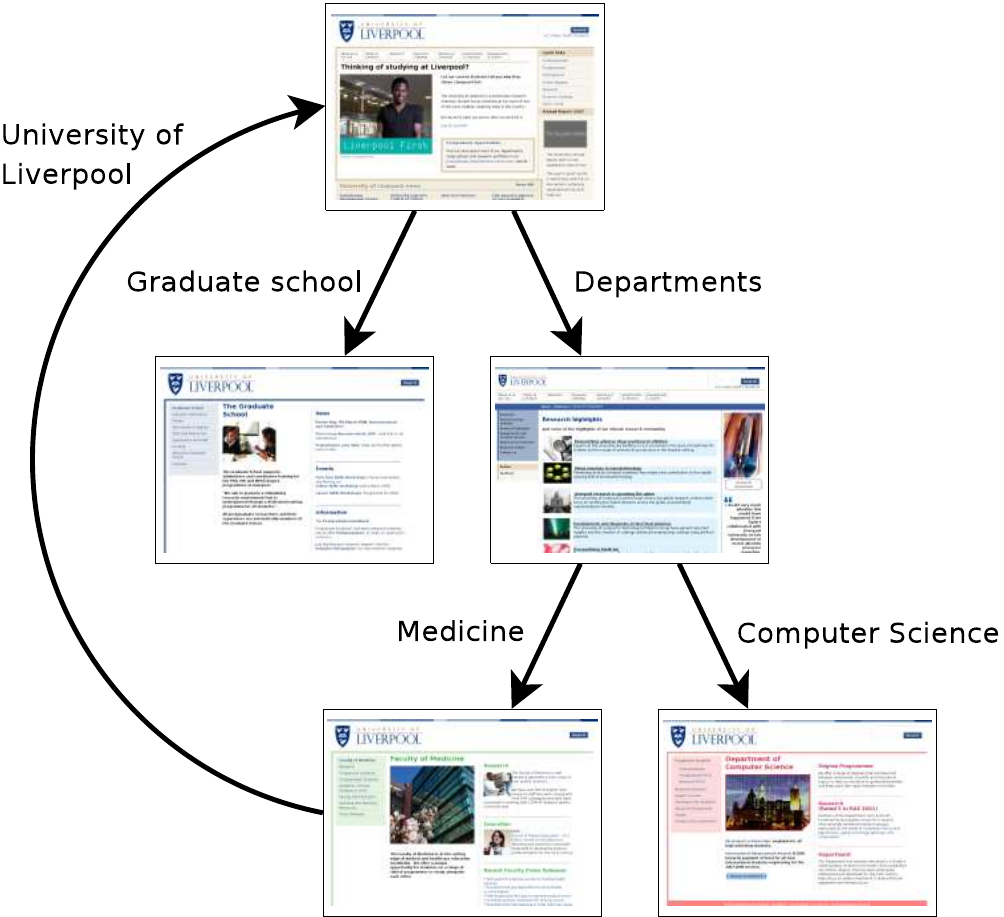}
\caption{Browsing of hyperlinked HTML documents on the University of Liverpool website.}
\label{fig:real_WDB}
\end{figure}

The World-Wide Web (WWW) can, in principle, be considered as a large semi-structured database, consisting of an arbitrarily organised collection of hyperlinked HTML documents. Each HTML document has a corresponding URL (WWW address), and contains textual data with markup tags denoting visualisation and hyperlink information. The following fragment of HTML code is an example of a hyperlink,
\[
\texttt{<a href="}\mbox{\url{http://www.liv.ac.uk/}}\texttt{">University of Liverpool</a>}
\]

\noindent
what in our symbolism of labelled elements can be represented as 
\[
\texttt{University of Liverpool}\,:\,\mbox{\url{http://www.liv.ac.uk/}}
\]

\noindent
and visually (in Web browser) this hyperlink would appear as ``clickable'' fragment of text 
\[
\underline{\texttt{University of Liverpool}}
\]

\noindent
with the URL hidden. Hiding of URLs corresponds to the idea mentioned above that set names (names of graph nodes) actually do not matter from the point of view of the proper information. Only labels on edges or the ``clickable'' links (and other text and visual content) on Web pages carry information, plus, of course, the graphical structure. That is, URLs play a different role than proper information in the WWW. In Figure~\ref{fig:real_WDB} we consider browsing between hyperlinked HTML documents by ``clicking'' on such links. It is evident from this example that hyperlinked HTML documents can express arbitrary relationships, for example the cycle when browsing by ``clicking'' on the links, \verb+Departments+, \verb+Medicine+, \verb+University of Liverpool+, and so on.

\medskip

Thus, any hyperlink can be denoted by the labelled edge $url_i \stackrel{label}{\longrightarrow} url_j$, suggesting the intuitive understanding of hyperlinking as arbitrary labelled directed graph. Therefore, systems of set equations or equivalently labelled direct graphs, can 
be more generally named by the analogy \emph{Web-like Databases} (WDB) \cite{CDQT02,LS97,S01,S06}. Furthermore, our approach also considers WDB as Web-like with distribution over the Internet (in a similar manner to hyperlinks), however, it is intended to be smaller, simpler and better organised than the WWW. Such WDB graphs can, in principle, be quite arbitrary but in real applications it is assumed to be governed by some organisation or company, and possibly not allowed to be arbitrarily extended by anybody in the world (like typical databases). Additionally, WDB (or semi-structured data) can also have a schema restricting the shape of the WDB, but not necessarily so rigid like in the case of relational databases, see for example \cite{BDFS97,LS97,S93}. However, we will not go further into these details.

\subsection{Distributed WDB}\label{sec:prelim_distributed_WDB}

Any WDB represented as a system of set equations $\system$ can be quite big, and naturally divided into subsystems of set equations. Each subsystem corresponds to a XML-WDB file (see Chapter~\ref{chap:xml-wdb} for details of the XML-WDB representation) containing only some of the equations  (desirably closely interrelated by a subject matter). Moreover, these files could be distributed between various servers over the world, like HTML files on the World-Wide Web. It may happen that set equations defined in some WDB file may involve set names defined by equations in other (non-local) WDB files.

\medskip

Furthermore, when considering the real application of WDB distribution proves useful in the creation and management of (potentally large) databases, such as the plausible distribution of the University WDB. Let us consider that in the case of the University WDB, set equations might be distributed between many WDB files, let us say by department. Therefore, the WDB file \url{http://www.liv.ac.uk/ChemistryDepartment.xml} could contain the following subsystem of set equations%
\footnote{This is still not very realistic situation to assume that the file \texttt{ChemistryDepartment.xml} contains all set equations related with this department (on students, lecturers, etc.). These set equations should be further divided into natural fragments (WDB files).
}%
:
\begin{small}
\begin{verbatim}
    DeptChemistry = { ..., enrolled:ChemStud, ... }
    ChemStud = { student:st1, ... }
\end{verbatim}
\end{small}

\noindent
Likewise, the WDB file \url{http://www.liv.ac.uk/BiologyDepartment.xml} could contain the subsystem of set equations:

\begin{small}
\begin{verbatim}
    DeptBiology = { ..., enrolled:BiolStud, ... }
    BiolStud = { student:st2, ... }
\end{verbatim}
\end{small}

\noindent
Moreover, there could also be the WDB file \verb+Students.xml+ containing the set equations \verb+st1 = {...}+ and \verb+st2 = {...}+. Thus, the set names \verb+st1+, \verb+st2+, etc.\ participating, respectively, in \verb+ChemistryDepartment.xml+ and \verb+BiologyDepartment.xml+ would now be described as sets in another file. In this case, we should consider the full versions of the simple set names, \verb+st1+, \verb+st2+, etc., described in \url{http://www.liv.ac.uk/Students.xml}, as discussed below.

\subsubsection{Full versus simple set names}

Taking into account the above example, any given set name should be considered as a \emph{full set name}, consisting of WDB file URL and \emph{simple set name} (with the simple set name described within the WDB file). For example, in the distributed University WDB considered above, the full set name of the biology student \verb+st2+ would be
\begin{small}
\begin{verbatim}
    http://www.liv.ac.uk/Students.xml#st2
\end{verbatim}
\end{small}

\noindent
with the WDB file URL and simple set name delimited by \verb+#+ symbol. However, in practice it suffices to use simple set names in the left-hand side of set equations, and also for those occurrences of set names appearing in the right-hand side of set equation definitions if they are defined in the same WDB file. In particular, the author of a WDB file can freely use any simple set name (as such or as part of full set names) without the danger of clashes with simple names participating in the other WDB files.

\medskip

However, there is one subtle point: if a simple set name \verb+set_name+ occurs twice in some WDB file, once as a simple set name and again as part of a full set name \verb+url#set_name+ (with \verb+url+ referring to some different WDB file). Then in the latter case it refers to another file where the corresponding equation is defined, even if the current file already contains the equation \verb+set_name = {...}+. Thus, these two occurrences are actually different set names because their corresponding full set names are indeed different. Of course, each set name must be defined either in the same or some other WDB file. Otherwise it is considered as syntactical error. Thus, it is necessary to download some WDB files whose URLs appear in full set names of the given file to confirm the existence of defining equations of the referenced set names.

\section{Hyperset data considered abstractly}\label{sec:hypersets_abstractly}

The notion of WDB as a system of set equations presents a low level, syntactical understanding of semi-structured data. 
However, conceptually (and semantically) WDB is understood as consisting of \emph{abstract} hypersets (like relational database consists of abstract relations). The hyperset approach considers WDB as an arbitrary finite system of set equations, each set equation consisting of set name equated to corresponding bracket expression. But the intended meaning of such a syntactical expression is a set of labelled elements, \emph{not} an ordered sequence. Therefore according to this (hyper)set theoretic approach ordering and repetition of elements in a bracket expression should be completely ignored. That is, ignoring ordering and repetitions has some both \emph{operational} and \emph{conceptual} consequences.

\medskip

This can possibly lead to equality between different set names $s_i$ and $s_j$ denoted as $s_i = s_j$ and meaning that $s_i$ and $s_j$ denote the same abstract hyperset, or strictly denoted as
\linebreak
$s_i \bis s_j$ (to avoid possible misunderstanding of $s_i = s_j$ as the assertion that these set names are identical, and to stress on the particularly important role of this concept of equality). In fact, $\bis$ is the well known concept in the context of graphs called \emph{bisimulation relation} between graph nodes or, in our case, between set names \cite{A88,BM96,S06}. As the role of this relation is crucial for the hyperset approach to semi-structured databases, this approach is therefore more than pure graph theoretic, as considered in the approaches to semi-structured databases as
%
graphs e.g.\ in \cite{ABS00,AQMWW97lorel,BFS00,CM90,CDQT02,GPBG94,MAGQW97lore} 
%
or as XML tree-like data e.g.\ in \cite{DFFLS99,GMW99}. Note that, however, \cite{BFS00} is also heavily based on the bisimulation relation, it is rather a graph than a hyperset approach as was argued in \cite{S06}.

\subsection{Bisimulation -- preliminary considerations}\label{sec:preliminary_bisimulation}

In general, the bisimulation relation between set names (graph nodes) of a WDB, i.e.\ a system of set equations, and the corresponding recursive algorithm is based on the idea that any two sets are equal if for each (labelled) element of the first set there exists an equal (bisimilar) element in the second set (and vice versa). Bisimilar set names are said to denote the same abstract (hyper)set. The bisimulation relation will be further described in Chapter~\ref{chap:bisimulation}, with formal theoretical definition, and practical considerations for its implementation. We consider that this hyperset approach to WDB is worth implementing as it suggests a clear and mathematically well-understood view on querying such semi-structured data.
\medskip

A WDB is called \emph{strongly extensional} \cite{A88} or non-redundant, if different set names (nodes) are non-bisimilar i.e.\ denote different hypersets. In the case of strongly extensional WDB, equality between set names (nodes) trivially becomes the syntactical identity relationship. Otherwise, even the simplest queries like $x=y$ or $x\in y$ can be quite expensive to evaluate, especially in the case of distributed WDB. Therefore, we devote Part~\ref{part:local_global} to some approach of dealing with this problem practically.

\subsubsection{Example}

Consider the set equations below, where trivially $x \bis x'$ holds because our (hyper)set approach 
ignores the ordering and repetition of elements:
\begin{align*}
&x = \setof{y,z} \\
&x' = \setof{z,y,z}.
\end{align*}
However, set names (or graph nodes) may be equal (bisimilar) for some 
``deeper'' reason than for $x$ and $x'$ above.
Let us consider the above example extended with the (recursive) definitions of the sets $z$, $y$ and $y'$:
\begin{align*}
&z = \setof{} \\
&y = \setof{x} \\
&y' = \setof{x'}.
\end{align*}
The sets $y$ and $y'$ both contain one element of syntactically differing set names ($x$ and $x'$ respectively), 
thus suggesting that $y$ and $y'$ might not be equal.
However, the bisimulation relation defines two sets as equal if for each element of the first set there 
exists an equal (or bisimilar) element in the second set, and vice versa.
In the case above we already know that $x \bis x'$ holds, and according to this informal 
definition of bisimulation all of the elements of $y$ are bisimilar to
the elements of $y'$, and vice versa. Therefore we can deduce that, in fact, $y \bis y'$ holds.

\medskip

Let us now consider the strongly extensional version of this
system of set equations obtained by eliminating 
the redundant set names $x'$ and $y'$, and omitting repetitions.
Thus, after ``collapsing'' the bisimilar nodes $x'$ to $x$ and $y'$ to $y$, and omitting element repetitions,
the resulting system of set equations is
\begin{align*}
&x = \setof{y,z} \\
&y = \setof{x} \\
&z = \setof{}.
\end{align*}
Thus, the elimination of redundancies (in the above system of set equations) is visualised by
Figure~\ref{fig:bisimulation_strong_extensionality}.

\newsavebox{\bsmltnexplbox}
\begin{figure}[ht]%
\centering%
\sbox{\bsmltnexplbox}{\includegraphics[]{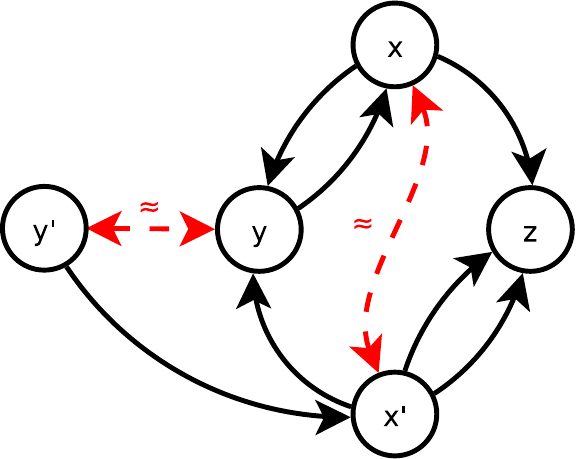}}%
\subfloat[Redundant version, with red dashed edges relating bisimilar nodes (or sets)]{\usebox{\bsmltnexplbox}}%
\qquad%
\subfloat[Non-redundant (strongly extensional) version]{%
\vbox to \ht\bsmltnexplbox{%
\vfil%
\hbox to 5cm{%
\hfil%
\includegraphics[]{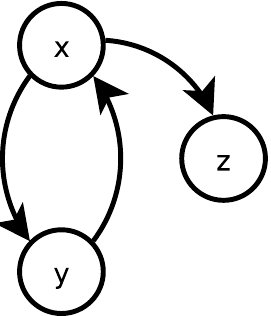}%
\hfil%
}
\vfil}}%
\caption{Graphical representation of a trivial WDB (cf. corresponding set equations above).}%
\label{fig:bisimulation_strong_extensionality}%
\end{figure}

\subsection{Redundancies in WDB}\label{sec:redundancies}

The above example, although artificial, demonstrates that bisimilarity between set names introduces redundancies into WDB. However, the crucial question in implementing the hyperset approach to WDB is whether the bisimulation relation ($\bis$) can be computed in any reasonable and practical way. Some possible approaches and views are outlined below.

\medskip

In principle, the occurrence of bisimilar nodes in a realistic WDB (i.e.\ redundancies) should be infrequent. Therefore, such rare redundancies can be eliminated by supporting WDB in a \emph{strongly extensional} state, with redundancies detected or even eliminated instantly as soon as they might potentially appear. Trivially, after eliminating redundancies equality between sets (i.e.\ bisimulation relation between set names or graph nodes) becomes the identity relation. However, eliminating redundancies is more expensive than only detecting them i.e.\ just computing bisimulation relation on the WDB. Thus, supporting WDB in strongly extensional form may be reasonable option when WDB is not large.

\medskip

WDB should not be assumed to be just another version of WWW, freely extensible by anybody in the world. That is, an appropriate discipline of working with WDB could make the problem of bisimulation practically resolvable. Let us now consider several ways by which redundancies can appear.

\subsubsection{Redundancies arising during query execution} 

Execution of queries leads to the temporary extension $\WDB'$ of the original WDB (as detailed later in Section~\ref{sec:operational_semantics}), with the addition of new set names and set equations locally. Such extensions $\WDB'$ may potentially give rise to new redundancies, so that equality subqueries applied to these newly generated sets becomes non-trivial. Note that the set names in original WDB do not refer to new ones in $\WDB'$, thus WDB remains self-contained. Therefore, the new bisimulation relation ($\bis'$) on $\WDB'$ restricted to those set names in WDB coincides with the identity relation on WDB. Moreover, the algorithm of query execution could be amended in such a way that as soon as new (auxiliary) set names are generated (like $res$ in Section~\ref{sec:operational_semantics}) any possible redundancies will be eliminated immediately. It should also be taken into account that the extensions $\WDB'$ arising during query execution have several specific types, and are sufficiently simple and small, thus making the process of detecting/eliminating redundancies easier, see also \cite{LS01,LS97logspace}, but we will not go into the details here.

\subsubsection{Redundancies which can appear during a local update} 

Local updates of WDB files are more problematic because previously non-bisimilar nodes outside this file may become bisimilar due to possible links (or paths) to the local nodes with changed/added meaning. The appropriate (more efficient than the standard) strategy of detecting/removing all such redundancies is not so straightforward and needs to be developed yet. However, taking into account the locality of changes, this task does not seem to be unrealistic.

\subsubsection{Deliberate redundancies}

Deliberate redundancies in WDB can also appear with the same aim as mirroring in WWW. But, if there is a requirement to officially registered such mirroring in the WDB, then such deliberate redundancies should most plausibly be dealt with in a quite feasible way.

\subsubsection{Local versus global bisimulation}

Unlike the other considerations above, we will consider the ``local/global'' approach and its implementation for supporting bisimulation relation on WDB (in background time) in more detail (see Part~\ref{part:local_global}). Now we present only some general introductory comments on this idea.

\medskip

Assume that all WDB nodes are divided into classes $L_i$ according to their sites (WDB servers) or even files. There is a quite natural definition of local (i.e. computed locally) lower and upper approximations 
($\bis_-^L, \bis_+^L$) to the global bisimulation relation ($\bis$) on the whole WDB:
\begin{align*}
n_1 \bis_-^L n_2 \Rightarrow n_1 \bis n_2 \Rightarrow n_1 \bis_+^L n_2
\end{align*}
These approximations can help to compute and to permanently support global bisimulation in a distributed way in background time. Moreover, we could require \emph{local independence} (${\bis^L_-} = {\bis^L_+}$, and hence ${}={\bis\upharpoonright L}$) and additionally \emph{local non-redundancy} (${\bis^L_-} = {\bis^L_+} = {=^L}$).

\subsection{Bisimulation invariance}

The hyperset approach assumes considering WDB (graphs or systems of set equations) up to bisimulation. Therefore, it is an important requirement for set theoretic operations and relations to be \emph{bisimulation invariant}, that is to preserve the bisimulation relation. Although not fully proven here, it can be shown \cite{S95} that all definable queries $q$ of the hyperset $\Delta$-query language%
\footnote{
The operational meaning of $\Delta$-queries are defined graph theoretically or in terms of set equations.
}
(see Chapter~\ref{chap:theoretical_language}) are bisimulation invariant:
\begin{align*}
&\bar{x} \bis \bar{y} \Longrightarrow q(\bar{x}) \bis q(\bar{y})\quad\textrm{(for set valued queries)} 
\\
&\bar{x} \bis \bar{y} \Longrightarrow q(\bar{x}) \Leftrightarrow q(\bar{y})\quad\textrm{(for boolean queries).}
\end{align*}
For example, in the case of the set theoretic operation union we have:
\begin{align*}
x_1 \bis y_1 \; \& \; x_2 \bis y_2 \Rightarrow ( x_1 \cup x_2 ) \bis ( y_1 \cup y_2 ).
\end{align*}
\noindent
This actually means that we work with (abstract) hypersets rather than just with graph nodes or set names, however the operational semantics of the language $\Delta$ is based on the syntactical manipulations of set equations \cite{S06}. The point is that the semantics of the language $\Delta$ respects bisimulation and completely agrees with the hyperset theory \cite{A88,BM96}.

\medskip

In particular, $x_1\cup x_2$ is defined as a new set name, say $u$, with corresponding new set equation $u=\setof{\ldots,\ldots}$, where the first ``$\ldots$'' is the content of the right-hand side of the equation $x_1=\setof{\ldots}$ from the given WDB, and similarly for the second ``$\ldots$'' and the equation $x_2=\setof{\ldots}$. The union $y_1\cup y_2$ is computed in the same way from set equations for $y_1$ and $y_2$ giving rise to new set name, $u'$, and the corresponding set equation $u'=\setof{\ldots,\ldots}$. Then the conclusion of the above bisimulation invariance condition for $\cup$ actually means $u\bis u'$, and can evidentially be shown. 

\medskip

Note that the membership relation $x\in y$ for two sets (considering the unlabelled case for simplicity) is defined to be true if the set equation for $y$ involves some set name $x'$, where $y=\setof{\ldots,x',\ldots}$ and,  moreover, $x\bis x'$. Additionally, it can be shown that the membership relation is also bisimulation invariant:
\begin{align*}
x_1 \bis y_1 \; \& \; x_2 \bis y_2 \implies  x_1 \in x_2 \iff y_1 \in y_2
\end{align*}
For all other constructs of the $\Delta$-language the operational semantics maybe more complicated, however, it follows that they also agree with this intuitive (abstract) set theoretical meaning. The syntax and semantics of the $\Delta$-query language will be further detailed in Sections~\ref{sec:syntax} and~\ref{sec:denotational_semantics}, with some further indications of the operational semantics in terms of set equations detailed in Section~\ref{sec:operational_semantics}.

\subsection{Anti-Foundation Axiom}\label{sec:AFA}

Finally, we do not go into full mathematical details on hypersets, however, we could assert the following form of {\bf Anti-Foundation Axiom} (AFA) \cite{A88,BM96}, which holds in the universe of abstract (in our case finite) hypersets:
\begin{quote}
\emph{Any system of set equations $\system$ has a unique abstract hyperset solution for set names 
$\bar{s}$ making these equations true.}
\end{quote}
Therefore, set names of any WDB (as system of set equations) denote quite concrete, uniquely defined abstract hypersets. In this sense each set name (in a $\Delta$-query) serves as a set constant (relative to the given WDB) denoting a unique hyperset. Note that, the $\Delta$-language also has set variables which can be quantified unlike constants.

\medskip

Strictly speaking all of this makes precise mathematical sense only in context of Chapter~\ref{chap:bisimulation}, which further details the bisimulation relation (with some additional mathematical considerations) beyond the general informal description of bisimulation relation so far.


\chapter{Query language $\Delta$}\label{chap:theoretical_language}

\section{The syntax}\label{sec:syntax}

There has already been much theoretical considerations on (some versions of) the $\Delta$ (Delta) query language to hyperset/WDB databases \cite{LS01,LS97,LS99,S93,S06}. The two main syntactical categories of $\Delta$ are:
\begin{itemize}
\item {
$\Delta$-\emph{terms} representing set valued operations over hypersets (\emph{set queries}), and 
}
\item {
$\Delta$-\emph{formulas} representing truth valued operations (\emph{boolean queries}).
}
\end{itemize}

\noindent
Note that the denotation $\Delta$ bears partly from the well-known class $\Delta_0$ of bounded formulas introduced by Levy, although $\Delta$, as defined here, denotes a wider language. It is based on the \emph{basic} or \emph{rudimentary} set theoretic languages of Gandy~\cite{G74} and Jensen~\cite{J72}. Moreover, inclusion of set theoretic operators: transitive closure (\TC), recursion (\Rec) and, for the case of hypersets, decoration (\Dec) (the latter due to Forti and Honsell \cite{FH83} and Aczel~\cite{A88}), allows to define in $\Delta$ exactly all polynomial time computable operations over hypersets represented as WDB, thus demonstrating and characterising theoretically its rich expressive power (assuming that a linear order on labels is given) \cite{LS99,S87,S93,S95}. The operators of $\Delta$ are defined as follows:
\begin{align*}
  \langle\mbox{$\Delta$-term}\rangle ::= \; & 
		\langle\mbox{set variable or constant}\rangle \MID
		\emptyset \MID
		\{l_1:a_1,\ldots,l_n,a_n\} \MID
		\bigcup a \MID
		\TC(a) \MID																							\\
  &	
  	\{l:t(x,l) \mid l:x \in a \AND \varphi(x,l)\} \MID 
  	\Rec\; p.\{l:x \in a \mid \varphi(x,l,p)\}  \MID							
  	\Dec(a,b) \\
  %
  \langle\mbox{$\Delta$-formula}\rangle ::= \; &
		a=b \MID
  	l_1=l_2 \MID
		l_1<l_2 \MID 
		l_1\mathrel{R}l_2 \MID
		l:a \in b \MID
		\varphi \AND \psi \MID
		\varphi \vee \psi \MID
		\neg \varphi \MID																				\\
  &	\forall l:x\mathrel{\in}a.\varphi(x,l) \MID
  	\exists l:x\mathrel{\in}a.\varphi(x,l)
\end{align*}

\noindent
The intuitive set theoretic semantics of the majority of the above constructs should be
\linebreak
well-understood by anyone with the minimal mathematical background in set theory and logic. In the above constructs we denote: $a,b,\ldots$ as (set valued) $\Delta$-terms; $x,y,z,\ldots$ as set variables; $l,l_i$ as label values or variables (depending on the context); $l:t(x,l)$ is any  $l$-labelled $\Delta$-term $t$ possibly involving the label variable $l$ and the set variable $x$; and $\varphi,\psi$ as (boolean valued) $\Delta$-formulas. Note that labels $l_i$ participating in the $\Delta$-term $\setof{ l_1\!:\!a_1,\ldots,l_n\!:\!a_n }$ need not be unique, that is, multiple occurrences of labels are allowed. This means that we consider arbitrary sets of labelled elements rather than records or tuples 
of a relational table where $l_i$ serve as names of fields (columns).

\medskip

The binding label and set variables $l,x,p$ of quantifiers, collect, and recursion  constructs should not appear free in the bounding term $a$ (denoting a finite set). Otherwise, these operators may become unbounded and thus, in general, non-computable. For example, let us consider the universal quantifier $\forall l\!:\!x\mathrel{\in} \setof{ \ldots, l\!:\!x ,\ldots } .\varphi(x,l)$ which becomes unbounded due to the quantified variables $l\!:\!x$ participating in the bounding term $\setof{ \ldots, l\!:\!x ,\ldots }$. In fact, as $l\!:\!x\in \setof{ \ldots, l\!:\!x ,\ldots }$ is always true the above quantified formula proves to be equivalent to unbounded one: $\forall l\!:\!x .\varphi(x,l)$.

\section{Intuitive denotational semantics}\label{sec:denotational_semantics}

Any $\Delta$-query without free variables has either: i) (hyper)set value in the case of $\Delta$-terms, or ii) boolean value in the case of $\Delta$-formulas. Those participating set variables or set constants represent abstract hypersets (and thus correspond to set names in WDB), whereas participating label variables or label constants represent label values  (corresponding to strings of symbols).

\medskip

The intuitive meaning of $\Delta$-queries is described by the \emph{denotational semantics}, that is what any expression denotes%
\footnote{
There is a deep mathematical theory of denotational semantics of programming languages based on Domain Theory  \cite{SSt71,St67} (also see the contemporary reference \cite{FJMORRS96}) to represent denotational values of a programming language expressions. The language $\Delta$, where all computations evaluating queries are finite, does not require this theory which is based on the idea of potentially infinite computations (embodied in the so called ``undefined'' element $\perp$). Anyway, it makes sense to use the term denotational semantics, although we will describe this semantics on a very intuitive level by reference to the ``domain'' of sets and hypersets. 
}%
. For the purposes of implementation $\Delta$-queries are also described by means of their \emph{operational} or computational semantics (see Section~\ref{sec:operational_semantics}) which must be coherent with our intuitive denotational semantics. Here we will also rely on intuition, without presenting any precise argument. In fact, the required coherence will be pretty much evident. So, we can concentrate on examples of queries and implementation aspects.

\subsection{Boolean valued expressions --- $\Delta$-formulas}\label{sec:denotational_semantics_formulas}

\emph{Equality} ($=$) and the \emph{alphabetic ordering} ($<$) between labels is understood standardly. In the theoretical $\Delta$-language the relation $\mathrel{R}$ over labels is any easily computable relation over labels, however, in the implemented $\Delta$-language described in this thesis we consider $R$ as any of the following \emph{substring} relations

\begin{align*}
*l_1=l_2 \MID l_1*=l_2 \MID *l_1*=l_2  
\end{align*}

\noindent
where the wildcard $*$ represents any string of symbols. In principle we could include into the language more relations over labels, but in the implementation there are only $<$ and substring relations, and the user currently has no way to define more primitive relations over labels. It should be noted that equality between $\Delta$-terms, $a=b$ or, for technical reasons, $a\bis b$, is understood as the equality of abstract hypersets denoted by these terms and, as such, is computed by the bisimulation algorithm discussed in Chapter~\ref{chap:bisimulation}. That is, when we discuss hypersets abstractly, we use $=$. But when considering bisimulation algorithm to determine whether two set names or graph nodes denote the same abstract hyperset, we use $\bis$. In the implemented version of the language we have only $=$ which, of course, involves calling the bisimulation algorithm, but this is hidden from the user who, therefore can think on hypersets abstractly. Moreover, bisimulation is implicitly involved in the (computational) meaning of the \emph{membership} relation according to the equivalence

\begin{align*}
l\!:\!a\in b\iff\exists m\!:\!x\in b.(m\!=\!l\AND x\!\bis\!a)
\end{align*}

\noindent
informally having the meaning: find an outgoing $l$-labelled edge from $b$ which leads to some node $x$ bisimilar to $a$. But, thinking abstractly, $l\!:\!a\in b$ says simply that $a$ is an $l$-labelled element of $b$. 

\medskip

The \emph{logical operators} ($ \AND, \vee, \neg $) have the usual meaning from propositional logic and can be used to form logical sentences from $\Delta$-formulas. \emph{Universal quantification} can be understood in terms of conjunction:
\begin{align*}
\forall l\!:\!x\in a.\varphi(x,l) & \iff \bigwedge_{l_i:x_i\in a} \varphi(x_i,l_i)
\end{align*}

\noindent
and \emph{existential quantification} in terms of disjunction:
\begin{align*}
\exists l\!:\!x\in a.\varphi(x,l) & \iff \bigvee_{l_i:x_i\in a} \varphi(x_i,l_i)
\end{align*}

\noindent
assuming that $a=\setof{l_1:x_1,\ldots,l_n:x_n}$. It is evident from this definition that quantification occurs over those elements of the set denoted by $a$ which satisfy the formula $\varphi$. That is, quantification is bounded by (elements of) the set $a$, with the $\Delta$ formula $\varphi$ being called the scope of the quantifier.

\medskip

Note that when a quantified formula participates as a subformula of a bigger formula or of a term the technical problem arises where exactly this (sub)formula is finished, that is what is the scope of the quantifier. In the implemented $\Delta$-language (Appendix~\ref{app:BNF}) there is a discipline of using parentheses to find unambiguously the scope of quantifiers, both intuitively and by the implemented parser (and contextual analysis algorithm). Say, in 
\begin{align*}
\forall l:x\in a\,.\, (\varphi\; \&\; \psi\; \&\; \chi)
\end{align*}

\noindent
the scope of the quantifier is the whole expression in the parentheses. But the general informal rule is: the scope of any quantifier is as small as possible. For example, in 
\begin{align*}
(\forall l:x\in a\,.\, \varphi\; \&\; \psi\; \&\; \chi)
\end{align*}

\noindent
the multiple conjunctions requires some compulsory external parentheses (exactly as shown), and then the scope of the quantifier is either $\varphi$ (excluding $\psi$ and $\chi$) or some initial part of $\varphi$, if syntactically meaningful at all. We will not give the formal definition which is usually widely known and intuitively evident. For the precise definition of the scope of quantifiers, declarations, etc.\ the reader should, first,  inspect the relevant part of the $\Delta$-language syntax in Appendix~\ref{app:BNF} and, most importantly, read the Section~\ref{sec:well-typed_delta_queries} on contextual analysis which, in fact, served as a rigorous conceptual guidance for us to implement 
the language correctly.

\subsection{Set valued expressions --- $\Delta$-terms}\label{sec:denotational_semantics_terms}

The set constant \emph{empty set} ($\emptyset$) denotes the set $\setof{}$ having no elements. In general, set values are represented symbolically by either: set constants, set variables or $\Delta$-terms. Furthermore, ``literal'' set values can be introduced with the \emph{enumeration} expression $\setof{l_1\!:\!a_1, ..., l_n\!:\!a_n}$ which can create new sets, possibly with nesting if some $a_i$ are also enumeration expressions, however, $a_i$ may also be arbitrary $\Delta$-terms.

\medskip

The \emph{collection} operation $\setof{l\!:\!t(x,l) \mid l\!:\!x \in a \AND \varphi(x,l)}$ denotes the set of labelled elements $l\!:\!t(x,l)$ with $t(x,l)$ a $\Delta$-term depending on the set and label variables $l$ and $x$, where $l\!:\!x$ ranges over the set $a$, for which the $\Delta$-formula $\varphi(x,l)$ holds. We can also consider the more special case of collection called the \emph{separation} operation $\setof{l\!:\!x \in a \mid \varphi(x,l)}$ which denotes the set of labelled elements $l\!:\!x$ in $a$ for which $\varphi(x,l)$ holds.

\medskip

The (unary) \emph{union} operation $\bigcup a$ is understood as the (multiple) ordinary union over the elements of $a$. Let us assume $a = \setof{l_1\!:\!a_1, \ldots , l_n\!:\!a_n}$ then

\begin{align*}
\bigcup a=a_1 \cup \ldots \cup a_n
\end{align*}

\noindent
with the ordinary union used in the right-hand side of equality. In particular, this also shows that the ordinary union is definable by means of the unary union and enumeration operators. This is only the simplest example of expressibility in $\Delta$. As we mentioned, this language has, in fact, very high expressive power exactly corresponding to polynomial time computability over hereditarily-finite hypersets%
\footnote{
Any hyperset set is hereditarily-finite if and only if it contains a finite number of elements, and these elements are also hereditarily-finite hypersets, etc. Moreover, it is required that the transitive closure of this hyperset 
is also finite.
}%
.

\medskip

The \emph{transitive closure} $\TC(a)$ denotes the set of (labelled) elements of elements, $\ldots$ , of elements of $a$ including $a$ itself. This can also be written (not fully formally, say, due to $\ldots$ present)~as:

\begin{align*}
l\!:\!x \in \TC(a) \iff &l\!:\!x \in x_0 \in \ldots \in x_n = a \; \vee \\
                        &(l=\Box \AND x=a)
\end{align*}

\noindent
with $x_i$ some intermediate elements in the membership chain, each belonging to the next $x_{i+1}$ with some label $l_i$ whose value is not important. In particular, we let $\Box:a\in\TC(a)$.

\bigskip

The above core constructs of the $\Delta$-language extended with the two additional constructs recursion and decoration (introduced below) define all polynomial time computable operations and relations over hypersets (represented as WDB); see the precise formulations in \cite{LS97,LS99,S93}.

\subsubsection{Recursion operation}

The \emph{recursion} operator $\Rec\; p.\{l\!:\!x \in a \mid \varphi(x,l,p)\}$ defines a subset $\pi$ of the set denoted by (the $\Delta$-term) $a$, obtained as the result of stabilising (due to finiteness of $a$) the inflating sequence of subsets of $a$ defined iteratively as:
\begin{align*}
p_0     & =\emptyset \\
p_{1}   & =p_0\cup\{l\!:\!x \in a \mid \varphi(x,l,p_0)\}\\
p_{2}   & =p_1\cup\{l\!:\!x \in a \mid \varphi(x,l,p_1)\}\\
        & \ldots \\
p_{k+1} & =p_k\cup\{l\!:\!x \in a \mid \varphi(x,l,p_k)\}.
\end{align*}
Evidently, all $\emptyset=p_0\subseteq p_1\subseteq\ldots$ are subsets of $a$. As $a$ is finite, $p_k = p_{k+1}=p_{k+2},\ldots$ for some $k$, and this  stabilised value, denoted above as $\pi$, is taken as the value of the recursion operator.

\subsubsection{Decoration operation}\label{sec:dec_operation}

Recall that in Chapter~\ref{chap:ssd_wdb} graph nodes were shown to denote (hyper)sets, and vice versa, arbitrary hereditarily-finite hyperset can be represented in this way.

\medskip

Now, we shall consider finite graphs in set theoretic terms. Traditionally, this is done by defining a graph as a set of ordered pairs where ordered pairs represent graph edges, for example $\orderedsetof{a,b}$ denoting the edge $a \rightarrow b$. Here (the arbitrary sets) $a$ and $b$, play the role of the source and target vertices of the edge $a \rightarrow b$. Thus, any set $g$ of ordered pairs can be treated as a graph. Formally such ordered pairs are represented as the sets containing two elements labelled by $fst$ and $snd$ respectively, such as $\setof{fst\!:\!a, snd\!:\!b}$. That is, we define $\orderedsetof{a,b}=\setof{fst\!:\!a, snd\!:\!b}$. Any labelled ordered pair $l:\setof{fst\!:\!a, snd\!:\!b}$ represents a labelled edge $a\stackrel{l}{\rightarrow}b$. In general, we can consider absolutely arbitrary hyperset $g$ as representing a graph. Indeed, we can take into account only those elements of $g$ which happen to be ordered pairs, and ignore the other non-pair elements. This will make the operation of decoration defined below applicable to the arbitrary hyperset $g$ what is convenient. Otherwise the formulation of the language $\Delta$ would be more complicated. Also, the arbitrary set $v$ may either participate as an element of the ordered pairs of $g$, i.e.\ serving as a $g$-vertex, or, otherwise, it is considered as an isolated vertex of the graph $g$. In this sense each set $v$ serves as a $g$-vertex.

\begin{definition}
The abstract set theoretic \emph{decoration operator} $\Dec(g,v)=d$ takes two arbitrary input sets $g$ and $v$ where the former represents a graph as a set of ordered pairs, and the latter represents some vertex $v$ of this graph. It outputs a new (hyper)set $d$ corresponding to the $v$-rooted graph $g$ according to the first paragraph of this section.
\end{definition}

Note that decoration is the only operator in $\Delta$ which allows for the construction of cyclic hypersets, like $\Omega=\setof{\Omega}$, from the ordinary ``uncycled'' sets (of sets of sets,\ldots) of finite depth. For example, consider the \emph{trivial cyclic} graph $g$ defined by the following system of set equations,
\begin{align*}
g & = \setof{ \; \setof{ fst\!:\!a, snd\!:\!a} \; } \\
a & = \setof{} \\
\end{align*}

\noindent
The result of applying decoration to the graph $g$ and the participating vertex $a$ would be,
\begin{align*}
\Omega & = \setof{ \Omega }
\end{align*}

\noindent
where $\Omega$ denotes the result $\Dec(g,a)$. Indeed this leads to the construction of the cyclic membership represented by the unique $g$-edge 
$a \rightarrow a$. In fact, here the Anti-Foundation Axiom from Section~\ref{sec:AFA} guarantees that $\Omega$ is a unique hyperset denoted by $\Dec(g,a)$ (and the same for arbitrary $g$ and $a$).

\medskip

This operator can also be reasonably called the \emph{plan performance operator} \cite{S06} because its input(s) can be considered as a graphical plan for the construction of a hyperset with the output being the resulting abstract hyperset. Imagine that we have a plan of a Web site (i.e. of a system of hyperlinked Web pages) and that $\Dec$ is a tool (or query) which automatically creates all the required Web pages. See also Section~\ref{sec:restructuring-query} for a more involved example of using the decoration operation for defining a restructuring query.

\section{Operational semantics}\label{sec:operational_semantics}

Consider any set or boolean query $q$ which involves no free variables and whose participating set names (constants) are taken from the given WDB system of set equations. Resolving $q$ consists in the following two macro steps: 
\begin{itemize}
\item{
{\bf Extending} this system by new equation $res = q$ with $res$ a fresh (i.e.\ unused in WDB) set or boolean name, and
}
\item{
{\bf Simplifying} the extended system:
\begin{align*}
\WDB_0=\WDB+(res = q)
\end{align*}
until it will contain only flat bracket expressions as the right-hand sides of the equations or the truth values \emph{true} or \emph{false} (if the left-hand side is boolean name).
}
\end{itemize}

\noindent
After simplification is complete, these set equations will contain no complex set or boolean queries (like $q$ above). In fact, the resulting version $\WDB_{res}$ of WDB will consist (alongside the old equations of the original WDB) of new set equations (new set names equated to flat bracket expressions) and boolean equations (boolean names equated to boolean values, \emph{true} or \emph{false}). This process of computation by \emph{extension} and \emph{simplification} was described in \cite{S06} as reduction steps

\begin{align*}
WDB_0 \rhd WDB_1 \rhd \ldots \rhd WDB_{res}
\end{align*}

\noindent
where $WDB_0$ is the initial state of $WDB$ extended by the equation $res = q$, and $WDB_{res}$ is the final step of reduction consisting of only flat set equations including the flattened version of set equation $res=q$ (or boolean equation, if $q$ is a $\Delta$-formula). Each reduction step represents simplification by applying rewrite rules which transform set equations involving complicated $\Delta$ expressions into simpler, semantically equivalent, equations. Note that the rewrite rules described here are based on those in \cite{S06} but extended to the labelled case as considered in this thesis. In general, rewrite steps are denoted by the $\rhd$ symbol which means ``transforms to''. Firstly, let us assume participation of the set names $s,p,r$ in the rewrite rules below, which correspond to the set equations

\begin{align*}
s & = \setof{ l_1\!:\!s_1, ..., l_a\!:\!s_a },\\
p & = \setof{ m_1\!:\!p_1, ..., m_b\!:\!p_b },\\
  &\ldots \\
r & = \setof{ n_1\!:\!r_1, ..., n_c\!:\!r_c }
\end{align*}

\noindent
existing either in the initial $WDB$ or in the current reduction $WDB_i$. The operational semantics for the $\Delta$ operators (except for recursion, decoration, transitive closure, bisimulation and label relation operators) are described as the reduction rules

{\allowdisplaybreaks 
\begin{flalign*}
res & = t(t_1, \ldots, t_a) \rhd \begin{cases}res & = t(res_1, \ldots, res_a),\\res_1 & = t_1,\\    & \ldots\\res_a & = t_a.\end{cases}\\
res & = \setof{ l\!:\!s, m\!:\!p,\ldots, n\!:\!r } \mbox{ -- no further reduction required once $s,p\ldots,r,$ are set names},\\
res & = s \cup p \cup\ldots\cup r \rhd res=\setof{ l_1\!:\!s_1, ..., l_a\!:\!s_a, \; m_1\!:\!p_1, ..., m_b\!:\!p_b, \; \ldots, n_1\!:\!r_1, ..., n_c\!:\!r_c }, \\
res & = \bigcup s \rhd res=s_1 \cup \ldots \cup s_a,\\
res & = \TC(p) \mbox{ -- operational semantics described in Section~\ref{sec:impl_tc}},\\
res & = \setof{ l:x \in p \mid \varphi(l,x) } \rhd res=\setof{ m_{i_1}\!:\!p_{i_1}, \ldots, m_{i_{b'}}\!:\!p_{i_{b'}} } \\
    & \mbox{where } m_{i_j}\!:\!p_{i_j} \mbox{ are all those }m_{i}\!:\!p_{i}\in p \mbox{ for which } res_i = \varphi( m_{i},p_{i}) \rhd res_i = \TRUE,\\
res & = \setof{ t(l,x) \mid l\!:\!x \in p \; \& \; \varphi(l,x) } \rhd res=\setof{ t(m_{i_1}\!:\!p_{i_1}), \ldots, t(m_{i_{b'}}\!:\!p_{i_{b'}}) } \\
    & \mbox{where } m_{i_j}\!:\!p_{i_j} \mbox{ are all those }m_{i}\!:\!p_{i}\in p \mbox{ for which } res_i = \varphi( m_{i},p_{i}) \rhd res_i = \TRUE,\\
res & = \Rec\; p.\{l:x \in a \mid \varphi(l,x,p)\} \mbox{ -- operational semantics described in Section~\ref{sec:impl_rec_sep}},\\
res & = \Dec(a,b) \mbox{ -- operational semantics described in Section~\ref{sec:impl_dec}},\\
res & = \forall l\!:\!x \in p \; . \; \varphi(l,x) \rhd res = \varphi(m_1,p_1) \; \& \; ... \; \& \; \varphi(m_n,p_n),\\
res & = \exists l\!:\!x \in p \; . \; \varphi(l,x) \rhd res = \varphi(m_1,p_1) \vee ... \vee \varphi(m_n,p_n),\\
res & = \TRUE \; \& \; \TRUE \rhd res=\TRUE,\\
res & = \FALSE \; \& \; \varphi \rhd res=\FALSE,\\
res & = \varphi \; \& \; \FALSE \rhd res=\FALSE,\\
res & = \varphi \vee \psi \rhd res=\neg(\neg\varphi \; \& \; \neg\psi),\\
res & = \neg\FALSE \rhd res=\TRUE,\\
res & = \neg\TRUE \rhd res=\FALSE,\\
res & = l\!:\!s \in p \rhd res=\exists m\!:\!x \in p \; . \; (s=x \; \& \; l=m),\\
res & = x=y \rhd x \approx y \mbox{ -- operational semantics described in Section~\ref{sec:impl_bisim_algo}},\\
res & = l \mathrel{R} m \mbox{ -- operational semantics described in Section~\ref{sec:denotational_semantics_formulas}}.
\end{flalign*}
}

\noindent
The implementation of $\Delta$-query execution is based on this process of reduction except for the $\Delta$-terms: recursion, decoration, transitive closure described in Section~\ref{sec:impl_rec_sep}, Section~\ref{sec:impl_dec} and Section~\ref{sec:impl_tc} respectively; and the $\Delta$-formulas: set equality (bisimulation) and label relation operators described in Section~\ref{sec:impl_bisim_algo} and Section~\ref{sec:denotational_semantics_formulas} respectively.

\subsection{Examples of reduction}

The above process of computation by \emph{reduction} is quite natural as shown in the following examples.

\subsubsection{Example elimination of complicated subterms}

Let us consider the reduction of the query $q=\bigcup q_1$ containing the complex subquery $q_1$. In general, any complicated term $t(t_1, \ldots, t_n)$ can be simplified by invoking the splitting rule which transforms the equation $res=t(t_1, \ldots, t_n)$ to the resultant equations

\begin{align*}
res      & = t(res_1, \ldots, res_n) \\ 
res_1	 & = t_1 \\ 
	         & \ldots \\ 
res_n	 & =  t_n
\end{align*}

\noindent
Therefore, the complicated query $res=\bigcup q_1$ can be split into two subqueries, $res =\bigcup res_1$ and $res_1=q_1$ where $res_1$ is a new set name.

\subsubsection{Example reduction of union}

In the case of our union query having the particular form $q = \bigcup \setof{l\!:\!s, m\!:\!p, n\!:\!r}$ where $s,p,r$ represent set names, it follows that the equation $res=q$ is reduced by the following steps:

\begin{enumerate}
\item{
Split the complicated equation $res = \bigcup \setof{l\!:\!s, m\!:\!p, n\!:\!r}$ resulting in the equations:
\begin{align*}
res     & = \bigcup res_1 \\
res_1 & = \setof{l\!:\!s, m\!:\!p, n\!:\!r}
\end{align*}
where $s,p,r$ are set names, and hence do not require further splitting.
}
\item{Reduce unary union $res = \bigcup res_1$ to multiple union resulting in the equation:
\begin{align*}
res = s \cup p \cup r
\end{align*}
with the unary union reduced to multiple unions over the elements of the set $res_1$ (the set names $s,p,r$).
}
\item{Reduce multiple union $res = s\, \cup\, p\, \cup\, r$ to the bracket expression resulting in the equation:
\begin{align*}
res = \setof{ l_1\!:\!s_1, ..., l_i\!:\!s_i, \; m_1\!:\!p_1, ..., m_j\!:\!p_j, \; n_1\!:\!r_1, ..., 
n_k\!:\!r_k }
\end{align*}
assuming that the current extension of the original WDB already contains the simplified equations $s = \setof{l_1\!:\!s_1, ..., l_i\!:\!s_i}$, $p = \setof{m_1\!:\!p_1, ..., m_j\!:\!p_j}$ and $q = \setof{n_1\!:\!q_1, ..., n_k\!:\!r_k}$. Here multiple union over the sets $s,p,r$ is reduced to the bracket expression containing the elements of these sets.
}
\end{enumerate}

\noindent
In general, most of the $\Delta$ operators can be resolved using the above reduction rules except for recursion, decoration, transitive closure, bisimulation and label relation operators. In fact, there is no common framework for describing the operational semantics for all the $\Delta$ operators, with the latter exceptions described as lower-level algorithms in Chapters~\ref{chap:bisimulation}~and~\ref{chap:exec}.

\medskip

The main conclusion is that after reduction we will have the equation $res=\setof{\ldots}$ of the required form whose right-hand side should involve no complicated terms or formulas, only set names either from the original WDB or new set names introduced during reduction (like $res_1$ above) together with the corresponding equations of the required form. Thus, execution of a query extends the original WDB to $\WDB_{res}$ (simplification of $\WDB_0$ above). This extension with the set name $res$ as an ``entrance point'' to the result of the query can be considered as a temporary one until we need this result.

\bigskip

\noindent
In principle, we could also consider \emph{update queries} which would change the original WDB (not only extend it as above), but this is beyond the scope of this work.

\section{Implemented $\Delta$-query language}\label{sec:implemented_delta__language}

The implemented $\Delta$-query language can express all operations definable in the original (as described above). For the purpose of writing queries the grammar of this language is expressed as BNF (see Appendix~\ref{app:BNF}) which the reader should take into consideration whilst reading the current section. (See Chapter~\ref{chap:exec} for technical details of the implementation of the $\Delta$-query language.) Note that, not every computable set theoretic operation is definable within the
\linebreak
\mbox{$\Delta$-language} but everything which is polynomial time computable (and generic; cf.\ \cite{LS97}) is already definable in the original language.

\medskip

Additional features (not present in the theoretical version of the language) have also been included in the implemented language making the language more practically convenient, but not increasing its theoretical expressive power. These additions, however important practically, are just ``syntactic sugaring'' of the above theoretical version of~$\Delta$.

\subsection{Queries with declarations}

Like in many programming languages allowing procedure declarations and calls we also introduce in the language $\Delta$ query declarations and calls. Thus, a query once declared can be invoked as many times as we want by using its name with various parameters. Besides queries, we allow also constant declarations. Each declaration has its own scope especially delimited (unlike quantifiers) by the keywords \verb+in+ and \verb+endlet+ where the declared queries or constants can be used (called). For example, let us show how full set names%
\footnote{
Recall that full set name consists of XML-WDB file URL extended by simple set name (delimited by \texttt{\#} symbol).
}
(which can be quite long and unmanageable) can be declared and then used as set constants. The following query declares the set constant \texttt{BibDB} as an abbreviation of the corresponding full set name:

\begin{small}
\begin{verbatim}
    set query 
      let set constant BibDB be
        http://www.csc.liv.ac.uk/~molyneux/t/BibDB.xml#BibDB
      in QUERY( BibDB )
    endlet;
\end{verbatim}
\end{small}

\noindent
Here \texttt{QUERY} denotes any subquery (according to the syntax in Appendix~\ref{app:BNF}) which may involve (possibly many times) the set constant \texttt{BibDB} declared once in the \verb+let+ declaration at the beginning of the whole query. However in general \verb+let+ declarations of constants and queries can appear at any depth of a query.

\medskip

Let us now consider the more useful case of the query declaration \verb+getBooks+, which in the following example gives the set of all books in the bibliography database illustrated by the graph in Figure~\ref{fig:WDB-bibdb} in Section~\ref{sec:example_queries} below. We first declare the query \texttt{getBooks} with one set variable argument \texttt{input} and then call it with the argument value \texttt{BibDB}:

\begin{small}
\begin{verbatim}
    set query 
      let set constant BibDB be
            http://www.csc.liv.ac.uk/~molyneux/t/BibDB-f1.xml#BibDB,
          set query getBooks (set input) be
            separate {
                pub-type:pub in input
                where pub-type='book'
            }
      in call getBooks(BibDB)
    endlet;
\end{verbatim}
\end{small}

\noindent
Here the keyword \verb+call+ means that we invoke the set query \verb+getBooks+ defined above. In general, any query can be declared once and invoked many times, e.g. \texttt{getBooks(BibDB1)}, \texttt{getBooks(BibDB2)}, etc., each time with various \texttt{<parameters>} which may be either any \texttt{<delta-term>} or \texttt{<label>} according to the BNF. Those relevant parts of the BNF for this set query are as follows,

\begin{small}
\begin{verbatim}
    <delta-term with declarations> ::= 
         "let" <declarations> "in"  <delta-term> "endlet"

    <set constant declaration> ::=
          "set constant" <set constant> ("be"|"=") <delta-term>

    <set query declaration> ::=		
         "set query" <set query name> "(" <variables> ")" 
         ("be"|"=") <delta-term>

    <set query call> ::=
         "call" <set query name> "(" <parameters> ")"
\end{verbatim}
\end{small}

\noindent
In general, there are also \texttt{<label constant declaration>} and \texttt{<boolean query declaration>} syntactical categories. Note that in the syntactic category \texttt{<delta-term with declarations>} the keyword \verb+in+ evidently does not play the role of the membership relation such as in the case of the other contexts of the $\Delta$-language. Recursive calls are not allowed in query declarations, that is the declared query name or constant should not occur in the scope of the declaration. For \texttt{<recursion>} (see the syntax in Appendix~\ref{app:BNF}) we have the special construct recursive separation already discussed above and illustrated below in Section~\ref{sec:horizontal_tc}.

\subsection{Library}\label{sec:libraries}

The library allows to create query or constant declarations independent of a query. Library commands allow creation and modification of user defined queries and constants. Predefined and also user defined queries and constants can then be used, i.e.\ called, (globally) in any query. For example, the following library command adds the set constant \texttt{some-book} for the appropriate full set name to the library:
\begin{small}
\begin{verbatim}
    library add set constant some_book =
      http://www.csc.liv.ac.uk/~molyneux/t/BibDB-f1.xml#b1;
\end{verbatim}
\end{small}

\noindent
where the identifier \verb+some-book+ may now participate in any subsequent queries in the current query session%
\footnote{
Query session is the period of time between opening the query system (for running queries and library commands) and closing it. When query system is restarted, only build in query and constant declarations (see the current list in the Appendix~\ref{app:predefined}) can be used.
}%
. Queries and constants can be modified or redeclared by rerunning the library \verb+add+ command. For example, the set constant \verb+some_book+ (above) could be redeclared as follows:
\begin{small}
\begin{verbatim}
    library add set constant some_book = 
      http://www.csc.liv.ac.uk/~molyneux/t/BibDB-f1.xml#b2;
\end{verbatim}
\end{small}

\noindent
Predefined and user defined%
\footnote{
added in the current query session
}
library queries/constants can be listed, in brief without the full declarations, with the command,
\begin{small}
\begin{verbatim}
    library list;
\end{verbatim}
\end{small}
\noindent
with result of this command (including predefined queries/constants) being,

\begin{small}
\begin{verbatim}
    Library command is well-formed and well-typed, but not 
    executable

    Warning, library command successful but no query executed.

    Warning, in the case of duplicate declaration names those 
    declarations at the bottom of the list have precedence.

    List of library declaration(s):

      set query Pair (set x,set y),
      boolean query isPair (set p),
      set query First (set p),
      set query Second (set p),
      set query CartProduct (set x,set y),
      set query Square (set z),
      set query LabelledPairs (set v),
      set query Nodes (set g),
      set query Children (set x,set g),
      set query Regroup (set g),
      set query CanGraph (set x),
      set query Can (set x),
      set query TCPure (set x),
      set query HorizontalTC (set g),
      set query TC_along_label (label l,set z),
      set query SuccessorPairs (set L),
      boolean query Precedes5 (set R,label l,set x,label m,set y),
      set query StrictLinOrder_on_TC (set z),
      set constant some_book,
      set constant some_book
\end{verbatim}
\end{small}

\noindent
The order of query/constant declarations depends on the order in which the corresponding \texttt{library add} commands were executed. Note that, the duplicate declarations named \verb+some_book+ is the result of running above the \texttt{library add} commands, and those declarations appearing at the bottom of the list have precedence over those at the top of the list. Thus, the set constant \verb+some_book+ appearing globally in any query would, in fact, have the redeclared set name \url{http://www.csc.liv.ac.uk/~molyneux/t/BibDB-f1.xml#b2}. However, there is one subtle point: if a query $q$ is declared in the library which calls another library query $q_1$ (or constant), then $q$ will invoke the latest declaration of $q_1$ \emph{preceding} this declaration of $q$ even if $q_1$ is redeclared again after $q$. Note that the modification or deletion of user defined declarations is not yet implemented, but it can be done easily.

\medskip

Also, the full declarations of user defined and predefined queries/constants can be listed with the command,
\begin{small}
\begin{verbatim}
    library list verbose;
\end{verbatim}
\end{small}

\noindent
with the result being,
\begin{small}
\begin{verbatim}
    Library command is well-formed and well-typed, but not
    executable

    Warning, library command successful but no query executed.

    List of library declaration(s):

      set query Pair (set x,set y) be
        { 'fst':x, 'snd':y },

      boolean query isPair (set p) be (
        exists l: x in p . (
          l='fst'
          and
          forall m:z in p . ( m='fst' => z=x )
        )
        and
        exists l:y in p . (
          l='snd'
          and
          forall m:z in p .( m='snd' => z=y )
        )
      ),

                      ...

      set constant some_book be
      http://www.csc.liv.ac.uk/~molyneux/t/BibDB-f1.xml#b1

      set constant some_book be
      http://www.csc.liv.ac.uk/~molyneux/t/BibDB-f1.xml#b2
\end{verbatim}
\end{small}

\noindent
Here the list of queries/constants follows as above, but including the full declaration for all other default library declarations (omitted here for brevity; see the full listing of predefined library declarations in Appendix~\ref{app:predefined}). Those relevant parts of the BNF for the library commands are as follows:
\begin{small}
\begin{verbatim}
    <library commands> ::= "add" <declarations> |
                           "list" [ "verbose" ]
\end{verbatim}
\end{small}

\noindent
Note that, only the predefined library declarations will remain in the library after finishing the query session. In principle the ability to work with several libraries (as well as user defined libraries) should also be implemented. The queries \texttt{Pair}, \texttt{isPair}, \verb+First+, \verb+Second+ will be
\linebreak
formally explained below; \texttt{CartProduct}, \texttt{Square} and \texttt{HorizontalTC} in Section~\ref{sec:horizontal_tc}; \texttt{LabelledPairs}, \texttt{CanGraph} and \texttt{Can} in Section~\ref{sec:query_optimisation_removing_redundancies}; \texttt{TC\_along\_label} in Section~\ref{sec:imitating_path_expressions}; \texttt{SuccessorPairs}, \texttt{Precedes5}, \texttt{TCPure}, \texttt{StrictLinOrder\_on\_TC} in Section~\ref{sec:lin-ord} and Appendix~\ref{app:lin-ord-declarations}; whereas \texttt{Nodes}, \texttt{Children} and \texttt{Regroup} in Section~\ref{sec:impl_dec_aux}.

\pagebreak

\subsubsection{The queries \texttt{Pair}, \texttt{isPair}, \texttt{First} and \texttt{Second}}\label{sec:pair_ispair_first_second_defs}

Thus, let us now define several auxiliary queries dealing with ordered pairs. According to the syntax in Appendix~\ref{app:BNF} query declarations have the general form:
\begin{align*}
&\mbox{\tt set query } \; q(\bar{x})  =  t(\bar{x}),\\
&\mbox{\tt boolean query } \; q(\bar{x})  =  \varphi(\bar{x}).
\end{align*}

\noindent
Here $q$ is either set or boolean query name, respectively, with query parameters defined by the list $\bar{x}$ of participating set or label variables.

\paragraph{\texttt{Pair}:}

Our first query defines the operation creating an ordered pair:
\begin{small}
\begin{verbatim}
    set query Pair(set x,set y) = {'fst':x,'snd':y}
\end{verbatim}
\end{small}

\noindent
where \verb+'fst'+ and \verb+'snd'+ are label values helping to distinguish the first element \verb+x+ from the second element \verb+y+ of the ordered pair, with \verb+x+,\verb+y+ as set variables denoting any (hyper)sets. Recall that the order of elements in a set is ignored, playing no role. But, labels of elements such as \verb+fst+ and \verb+snd+ add the required structure.

\medskip

\paragraph{\texttt{isPair}:}

Now we consider the boolean valued query \verb+isPair(p)+ which given a set \verb+p+ says whether it is an ordered pair \verb+p={'fst':x,'snd':y}+ for some sets \verb+x+ and \verb+y+:
\begin{small}
\begin{verbatim}
    boolean query
    isPair(set p) = 
        (exists l:x in p .
          ( l='fst' and forall m:z in p . (m='fst' implies z = x) )
        and
         exists l:y in p .
          ( l='snd' and forall m:z in p . (m='snd' implies z = y) )
        )
\end{verbatim}
\end{small}

\noindent
Note that the equalities \verb+z=x+ and \verb+z=y+ in this query are actually based on the bisimulation relation. It follows that \verb+isPair(p)+ can hold even if the set equation \verb+p={...}+ contains syntactically more than two elements between braces. It is required that there exists only one element in \verb+p+ labelled by \verb+'fst'+ and one labelled by \verb+'snd'+ only up to bisimulation.

\paragraph{\texttt{First} and \texttt{Second}:}\label{par:first_second}

Let us also define the set valued operations \verb+First(p)+ and \verb+Second(p)+ giving the first and the second elements of any pair $p$:
\begin{small}
\begin{verbatim}
    set query First(set p) =
        union separate {l:x in p where l='fst' }

    set query Second(set p) =
        union separate {l:x in p where l='snd' }
\end{verbatim}
\end{small}

\noindent
Note that the union operation is necessary here. Indeed, assuming that the input is an ordered pair \verb+p = {'fst':u,'snd':v}+, then we would get 
without union just singleton sets \verb+{'fst' : u}+ and \verb+{'snd' : v}+, respectively, generated by the separation operator whereas we need their elements \verb+u+ and \verb+v+, respectively. Therefore, we need to use the general set theoretic identity
\[
\bigcup\setof{l:u} = u
\]
where $u$ is any set. Of course, in the case of arbitrary set input \verb+p+ separation will not necessary generate a singleton set. Anyway, \verb+First(p)+ and \verb+Second(p)+ will give some set values so that these operations are always defined.

\subsubsection{Implementation of the library}
\label{sec:library-implementation}

Although general implementation issues will be postponed till Part~\ref{part:implementation}, we can easily comment here how implementation of the library can be reduced to the general \texttt{let-endlet} construct of the language. Thus, let us assume that the library contains a list of declarations
\begin{align*}
d_1, d_2, \ldots, d_n
\end{align*}

\noindent
already added by the \verb+add+ command. Then any query $q$ can use these declarations and thus can contain constants and query names which are not declared in $q$, but must be declared above in the library. In fact, any such query
\[
\texttt{set query $q$;} \quad\textrm{or}\quad \texttt{boolean query $q$;} 
\]

\noindent
is automatically transformed by the implemented query system, respectively, to the query
\begin{align}\label{eq:library-translation}
\texttt{set/boolean query let } d_1, d_2, \ldots, d_n \texttt{ in } q \texttt{ endlet;}
\end{align}

\noindent
Then this query is checked to be well-formed and well-typed and then executed as it is discussed formally in Chapters~\ref{chap:syntax}~and~\ref{chap:exec}. This way also the problem of dependency between library declarations $d_1, d_2, ..., d_n$, whose order may be essential%
\footnote{
A declaration $d_i$ can depend only on $d_j$ with $j<i$. Even if $d_i$ calls a constant or query name declared by $d_k$ with $i<k$, appropriate (rightmost) $d_j$ with $j<i$ should be really found and used. But this does not require any special or additional care for the library declarations because the contextual analysis algorithm in Section~\ref{sec:well-typed_delta_queries} will guarantee this automatically under translation (\ref{eq:library-translation}).
}%
, is resolved automatically. Also query declarations when added to the library are automatically checked simply by transforming them to the usual query

\begin{align*}
\texttt{set query let } d_1, d_2, \ldots, d_n \texttt{ in } \setof{} \texttt{ endlet;}
\end{align*}

\noindent
where the trivial version of $q=\setof{}$ is used. Well-formedness and well-typedness of the latter query is considered, by definition, as well-formedness and well-typedness of the declarations in the library.

\section{Example $\Delta$-queries}\label{sec:example_queries}

\begin{figure}[!ht]
\centering
\includegraphics[scale=1.00]{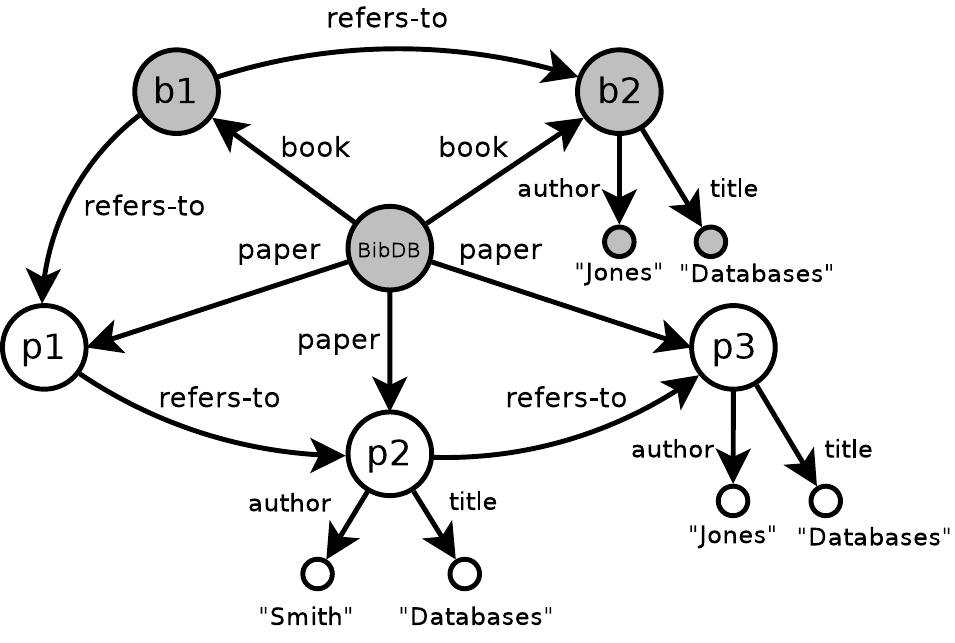}
\caption{Example distributed WDB of a small bibliographic database, distributed into two fragments.}
\label{fig:WDB-bibdb}
\end{figure}

Let us consider the following example queries based on the bibliographic WDB presented in \cite{MS07} and similar to the example in  \cite{ABS00}. This WDB is distributed (split into two fragments) as illustrated by the colouring of the graph in Figure~\ref{fig:WDB-bibdb}. Each fragment is given by a subsystem of set equations represented practically as an XML-WDB file (see Chapter~\ref{chap:xml-wdb} for the technical details of the XML-WDB representation). These files can be examined in the Appendix~\ref{app:XML-WDB_files}.

\noindent
Let us consider the corresponding subsystems of set equations represented practically as
\linebreak
\mbox{XML-WDB} files. Note that, full set names are denoted as  the concatenation of URL, \verb+#+, and simple set name; however, the URL and the delimiter \verb+#+ can be omitted for local set names. The subsystem of set equations represented by the XML-WDB file \url{http://www.csc.liv.ac.uk/~molyneux/t/BibDB-f1.xml} is as follows:

\medskip
\pagebreak[3]

\begin{small}
\begin{verbatim}
   BibDB = {
    'book':b1,
    'book':b2,
    'paper':http://www.csc.liv.ac.uk/~molyneux/t/BibDB-f2.xml#p1,
    'paper':http://www.csc.liv.ac.uk/~molyneux/t/BibDB-f2.xml#p2,
    'paper':http://www.csc.liv.ac.uk/~molyneux/t/BibDB-f2.xml#p3
   }

   b1 = {
    'refers-to':http://www.csc.liv.ac.uk/~molyneux/t/BibDB-f2.xml#b2,
    'refers-to':p1
   }

   b2 = {
    'author':"Jones",
    'title':"Databases"
   }
\end{verbatim}
\end{small}

\noindent
The XML-WDB file \url{http://www.csc.liv.ac.uk/~molyneux/t/BibDB-f2.xml} represents the subsystem

\begin{small}
\begin{verbatim}
   p1 = {
    'refers-to':p2
   }

   p2 = {
    'author':"Smith",
    'title':"Databases",
    'refers-to':p3
   }

   p3 = {
    'author':"Jones",
    'title':"Databases"
   }
\end{verbatim}
\end{small}

\noindent
Recall that single quotation marks are used to denote labels such as \verb+'author'+, whereas double quotation marks denote atomic values which are, strictly speaking, special singleton sets, e.g.\ \verb+"Jones"+ means \verb+{'Jones':{}}+.

\pagebreak

\subsection{Example of a non-well-typed query}

In our first example the query is non-well-typed because the identifiers \verb+BibDB+ and \verb+b2+  are formally undeclared within the following query, although intuitively corresponding to some graph nodes. The intended informal meaning of the query being: find all publications which refer to the book \verb+b2+.
\begin{small}
\begin{verbatim}
    set query collect {
        pub-type:pub 
        where pub-type:pub in BibDB
        and exists 'refers-to':ref in pub . ref=b2
    };
\end{verbatim}
\end{small}

\noindent
The result of running this query is the error messages:
\begin{small}
\begin{verbatim}
    Query is well-formed, but not well-typed

    Error at character 76,

    occurrence of identifier name BibDB not declared:
      set query collect { pub-type:pub 
      where pub-type:pub in BibDB <-------

      and exists 'refers-to':ref in pub .

    Error at character 127,

    occurrence of identifier name b2 not declared:
      and exists 'refers-to':ref in pub .
      ref=b2 <-------

      };
\end{verbatim}
\end{small}

\noindent
Here \texttt{well-typed} would intuitively mean that all identifiers and their types (\emph{set} or \emph{label}, etc.) in the query are appropriately described by declarations, quantifiers, etc., and used in other places of the query accordingly. But unfortunately the error messages show that it is not the case. The corrected version of this query is presented in Section~\ref{sec:valid-executable-query}, where the identifiers \verb+BibDB+ and \verb+b2+ are appropriately related to the WDB considered. We will pay much more attention to well-typedness of queries in  Chapter~\ref{chap:syntax} which is highly important for the correct implementation of $\Delta$.

\subsection{Example of valid and executable query}\label{sec:valid-executable-query}

After correction of the above query we have:
\begin{small}
\begin{verbatim}
    set query 
      let set constant BibDB be
            http://www.csc.liv.ac.uk/~molyneux/t/BibDB-f1.xml#BibDB,
          set constant b2 be
            http://www.csc.liv.ac.uk/~molyneux/t/BibDB-f1.xml#b2
      in collect { pub-type:pub
          where pub-type:pub in BibDB
          and exists 'refers-to':ref in pub . ref=b2
        }
      endlet;
\end{verbatim}
\end{small}

\noindent
Evidently the result of this query contains the book \verb+b1+ (which refers to \verb+b2+) and, not so obviously, the paper \verb+p2+ which refers to \verb+p3+, the latter being formally bisimilar to \verb+b2+ with the same \verb+title+ and \verb+author+ elements. The result of the modified query is,
\begin{small}
\begin{verbatim}
    Query is well-formed, well-typed and executable

    Result = { 
      'paper':http://www.csc.liv.ac.uk/~molyneux/t/BibDB-f2.xml#p2,
      'book':http://www.csc.liv.ac.uk/~molyneux/t/BibDB-f1.xml#b1
    }

    Finished in: 398 ms
\end{verbatim}
\end{small}

\noindent
This result might seem strange, but formally it is correct taking into account our hyperset theoretic approach to WDB. The question here is to the designer(s) of this bibliographic database who overlooked that essentially the \emph{same} publication is presented in the database both as a book and as a paper. If these are really different publications then they should be represented in the database accordingly (as discussed in the considerations below). Note that the incoming edges labelled by \texttt{book} or \texttt{paper} do not count when determining bisimilarity of the nodes \verb+p3+ and \verb+b2+ --- only outgoing edges play a role. Such fundamental flaws can be introduced accidentally when possibly many users create distributed WDB. Evidently, this WDB was poorly designed, therefore, better understanding of the structural design of WDB would make this process less error-prone. Anyway, even with the (traditional) relational approach database design is a crucial step.

\subsubsection{Query semantics versus WDB design}

If we really want to include only references to the book \verb+b2+ (without redesigning this WDB), then it might seem that the solution is to replace the equality \verb+ref=b2+ by the formula
\begin{small}
\begin{verbatim}
    (ref=b2 and 'book':ref in BibDB)
\end{verbatim}
\end{small}

\noindent
in the above query. However, this would not really help because in any case \verb+p3=b2+ (these set names / graph nodes are bisimilar) in the above WDB. Equality of (hyper)sets is defined by their elements, elements of elements, etc., i.e. by outgoing edges, and not by incoming edges. So, after formally removing redundancies (say, omitting \verb+p3+) we should have one joint node \verb+b2+ with two incoming edges \verb+BibDB+ $\stackrel{\texttt{book}}{\arrr}$ \verb+b2+ and \verb+BibDB+ $\stackrel{\texttt{paper}}{\arrr}$ \verb+b2+  (besides two more incoming \verb+refers-to+ edges from \verb+b1+ and \verb+p2+ and the evident two outgoing edges). This is probably not what the designer(s) of this distributed WDB had in mind. Anyway, we will continue using this example as a good and simple illustration of the (hyper)set theoretic approach. In principle, we could imagine that the creators of this WDB really wanted to have publications classified both as a book and a paper. This is not a contradiction, as anything is possible in semi-structured data. In fact, the problem is only to decide what we really want and whether this intuition is reflected correctly by the given WDB design.

\medskip

This example emphasises the real meaning of set theoretic versus pure graph approaches to semi-structured databases, and the role of removing redundancies on the level of the design. The right approach here should be based on a well-chosen discipline, for example:
\begin{itemize}
\item[(i)]{
\emph{Reconstruct} this database by replacing labels \verb+book+ and \verb+paper+ by \verb+publication+ and adding outgoing edges from each publication showing its \verb+type+ (\verb+'book'+ or \verb+'paper'+; see Figure~\ref{fig:restructured_bibdb}
\footnote{
Strictly speaking, Figure~\ref{fig:restructured_bibdb} reflects this idea only partially because it is devoted to illustrate a related but formally different example of restructuring query in the $\Delta$-language. It still has a publication which is characterised as both book and a paper, however, this is more noticeable ``locally'' reducing accidental user error.
}), 
or alternatively
}
\item[(ii)]{
Enforce some WDB \emph{schema} during the design of WDB e.g.\ requiring that there is only one \verb+book+ or \verb+paper+ edge from \verb+BibDB+ leading to any given publication considered up to bisimulation.
}
\end{itemize}

\noindent
Here the term ``up to bisimulation'' means that if two children of \verb+BibDB+ are bisimilar then they, in fact, have identical labelling. But it is not our goal here to go into details of such kind of discipline and consider WDB schemas. In any case, we should be precise and accurate with the design of WDB, and in formulating both formal and intuitive versions of our queries. The mathematical ground of hyperset theory is quite solid and sufficient for that. 

\medskip

The main point is that any formal query has a unique (up to bisimulation) answer -- in fact, either a hyperset or boolean value -- and all the queries are \emph{bisimulation invariant} and can  be computed in polynomial time (with respect to the size of WDB). Vice versa, any P-time computable and bisimulation invariant (and also ``generic'' \cite{LS97,S93}) query is definable in $\Delta$. In fact, this also means that the language $\Delta$ has full P-time computable power of \emph{restructuring}, not only simple retrieval of already existing elements in the WDB. For example the query restructuring the \verb+BibDB+ database as is essentially described in (i) above could be written in $\Delta$ using the plan performance operator $\Dec$. 

\subsection{Restructuring query}\label{sec:restructuring-query}

The ability to define queries arbitrarily restructuring any given data is the most essential requirement of any database query language. Here we will consider one simple example which could hopefully convince the reader that $\Delta$ has a very strong 
restructuring power.

\medskip

Firstly, let us recall the informal meaning of the following useful query declarations in the default library (with the formal meaning fully described in Section~\ref{sec:pair_ispair_first_second_defs}) and introduce \mbox{semi-formally} one more query \texttt{CanGraph} to be formally defined in Section~\ref{sec:query_optimisation_removing_redundancies}:

\begin{itemize}
\item{
\verb+Pair(x,y)+ -- denoting the ordered pair $\orderedsetof{x,y}$, in fact the two element set of the form \verb+{'fst':x,'snd':y}+ allowing to distinguish between the first and second elements.
}
\item{
\verb+First(p)+ -- first element of $p$ if $p$ is an ordered pair.
}
\item{
\verb+Second(p)+ -- second element of $p$ if $p$ is an ordered pair.
}
\item{
\verb+CanGraph(x)+ -- denoting the set of labelled pairs $l:\orderedsetof{u,v}$ where $l\!:\!v \in u$ holds in the transitive closure $\TC(x)$.
}
\end{itemize}

\noindent
Then the required restructuring query (described informally in (i) above) is defined as follows:
\begin{small}
\begin{verbatim}
    set query
      let set constant BibDB =
            http://www.csc.liv.ac.uk/~molyneux/t/BibDB-f1.xml#BibDB,
          set constant restructuredBibDB be
            (U collect{
              'null':if (L='paper' or L='book')
                     then { 'publication':X,
                            'type':call Pair(call Second(X),{L:{}}),
                            L:call Pair({L:{}}, {}) }
                     else {L:X}
                     fi
               where L:X in call CanGraph(BibDB)
               }
            )
      in
        decorate ( restructuredBibDB, BibDB )
      endlet;
\end{verbatim}
\end{small}

\noindent
Here \verb+CanGraph(BibDB)+ is essentially the bibliography graph in Figure~\ref{fig:WDB-bibdb}, but represented in the traditional set theoretic way as the set of labelled ordered pairs, each denoted in the query as \texttt{L:X} with \texttt{L} the label and \texttt{X} the ordered pair in question. The required restructuring {in terms of ordered pairs consists in relabelling of labels \verb+'book'+ and \verb+'paper'+ as \verb+'publication'+, and creating additional leaf edges with the publication type is done essentially by the following fragment
\begin{small}
\begin{verbatim}
    'null':if (L='paper' or L='book') 
           then { 'publication':X,
                  'type':call Pair(call Second(X),{L:{}}),
                   L:call Pair({L:{}}, {}) 
                 }
           else {L:X}
           fi 
\end{verbatim}
\end{small}

\noindent
generating appropriate sets of labelled ordered pairs. Then these sets%
\footnote{
where the value of the label \texttt{'null'} is not important
}
are \texttt{collect}ed, and taking the union gives rise to the required restructured set of labelled ordered pairs denoted as \texttt{restructuredBibDB}. But abstractly, we need a hyperset rather than this graph (a set of pairs). Thus, finally, the decoration operation applied to the graph \verb+restructuredBibDB+ and the vertex \verb+BibDB+ generates the required abstract hyperset (as described in general in Section~\ref{sec:dec_operation}). The result of this query is,
\begin{small}
\begin{verbatim}
    Query is well-formed, well-typed and executable

    Result = {
      'publication':res2,
      'publication':res0,
      'publication':res1,
      'publication':{
        'type':"book",
        'refers-to':res1,
        'refers-to':res2
      }
    }

    res0 = {
      'type':"paper",
      'author':"Smith",
      'title':"Databases",
      'refers-to':res1
    }

    res1 = {
      'type':"paper",
      'type':"book",
      'author':"Jones",
      'title':"Databases"
    }

    res2 = {
      'type':"paper",
      'refers-to':res0
    }

    Finished in: 1646 ms (query execution is 1643 ms, and 
    postprocessing time is 3 ms)
\end{verbatim}
\end{small}

\noindent
As we discussed formerly, atomic values, strictly speaking, denote corresponding singleton sets, for example \verb+"Smith"+, denotes \verb+{'Smith':{}}+. The (new) set names \verb+res0+, \verb+res1+ and \verb+res2+ correspond, respectively, to the ``restructured'' publications \texttt{p2'}, \texttt{p3'/b2'} and \texttt{p1'}. Note that, the query system replaces some set names on the right-hand side by the corresponding bracket expression where suitable, thereby presenting the result in a ``nested'' form. For example the publication \texttt{b1'} is implicitly nested in the \verb+Result+ set equation.

\medskip

This result can be more conveniently visualised by Figure~\ref{fig:restructured_bibdb} with the set name \texttt{Result} replaced by \texttt{BibDB'}, and new set names replaced by corresponding names revelant to the restructured publications (as was discussed above).

\begin{figure}[ht!]
\centering
\includegraphics[scale=1.00]{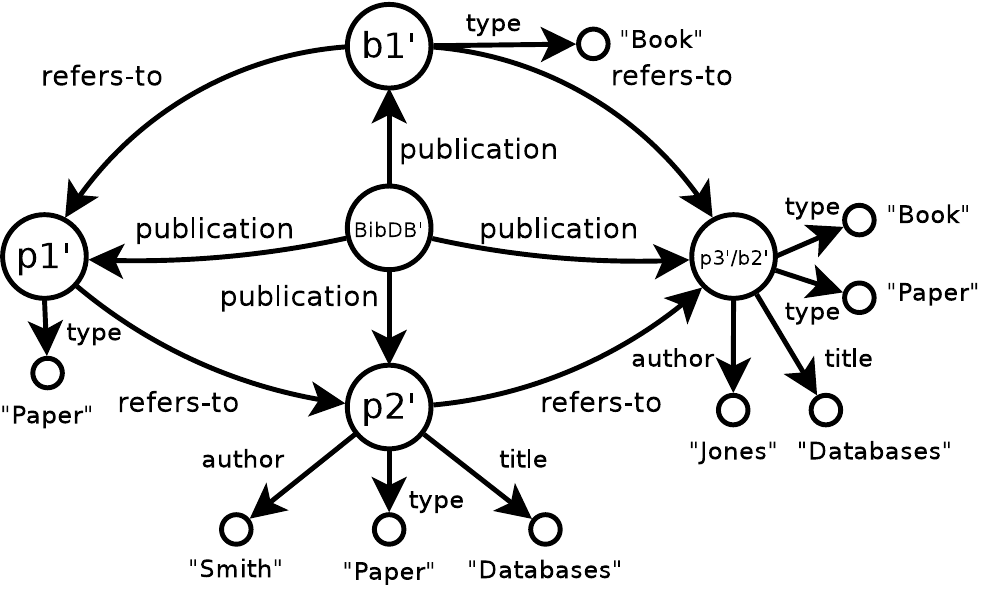}
\caption{The result of the restructuring query.}
\label{fig:restructured_bibdb}
\end{figure}

\noindent
Note that the publication \texttt{p3'/b2'}%
\footnote{
denoted by the new set name \texttt{res1} (see query result above)
}
has both the type \verb+book+ and \verb+paper+, and that this unusual feature is the result of the initial design of \verb+BibDB+ and not a failure of the above query. Anyway, in principle this graph suggests a potentially better (less semantically error prone) design for the bibliography database.

\subsection{Horizontal transitive closure}\label{sec:horizontal_tc}

Let us now consider the query which can generate the ``horizontal'' transitive closure%
\footnote{
This should not be mixed with the set theoretic meaning of the $\Delta$-term operator transitive closure $\TC$ which can be understood intuitively as ``vertical'' transitive closure, that is $\TC(x)$ represents the set of (labelled) elements of element of elements, etc.\ of $x$ (including $x$ itself) as defined in Section~\ref{sec:denotational_semantics_terms}. The point is that it is typically convenient to think of elements of a set as lying \emph{under} this set -- hence \emph{vertical} view.
}
of any graph $g$ (a set of ordered pairs). Consider the trivial example graph $g$ represented as the nodes $a,b,c$ with edges $\orderedsetof{a,b}$ and $\orderedsetof{b,c}$ depicted by solid black edges in Figure~\ref{fig:WDB-horizontal_tc_example}%
\footnote{
We should not mix this graph, which is only a visual representation of a \emph{set of ordered pairs}, with any other graphs depicted before and having rather a visual representation of a \emph{system of set equations}.
}%
. The result of applying horizontal transitive closure to the graph $g$ is shown by the original edges (in solid black) and the additional edges $\orderedsetof{a,c}$, $\orderedsetof{a,a}$, $\orderedsetof{b,b}$ and $\orderedsetof{c,c}$ highlighted in Figure~\ref{fig:WDB-horizontal_tc_example} as red dashed edges.

\begin{figure}[!ht]
\centering
\includegraphics[scale=1.00]{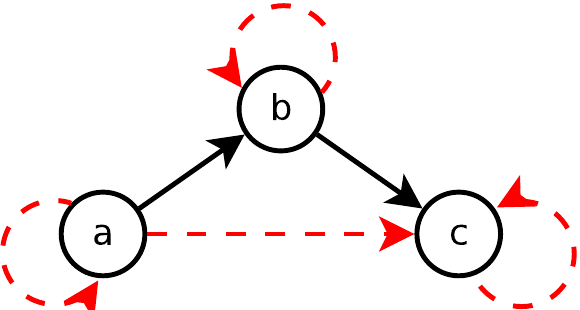}
\caption{The result of ``horizontal'' transitive closure applied to the abstract graph $g$.}
\label{fig:WDB-horizontal_tc_example}
\end{figure}

\noindent
The result is also a graph denoted as $g^*$ which extends $g$ by new ordered pairs ($g\subseteq g^*$) such that for each edge $\orderedsetof{x,y}\in g^*$ there exists  a path from $x$ to $y$ belonging to the original graph $g$, and vice versa. This can be recursively defined as follows:
\begin{align*}
\orderedsetof{x,y}\in g^* \iff x=y \vee \exists z . ( \orderedsetof{x,z}\in g^* \wedge 
\orderedsetof{z,y}\in g ) 
\end{align*}

\noindent
or as 
\begin{align}\label{eq:horizontalTC}
g^* = \setof{\orderedsetof{x,y}\in|g| \mid x=y \vee \exists z\in|g| . ( \orderedsetof{x,z}\in g^* \wedge 
\orderedsetof{z,y}\in g )} 
\end{align}

\noindent
where $|g|$ is the set of all $g$-nodes. It is assumed that $g^*$ is the least set of pairs satisfying the above equivalence. This operation could prove useful complementing ``vertical'' transitive closure $\TC(x)$ in the original $\Delta$-language, whose result is the set of elements of elements, etc.\ for any given set $x$ (including $x$ itself).

\pagebreak

Thus, let us implement $g^*$ (denoted below as \verb+HorizontalTC(g)+) in the following straightforward way based on the above formula (\ref{eq:horizontalTC}). Firstly, let us add to the library the set query declaration \texttt{Nodes(g)} (formally described in Section~\ref{sec:impl_dec_aux}), denoted above as $|g|$ and extracting from the set of ordered pairs \texttt{g} the set of elements participating in these ordered pairs.

\paragraph*{\texttt{Nodes:}}

\begin{small}
\begin{verbatim}
    set query Nodes (set g) =
        union separate { m : p in g  | call isPair ( p ) }
\end{verbatim}
\end{small}

\noindent
We will also need the ordinary and very important (not only for defining the horizontal transitive closure) set theoretic operations of

\paragraph*{\texttt{CartProduct} and \texttt{Square:}}

%
\begin{small}
\begin{verbatim}
    set query CartProduct(set X,set Y) =
          U collect {'null':collect {'null':call Pair(x,y)
                                     where l:y in Y
                                     }
                     where m:x in X
                     }

    set query Square(set X) = call CartProduct(X,X)
\end{verbatim}
\end{small}

\noindent
Finally, the set query \verb+HorizontalTC(g)+ can be easily defined using the recursion operator as follows.

\paragraph*{\texttt{HorizontalTC:}}

%
\begin{small}
\begin{verbatim}
    set query HorizontalTC(set g) be
        recursion p {
            'null':pair in call Square(call Nodes(g)) where (
                call First(pair)=call Second(pair)
                or
                exists m:z in call Nodes(g) . ( 
                    'null':call Pair(call First(pair),z) in p
                    and
                    'null':call Pair(z,call Second(pair)) in g
                )
            )
        }
\end{verbatim}
\end{small}


\noindent
Let us now execute \texttt{HorizontalTC} applied to the graph \texttt{g} (see above),
\begin{small}
\begin{verbatim}
    set query
      let set constant g be {
                       'null':call Pair("a","b"),
                       'null':call Pair("b","c")
                     }
      in
      call HorizontalTC(g)
    endlet;
\end{verbatim}
\end{small}


\noindent
and see that the result is as expected, although with many repetitions which witness that the implementation is currently not optimal. However, all the repetitions in the query result can be easily eliminated by \emph{canonisation} (to be discussed in Section~\ref{sec:query_optimisation_removing_redundancies} below). First note that the canonisation set query declaration (\texttt{Can}) is already added to the default library
\begin{small}
\begin{verbatim}
    set query Can(set x) be decorate(call CanGraph(x),x)
\end{verbatim}
\end{small}

\noindent
and that the above query can be rewritten using \texttt{Can} as follows:
\begin{small}
\begin{verbatim}
    set query
      let set constant g be {
                       'null':call Pair("a","b"),
                       'null':call Pair("b","c")
                     }
      in
      call Can(call HorizontalTC(g))
    endlet;
\end{verbatim}
\end{small}

\noindent
Now, by running the amended query, we see that all repetitions have been eliminated.

\subsection{Dealing with proper hypersets }\label{sec:querying_hypersets}

The hyperset theoretic approach to WDB can represent and query semi-structured databases possibly involving arbitrary cycles (see  Chapter~\ref{chap:ssd_wdb}). For example let us consider the WDB graph in Figure~\ref{fig:hyperset_example} with the cycle between the vertices $a$ and $b$ (edges $a \longrightarrow b$ and $b \longrightarrow a$).
\begin{figure}[!ht]
\centering
\includegraphics{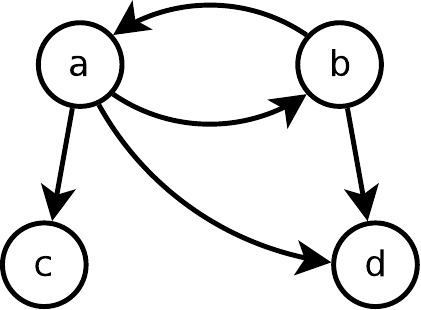}
\caption{WDB graph with cycle.}
\label{fig:hyperset_example}
\end{figure}

\noindent
It is easy to see that $a\bis b$ and $c\bis d$ are the only positive bisimulation facts, and hence $a$ and $b$, and also $c$ and $d$ actually denote the same hypersets (the latter two denote $\emptyset$). The strongly extensional version of this WDB with all redundancies removed is shown in Figure~\ref{fig:hyperset_example_strong_ext}.


\begin{figure}[!ht]
\centering
\includegraphics{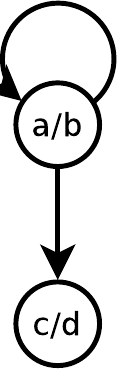}
\caption{Strongly extensional version of the WDB in Figure~\ref{fig:hyperset_example}.}
\label{fig:hyperset_example_strong_ext}
\end{figure}

\noindent
Let us show how to define in $\Delta$ the hyperset denoted by the vertex $a$. It can be done with the help of decoration operation as follows:

\begin{small}
\begin{verbatim}
   set query let
      set constant g = {
            'null':call Pair("a","b"), 'null':call Pair("b","a"),
            'null':call Pair("a","c"), 'null':call Pair("a","d"),
            'null':call Pair("b","d")
      }
   in 
     decorate (g, "a")
   endlet;
\end{verbatim}
\end{small}

\noindent
The result of this query exactly corresponds to the graph in Figure~\ref{fig:hyperset_example}:
\begin{small}
\begin{verbatim}
    Query is well-formed, well-typed and executable

    Result = {
      'null':{
        'null':Result,
        'null':{}
      },
      'null':{},
      'null':{}
    }

    Finished in: 20 ms (query execution is 20 ms, and 
    postprocessing time is 0 ms)
\end{verbatim}
\end{small}

\noindent
In the next section we will show how the strongly extensional result (corresponding to Figure~\ref{fig:hyperset_example_strong_ext}) can be obtained.
In fact, without using decoration it would be impossible to define this cyclic set \verb+Result+ corresponding to the vertex $a$. Further, let us consider the query to compute equality (bisimulation) between the sets denoting the vertices $a$ and $b$ as
\begin{small}

\begin{verbatim}
   boolean query let
      set constant g = {
            'null':call Pair("a","b"), 'null':call Pair("b","a"),
            'null':call Pair("a","c"), 'null':call Pair("a","d"),
            'null':call Pair("b","d")
      }
   in
     decorate (g, "a") = decorate (g, "b")
   endlet;
\end{verbatim}
\end{small}

\noindent
where the evident result \texttt{true} of this query corresponds to the intuitive observation that, in fact, \verb+"a"+ and \verb+"b"+ denote bisimilar graph $g$-nodes.

\subsection{Query optimisation by removing redundancies}\label{sec:query_optimisation_removing_redundancies}

The following example demonstrates the general task of removing redundancies by  a particular set query \texttt{Can} (for ``canonisation'') on the above graph in Figure~\ref{fig:hyperset_example} (in Section~\ref{sec:querying_hypersets}). Here we use our knowledge%
\footnote{
This solution may not be so intuitively evident yet to those users who are unfamiliar with the set theoretic meaning of decoration and the details of \emph{how} this operation was implemented (see {Section~\ref{sec:impl_dec}}). But running queries with \texttt{Can} can nevertheless clearly demonstrate its usefulness.
}
on the implementation of the decoration operation (see Section~\ref{sec:impl_dec}) to remove the redundancies in the original graph (see the result of the set query above) by applying the decoration operator to the canonical form of this graph (as a set of pairs representing graph edges) and the participating vertex $a$.

\medskip

\noindent
First, let us define the set query declaration

\paragraph*{\texttt{LabelledPairs:}}

%
\begin{small}
\begin{verbatim}
    set query LabelledPairs (set v) be 
        collect {
            l:{ 'fst':v , 'snd':u }
            where l:u in v
        } 
\end{verbatim}
\end{small}

\noindent
with the result of \verb+LabelledPairs(v)+ being the set of labelled pairs $l\!:\!\orderedsetof{v,u}$ denoting labelled edges $v \stackrel{l}{\longrightarrow} u$ corresponding to the set memberships \verb+l:u+ in the set \verb+v+. This set query declaration participates in another important library set query

\pagebreak

\paragraph*{\texttt{CanGraph:}}

%
\begin{small}
\begin{verbatim}
    set query CanGraph(set x) be
        union
            collect {
                'null':call LabelledPairs ( v )
                where m:v in TC(x)
            }
\end{verbatim}
\end{small}

\noindent
whose output is the set of labelled pairs $l\!:\!\orderedsetof{u,v}$ corresponding to those labelled elements $l:v\in u$ with $u$ ranging over the elements of transitive closure $\TC(x)$. Here \verb+'null'+ is a label whose value is not important. Indeed, the \verb+union+ operation unifies the labelled pairs from \verb+LabelledPairs(v)+. The third library query we need is the set query \verb+Can(set x)+ (invoking \verb+CanGraph+ above) which takes any set $x$ and returns the same abstract set $x$, but in its strongly extensional form.


\paragraph*{\texttt{Can:}}

%
\begin{small}
\begin{verbatim}
    set query Can(set x) be
        decorate (call CanGraph(x), x)
\end{verbatim}
\end{small}

\noindent
In fact, we should always have \verb+Can(x)=x+ because \verb+CanGraph(x)+ is evidently the canonical graph whose node \verb+x+ represents the set \verb+x+ itself, and, in this sense, the set query \verb+Can+ does nothing. It follows also that \verb+Can+ and \verb+decorate+ are essentially inverse operations. Thus, \verb+Can+ changes nothing in the abstract set theoretical sense. But due to applying decoration to get \verb+Can(x)+ and taking into account both strong extensionality of \texttt{CanGraph(x)} and the way \texttt{decoration} used in \texttt{Can} is implemented in  Section~\ref{sec:impl_dec}, the resulting system of set equations generated by \texttt{Can(x)} is always non-redundant (strongly extensional). 

\medskip

Therefore the result of \verb+Can(a)+ for the example in Figure~\ref{fig:hyperset_example} consists of one set equation for the node $a/b$ of the graph shown in Figure~\ref{fig:hyperset_example_strong_ext}. Indeed, running the query:

\begin{small}
\begin{verbatim}
   set query let
      set constant g = {
            'null':call Pair("a","b"), 'null':call Pair("b","a"),
            'null':call Pair("a","c"), 'null':call Pair("a","d"),
            'null':call Pair("b","d")
      }
   in 
     call Can ( decorate (g, "a") )
   endlet;
\end{verbatim}
\end{small}

\noindent
gives the result:
\begin{small}
\begin{verbatim}
    Query is well-formed, well-typed and executable

    Result = {
      'null':Result,
      'null':{}
    }

    Finished in: 35 ms (query execution is 35 ms, and
    postprocessing time is 0 ms)
\end{verbatim}
\end{small}

\noindent
with the set \verb+Result+ denoting $a/b$. From the abstract hyperset view this is exactly the same result as without using \texttt{Can}, but represented in a better, non-redundant way.

\medskip

Note that \verb+Can+ can be used for the more general purpose of query optimisation (not only for optimisation of query results by removing redundancies). Of course, using \verb+Can(t)+ instead of \verb+t+ will require some time to compute \texttt{TC(t)} and then decoration (which in fact requires computation of many bisimulation facts). But the benefit is that \verb+Can(t)+ will be represented without any redundancies at all, in contrast to the set \verb+t+ which could contain a large number of equal elements due to possible redundancies and thus would be much smaller after eliminating them. Then, for example, \verb+Square(t)+ (the Cartesian product of \verb+t+) would also be represented without any unnecessary repetitions, and thus possibly much smaller. In particular, if we want to have recursion over this \texttt{Square} (like in the case of recursive definition of \verb+HorizontalTC+), it would be computed much more efficiently, also with smaller number of iteration steps, assuming \verb+Can(t)+  instead of \verb+t+.

\medskip

In principle, we could extend the language by adding \emph{literal equality} \texttt{eq(x,y)} for set names (object identities). This, of course, would change the set theoretic character of the language as queries using such equality will not necessarily be bisimulation invariant. But if we would use this equality only over the elements of sets represented as \verb+Can(t)+, then this can work as an additional optimisation. In principle, the query system could recognise the expressions \verb+Can(t)+ and automatically replace bisimulation over this set by literal equality.

\medskip

Finally, note that the above optimisation was given for the current implementation of the $\Delta$-language so that users can exploit  canonisation to optimise some queries. In principle, this optimisation could be build into the implementation, so that, any possible redundancies are removed during query execution. In fact, the query system, while executing a query, supports a list of currently known positive bisimulation facts (see Chapter~\ref{chap:bisimulation}) which can be used in background time to remove at least some redundancies in set equations stored in local memory.

\section{Imitating path expressions}\label{sec:imitating_path_expressions}

The ability to select nodes of a WDB graph to arbitrary depth can be elegantly achieved using path expressions. As shown in \cite{S06}, the action of a rich class of path expressions is definable in the original $\Delta$, itself having no path expressions at all, with the help of $\TC$ and $\Rec$. In spite of this fact, an important goal for the future work is to implement the extension of $\Delta$ by such user friendly path expressions like~in the following example query%
\footnote{
The keyword \texttt{path} is added to aid reading.
}
(for simplicity only involving set constants for full set names from the bibliographic WDB):

\begin{small}
\begin{verbatim}
    set query
    separate {
        pub-type:x in BibDB
        where exists path <b1>refers-to*<x>refers-to<b2> .
              'author':"Smith" in x
    };
\end{verbatim}
\end{small}

\noindent
The result of this query would be:

\begin{small}
\begin{verbatim}
    Result = {
        paper:p2
    }
\end{verbatim}
\end{small}

\noindent
Quantification goes over paths from \verb+b1+ to \verb+b2+ having an appropriate intermediate set (or node for a publication) \verb+x+ which is required to have the element \verb+author:"Smith"+, but it appears that there does not exists such an explicit path. Nevertheless, the required path does exist, as shown in Figure~\ref{fig:WDB-bibdb_path_expression} by the dashed edges labelled \verb+refers-to+ leading from \verb+b1+ to \verb+p3+, where \verb+p3+ is equal (bisimilar) to \verb+b2+ (\verb+p3+${}\bis{}$\verb+b2+) as we already know. In strongly extensional graphs (where there are no bisimilar nodes) path expressions would be understood quite straightforwardly. Our hyperset approach leads to such kind of complications, but this is the compromise for having a natural language with clear semantics and strong (also precisely characterised) expressive power.

\medskip

Note that the result of the above query would be the empty set if the Kleene star ``\verb+*+'' was removed from the path expression. Indeed, there are no paths of length two from b1 to b2, even up to bisimulation.

\pagebreak

\begin{figure}[!ht]
\centering
\includegraphics[scale=1.00]{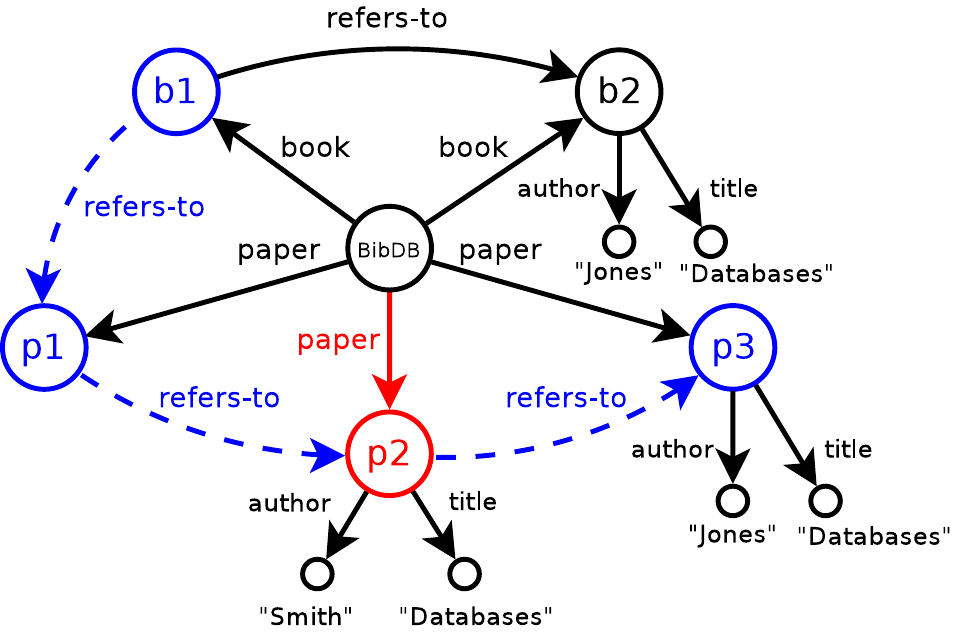}
\caption{Visualisation of the path expression \texttt{<b1>refers-to*<x>refers-to<b2>} applied to the bibliographic WDB.}
\label{fig:WDB-bibdb_path_expression}
\end{figure}

\noindent
The action of the path expression \verb+<b1>refers-to*<x>refers-to<b2>+ can, in fact, be ``rewritten'' into $\Delta$ (in its present form) by the following steps. Firstly, consider the subexpression \verb+<x>refers-to<b2>+ denoting a path from the candidate publication \verb+x+ to \verb+b2+ labelled by \verb+'refers-to'+. This can be expressed as the $\Delta$-formula:
\begin{small}
\begin{verbatim}
    'refers-to':b2 in x
\end{verbatim}
\end{small}

\noindent
where \verb+b2+ is set constant and \verb+x+ is set variable. Secondly, the subpath expression 
\linebreak
\verb+<b1>refers-to*<x>+ denotes set of candidate publications \verb+x+ which can be reached from \verb+b1+ by navigating zero or more \verb+refers-to+ labelled edges. Thus, let us include in the library the general set query which will give the set of graph nodes (of a graph representing a hyperset~\verb+z+) reachable by navigating zero or more \verb+l+-labelled edges.

\paragraph*{\texttt{TC\_along\_label}:}


\begin{small}
\begin{verbatim}
    set query TC_along_label(label l, set z) be
        recursion p { k:x in TC(z)
                      where (
                              ( x=z and k='null' )
                             or
                              ( k=l and exists m:y in p . l:x in y )
                             )
        };
\end{verbatim}
\end{small}

\noindent
Here \texttt{p} is a recursion set variable to representing the set \verb+T=TC_along_label(l,z)+ of nodes lying on potentially all the \texttt{l}-labelled paths outgoing from \texttt{z}. All elements of \texttt{T} are \texttt{l}-labelled, except possibly \texttt{z}. If \texttt{l:z} is in \texttt{z} then \texttt{l:z} will be added to \texttt{T}. But in any case \texttt{'null':z} will appear in \texttt{T} at the first stage of iteration. Hence the query call
\begin{small}
\begin{verbatim}
    TC_along_label('refers-to', b1)
\end{verbatim}
\end{small}

\noindent
represents the path expression \verb+<b1>refers-to*<x>+ where \verb+'refers-to'+ is label value and \verb+b1+ is set constant.

\medskip

Finally the path expression \verb+<b1>refers-to*<x>refers-to<b2>+, understood as the set of all \verb+x+ lying on the paths matching this 
path expression, is expressed as:
\begin{small}
\begin{verbatim}
    set query
          separate {
              n:xx in call TC_along_label('refers-to', b1)
              where 'refers-to':b2 in xx
          };
\end{verbatim}
\end{small}

\noindent
Now, the fragment 
\begin{small}
\begin{verbatim}
    exists path <b1>refers-to*<x>refers-to<b2> .
    'author':"Smith" in x
\end{verbatim}
\end{small}

\noindent
of our path expression query can be rewritten as
\begin{small}
\begin{verbatim}
    exists m:y in separate 
                  {n:xx in call TC_along_label('refers-to',b1)
                   where 'refers-to':b2 in xx
                   } .
    (x=y and 'author':"Smith" in x)
\end{verbatim}
\end{small}

\noindent
so that we can insert it in the full query

\begin{small}
\begin{verbatim}
    set query
    let
    set constant BibDB =
        http://www.csc.liv.ac.uk/~molyneux/t/BibDB-f1.xml#BibDB,
    set constant b1 =
        http://www.csc.liv.ac.uk/~molyneux/t/BibDB-f1.xml#b1,
    set constant b2 =
        http://www.csc.liv.ac.uk/~molyneux/t/BibDB-f1.xml#b2
    in
    separate {
        pub-type:x in BibDB
        where
            exists m:y in separate {
                n:xx in call TC_along_label('refers-to',b1)
                where 'refers-to':b2 in xx
            } .
            ( x=y and 'author':"Smith" in x)
          }
    endlet;
\end{verbatim}
\end{small}

\noindent
and run it to see the required result:

\begin{small}
\begin{verbatim}
    Query is well-formed, well-typed and executable

    Result = {
      'paper':http://www.csc.liv.ac.uk/~molyneux/t/BibDB-f2.xml#p2
    }

    Finished in: 5766 ms (query execution is 5764 ms, and 
    postprocessing time is 2 ms)
\end{verbatim}
\end{small}

\noindent
Despite this example of successfully imitating path expressions it would be more useful to also include path expressions directly within the implementation language. Although much more general path expressions can be imitated by $\Delta$-queries in the current version \cite{S06}, this  imitation can be quite complicated in general and is not a particularly efficient way of implementing and executing queries with path expressions. Anyway, the $\Delta$-language, as it is implemented now, is very expressive.

\section{Linear ordering query}\label{sec:lin-ord}

The query example considered in this section has mainly theoretical interest, although it might be useful in practice. The point is that we can define in $\Delta$ linear ordering on the transitive closure of any hyperset by using the lexicographical linear ordering we have on labels. In fact, the resulting linear ordering on hypersets is itself, in a sense, lexicographical. Having  defined linear ordering, we can further define any (``generic'' polynomial-time) computable operation over hypersets by simulating any given Turing Machine (as shown in descriptive complexity theory \cite{G83,I82,S80,V82}). This is the key point of the main result in \cite{S93} (for well-founded sets) and in \cite{S95,LS97,LS99} (for hypersets) on the expressive power of $\Delta$ coinciding with polynomial time computability over (hyper)sets. (We omit precise formulation which is more subtle in the case of hypersets having labelled elements; see \cite{S93,LS97}).

\medskip

Let us consider the set query declaration \verb+StrictLinOrder_on_TC(set z)+ (and other associated declarations) which can be found in  Appendix~\ref{app:lin-ord-declarations}%
\footnote{
It is based on formula (22) and Theorem~2 in \cite{LS99}. We leave this for the reader to realise how this query below is related with this formula and why it gives a strict linear ordering (see \cite{LS99}).
}%
. In fact, the rather complicated query \verb+StrictLinOrder_on_TC+ serves as additional witness demonstrating that everything is implemented correctly, and to check whether and where any optimisation of the implementation is required. Note that \verb+StrictLinOrder_on_TC+ invokes \texttt{Can} and without this canonisation the transitive closure
\begin{small}
\begin{verbatim}
    TCPure(BibDB)
\end{verbatim}
\end{small}

\noindent
participating in the query below (according to Appendix~\ref{app:lin-ord-declarations}) would have too many repetitions, and, hence, \texttt{Square} would have even more repetitions so that the recursion in the set query 
\linebreak
\verb+StrictLinOrder_on_TC+ over this \texttt{Square} would take many hours. Now let us run
\begin{small}
\begin{verbatim}
    set query 
    let
    set constant BibDB =
        http://www.csc.liv.ac.uk/~molyneux/t/BibDB-f1.xml#BibDB
    in 
        call SuccessorPairs( 
            call StrictLinOrder_on_TC(BibDB)
        )
    endlet;
\end{verbatim}
\end{small}

\noindent
Note that \verb+SuccessorPairs+ (defined in Appendix~\ref{app:lin-ord-declarations}) makes the result more concise. We see that our database \verb+BibDB+ becomes linear ordered (with corresponding simple set names from the bibliographic database substituted in the place of new set names generated by the query system):
\begin{small}
\begin{verbatim}
    Query is well-formed, well-typed and executable

    Result = {
      'null':{'fst':{},         'snd':"Databases"},
      'null':{'fst':"Databases",'snd':"Jones"},
      'null':{'fst':"Jones",    'snd':"Smith"},
      'null':{'fst':"Smith",    'snd':BibDB},
      'null':{'fst':BibDB,      'snd':p1},
      'null':{'fst':p1,         'snd':b1},
      'null':{'fst':b1,         'snd':b2/p3},
      'null':{'fst':b2/p3,      'snd':p2}
    }

    p2 = {'author':"Smith",'title':"Databases",'refers-to':b2/p3}
    b2/p3 = {'author':"Jones",'title':"Databases"}
    p1 = {'refers-to':p2}
    b1 = {'refers-to':b2/p3,'refers-to':p1}
    BibDB = {'paper':p1,'paper':p2,'paper':b2/p3,'book':b1,
             'book':b2/p3}

    Finished in: 270500 ms (~ 4 minutes and 30 seconds) 
\end{verbatim}
\end{small}

\noindent
The correspondence of set names with those nodes in the graph in Figure~\ref{fig:WDB-bibdb} is explicitly shown in the above result. Thus, the resulting linear ordering on the transitive closure of \verb+BibDB+ is:
\begin{small}
\begin{verbatim}
    {}, "Databases", "Jones", "Smith", BibDB, p1, b1, b2/p3, p2.
\end{verbatim}
\end{small}

\noindent
Here it is important that recursion in \verb+StrictLinOrder_on_TC+ does not use bisimulation for comparison iteration steps (see Chapter~\ref{chap:bisimulation}). This crucially optimises recursion, and in particular the query \verb+StrictLinOrder_on_TC+ which also uses \texttt{Can} in its library declaration. Without the first optimisations this query would take about 30 minutes, and without also using \texttt{Can} even hours. Of course, several minutes for such a small 
database (with \texttt{TC(BibDB)} containing 9 sets) is also quite long, and thus the query system implementation 
needs to be further optimised. But the query is rather complicated  (see Appendix~\ref{app:lin-ord-declarations}), and recursion 
actually uses $81=9^2$ steps of iteration if \texttt{Can} is involved. This means in the average 
about 3.3 seconds per iteration step.


\chapter{Bisimulation}\label{chap:bisimulation}

Before discussing the theoretical and practical issues surrounding  bisimulation, let us recall some relevant details of the hyperset approach to WDB. As previously described in Chapter~\ref{chap:ssd_wdb} WDB is represented as a system of set equations  $\bar{x}=\bar{b}(\bar{x})$ where $\bar{x}$ is a list of set names $x_1,\ldots,x_k$ and $\bar{b}(\bar{x})$ is the corresponding list of bracket expressions (for simplicity, ``flat'' ones). Visually equivalent representation can be done in the form of labelled directed graph, where labelled edges $x_i\stackrel{label}{\longrightarrow} x_j$ correspond to the set memberships $label\!:\!x_j \in x_i$  meaning that the equation for $x_i$ has the form $x_i=\{\ldots,label\!:\!x_j,\ldots\}$. In this case we also call $x_j$ a child of $x_i$. Note that, our usage of the membership symbol ($\in$) as relation between set names or graph nodes is non-traditional but very close to the traditional set theoretic membership relation between abstract (hyper)sets.  Of course this analogy is very important for us and it is indeed highly natural, hence we decided not to introduce a new kind of membership symbol here. For the purposes of our description below labels can be ignored, as inclusion of labels will not affect the nature of our discussion. We will also apply the transitive closure operator $\TC(x)$ to a set name $x$. The essential point is that in this context $\TC(x)$ is understood as a set of set names (or graph nodes) rather than of abstract sets denoted by these names. Again, we do not bother with introducing a new denotation for such $\TC$.

\section{Hyperset equality and the problem of efficiency}

One of the key points of our approach is the interpretation of WDB-graph nodes as set names $x_1,\ldots,x_k$ where different nodes $x_i$ and $x_j$ can, in principle, denote the same (hyper)set,
\linebreak
\mbox{$x_i=x_j$}. This notion of equality between nodes is defined by the bisimulation relation denoted also as $x_i\bis x_j$ (to emphasise that set names can be syntactically different, but denote the same set) which can be computed by the appropriate recursive comparison of child nodes or set names. Thus, in outline, to check bisimulation of two nodes we need to check bisimulation between some children, grandchildren, and so on, of the given nodes, i.e. many nodes could be involved. If the WDB is distributed amongst many WDB files and remote sites, downloading the relevant WDB files might be necessary in this process and will take significant time. There is also the analogous problem with the related transitive closure operator ($\TC$) whose efficient implementation in the distributed case requires additional considerations  not discussed here. So, in practice the equality relation for hypersets seems intractable, although theoretically it takes polynomial time with respect to the size of WDB. Nevertheless, we consider that the hyperset approach to WDB based on bisimulation relation is worth implementing because it suggests a very clear and mathematically well-understood view on semi-structured data and the querying of such data. Thus, the crucial question is whether the problem of bisimulation can be resolved in any reasonable and practical way. Some possible approaches and strategies related with the possible distributed nature of WDB and showing that the situation is manageable in principle are outlined below.

\medskip

Although for the general database perspective we should consider graphs with labels on edges and hypersets with labelled elements, the majority of our considerations in this chapter will be devoted to the pure case without any labels. Extension to the labelled case is quite straightforward and is not explicitly considered, except in Definition~\ref{def:bis}~(b). Of course, our implementation of bisimulation relation considers the labelled case.

\subsection{Bisimulation relation}

Equality between set names (or graph nodes) of any WDB is determined by bisimulation relation defined according to \cite{A88} (see also \cite{M80,P81}).
\begin{definition}\label{def:bis}
(a) \emph{Bisimulation relation} $\bis$ (or $\bis_{\rm WDB}$) on a WDB without labels (the pure case) is the largest one such that for all set names $x,y$ the following implication holds:
\begin{equation}
x\bis y\Rightarrow 
\forall x'\in x\exists y'\in y(x'\bis y')\;\&\;
\forall y'\in y\exists x'\in x(x'\bis y').
\end{equation}
(b) In the general labelled case, it should satisfy the implication 
\begin{align}
  x\bis y\Rightarrow &	\;\forall l:x'\in x\exists m:y'\in y(l=m \wedge x'\bis y')\;\&\; \nonumber \\
	& \;\forall m:y'\in y\exists l:x'\in x(l=m \wedge x'\bis y').
\end{align}
\end{definition}
\noindent
It is well-known that the largest such relation does exist. Indeed, the class $\cal R$ of relations $R$ satisfying any of the above formulas (in place of $\bis$) is evidently closed under taking unions, so the union of all of them is the required largest one $\bis$. In fact, for $\bis$ the implication $\Rightarrow$ above can be replaced by $\iff$. Moreover, the class $\cal R$ evidently contains the identity relation $=$ and is closed under taking compositions $R\circ S$ and inverse relations $R^{-1}$. It follows that the largest such relation $\bis$ is reflexive, transitive and symmetric, that is, an equivalence relation. The bisimulation relation is completely coherent with hyperset theory as it is fully described in the books of Aczel~\cite{A88}, and Barwise and Moss~\cite{BM96} for the pure case, and this fact extends easily to the labelled case. It is by this reason that the bisimulation relation $\bis$ between set names can be considered as equality relation $=$ between corresponding abstract hypersets. So, we will not go into further general theoretical details concerning the bisimulation relation (except for the concept of local bisimulation in Chapter~\ref{chap:local} below), paying the main attention to implementation aspects.

\section{Computing bisimulation over WDB}\label{sec:bisim_algo}

Bisimulation relation is computed in our implementation by recursively deriving bisimulation facts. Both positive ($\bis$) and negative ($\not\bis$) bisimulation facts can be derived with the following rules:
\begin{equation}\label{eq:bis-posi}
x\bis y\derivablefrom 
\forall x'\in x\exists y'\in y(x' \bis y') \;\&\;
\forall y'\in y\exists x'\in x(x' \bis y').
\end{equation}
\begin{equation}\label{eq:bis-negi}
x\not\bis y\derivablefrom 
\exists x'\in x\forall y'\in y(x'\not\bis y')\vee
\exists y'\in y\forall x'\in x(x'\not\bis y').
\end{equation}
\noindent
In principle, using the rule (\ref{eq:bis-posi}) for deriving positive facts is unnecessary. They will be obtained, anyway, at the moment of stabilisation in the derivation process by using only~(\ref{eq:bis-negi}) (see below). Derivation of bisimulation facts using the above rules (\ref{eq:bis-posi} and \ref{eq:bis-negi}) occur after initial facts have been derived. The rules for deriving these initial facts are partial cases of the main rules (\ref{eq:bis-posi} and \ref{eq:bis-negi}):
\begin{equation}\label{eq:bis-pos-initial}
x\bis y\derivablefrom 
(x = \emptyset \;\&\; y = \emptyset)
\end{equation}
\begin{equation}\label{eq:bis-neg-initial}
x\not\bis y\derivablefrom 
(x = \emptyset \;\&\; y \not= \emptyset)\vee
(y = \emptyset \;\&\; x \not= \emptyset)
\end{equation}
\begin{equation}\label{eq:bis-pos-reflexivity}
x\bis x
\end{equation}
\noindent
After the derivation of initial facts, rules \ref{eq:bis-posi} and \ref{eq:bis-negi} can be recursively applied. Since it is known that bisimulation is an equivalence relation, the following transitivity and symmetry rules can be used alongside the above rules:
\begin{equation}\label{eq:bis-pos-trans}
x\bis z\derivablefrom 
x \bis y \;\&\; y \bis z
\end{equation}
\begin{equation}\label{eq:bis-pos-sym}
x\bis y\derivablefrom y\bis x
\end{equation}
\noindent
All these rules should be applied until stabilisation, the stage when no more new $x\bis y$ or $x\not\bis y$ facts can be derived by the above rules. Evidentially, stabilisation is inevitable because there are only finitely many set names in the original WDB, i.e.\ in the corresponding system of set equations. All remaining non-resolved bisimulation questions ($\question{x}{y}$) can now be concluded as resolved positively as $x\bis y$.

\subsection{Implemented algorithm for computing bisimulation over distributed WDB}\label{sec:impl_bisim_algo}

The deeply recursive nature of the bisimulation algorithm seems to suggest that it maybe necessary to effectively compute the transitive closure of the two set names participating in any bisimulation question. For example in the case of the bisimulation question $\question{x}{y}$, stabilisation is sufficient to establish only for the facts between set names in $\TC(x)$ and $\TC(y)$. In general, it may happen that the full transitive closures will be involved. However, in an optimistic approach, derivation rules (described in Section~\ref{sec:bisim_algo}) may be applied to the partial transitive closures, with a ``progressive'' transitive closures computed as necessitated by the derivation rules to facilitate the resolution of a bisimulation question.

\subsection*{Bisimulation algorithm $Bis(x,y)$:}

\begin{enumerate}
\item[]{\bf START with resolving the bisimulation question $\question{x}{y}$.}
\item{
{\bf Create two (initially empty) lists $Q$ and and $Eq$}. $Q$ will consist of bisimulation questions $\question{u}{v}$ or their possible answers, and $Eq$ of (downloaded) set equations.

{\bf Note:}
\emph{During the computation, some bisimulation questions $\question{u}{v}$ from the list $Q$ can be resolved -- replaced by either $u\bis v$ (positive) or $u\not\bis v$ (negative) facts. Thereby $Q$ will contain both non-resolved questions, and positive or negative facts. The process will continue until $Q$ will stabilise}%
\footnote{In the case of using the Oracle, as described later in Chapter~\ref{chap:oracle}, the questions already asked to the Oracle should be appropriately labelled to avoid asking them again.
}%
.
}
\item{
{\bf Initialise populating $Q$ by inserting the bisimulation question $\question{x}{y}$.}
}
\item{\label{step:acquire}
{\bf Acquire set equations} corresponding to those set names involved in all non-resolved bisimulation questions in $Q$ by downloading appropriate WDB files containing these equations. That is, for the question $\question{u}{v}$ in $Q$, download the uniquely defined WDB files (by full set names $u,v$) containing equations $u=\{\ldots\}$ and $v=\{\ldots\}$ (if they have not been downloaded yet).

{\bf Add} these equation into the (originally empty) list of set equations $Eq$ (acquired from the WDB).

{\bf Extend} $Q$ by all new bisimulation questions (more precisely, those not yet included in $Q$ neither as questions nor as positive or negative answers) for all set names participating in $Q$ plus set names in the right hand side of the (downloaded) set equations from~$Eq$.

{\bf Note:} 
{\em Not all the downloaded equations (from the downloaded files) will likely participate in $Eq$ and in the generation of transitive closure $\TC(x)\cup \TC(y)$ for the initial question $\question{x}{y}$, and in this case they may be ignored when generating new questions (to be added in $Q$). But they could probably be useful in future computations and could save time on downloading if some equations to be downloaded as prescribed by the current stage have been already downloaded earlier. Thus, all downloaded equations (in fact, WDB files) should be saved in a cache of WDB (in memory) for possible future use. Therefore, before making the quite expensive step of downloading a WDB file the system should check whether it has already been downloaded. This WDB cache should be initialised when beginning general query execution and used by both the general query evaluation procedure and algorithm described here for evaluating bisimulation (or equality) subqueries $\question{u}{v}$.

Similarly to the cache of WDB, the current versions of $Q$ and $Eq$ should not be discarded from the memory till the end of executing a given query, involving the subquery $\question{x}{y}$ considered in the current algorithm, because some other bisimulation questions might be involved which could be easily answered with already known $Q$ and $Eq$. 
}
}
\item{\bf Iteratively apply derivation rules (\ref{eq:bis-posi}) and (\ref{eq:bis-negi})} (thereby resolving some questions in $Q$) until the initial bisimulation question $\question{x}{y}$ becomes a resolved fact or, otherwise, until exhaustion by using the currently downloaded (probably incomplete) list $Eq$ of set equations.

{\bf Note:} \emph{Some enumerated in $Q$ questions could still remain unresolved.}

\item{{\bf Recursive jump}:
	\begin{enumerate}
	\item{{\bf Is the initial bisimulation question $\question{x}{y}$ now a resolved fact in $Q$?}\\ \\{\bf Yes} -- The original bisimulation question has now been resolved (end of algorithm).\\ \\{\bf No } -- Move to step~\ref{step:downloaded} to continue trying to resolve initial bisimulation question and other non-resolved questions in $Q$.}
	\item{{\label{step:downloaded}\bf Are there set names $u$ participating in non-resolved questions in $Q$ 
	for which set equations $u=\{\ldots\}$ have not yet been downloaded?}\\ \\{\bf Yes} -- Then move to step~\ref{step:acquire} by which further facts may be derived once the relevant set equations have been downloaded. \\ \\{\bf No} -- Then the full transitive closure 
	$\TC(x)\cup \TC(y)$ of the initial bisimulation question 
	$\question{x}{y}$ has been completed, therefore there are no further possibilities to derive/resolve new facts, and stabilisation of the list $Q$ has been achieved.
	Postulate all non-resolved bisimulation questions as {\bf true} facts. In particular, the original bisimulation question $\question{x}{y}$ has now been resolved positively as $x\bis y$ (end of algorithm).
        }
	\end{enumerate}
}
\item[]{{\bf END with the bisimulation question $\question{x}{y}$ resolved positively $x\bis y$ or negatively $x\not\bis y$.}}
\end{enumerate}

\noindent
The essential point of the above algorithm for computing bisimulation is that downloading of WDB files is done in a ``lazy'' way -- only when no derivation step is possible. This strategy is chosen because downloading WDB files is the most expensive process of the general implemented bisimulation algorithm. Therefore only in the worst case downloading all the necessary set equations (generating the full transitive closure of the original bisimulation question) will be necessary. Usually this should save a lot of time and memory.


%
%


\part{Local/global approach to optimise bisimulation and querying}\label{part:local_global}


\chapter{The Oracle}\label{chap:oracle}

\section{Computing bisimulation with the help of the Oracle}\label{sec:oracle_description}

The concept of the Oracle for Web-like databases is somewhat similar to that of an Internet search engine, such as Google, where the
Oracle will attempt to provide bisimulation facts to the $\Delta$-query system when requested and thereby to facilitate the easier computation of set equality. Furthermore, the Oracle should work in background time independently (as well as by requests from the $\Delta$-query system) to derive bisimulation facts.

\medskip

We assume that to the bisimulation question $\question{x}{y}$ the Oracle should give one of three answers \emph{``Yes''}, \emph{``No''} or \emph{``Unknown''}%
\footnote{\label{foot:full-answers}More precisely, to know which question is answered, full answers should be given: ``$x\bis y$'', ``$x\not\bis y$'' or ``$\question{x}{y}$''.
}%
. In the latter case \emph{``Unknown''} should consequently be replaced by the Oracle (after resolving the question itself, probably resulting in some delay) with either \emph{``Yes''} or \emph{``No''}. The answers \emph{``Yes''} or \emph{``No''} must be correct. 
%
%
In fact, asking the Oracle is a way to resolve bisimulation questions, just like applying derivation rules. However, it is likely that the Oracle only provides a partial bisimulation relation (depending on the current state of its work) because of possible updates to WDB forcing the Oracle to redo at least some of its work and the time required to compute bisimulation. Thus, those bisimulation questions answered \emph{``Unknown''} should invoke an initial attempt by the query system to resolve the question locally, hence downloading WDB files with those set equations corresponding to the set names participating in the question(s), etc., as in the algorithm of Section~\ref{sec:impl_bisim_algo} above. If during the process of local computation the Oracle will replace \emph{``Unknown''} by \emph{``Yes''} or \emph{``No''} then this local attempt to resolve the bisimulation question will be automatically halted due to replacing this question by its answer, however, downloaded WDB files may prove to be useful in future derivation steps of other possible bisimulation questions and should not be discarded from the local cache.

\pagebreak

For example, let us consider the Oracle attempting to resolve a bisimulation question posed by the $\Delta$-query system as shown below:

\begin{itemize}
\item[]{{\bf $\Delta$-query system: }$\question{x}{y}$ (is the set name $x$ bisimilar to the set name $y$?).}
\item[]{{\bf Oracle: }\emph{``Unknown''} (based on the current state of knowledge of the Oracle).}
\item[]{The Oracle works towards resolving various bisimulation questions, in particular $\question{x}{y}$.}
\item[]{500ms later...}
\item[]{{\bf Oracle: }\emph{``No''} ($x\not\bis y$ holds).}
\end{itemize}

\section{Imitating the Oracle for testing purposes}\label{sec:trivial_oracle}

As the first attempt, an Oracle which is able to answer bisimulation questions can be simulated with a single file containing a list of bisimulation facts with the states \emph{``Yes''} or \emph{``No''}. Further, those bisimulation questions initially answered as \emph{``Unknown''} 
can be also simulated as delayed answers of \emph{``Yes''} and \emph{``No''} by associating each bisimulation fact with number of milliseconds delay.

\medskip

For the purposes of our preliminary implementation the trivial Oracle (simulated as a file instead of a special Internet server) was implemented as an XML file%
\footnote{
which should not be mixed with XML-WDB files used to represent set equations
}%
. The trivial Oracle (XML file) contains all the necessary information to simulate the behaviour of the Oracle: bisimulation facts corresponding to all possible bisimulation questions. Also, to simulate those questions initially answered \emph{``Unknown''} by the Oracle (such as in the example above) each bisimulation fact has an associated delay time. These XML files are generated by one of the programs belonging to our suite of tools from a given WDB in such a way that all \emph{``Yes''}/\emph{``No''} facts presented there are automatically true, that is the bisimulation relation is computed by this program and presented as an XML file. Furthermore, arbitrary delay times (useful for the purposes of testing) are added manually to those XML files generated by this program.

\medskip

Each bisimulation fact (in the trivial Oracle) is represented as an XML tag with 
\linebreak
\verb+set_name+s, bisimulation {\verb+value+} and \verb+delay+ times as mandatory attributes. For example, let us consider the bisimulation fact $y\not\bis z$ with no delay time represented in the trivial Oracle as,

\begin{small}
\begin{verbatim}
  <facts set_name="y">
    <fact set_name="z" value="no" delay="0" />
  </facts>
\end{verbatim}
\end{small}

\noindent
where bisimulation facts are grouped, inside \verb+<facts>+ and \verb+<fact>+ tags, according to those set name participating in the WDB. The grouping of facts is a feature of the implementation used to generate these XML files. Let us consider the trivial Oracle for the bibliographic WDB (considered in Section~\ref{sec:example_queries}) represented as the XML file:

\begin{tiny}
\begin{verbatim}
    <oracle>

      <facts set_name="http://www.csc.liv.ac.uk/~molyneux/t/BibDB-f1.xml#BibDB">
        <fact delay="0"
            set_name="http://www.csc.liv.ac.uk/~molyneux/t/BibDB-f1.xml#b1" value="no"/>
        <fact delay="0"
            set_name="http://www.csc.liv.ac.uk/~molyneux/t/BibDB-f1.xml#b2" value="no"/>
        <fact delay="0"
            set_name="http://www.csc.liv.ac.uk/~molyneux/t/BibDB-f2.xml#p1" value="no"/>
        <fact delay="0"
            set_name="http://www.csc.liv.ac.uk/~molyneux/t/BibDB-f2.xml#p2" value="no"/>
        <fact delay="0"
            set_name="http://www.csc.liv.ac.uk/~molyneux/t/BibDB-f2.xml#p3" value="no"/>
      </facts>

      <facts set_name="http://www.csc.liv.ac.uk/~molyneux/t/BibDB-f1.xml#b1">
        <fact delay="0"
            set_name="http://www.csc.liv.ac.uk/~molyneux/t/BibDB-f1.xml#b2" value="no"/>
        <fact delay="0"
            set_name="http://www.csc.liv.ac.uk/~molyneux/t/BibDB-f2.xml#p1" value="no"/>
        <fact delay="0"
            set_name="http://www.csc.liv.ac.uk/~molyneux/t/BibDB-f2.xml#p2" value="no"/>
        <fact delay="0"
            set_name="http://www.csc.liv.ac.uk/~molyneux/t/BibDB-f2.xml#p3" value="no"/>
      </facts>

      <facts set_name="http://www.csc.liv.ac.uk/~molyneux/t/BibDB-f1.xml#b2">
        <fact delay="0"
            set_name="http://www.csc.liv.ac.uk/~molyneux/t/BibDB-f2.xml#p1" value="no"/>
        <fact delay="0"
            set_name="http://www.csc.liv.ac.uk/~molyneux/t/BibDB-f2.xml#p2" value="no"/>
        <fact delay="0"
            set_name="http://www.csc.liv.ac.uk/~molyneux/t/BibDB-f2.xml#p3" value="yes"/>
      </facts>

      <facts set_name="http://www.csc.liv.ac.uk/~molyneux/t/BibDB-f2.xml#p1">
        <fact delay="0"
            set_name="http://www.csc.liv.ac.uk/~molyneux/t/BibDB-f2.xml#p2" value="no"/>
        <fact delay="0"
            set_name="http://www.csc.liv.ac.uk/~molyneux/t/BibDB-f2.xml#p3" value="no"/>
      </facts>

      <facts set_name="http://www.csc.liv.ac.uk/~molyneux/t/BibDB-f2.xml#p2">
        <fact delay="0"
            set_name="http://www.csc.liv.ac.uk/~molyneux/t/BibDB-f2.xml#p3" value="no"/>
      </facts>

      <facts set_name="http://www.csc.liv.ac.uk/~molyneux/t/BibDB-f2.xml#p3">
      </facts>

    </oracle>
\end{verbatim}
\end{tiny}

\noindent
Note that only one value \verb+"yes"+ appears above as it is already known concerning our bibliography database that only the set names \verb+b2+ and \verb+p3+ are bisimilar. Information encoded within the such an XML file simulates the responses of the Oracle, i.e. the responses to bisimulation questions. These responses, i.e.\  the desired bisimulation facts (possibly delayed with the immediate temporary answer \emph{``Unknown''}) may assist the regular bisimulation algorithm. To simulate the Oracle, the bisimulation algorithm in  Section~\ref{sec:impl_bisim_algo} should be extended replacing step~\ref{step:acquire} as follows:

\begin{itemize}
\item[\ref{step:acquire}.]{{\bf Acquiring set equations} $u=\{\ldots\}$ and $v=\{\ldots\}$ corresponding to all those unresolved questions $\question{u}{v}$ in $Q$ should now begin with asking the Oracle all these questions (which have not already been asked), and the necessary downloads should follow only in the case where the Oracle answers with \emph{``Unknown''}.

{\bf Note:}
\emph{According to Footnote~\ref{foot:full-answers} (on page~\pageref{foot:full-answers}), the answer \emph{``Unknown''}, in fact, means that the Oracle returns back to the query system the question ``$\question{u}{v}$'', and similarly for the answers 
\emph{``Yes''} and \emph{``No''} in which case the full answers ``$x\bis y$'' and ``$x\not\bis y$'', 
respectively, should be returned. Otherwise, because of delays, the system will not know how 
to treat \emph{``Yes''}, \emph{``No''} and \emph{``Unknown''}.}

}
\end{itemize}

\noindent
Evidentially, whilst resolving bisimulation questions (the modified version of) Step~2 will pose many bisimulation question to the Oracle, which will be answered either ``Yes'' ($u\bis v$) or ``No'' ($u\not\bis v$) possibly with delays. In fact, the behaviour of the modified bisimulation algorithm can be characterised as follows, depending on the Oracle's responses:

\begin{itemize}
\item{
{\bf Bisimulation questions ($\question{u}{v}$) to the Oracle directly answered \emph{``Yes''} ($u\bis v$) or \emph{``No''} ($u\not\bis v$):} 
In this case, the answer from the Oracle should immediately replace the unresolved question in $Q$, and the modified bisimulation algorithm will resume its work resolving other non-resolved bisimulation questions from~$Q$.
}
\item{
{\bf Bisimulation questions ($\question{u}{v}$) to the Oracle initially answered \emph{``Unknown''} ($\question{u}{v}$):} In this case, the modified bisimulation algorithm will, in fact, resume its work resolving $\question{u}{v}$ and other non-resolved bisimulation questions from~$Q$. Thus, the question will either be resolved locally or the Oracle will replace its answer \emph{``Unknown''} ($\question{u}{v}$) by either \emph{``Yes''} ($u\bis v$) or \emph{``No''} ($u\not\bis v$) possibly with some delay.

Note that, if the Oracle answers the question positively or negatively before being resolved locally then this answer should replace the question in $Q$ and the modified bisimulation algorithm should continue its work (taking into account the newly resolved question -- it does not matter in which way the question is resolved, by the Oracle or by the query system)%
\footnote{
A question answered \emph{``Unknown''} does not require asking the Oracle again. In general, Oracle (as a special Internet server) should remember all questions and reply to the appropriate client accordingly when the answer will be ready.
}%
.
}
\end{itemize}

\noindent
Note that, step~2 in the present modified form plays a crucial role in performance: resolution of bisimulation questions by the Oracle will save costly downloading of WDB files.

\section{Empirical testing of the trivial Oracle}\label{sec:testing_simulated_oracle}

In principle, with the help of the Oracle those $\Delta$-queries which involve set equality (bisimulation) should be executed quicker. The aim of the following empirical testing is to measure the improvement in query performance with the help of the Oracle, in addition to demonstrating the effects of delayed answers to bisimulation questions (those initially answered \emph{``Unknown''}) by the Oracle.%
\footnote{Even more optimal would be to postpone local resolution of bisimulation questions in favour of some other independent subqueries of the given query with the hope that the Oracle will give a definite answer before starting local resolution. There are many ways to optimise our implementation, but we can consider only a limited range of such possibilities.}

\medskip

The distributed bibliographic WDB considered in Section~\ref{sec:example_queries} (see Figure~\ref{fig:WDB-bibdb}) is fragmented into two XML-WDB files, thus making computation of bisimulation more dependent on the time taken to download these files. The following example query (already considered in Section~\ref{sec:valid-executable-query}) involves set equality to determine which publications belonging to \texttt{BibDB} refer to the publication (possibly bisimilar to) \texttt{b2}. The requirement to compute bisimulation across the distributed bibliographic WDB makes this simple example particularly suitable for empirical testing of the Oracle:

\begin{small}
\begin{verbatim}
    set query 
      let set constant BibDB be
            http://www.csc.liv.ac.uk/~molyneux/t/BibDB-f1.xml#BibDB,
          set constant b2 be 
            http://www.csc.liv.ac.uk/~molyneux/t/BibDB-f2.xml#b2
      in collect { pub-type:pub 
          where pub-type:pub in BibDB
          and exists 'refers-to':ref in pub . ref=b2
         }
      endlet;
\end{verbatim}
\end{small}

\noindent
The execution time of this example query under various experimental conditions can be seen in the Table~\ref{results_table_trivial_oracle}. 
The results suggest a marked improvement in performance with help of the Oracle, and only a slight improvement in performance when the Oracle returns an answer after delay 50ms or 75ms. However, when the Oracle provided a greatly delayed answer ($\ge$~100ms) query execution occurs with no real help by the Oracle, and bisimulation is computed locally without any real help from the Oracle. Thus, under this circumstance, query execution time increases, and the optimal approach appears to be query execution without invoking the Oracle. This result may be explained by the numerous (and seemingly futile) bisimulation questions posed to the Oracle (all of which are answered \emph{``Unknown''} and never improved) which provide no real help.

\medskip

In summary, these results were based on experiments with the trivial Oracle (simulated as an XML file instead of an Internet server). Additionally, the example WDB is too small and, crucially, only distributed into two fragments. In principle, invoking the help of the Oracle should improve query performance considerably when the WDB is distributed into a large number of fragments.

\begin{table}[!h]
\begin{center}
\begin{tabular}{ | l | l | }
\hline
Strategy&Query execution time [ms]\\
\hline
Bisimulation algorithm without invoking the Oracle&588\\
with help of the Oracle (no delay time per question)&390\\
with help of the Oracle (50ms delay time per question)&500\\
with help of the Oracle (75ms delay time per question)&500\\
with help of the Oracle (100ms delay time per question)&608\\
with help of the Oracle (125ms delay time per question)&608\\
\hline
\end{tabular}
\end{center}
\caption{Experimental results showing query execution time [ms] corresponding to each strategy for computing bisimulation.}
\label{results_table_trivial_oracle}
\end{table}

\noindent
In a more realistic situation, the Oracle should be implemented as an Internet service (called the bisimulation engine) for large distributed WDB, working in background time to derive all possible bisimulation facts on the current state of WDB. The goal of the bisimulation engine consists in answering bisimulation questions $\question{x}{y}$ from the $\Delta$-query system (possibly with a delay%
\footnote{
In principle, the Oracle, when asked the question $\question{x}{y}$, could change its regular behaviour, and try to resolve such questions (with appropriate strategy of priority) from one or more querying clients.
}%
). The Oracle should be based on the bisimulation algorithm described in Section~\ref{sec:impl_bisim_algo} and, additionally, on the idea of local/global bisimulation considered in Chapter~\ref{chap:local}. We will consider implementation (still rather an imitation) of the Oracle in Chapter~\ref{chap:oracle_local} and some further advanced experiments.


\chapter{Local/global bisimulation}\label{chap:local}

Let a proper set%
\footnote{$\LLL\ne\emptyset$ and $\LLL\ne \SNames$}
$\LLL\subseteq \SNames$ of ``local'' vertices (set names) in a graph WDB (a system of set equations) be given, where $\SNames$ is the set of all WDB vertices (set names). Let us also denote by $\LLL'\supseteq\LLL$ the set of all set names participating in the set equations for each set name in $\LLL$ both from left and right-hand sides. Considering the graph as a WDB distributed among many \emph{sites}, $\LLL$ plays the role of (local) set names defined by set equations in some (local) WDB files of one of these sites. Then $\LLL'\setminus \LLL$ consists of non-local set names which, however, participate in the local WDB files, have defining equations in other (possibly remote) sites of the given WDB. Non-local (full) set names can be recognised by their URLs as different from the URL of the given site. Set names (or vertices) from $\LLL'$ can be reasonably called ``almost local''.

\medskip	

We will consider \emph{derivation rules} of the form $xRy\derivablefrom\ldots R\ldots$ 
for three relations over $\SNames$:
\[
{\bis^{\LLL}_{-}}\subseteq{\bis}\subseteq{\bis^{\LLL}_{+}}
\quad\textrm{or, rather, their negations}\quad
{\not\bis^{\LLL}_{+}}\subseteq{\not\bis}\subseteq{\not\bis^{\LLL}_{-}}
\]

\noindent
defined on the whole WDB graph (however, we will be mainly interested in the behaviour of $\bis^{\LLL}_-$ and $\bis^{\LLL}_+$ on $L$). We will usually omit the superscript $\LLL$ when it is clear from the context. In particular, this chapter deals mainly with one $\LLL$, so no ambiguity can arise.

\section{Defining the ordinary bisimulation relation $\bis$}

Recall the derivation rule defining $\not\bis$:
\begin{equation}\label{eq:bis-neg}
x\not\bis y\derivablefrom 
\exists x'\in x\forall y'\in y(x'\not\bis y')\vee
\exists y'\in y\forall x'\in x(x'\not\bis y').
\end{equation}

\noindent
If $u\not\bis v$ is underivable for some vertices/set names $u,v$ then we assume $u\bis v$ to be true (indistinguishable sets are considered equal), and similarly in other cases below. Equivalently, $\not\bis$ is the least relation satisfying (\ref{eq:bis-neg}), and its positive version $\bis$ is the largest relation satisfying
\begin{equation}\label{eq:bis}
x\bis y\Rightarrow 
\forall x'\in x\exists y'\in y(x'\bis y')\;\&\;
\forall y'\in y\exists x'\in x(x'\bis y').
\end{equation}

\noindent
The relation $\bis$ is called \emph{bisimulation} relation which is also known to be an equivalence relation on the whole graph. Below are defined its upper and lower (relativised to $\LLL$) approximations $\bis_{+}$ and $\bis_{-}$. 

\section{Defining the local upper approximation $\bis^\LLL_{+}$ of $\bis$}

\noindent
Let us define the relation ${\not\bis_{+}}\subseteq \SNames^2$ by derivation rule
\begin{equation}\label{eq:bis-neg-+}
x\not\bis_{+}y\derivablefrom x,y\in \LLL\;\&\;[\exists x'\in x\forall y'\in y
         (x'\not\bis_{+} y')\vee\ldots].
\end{equation}

\noindent
Here and below ``$\ldots$'' represents the evident symmetrical disjunct (or conjunct). Thus the premise (i.e.\ the right-hand side) of (\ref{eq:bis-neg-+}) is a \emph{restriction} of that of (\ref{eq:bis-neg}). It follows by induction on the length of derivation of the $\not\bis_{+}$-facts that,
\begin{align}
{\not\bis_{+}}\subseteq{\not\bis}, \quad {\bis}\subseteq{\bis_{+}} \label{eq:+approx} \\
x\not\bis_{+}y\Rightarrow x,y\in \LLL \label{eq:loc} \\
x\not\in \LLL\vee y\not\in \LLL\Rightarrow  x\bis_{+}y. \label{eq:nonloc-bis}
\end{align}


\noindent
As $\LLL\ne \SNames$, the set of all vertices,
it follows from (\ref{eq:nonloc-bis}) that $\bis_{+}$ can be an equivalence relation 
on the whole graph \emph{only} if it is trivial, 
making all vertices equivalent. But we will show below that 
it is an equivalence relation locally, that is on $\LLL$. 

\medskip

Let us also consider another, ``more local'' version of the rule~(\ref{eq:bis-neg-+})
\begin{equation}\label{eq:bis-neg-+LL}
x\not\bis_{+}y\derivablefrom x,y\in \LLL\;\&\;[\exists x'\in x\forall y'\in y
         (x',y'\in\LLL\;\&\;x'\not\bis_{+} y')\vee\ldots].
\end{equation}

\noindent
It defines the same relation $\not\bis_{+}$ because in both cases (\ref{eq:loc}) holds implying that the right-hand side of (\ref{eq:bis-neg-+LL}) is equivalent to the right-hand side of (\ref{eq:bis-neg-+}). The advantage of (\ref{eq:bis-neg-+}) is its formal simplicity whereas that of (\ref{eq:bis-neg-+LL}) is its ``local'' computational meaning. From the point of view of distributed WDB with $\LLL$  one of its local sets of vertices/set names (corresponding to one of the sites of the distributed WDB), we can derive $x\not\bis_{+}y$ for local $x,y$ via (\ref{eq:bis-neg-+LL}) by looking at the content of local WDB files only. Indeed, participating URLs (full set names) $x'\in x$ and $y'\in y$,  although likely non-local names ($\in \LLL'\setminus\LLL$), occur in the locally stored WDB files with local URLs $x$ and $y\in L$. However, despite the possibility that $x'$ and $y'$ can be in general non-local, we will need to use in (\ref{eq:bis-neg-+LL}) the facts of the kind $x'\not\bis_{+}y'$ derived on the previous steps for local $x',y'\in\LLL$ only. Therefore,

\begin{note}[Local computability of $x\not\bis_{+}y$]\label{note:upper-approx}
For deriving the facts $x\not\bis_{+}y$ for $x,y\in\LLL$ by means of the rule (\ref{eq:bis-neg-+}) or (\ref{eq:bis-neg-+LL}) we will need to use the previously derived facts $x'\not\bis_{+}y'$ for set names $x',y'$ from $\LLL$ only, and additionally we will need to use set names from a wider set $\LLL'$ (available, in fact, also locally)%
\footnote{
This is the case when $y=\emptyset$ but there exists according to (\ref{eq:bis-neg-+LL}) an $x'$ in $x$ which can be possibly in $\LLL'\setminus\LLL$ (or similarly for $x=\emptyset$). When $y=\emptyset$ then, of course, there are no suitable witnesses $y'\in y$ for which $x'\not\bis_{+}y'$ hold. Therefore, only the existence of some $x'$ in $x$ plays a role here.
}.
In this sense, the derivation of all facts $x\not\bis_{+}y$ for $x,y\in\LLL$ can be done locally and does not require downloading of any external WDB files. (In~particular, facts of the form $x\not\bis_{+}y$ or $x\bis_{+}y$ for set names $x$ or $y$ in $\LLL'\setminus\LLL$ present no interest in such derivations.)
\end{note}

\noindent
The upper approximation $\bis_{+}$ (on the whole WDB graph) can be equivalently characterised as the largest relation satisfying 
any of the following (equivalent) implications for all graph vertices $x,y$:
\begin{align}
 x\bis_{+}y\Rightarrow 
              x\not\in \LLL\vee 
              y\not\in \LLL\vee 
              [\forall x'\in x\exists y'\in y
         (x'\bis_{+} y')\;\&\;\ldots] \nonumber \\
x\bis_{+}y\;\&\;x,y\in \LLL 
        \Rightarrow 
              [\forall x'\in x\exists y'\in y
         (x'\bis_{+} y')\;\&\;\ldots] \label{eq:bis+LL}
\end{align}


\noindent
The set of relations $R\subseteq \SNames^2$ satisfying (\ref{eq:bis+LL}), in place of $\bis_{+}$, evidently: {\bf(i)} contains the identity relation $=$ and is closed under {\bf(ii)} unions (thus the largest $\bis_+$ does exist), and {\bf(iii)} taking inverse.

\medskip

Evidently, any ordinary (global) bisimulation relation $R\subseteq \SNames^2$ (that is, a relation satisfying (\ref{eq:bis})) satisfies (\ref{eq:bis+LL}) as well%
\footnote{This imples (\ref{eq:+approx}) again because $\bis_+$ is the largest relation satisfying~(\ref{eq:bis+LL}).
}%
. For any $R\subseteq \LLL^2$ the converse also holds: if $R$ satisfies (\ref{eq:bis+LL}) then it is actually a global bisimulation relation (and $R\subseteq{\bis}$). It is easy to check that {\bf(iv)} relations $R\subseteq \LLL^2$ satisfying (\ref{eq:bis+LL}) are closed under compositions.

\medskip

It follows from {\bf(i)} and {\bf(iii)} that $\bis_{+}$ is reflexive and symmetric. Over $\LLL$, the relation $\bis_{+}$ (that is the restriction $\bis_{+}\upharpoonright \LLL$) is also transitive due to {\bf(iv)}. Therefore, $\bis_{+}$ is an \emph{equivalence relation}. (In general, as we noticed above, $\bis_{+}$ cannot be equivalence relation on the whole graph, due to (\ref{eq:nonloc-bis}).) Moreover, any $x\not\in \LLL$ is $\bis_{+}$ to all vertices (including itself).

\section{Defining the local lower approximation $\bis^\LLL_{-}$ of $\bis$}

Consider the derivation rule for the relation ${\not\bis_{-}}\subseteq \SNames^2$:
\begin{eqnarray*}
x\not\bis_{-}y & \derivablefrom & (x,y\not\in \LLL\;\&\;x\ne y)\vee
                             (x\in \LLL\;\&\;y\not\in \LLL)\vee
                             (y\in \LLL\;\&\;x\not\in \LLL)\vee {}
\\
 & & \qquad\qquad
               [\exists x'\in x\forall y'\in y
         (x'\not\bis_{-} y')\vee\ldots].
\end{eqnarray*}

\noindent
The following is an equivalent simplified rule:
\begin{eqnarray}
\nonumber
x\not\bis_{-}y & \derivablefrom & ((x\not\in \LLL\vee y\not\in \LLL)
                             \;\&\;x\ne y)\vee
\\ \label{eq:bis-neg--}
 & & \qquad\qquad
              [\exists x'\in x\forall y'\in y
         (x'\not\bis_{-} y')\vee\ldots]
\end{eqnarray}

\noindent
which can also be equivalently replaced by two rules:
\begin{eqnarray}
\label{eq:a-priori-knowledge}
x\not\bis_{-}y & \derivablefrom & (x\not\in \LLL\vee y\not\in \LLL)
                             \;\&\;x\ne y\textrm{ -- ``a priori knowledge''}, 
\\ \label{eq:bis-neg---}  
x\not\bis_{-}y & \derivablefrom & \exists x'\in x\forall y'\in y
         (x'\not\bis_{-} y')\vee\ldots\;.
\end{eqnarray}

\noindent
Thus, in contrast to (\ref{eq:bis-neg-+}), this is a \emph{relaxation}, or, an \emph{extension} of the rule (\ref{eq:bis-neg}) for $\not\bis$. It follows that
\begin{align}\nonumber
&{\not\bis}\subseteq{\not\bis_{-}}\ 
({\bis_{-}}\subseteq{\bis}),
\\
\nonumber
&x\bis_{-}x
\textrm{ for all }x\in \SNames \textrm{ --- reflexivity}.
\end{align}

\noindent
The former is trivial, and the latter means that $x\not\bis_{-}x$ is not derivable. (Indeed, $x\not\bis_{-}x$ can be derivable only if $x'\not\bis_{-}x'$ is derivable for some $x'\in x$ on an earlier stage; thus, there cannot exists a first such derivable fact.) It is also evident that
\begin{align}\nonumber
&\textrm{any }x\not\in \LLL\textrm{ is }\not\bis_{-}
\textrm{ to all vertices different from }x,
\\
\nonumber
&x\bis_{-}y\;\&\;x\ne y\Rightarrow (x,y\in \LLL).
\end{align}

\noindent
The latter means that $\bis_{-}$ (which is an equivalence relation on $\SNames$ and hence on $\LLL$ as it is shown below) is non-trivial only on the local set names. Again, like for $\not\bis_{+}$, we can conclude from the above considerations that,

\begin{note}[Local computability of $x\not\bis_{-}y$]\label{note:lower-approx}
We can compute the restriction of $\not\bis_{-}$ on $\LLL$ locally: to derive $x\not\bis_{-}y$ for $x,y\in\LLL$ 
with $x\ne y$ {\em (taking into account reflexivity of $\bis_{-}$)}
by (\ref{eq:bis-neg--}) we need to use only $x',y'\in\LLL'$ 
(by $x'\in x$ and $y'\in y$) and already 
derived facts $x'\not\bis_{-}y'$ for $x',y'\in\LLL, x\ne y$, 
as well as the facts $x'\not\bis_{-}y'$ for $x'$ or $y'\in\LLL'\setminus\LLL$, $x'\ne y'$ 
following from the ``a priori knowledge'' (\ref{eq:a-priori-knowledge}). 
\end{note}

\noindent
The lower approximation $\bis_{-}$ can be equivalently characterised 
as the largest relation satisfying 
\[
x\bis_{-}y\Rightarrow (x,y\in \LLL
                             \vee x=y)
                  \;\&\;
              (\forall x'\in x\exists y'\in y
         (x'\bis_{-} y')\;\&\;\ldots).
\]
Evidently, $=$ (substituted for $\bis_{-}$) satisfies this implication. 
Relations $R$ satisfying this implication are also closed 
under unions and taking inverse and compositions. 
It follows that 
$\bis_{-}$ is reflexive, symmetric
and transitive, and therefore an 
\emph{equivalence relation over the whole WDB graph}, and therefore \emph{on its local part} $L$.

\medskip

Finally, we summarise that both upper and lower approximations $\bis^\LLL_{+}$ and $\bis^\LLL_{-}$ to $\bis$ restricted to $\LLL$ are computable ``locally''. Each of them is defined in a trivial way outside of $\LLL$, and the computation requires only knowledge at most on the $\LLL'$-part of the graph. In fact, only edges from $\LLL$ to $\LLL'$ are needed, everything being available locally. 

\section{Using local approximations to aid computation of the global bisimulation}\label{sec:local_aid_global}

The point of previous considerations of this chapter was that, given any set $\LLL$ of ``local'' set names (or WDB graph vertices), we defined two (local to $\LLL$) approximations $\bis^\LLL_{+}$ and $\bis^\LLL_{-}$ to the global bisimulation relation $\bis$. Now, assume that the set \SNames\ of all set names (nodes) of a WDB is disjointly divided into a family of local sets $\LLL_i$, for each ``local'' site $i\in I$ (so that \SNames\ is the disjoint union $\SNames=\bigcup_{i\in I}\LLL_i$). Then we have many local approximations $\bis^{\LLL_i}_{+}$ and $\bis^{\LLL_i}_{-}$ to the global bisimulation relation $\bis$. As we discussed above, these relations can be easily computed locally by each site $i$ using the derivation algorithms described in Notes~\ref{note:upper-approx} and \ref{note:lower-approx}, respectively.

\bigskip

\noindent
Now the problem is how to compute the global bisimulation relation $\bis$ with the help of many its local approximations $\bis^{\LLL_i}_{+}$ and $\bis^{\LLL_i}_{-}$ in all sites $i$.

\subsection{Granularity of sites}\label{sec:sites-files}

However, for simplicity of implementation and testing the above idea (and also because it is more problematic to create many sites with their servers) we will redefine the scope of~$i$ to a smaller granularity. Instead of taking $i$ to be a site, consisting of many WDB files, we will consider that each $i$ itself is a name of a single WDB file $\textit{file}_i$. More precisely, $i$ is considered as the URL of any such a file. This will not change the main idea of implementation of the Oracle on the basis of using local information for each $i$. That is, we reconsider our understanding of the term local -- from being \emph{local to a site} to \emph{local to a file}%
\footnote{
Moreover, this idea of locality to files (described below in detail) belonging to each such a site $i$ is useful for computing $i$-th site's local upper and lower approximations of bisimulation as an intermediate step. Then these $i$-th approximations could be used in implementation of the global Oracle. That is, the idea of locality can be fruitfully used on various levels of granularity to optimise performance of the bisimulation engine (the Oracle).
}
-- as shown in Figure~\ref{fig:site_vs_file}.
%
%
Then $\LLL_i$ is just the set of all (full versions of) set names defined in file $i$ (left-hand sides of all set equations in this file). Evidently, so defined sets $\LLL_i$ are disjoint and cover the class $\SNames$ of all (full) set names from the WDB considered. Recall that $\bis^{\LLL_i}_{+}$ and $\bis^{\LLL_i}_{-}$ are formally defined on the whole WDB (not only on $\LLL_i$). Their restrictions to $\LLL_i$ are also equivalence relations (on~$\LLL_i$) denoted, for brevity and when it is clear from the context, also as $\bis^{\LLL_i}_{+}$ and $\bis^{\LLL_i}_{-}$.

\begin{figure}[!ht]
  \center
  \subfloat[Local to files]{
    \includegraphics[width=0.43\textwidth]{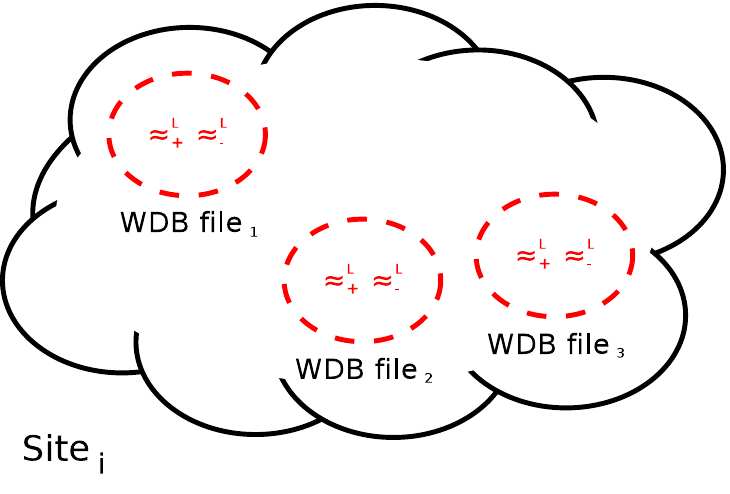}
    \label{subfig:file}
  }\qquad
  \subfloat[Local to sites]{
    \includegraphics[width=0.43\textwidth]{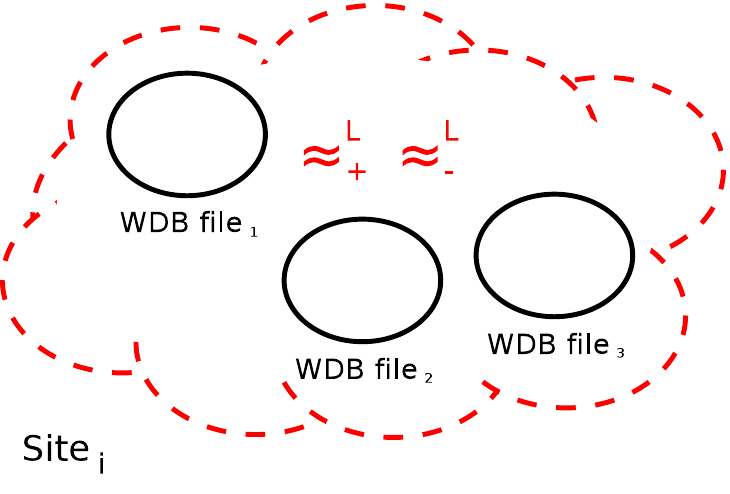}
    \label{subfig:site}
  }
\caption{Summary of a distributed WDB showing the difference between interpretation of local as: local to a file, or local to a site.}
\label{fig:site_vs_file}
\end{figure}

\noindent
The relations $\bis^{\LLL_i}_{+}$ and $\bis^{\LLL_i}_{-}$ should be automatically computed, saved as file and maintained as the current local approximations for each WDB file $i$. In principle a suitable tool is necessary for editting (and maintaining) WDB, which would save a WDB file $i$ and thereby generate the approximation relations $\bis^{\LLL_i}_{+}$ and $\bis^{\LLL_i}_{-}$ (file) automatically.

\subsection{Local approximations giving rise to global bisimulation facts}
\label{sec:local-approx}

We know that these approximations satisfy,
\[{\bis^{\LLL_i}_{-}}\subseteq{\bis}\subseteq{\bis^{\LLL_i}_{+}},
\]

\noindent
or, equivalently,
\[{\not\bis^{\LLL_i}_{+}}\subseteq{\not\bis}\subseteq{\not\bis^{\LLL_i}_{-}}.
\]

\noindent
It evidently follows that,
\begin{itemize}
\item{\em
each positive local fact of the form $x\bis^{\LLL_i}_{-}y$ 
is a positive fact about $\bis$, i.e.\ gives rise to the fact $x\bis y$, and}
\item {\em
each negative local fact of the form $x\not\bis^{\LLL_i}_{+}y$ 
is a negative fact about $\bis$, i.e.\ gives rise to the fact $x\not\bis y$.}
\end{itemize}

\noindent
Let $\bis^{\LLL_i}$ (without subscripts $+$ or $-$) denote the set of positive and negative facts for set names in $\LLL_i$ on the global bisimulation relation $\bis$ obtained by these two clauses. This set of facts $\bis^{\LLL_i}$ is called the \emph{local simple approximation set} to $\bis$ for the file (or site)~$i$. Let the \emph{local Oracle} $LO_i$ just answer \emph{``Yes''} (``$x\bis y$''), \emph{``No''} (``$x\not\bis y$'') or \emph{``Unknown''} to questions $\question{x}{y}$ for $x,y\in\LLL_i$ according to~$\bis^{\LLL_i}$.

\medskip

In the case of $i$ considered as a site (rather than a file) then $LO_i$ can have delays when answering \emph{``Yes''} (``$x\bis y$'') or \emph{``No''} (``$x\not\bis y$'') because $LO_i$ should rather compute $\bis^{\LLL_i}$ itself and find out in $\bis^{\LLL_i}$ answers to the questions asked which takes time. But, if $i$ is understood just as a file saved together with all the necessary information on local approximations at the time of its creation then $LO_i$ can submit the required answer and, additionally, all the other facts it knows at once (to save time on possible future communications).

\medskip

Therefore, a centralised Internet server (for the given distributed WDB) working as the (global) Oracle or \emph{Bisimulation Engine}, which derives positive and negative ($\bis$ and $\not\bis$) global bisimulation facts can do this by the algorithm of Section~\ref{sec:impl_bisim_algo}, in addition to asking (when required) various local Oracles $LO_i$ concerning $\bis^{\LLL_i}$. That is, the algorithm from Section~\ref{sec:impl_bisim_algo} extended to exploit local simple approximations $\bis^{\LLL_i}$ should, in the case of the question $\question{x}{y}$ in $Q$ with $x,y\in L_i$ from the same site/WDB file $i$%
\footnote{$x,y\in\LLL_i$ can be trivially checked by comparing the full set names $x,y$ with the URL $i$
}%
, additionally ask the oracle $LO_i$ whether it already knows the answer (as described in the above two items). If the answer is known, the algorithm should just use it. Otherwise (if $LO_i$ does not know the answer or $x,y$ do not belong to one $L_i$ -- that is, they are ``remote'' one from another), the global Oracle should work as described in Section~\ref{sec:impl_bisim_algo} by downloading set equations, making derivation steps, etc. Thus, local approximations serve as auxiliary local Oracles $LO_i$ helping the global Oracle.

\subsection{Practical algorithm for computation of local approximations}\label{sec:prac_algo_localappox}

The derivations rules for computing local approximations (described above by rules \ref{eq:bis-neg-+}, \ref{eq:bis-neg--} together with Notes~\ref{note:upper-approx},~\ref{note:lower-approx}) can be implemented in a very similar way to the practical algorithm for computing the global bisimulation described in Section~\ref{sec:bisim_algo}. Given a WDB file $i$ as the input, the algorithm will generate \emph{approximation files} $i^A$ and $i^{SA}$ containing local approximations $\bis^{\LLL_i}_{+}$, $\bis^{\LLL_i}_{-}$ and, respectively, local simple approximation set $\bis^{\LLL_i}$ (all three approximations restricted to~$\LLL_i$). The derivation rules (\ref{eq:bis-neg-+}, \ref{eq:bis-neg--}) were formulated to compute the relations $\bis^{\LLL_i}_{+}$ and $\bis^{\LLL_i}_{-}$ over all set names (both local and non-local). According to Notes~\ref{note:upper-approx},~\ref{note:lower-approx} on local computability of local approximations the computation of restricted relations can be also restricted to local set names in $\LLL_i$ (or to slightly wider set $\LLL'_i$). Additionally, the two clauses in Section~\ref{sec:local-approx} should be used.

\medskip

Unlike the practical algorithm for computing global bisimulations, the computation of local approximations $\bis^{\LLL_i}_+,\bis^{\LLL_i}_-$, and $\bis^{\LLL_i}$ (creation of approximation files  $i^{A}$ and $i^{SA}$) should be done after creating (and saving) WDB files $i$, therefore this operation does not require much attention towards optimisation.

\pagebreak

Local simple approximation files, $i^{SA}$, are represented as XML files (quite similar to those of the imitated Oracle; see Section~\ref{sec:trivial_oracle}) containing global bisimulation facts derived locally on the fragment $i$ ($\approx^{L_i}$). Each approximation fact is represented as an (XML) \verb+fact+ tag with boolean local approximation value and set name as mandatory attributes \verb+value+ and \verb+set_name+. These approximation facts are grouped (inside \verb+facts+ tag) corresponding to all local set names in $L_i$%
\footnote{
This is quite similar to the previous implemented tool to generate the (trivial) Oracle XML files.
}%
.

\medskip

For example, let us consider the simple approximation file $i^{SA}$, corresponding to the local simple approximation set $\approx^{L_i}$, for one particular fragment of the bibliographic WDB (see Section~\ref{sec:example_queries}) \url{http://www.csc.liv.ac.uk/~molyneux/t/BibDB-f1.xml}:

\medskip

\begin{tiny}
\begin{verbatim}
    <simple-approximation>

      <facts set_name="http://www.csc.liv.ac.uk/~molyneux/t/BibDB-f1.xml#BibDB">
        <fact set_name="http://www.csc.liv.ac.uk/~molyneux/t/BibDB-f1.xml#b1" value="no"/>
        <fact set_name="http://www.csc.liv.ac.uk/~molyneux/t/BibDB-f1.xml#b2" value="no"/>
      </facts>

      <facts set_name="http://www.csc.liv.ac.uk/~molyneux/t/BibDB-f1.xml#b1">
        <fact set_name="http://www.csc.liv.ac.uk/~molyneux/t/BibDB-f1.xml#b2" value="no"/>
      </facts>

      <facts set_name="http://www.csc.liv.ac.uk/~molyneux/t/BibDB-f1.xml#b2">
      </facts>

    </simple-approximation>
\end{verbatim}
\end{tiny}

\noindent
Note that all ``no'' values above correspond to negative bisimulation facts ($\not\bis$) resulting from the computation of the local simple approximation set $\bis^{L_{i}}$, where $i$ is the WDB file mentioned above. Simple approximation files are predictably named based on the name of the corresponding WDB file $i$ by concatenating the string ``\texttt{approximation}'' to the end of the WDB file name, for example the WDB file name ``\texttt{BibDB-f1.xml}'' will have corresponding simple approximation file with the name ``\texttt{BibDB-f1.approximation.xml}''.


\chapter{The Oracle based on the idea of local/global bisimulation}\label{chap:oracle_local}

\section{Description of the bisimulation engine (implementation of a more realistic Oracle)}

Empirical evidence from the implementation of the imitated Oracle in Section~\ref{sec:testing_simulated_oracle} concluded that a centralised service providing answers to bisimulation question would increase query performance (for those queries exploiting set equality) -- this service could be named \emph{bisimulation engine}. The goal of such bisimulation engine would be:

\begin{itemize}
\item{
{\bf Answer bisimulation queries} -- Answers bisimulation questions communicating via standardised protocol (as discussed in Section~\ref{sec:oracle_description}).
}
\item{
{\bf Compute bisimulation} -- Derive bisimulation facts in background time, and strategically prioritise bisimulation questions posed by the $\Delta$-query system by temporary changing the fashion of the background time work in favour of resolving these questions%
\footnote{
Although due to limitations of time, the current implementation is more basic and does not adopt this strategy of prioritising. (See more in Section~\ref{sec:strategies}.
}%
.
}
\item{{\bf Exploit local approximations} -- Exploit those local approximations corresponding to WDB files to assist in the computation of bisimulation.}
\item{{\bf Maintain cache of set equations} -- The Oracle (just like the $\Delta$-query system) should maintain a cache of the downloaded set equations in the previous steps. These set equations may later prove useful in deriving new bisimulation facts with saving time on downloading of already known equations.
}
\end{itemize}

\subsection{Strategies}\label{sec:strategies}

In principle, the bisimulation engine should give strategic prioritisation to resolving those bisimulation questions posed by clients -- favouring resolution of these bisimulation questions over background tasks (resolving all other bisimulation questions). Moreover, it is reasonable to make the query system adopt a ``lazy'' strategy while working on a query $q$. This strategy consists of sending bisimulation subqueries of $q$ to the Oracle but not attempting to resolve them in the case of the Oracle's answer ``Unknown'' (according to the standard algorithm). Instead of such attempts, the query system could try to resolve other subqueries of the given query $q$ until the resolution of the bisimulation question sent to the Oracle is absolutely necessary. The hope is that before this moment the bisimulation engine will have already given a definite answer.

\medskip

However these useful features have not yet been implemented. In the current version, we have only a simplified imitation of bisimulation engine 
which resolves all possible bisimulation questions for the given WDB in some predefined standard order without any prioritisation and answers 
these questions in a definite way when it has derived the required information. Thus the Oracle, while doing its main job in background time, should only remember all the pairs (client, question) for questions asked by clients and send the definite answer to the corresponding client when it is ready.

\subsection{Exploiting local approximations to aid in the computation of bisimulation}

For implementation of the Oracle we use again the algorithm for computing the bisimulation relation, as described in Section~\ref{sec:impl_bisim_algo}. But, this algorithm will be extended to exploit local approximations by adding an additional step after acquiring set equations (step~\ref{step:acquire}). This additional step (step~\ref{step:acquire}') is detailed below:

\begin{itemize}
\item[\ref{step:acquire}'.]{
{\bf Acquire local approximations} by (i) downloading the local approximation set $\bis^{\LLL_i}$ (consisting of some positive and negative bisimulation facts) represented as the simple approximations file $i^{SA}$ (cf.\ Section~\ref{sec:prac_algo_localappox}) for each WDB file $i$ retrieved during step~\ref{step:acquire}, and (ii) adding all the positive and negative bisimulation facts from $i^{SA}$ to the list $Q$ of questions and answers (replacing those questions in $Q$ which were thereby answered positively or negatively).
}
\end{itemize}

\noindent
Additionally, while computing global bisimulation by exploiting local approximations, the Oracle should always be ready to receive questions $\question{u}{v}$ from various, possibly remote $\Delta$-query systems and answer them immediately that the result is yet unknown (if it is so) and, when the result will become known either as $u\bis v$ or $u\not\bis v$, sending it back to the corresponding $\Delta$-query system.

\section{Empirical testing of the bisimulation engine}\label{sec:testing_real_oracle}

Preliminary results from testing of the simulated Oracle (described in Section~\ref{sec:testing_simulated_oracle}) indicated that, in principle, an Internet Service providing answers to bisimulation questions would decrease query execution time for those queries involving set equality. However, these preliminary tests were idealised situations and did not describe the \emph{relationship} between background work by the bisimulation engine and query performance. (In fact, the simulated Oracle did not work in background time, and only some intermediate result was represented.) Additionally, advantages of exploiting local approximations should be demonstrated.

\medskip

Let us consider empirical testing of the bisimulation engine by measuring the performance of the query client executing (with the help of the bisimulation engine) set equality queries of the form $\question{x}{y}$ where $x,y$ belong to a some suitable large WDB. To simplify our  considerations on measuring efficiency and to demonstrate some desirable effects we will consider rather artificial examples of WDB. As for WDB size, we will try to determine a threshold where the execution time becomes either unrealistically long or sufficiently reasonable. Note that, labels are ignored with just one (identical) label on all graph edges, as labels typically allow the bisimulation algorithm (see  Section~\ref{sec:impl_bisim_algo}) to derive more negative facts and, thus, possibly terminating too early (before the transitive closure of both set names involved in any bisimulation question will be fully explored).

\subsection{Determining the benefit of background work by the bisimulation engine on query performance}\label{sec:testing_without_approximations}

The aim of this experiment is to demonstrate the relationship between query execution time $t$ by the query system, and background work by the bisimulation engine. Background work by the bisimulation engine is simulated by delay time $d$, summarised briefly as follows:

\begin{enumerate}
\item{
The bisimulation engine should begin working with the goal of resolving all possible questions $\question{u}{v}$ for arbitrary set names of a given WDB. For the sake of the experiment, it should work uninterrupted (without being posed any questions by the query client) for the delay time $d$.
}
\item{
The query client should start executing the test query $\question{x}{y}$ after the delay time $d$ has expired, attempting resolution of the test question (and possibly other bisimulation questions which may arise during this process) with the help of the bisimulation engine. The bisimulation engine should continue its work, but now communicating with the query client.
}
\end{enumerate}

\pagebreak

\noindent
Thus, the query execution time $t(d)$ by the query client (working with the bisimulation engine starting from the delay time $d$) 
depends on $d$, and it is this dependence which we want to investigate experimentally. Evidently, $t(d)$ should be a decreasing function: the later the client starts its work after the bisimulation engine, the more help it can provide, and the smaller should be the client's working time $t(d)$. Note that this is still an idealised experiment, in practice, there could be many query clients communicating with the bisimulation engine at arbitrary times.

\begin{note}\em 
It should be noted that the current implementation of the hyperset language $\Delta$ does not use yet any bisimulation engine. These experiments were implemented separately and only to demonstrate some potential strategies for more efficient implementation of the most crucial concept of bisimulation relation underlying the hyperset approach.
\end{note}

In this experiment, the example WDB consists of 51 set names distributed over 10 WDB files, connected in chains as shown by the schematical graph in Figure~\ref{fig:test_bisimulation_engine_delay_time}. To increase the difficulty of computing bisimulation a copy WDB' of this WDB was made, changing only the URL part of full set names. Thus, the experiment is done over WDB + WDB'. Bisimulation between corresponding set names in WDB and WDB' under this circumstance is intuitively trivial (the answer being always ``true''). However, it is a non-trivial task when calculated by our algorithm which has no advance knowledge that WDB and WDB' are essentially identical (isomorphic).

\medskip

Further, our experimental procedure here was the measurement of execution time $t(d)$ by the query client executing the test query $\question{x}{x'}$ where $x,x'$ are corresponding set names in WDB and, its isomorphic copy, WDB'.

\begin{figure}[h]
\centering
\includegraphics[scale=0.95]{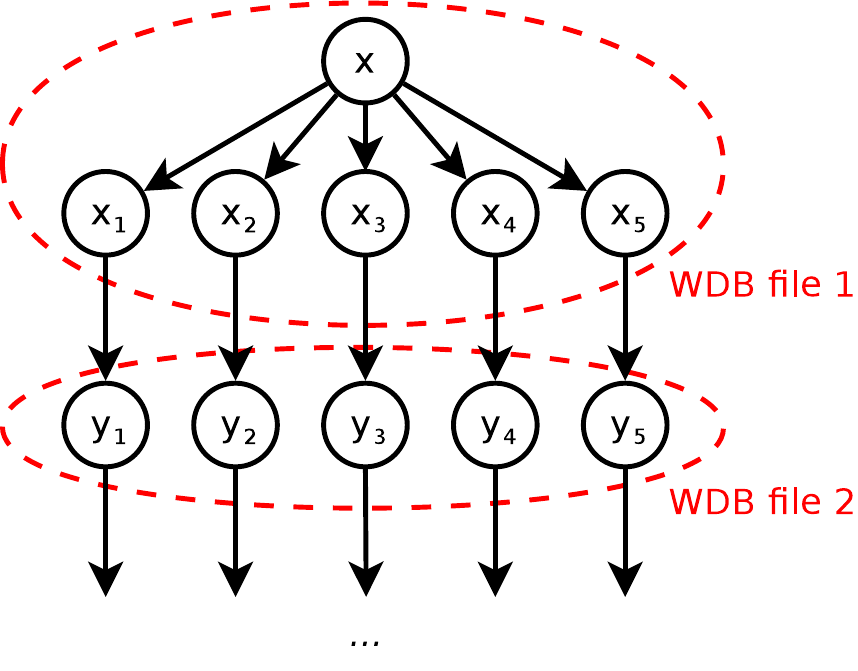}  
\caption{Schematical WDB graph divided into WDB files as shown by the red dashed ovals.}
\label{fig:test_bisimulation_engine_delay_time}
\end{figure}

\subsubsection{Experiment results}

On examination of the results graph in Figure~\ref{graph_bisimulation_engine_without_approx} the trend curve suggests an exponential decay relationship between partial work of the bisimulation engine and query performance. Moreover, this qualitative assessment by inspection of the graph is confirmed by examining the experimental values in Table~\ref{results_table_without_local_approximations}, which demonstrate that $t(d)$ approximately halves as $d$ increases by steps of 2500ms.

\medskip

Therefore, query performance benefits considerably even when the bisimulation engine has been working (in the background) for relatively short periods of time (say, 5 seconds or more), with an exponential decrease in $t(d)$ as $d$ increases. However, for sufficiently small delay time $d$, query performance suffers, as the bisimulation engine answers \emph{``Unknown''} to nearly all posed bisimulation questions. Thus, in this case, the bisimulation engine provides no real help, and the query client is forced to start resolving the bisimulation question itself. This suggests that in this circumstance that local computation of bisimulation by the query system without invoking the help of the bisimulation engine would be more efficient, as shown by the threshold on the graph (dashed horizontal line). In fact, here query execution time $t(d)$ with the help of the bisimulation engine is smaller than without the help of the bisimulation engine when delay $d$ is $> 2000$ms.

\begin{figure}[!ht]
\centering
\includegraphics[width=0.85\textwidth]{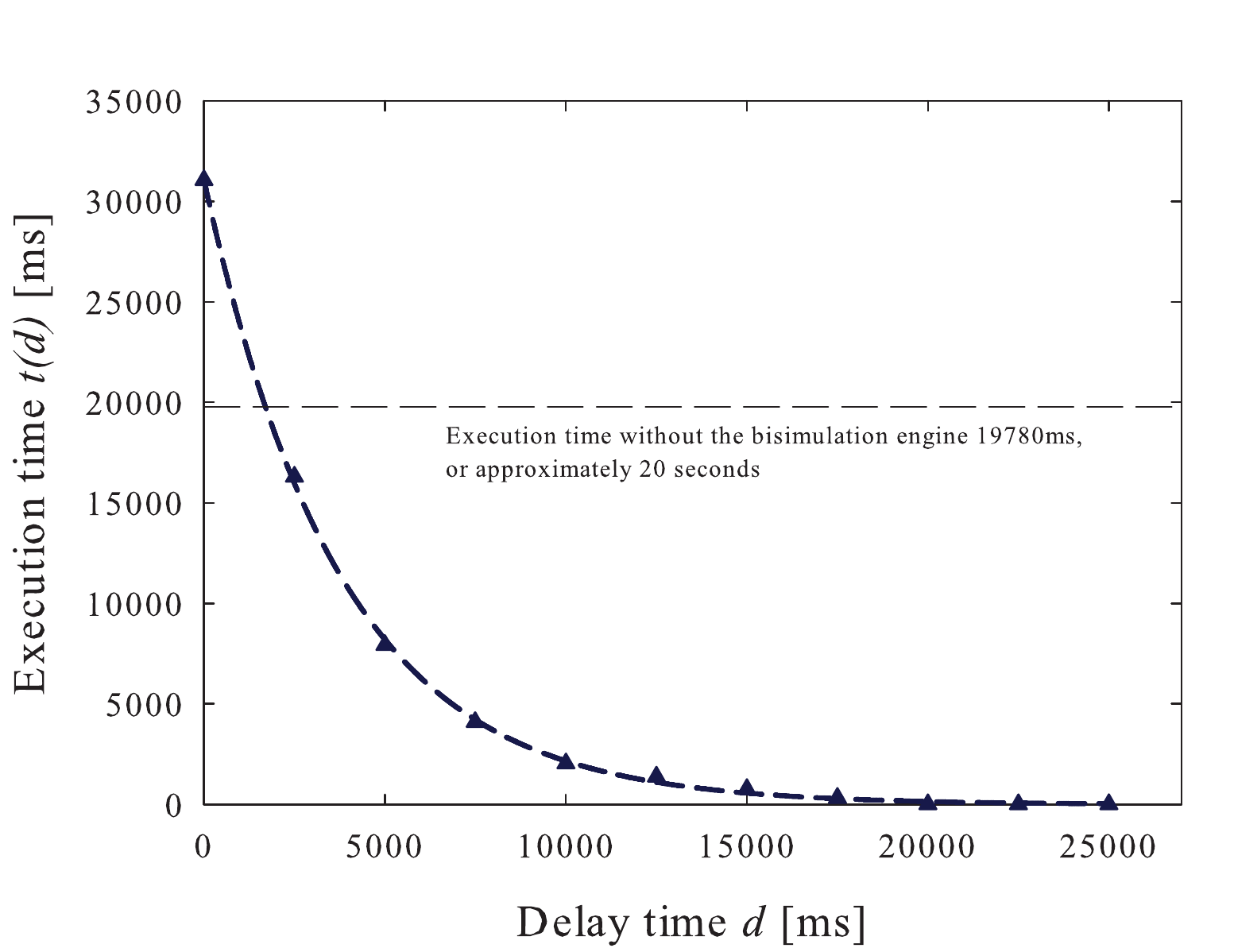}
\caption{Graph of experimental results (cf.\ Table~\ref{results_table_without_local_approximations} below) showing the dependence of query execution time $t(d)$ [ms] on delay time $d$ [ms]}
\label{graph_bisimulation_engine_without_approx}
\end{figure}

\begin{table}[h]
\begin{center}
\begin{tabular}{ | l | l | }
\hline
Delay time $d$ [ms]&Execution time $t(d)$ [ms]\\
\hline
0&31050\\
2500&16300\\
5000&7930\\
7500&4090\\
10000&2040\\
12500&1380\\
15000&770\\
17500&320\\
20000&10\\
22500&10\\
25000&10\\
\hline
\end{tabular}
\end{center}
\caption{Experimental results showing dependence of query execution time $t(d)$ [ms] on delay time $d$ [ms]}
\label{results_table_without_local_approximations}
\end{table}

\subsection{Determining the benefit of exploiting local approximations by the bisimulation engine on query performance}\label{sec:testing_with_approximations}

It seems plausible to expect that, in practice, each WDB file (or a group of closely related WDB files) should be sufficiently self-contained and have few links to the external files -- relatively small dependence on the ``external world''. Therefore, we should expect that the set of locally derived bisimulation facts should be sufficiently large (the majority of questions $\question{x}{y}$ for local set names should be resolved locally based on $\bis^L_+$ and $\bis^L_-$), and, hence, helpful for the work of bisimulation engine and improving its performance. 

\begin{figure}[!ht]
\centering
\includegraphics[scale=1.00]{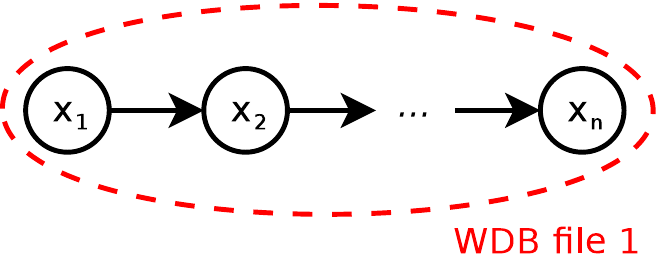}  
\caption{Schematical WDB graph consisting of one WDB file as shown by the red dashed oval.
}
\label{fig:test_bisimulation_local_global}
\end{figure}

\noindent
Taking this into account, our alternative example WDB for testing consists of one WDB file containing a variable number $n$ of set names (our experimental parameter as described below) connected in one chain, as shown by the schematical graph in Figure~\ref{fig:test_bisimulation_local_global}. Also, like the previous experiment, a copy WDB' of this WDB was made, changing only the URL part of full set name. Likewise, the experimental queries to follow are over WDB + WDB', that is over two files. This example represents an extreme, idealised case when each of these two files is fully self-contained, i.e.\ has no links to the ``external world''. As we wrote above, in more realistic situations we should rather expect a relatively small number of such external links.

\medskip

Recall that each of the WDB and WDB' files has a corresponding local approximations file, as described in Section~\ref{sec:prac_algo_localappox}, containing, respectively, local sets $\bis^L$ and $\bis^{L'}$ of (positive and negative) bisimulation facts which now will be available by demand to the bisimulation engine (as well as to the query system) which should considerably improve the performance. Thus, for our self-contained WDB file 1 (and similarly with its duplicate) the set of local set names is $L=\{x_1,\ldots,x_n\}$ and the corresponding local facts $\bis^L$ and $\not\bis^L$ obtained from the local approximations $\bis^L_+$ and $\bis^L_-$ trivially coincide with those global bisimulation facts $\bis$ and $\not\bis$ restricted to the set of names $L$.

\medskip

The aim of the experiment is to determine the relationship between the size of WDB (input size based on the parameter $n$) and query performance time comparing the three strategies: {\bf (i)} with the help of the bisimulation engine not exploiting local approximations; {\bf (ii)} with the help of the bisimulation engine, exploiting local approximations; and {\bf (iii)} without the help of the bisimulation engine%
\footnote{
That is, without the help of the bisimulation engine the query client running the test query is forced to compute bisimulation itself.
}%
. Similarly to the previous experiment we measure query performance for the test query $\question{x_1}{x_1'}$ where $x_1$,$x_1'$ are corresponding set names of the example WDB and its copy WDB'. But now there is \emph{no delay time} between the client and the bisimulation engine starting work. Delay time $d=0$ is the ``worst case'' for the bisimulation engine, as proved by the previous experiment.
(The case of variable $d$ for a fixed $n$ will be considered in another experiment later.)

\subsubsection{Experiment results}

The graph in Figure~\ref{graph_bisimulation_engine_with_approx} suggests a sufficiently close to linear trend between query performance and WDB size when the bisimulation engine exploits local approximations. Moreover, this looks almost like a horizontal line, and query execution seems practically viable ($\sim41$ seconds for $n=70$; see Table~\ref{results_table_with_local_approximations}). On the other hand, with help of the bisimulation engine not exploiting local approximations, as well as without help of the bisimulation engine at all, query performance with sufficiently large WDB ($n=70$) becomes intractable (more than one hour). In fact query performance improves at a threshold level of approximately $n=27$ (see Table~\ref{results_table_with_local_approximations}) with the bisimulation engine exploiting local approximations, with significant improvement in query performance for larger $n$ compared to the bisimulation engine not exploiting local approximations or without using bisimulation engine at all.

\pagebreak

Moreover, the absence of hyperlinks to other WDB files in our example WDB gives local approximations facts that coincide with those global bisimulation facts restricted to the set names in $L$ or $L'$. Thus, computing bisimulation requires fewer derivation steps, dramatically decreasing the time required to compute bisimulation. Furthermore, these results suggest that local approximations are more useful when the WDB is divided into larger almost
\linebreak
\mbox{self-contained} fragments. The latter is definitely the case when local is understood as \emph{local to a site}. However, in the latter case, local approximations to $\bis$ could take some time to compute at each site. This situation is somewhat different from saving a WDB file with its local approximation set $\bis^L$. Thus more experimentation is required.

\begin{figure}[!h]
\centering
\includegraphics[width=0.85\textwidth]{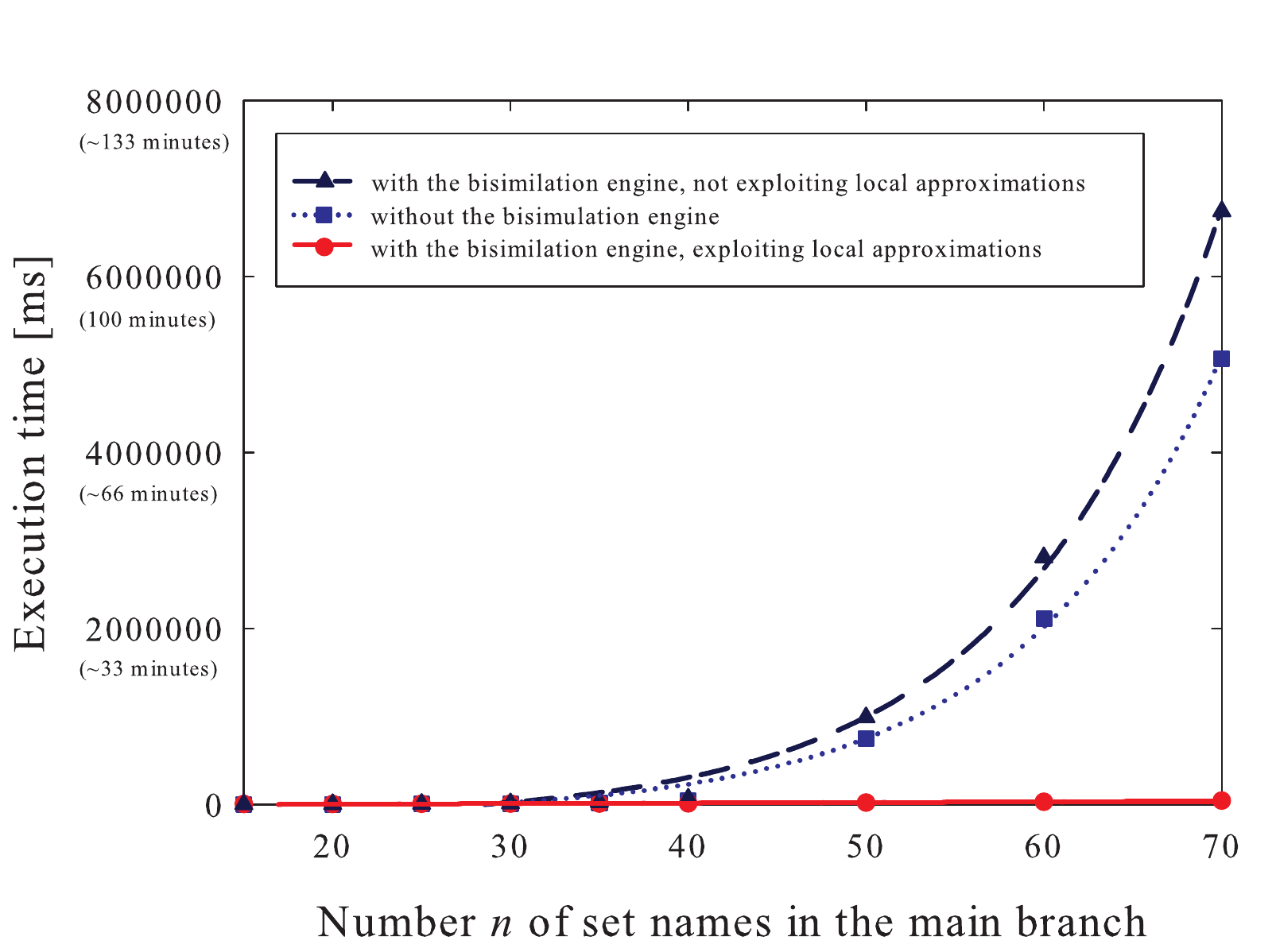}
\caption{Graph of experimental results (cf.\ Table~\ref{results_table_with_local_approximations} below) showing the relationship between query execution time [ms] and size of WDB (based on the parameter $n$) -- comparing the three strategies towards computing bisimulation}
\label{graph_bisimulation_engine_with_approx}
\end{figure}

\noindent
It might seem unexpectedly, but is actually quite natural that the results of this  experiment also demonstrate that query performance is worse with the help of the bisimulation engine not exploiting local approximations compared to without the help of the bisimulation engine. In fact, this experiment was conducted with no delay time ($d=0$), and we should recall the results of the experiments in Section~\ref{sec:testing_without_approximations} where a sufficiently small delay times decreased query performance with the help of the bisimulation engine (not exploiting local approximations) due to the additional expense of communication with the bisimulation engine.

\pagebreak

Note that the WDB considered in this and the following experiments was artificially created to make computation of bisimulation more difficult. In real situations, in particular where labels are used, it should be possible to derive non-bisimilarity of vertices without the need to go so deeply. However, only realistic application of the $\Delta$-query language can fully show its efficiency and where it should be improved.

\begin{table}[!ht]
\begin{center}
\begin{tabular}{ | p{1.70cm} | p{3.30cm} | p{3.30cm} | p{3.50cm} | }
\cline{2-4}
\multicolumn{1}{c}{} & \multicolumn{3}{|c|}{Query execution time (ms)} \\
\hline
Number of set names \newline $n$&with bisimulation \newline engine exploiting \newline local approximations&without bisimulation engine&with bisimulation \newline engine not exploiting local approximations\\
\hline
15&3422&1015&1340\\
20&4360&1781&2428\\
25&5500&3422&4585\\
30&7015&7781&10368\\
35&8547&19766&26309\\
40&10375&48422&64400\\
50&20063&746187 ($\sim13$ mins)&989750 ($\sim16$ mins)\\
60&27516&2113375 ($\sim35$ mins)&2810800 ($\sim47$ mins)\\
70&40983&5069797 ($\sim84$ mins)&6742890 ($\sim112$ mins)\\
\hline
\end{tabular}
\end{center}
\caption{Experimental results showing query execution time [ms] against WDB size (based on the parameter $n$) -- comparing the three strategies towards computing bisimulation.}
\label{results_table_with_local_approximations}
\end{table}

\subsection{Determining the benefits of background work by the bisimulation engine exploiting local approximations}\label{sec:testing_with_and_without_approximations}

Now let us consider the realistic case where the bisimulation engine is working in background time, comparing both strategies of working by the bisimulation engine: {\bf (i)} with exploitation of local approximations, and {\bf (ii)} without exploitation of local approximations. We shall adopt the same method of testing as previously in Section~\ref{sec:testing_without_approximations} with the aim to determine the relationship between query execution time against partial background work%
\footnote{Recall that, in Section~\ref{sec:testing_without_approximations} the experimental parameter, delay time $d$, simulated partial background work by the bisimulation engine.}
by the bisimulation engine for both strategies.

\medskip

The example WDB used in this experiment is based on notions described in Section~\ref{sec:testing_with_approximations} that WDB files (or groups of WDB files) should be relatively self contained with few external links. Thus, here the experimental WDB consists of one (main) WDB file with hyperlinks to two other (auxiliary) WDB files, describing 61 set names in total, as shown by the schematical graph in Figure~\ref{fig:test_bisimulation_comparison}. Note that, like those previous experiments in Section~\ref{sec:testing_without_approximations}~and~\ref{sec:testing_with_approximations}, the following experimental queries are over WDB and its identical copy WDB'.

\medskip

The aim of this experiment is to measure query execution time $t(d)$ by the query client with the help of the bisimulation engine for the test query $\question{x}{x'}$ where $x,x'$ are corresponding ``root'' set names of the example WDB and its copy WDB'. Our experimental parameter is the delay time $d$, as detailed in the previous experiment Section~\ref{sec:testing_without_approximations}.

\begin{figure}[!ht]
\centering
\includegraphics[scale=0.95]{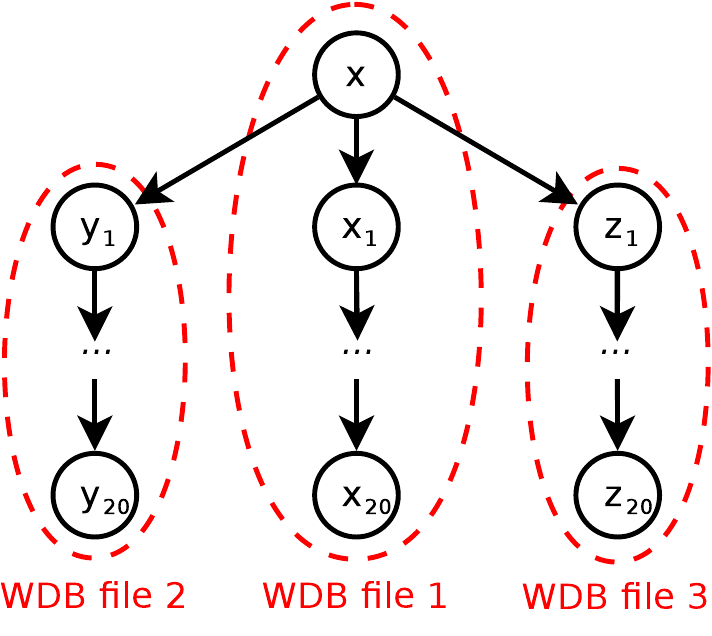}  
\caption{Schematical WDB graph divided into three WDB files as shown by the red dashed ovals.}
\label{fig:test_bisimulation_comparison}
\end{figure}

\subsubsection{Experiment results}

The results of the experiment in Table~\ref{results_table_comparison} extend previous results in Section~\ref{sec:testing_with_approximations} which suggested that exploitation of local approximation by the bisimulation engine increases query performance. However, comparing the influence of partial background work by the bisimulation engine, for both strategies of working, is somewhat difficult due to the difference in magnitude between the results (see Figure~\ref{graph_bisimulation_engine_comparison_both}). In fact, exploitation of local approximations (by the bisimulation engine) reduces query execution time from minutes to seconds, and hours to minutes.

\medskip

Note that in the case of exploitation of local approximations, the process of derivation is preceded%
\footnote{
Downloading approximation files can occur at any stage whilst resolving some bisimulation question.
}
by acquiring these approximations. The additional plot of data in Figure~\ref{graph_bisimulation_engine_comparison_with} shows threshold level, when $d$ is small, that background work by the bisimulation engine does not improve query performance whilst (the initial required) local approximations are being downloaded, as shown by the brown arrow in Figure~\ref{graph_bisimulation_engine_comparison_with}. Furthermore, when exploiting local approximations, a sufficiently large number of locally derived bisimulation facts (on the stage of creating WDB files) actually means in this example that fewer real derivation steps are required.

\newpage

\begin{figure}[!ht]
  \centering
  \subfloat[Comparison between bisimulation engines with and without exploiting local approximations]{
    \includegraphics[width=0.85\textwidth]{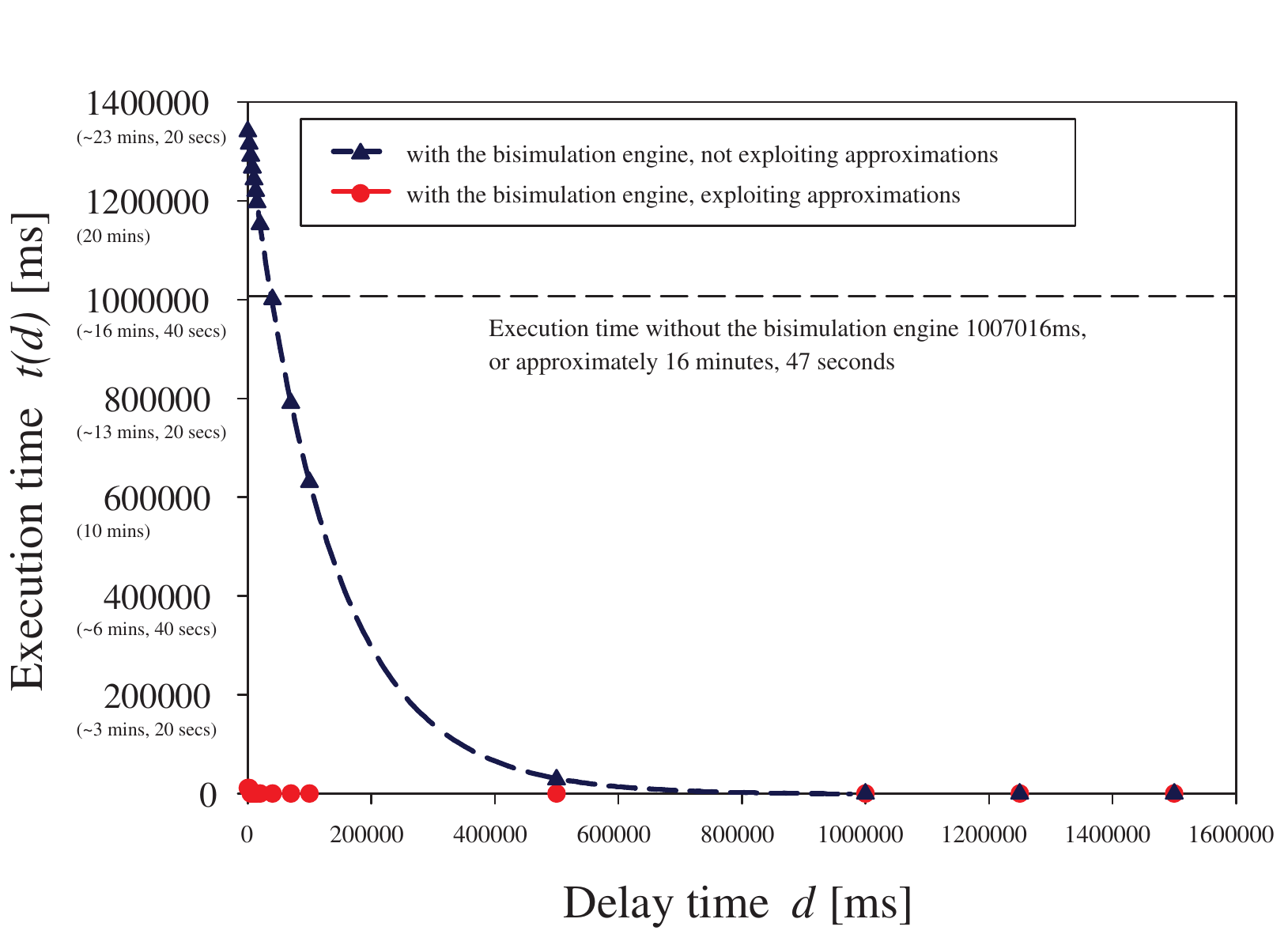}
    \label{graph_bisimulation_engine_comparison_both} 
  }\\
  \subfloat[Bisimulation engine exploiting local approximations]{
    \includegraphics[width=0.85\textwidth]{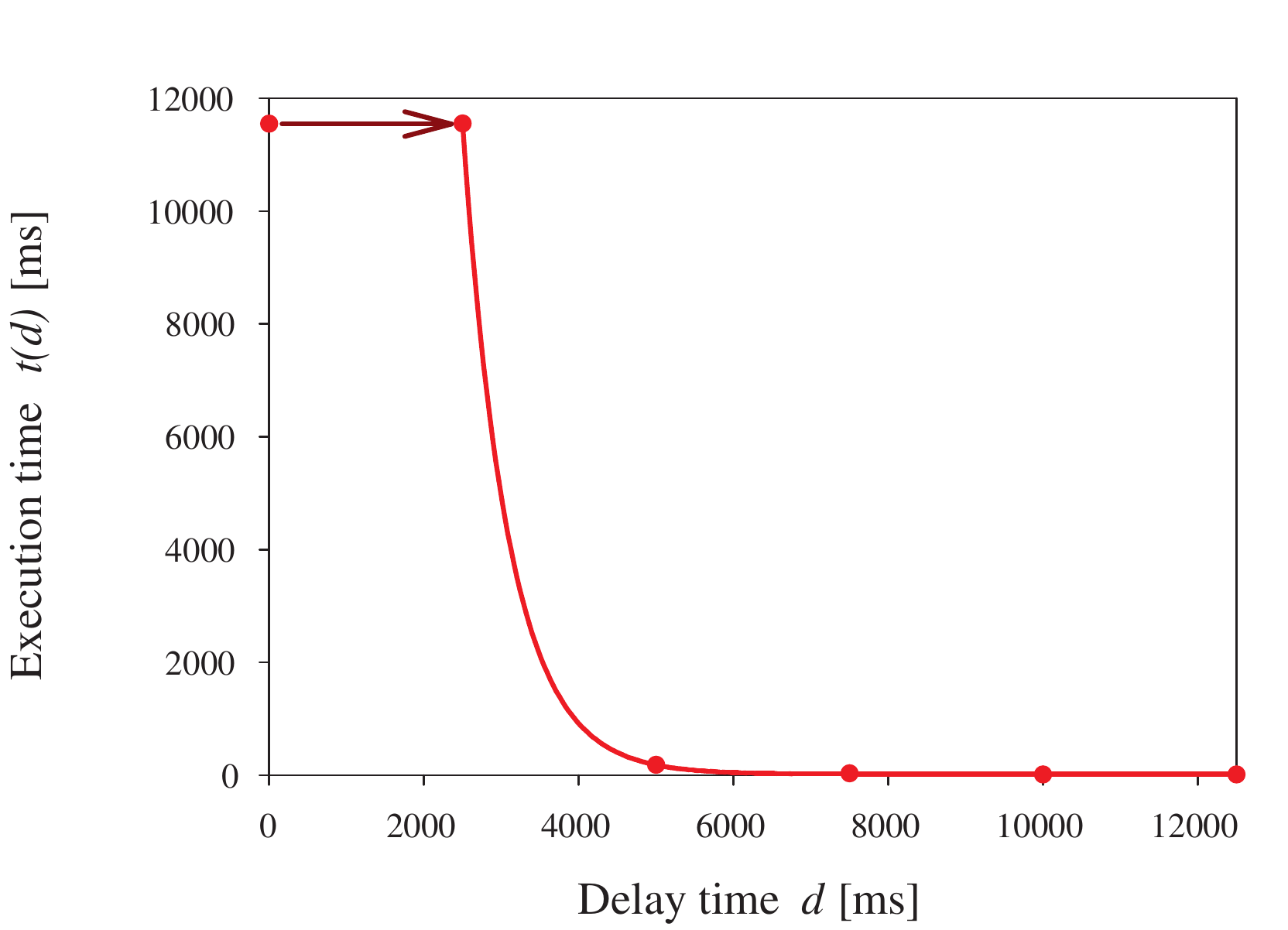}
    \label{graph_bisimulation_engine_comparison_with}  
  }
\caption{Graphs of experimental results demonstrating the relationship between query execution time [ms] and background work by the bisimulation engine simulated by delay time $d$ [ms]}
\label{graph_bisimulation_engine_comparison}
\end{figure}

\section{Overall conclusion}

In summary, here two strategies were suggested towards improving the performance of queries involving set equality (bisimulation), these strategies are: {\bf (i)} implementation of an Internet service, bisimulation engine, answering bisimulation questions; and {\bf (ii)} exploitation of local approximations (by the bisimulation engine) to facilitate the quicker computation of bisimulation. It was shown empirically that for an artificial  WDB that both strategies, and most dramatically {\bf (ii)}, improved query performance. In fact, the latter strategy demonstrates that querying of a medium sized example WDB could become practically viable.

\medskip

Note that other recent research into the efficient computation of the bisimulation relation was not considered here, for example the bisimulation algorithm described by Dovier et al \cite{DPP04} (which was intended to optimise the theoretical semi-structured query language G-log \cite{CDQT02}). However, the point of the approach presented here was to demonstrate some strategies for computing bisimulation in the case of distributed semi-structured data, unlike that by Dovier et al which did not consider distribution. There was not enough time to consider all possibilities for optimisation, and here we concentrated on those most novel and appropriate to our approach.

\begin{table}[!ht]
\begin{center}
\begin{tabular}{ | p{3.4cm} | p{2.7cm} | p{4.6cm} | }
\cline{2-3}
\multicolumn{1}{c}{} & \multicolumn{2}{|p{7cm}|}{Query execution time with help of the \newline bisimulation engine $t(d)$ (ms)} \\
\hline
Delay time $d$ [ms]&exploiting local \newline approximations&not exploiting local \newline approximations\\
\hline
0&11546&1340250 ($\sim22$ mins, 20 secs)\\
2500&11550&1315269 ($\sim21$ mins, 55 secs)\\
5000&180&1290715  ($\sim21$ mins, 31 secs)\\
7500&28&1266620  ($\sim21$ mins, 7 secs)\\
10000&10&1243000 ($\sim20$ mins, 43 secs)\\
12500&10&1219769 ($\sim20$ mins, 20 secs)\\
15000&10&1197025 ($\sim19$ mins, 57 secs)\\
20000&10&1152728 ($\sim19$ mins, 13 secs)\\
40000&10&1000520 ($\sim17$ mins)\\
70000&10&790760 ($\sim13$ mins)\\
100000&10&630772 ($\sim11$ mins)\\
500000 ($\sim8$ mins)&10&28765\\
1000000 ($\sim17$ mins)&10&118\\
1250000 ($\sim21$ mins)&10&10\\
1500000 ($25$ mins)&10&10\\
\hline
\end{tabular}
\end{center}
\caption{Experimental results showing query execution time $t(d)$ [ms] against partial background work by the bisimulation engine simulated by delay time $d$ [ms] -- comparing both strategies towards computing bisimulation, with and without exploiting local approximations.}
\label{results_table_comparison}
\end{table}

\subsection{Claims and limitations}

The main conclusion from the above experiments is that, although bisimulation (crucial to the hyperset approach to WDB and the $\Delta$-query language) presents some difficulty in efficient and realistic implementation, this problem appears to be resolvable in principle. Moreover, this assertion is somewhat supported by the empirical testing of artificial WDB examples described in Sections~\ref{sec:testing_without_approximations}--\ref{sec:testing_with_and_without_approximations}. In particular, these artificial WDB were chosen to simulate some specific worst case structural features of WDB similarly to physicists conducting some very specific experiments allowing to understand the most fundamental laws of the nature instead of dealing with something complicated as in the real life. On the other hand, those artificial WDB example presented here are intrinsically limited by their small size%
\footnote{
with the largest WDB considered here involving only 70 set names
}
and have restricted structural features%
\footnote{
which should involve not only nested chains but also nested tree structures
}%
, and, in principle, further comprehensive tests should be done to further characterise the usefulness of those practical strategies towards computing bisimulation suggested here. Also, empirical testing of some particular real-world WDB of sufficiently big size is important, but in this case a lot of further work should be done on optimisation of query execution which was outside of the scope of this work but deserves further investigation. We only considered one essential aspect of efficiency for the current version of the query system related with the idea of local/global bisimulation. However, in principle, the experiments done here suggest that these strategies show potential and merit further investigation.

\medskip

What has been demonstrated here is probably insufficient for a full-fledged implementation because in real-world circumstances using the $\Delta$-query language could be much more complicated. Anyway, only further work and practical experimentation can reveal problems with the current implementation, which is, of course, not fully perfect. However, it shows that the hyperset approach to databases looks promising and deserves further not only theoretical but also practical considerations --  and this was actually our main goal, as well as the goal to create a working implementation available to a wider range of users to realise practically what is the hyperset approach to WDB or semistructured databases.


%
%


\part{Implementation issues}\label{part:implementation}

\chapter*{Overview of Part~\ref{part:implementation}}

\noindent
In this part we discuss the most essential issues of implementing the $\Delta$-query language: (i) query execution (Chapter~\ref{chap:exec}), (ii) syntactical aspects (Chapter~\ref{chap:syntax}), and (iii) XML representation of WDB (Chapter~\ref{chap:xml-wdb}). These chapters can be read (almost) independently, however, logically their order should be the inverse. The chosen order rather reflects the importance of the material for the reader, who probably should be more interested in the principles of query execution than in the very technical details of implementation of the syntax (in particular related with the subtle points of well-formed vs.\ well-typed queries). But from the point of view of the actual implementation (including execution of queries) such syntactical aspects were very crucial and, in fact, such technical details serve as a guarantee that the whole implementation was done correctly. Indeed, the content of Chapter~\ref{chap:syntax} arose to overcome the problems of ensuring well-formed/well-typed queries encountered during the first attempt at implementation \cite{M04}. Finally, Chapter~\ref{chap:xml-wdb} details the XML representation of WDB, and has quite a separate role. We think and work exclusively in terms of hypersets and set equations, and any WDB could be represented adequately and straightforwardly in the latter form. However, we have chosen XML form (XML-WDB format) as a representation of set equations to make our approach potentially more closely related to the existing practice of using XML for semistructured data. The reader should choose the level of details he/she needs from this chapter for understanding examples of XML-WDB files we use when running $\Delta$-queries.


\chapter{$\Delta$ Query Execution}\label{chap:exec}

\section{Implementation of $\Delta$-query execution by reduction process}\label{sec:impl_delta_exe}

How to execute any $\Delta$-query was explained mostly in Section~\ref{sec:operational_semantics} as operational semantics (based on the general abstract mathematical approach described in \cite{S06}) and continued in Section~\ref{sec:bisim_algo} on computing bisimulation. Here we will finalise the operational semantics by considering the clauses omitted in Section~\ref{sec:operational_semantics} 
in the style more close to that of implementation. Recall that in this approach, any $\Delta$-term or $\Delta$-formula query $q$ should be equated, respectively, to a new set or boolean name $res$. Then this equation $res=q$ is reduced (in the context of all set equations of WDB) to an equation $res=V$,
\begin{equation}\label{eq:leadsto}
res = q \rhd res = V,
\end{equation}

\noindent
where $V$ is, respectively, either a
\begin{itemize}
\item{
set value -- flat bracket expression $\setof{l_1:v_1,\ldots,l_n:v_n}$ where $v_i$ are set names and $l_i$ label values, or
}
\item{
boolean value -- {\bf true} or {\bf false}.
}
\end{itemize}

\noindent
Note that this process of reduction can extend the original WDB by the auxiliary set equations $v_i=\setof{\ldots}$ defining those set names $v_i$ participating in $V$ which were not the original set names in the WDB, and, possibly, many others participating in equations for $v_i$, and so on. Thus, strictly speaking, the reducibility statement~(\ref{eq:leadsto}) only partially reflects this process of reduction as the whole WDB extended by the equation $res = q$ can be involved. In the case of distributed WDB, over which some query $q$ should be executed, this process of reduction also tacitly assumes downloading the (remote) WDB files with those required set equations participating in this process.

\bigskip

\noindent
Implementation of the $\Delta$-language should evidently follow the operational semantics in \cite{S06} or in Section~\ref{sec:operational_semantics}. In this chapter, we will give implementation details on four important
\linebreak
$\Delta$-language constructs: separation, quantification, recursion, decoration and transitive closure. Equality (bisimulation) was already discussed in detail. Other cases are sufficiently evident or do not add much to the operational semantics and by this reason are omitted.

\subsection{Separation construct}\label{sec:separate}

In the case of those queries which involve complex subqueries new equations will be created during the evaluation of the subquery (which was conceptually understood as the ``splitting'' rule; cf.\ Section~\ref{sec:operational_semantics}).

\medskip

\noindent
Consider the reduction process for $\Delta$-term separate $\setof{ l\!:\!x \in t \mid \varphi(l,x) }$:
\begin{align*}
res =\setof{ l\!:\!x \in t \mid \varphi(l,x) } \rhd & \; res = \setof{l_1\!:x_1, ..., l_n\!:x_n}
\end{align*}

\noindent
where $t$ is a set name with a flat set equation $t=\setof{l_1\!:x_1, ..., l_m\!:x_m}$ in the current version of WDB (possibly extended locally by the query system). In reality $t$ could be a complicated $\Delta$-term, but we may assume that the ``splitting'' rule from  Section~\ref{sec:operational_semantics} has already been applied so that we have here just a set name. In fact, $l_1\!:x_1, ..., l_n\!:x_n$ should be a sublist of
\linebreak
\mbox{$l_1\!:x_1, ..., l_m\!:x_m$} separated by the formula $\varphi(l,x)$ -- for simplicity of denotation some initial sublist (so that $n \le m$). Note that $l,x$ are label and set variables whereas $l_i,x_i$ are label values and set names participating in the current extended version of WDB. (See also the $\Delta$-language syntax in Appendix~\ref{app:BNF} on set names, and label and set variables.) The process of reduction is the quite evident iterative procedure,

\medskip

\noindent
{\bf Separation algorithm:}
\begin{enumerate}
\item[]{{\bf START with the current version of WDB and the separation term 
\[
\setof{ l\!:\!x \in t \mid \varphi(l,x) }
\] 
where $t$ is set name, and WDB contains flat set equation 
$t = \setof{l_1\!:\!x_1, ..., l_m\!:\!x_m}$.}
}
\item{
{\bf Extend current version of WDB} by the equation $res=\setof{ l\!:\!x \in T \mid \varphi(l,x) }$ where $res$ is a new set name. 
}
\item{
{\bf Create the new (temporary) set equation} $res = \setof{}$ (empty set) for the same set name $res$. (After populating the right-hand side by labelled set names, this equation will replace the above.)
}
\item{{\bf Iterate over the labelled elements} $l_i\!:\!x_i$ of $t$ where $t = \setof{l_1\!:\!x_1, ..., l_m\!:\!x_m}$.
  \begin{enumerate}
  \item{
  Call the corresponding reduction procedure for the $\Delta$-formula $\varphi(l_i,x_i)$,
\[
res_i=\varphi(l_i,x_i)\rhd res_i=\ldots,
\]
  for new set names $res_i$ resulting in the boolean equations $res_i = \TRUE$ or
  \linebreak
  $res_i = \FALSE$.%
\footnote{As the $\Delta$-language is bounded (quantifiers and other variable binding constructs are bounded by appropriately restricting the range of variables explicitly required by the language syntax) any such reduction process will inevitably halt (in fact, in polynomial time). In the current case either \TRUE\ or \FALSE\ will be obtained.
}
  \\
  \\{\bf Does $res = \varphi(l_i,x_i)\rhd res_i = \TRUE$?}
  \\
  \\{\bf Yes} -- Amend the equation for $res=\setof{\ldots}$ initiated in the step 2 as $res = \setof{\ldots, l_i\!:\!x_i}$ by adding the labelled element $l_i\!:\!x_i$. Move back to step 3 (iterate over next labelled element, if one exists).
  \\
  \\{\bf No} -- Move back to step 3 (iterate over next labelled element, if one exists).
  }
  \end{enumerate}
}
\item[]{{\bf END with the (simplified) set equation $res = \setof{l_1\!:\!x_1, ..., l_n\!:\!x_n}$} (with $res$ a subset 
of~$t$).}
\end{enumerate}

\subsection{Quantification}

Consider, for example, the reduction process for the quantified formula $\exists l\!:\!x \in t . \varphi (l,x)$ where $t$ is (for simplicity) a set name with a flat set equation $t = \setof{l_1\!:\!x_1, ..., l_m\!:\!x_m}$ (for $l_i,x_i$ label values and set names, like above). It starts by replacing the bounded existential quantifier with the disjunction:
\begin{align*}
res =\exists l\!:\!x \in t . \varphi (l,x) \rhd res = \varphi(l_1,x_1) \vee ... \vee \varphi(l_m,x_m)
\rhd\ldots .
\end{align*}

\noindent
By invoking the ``splitting'' rule it assumes the recursive subprocesses
\[
res_i =\varphi (l_i,x_i) \rhd\ldots 
\]

\noindent
(with new boolean names $res_i$) leading to truth values for $res_i$ from which an appropriate truth value for $res$ can  evidently be obtained.

\subsection{Recursive separation}\label{sec:impl_rec_sep}

Consider the recursion query:

\begin{align*}
\Rec\; p. \setof{l\!:\!x \in t \mid \varphi(x,l,p) }
\end{align*}

\noindent
where, as above, $t$ is considered as set name with a flat set equation $t = \setof{l_1\!:\!x_1,..., l_m\!:\!x_m}$ for $l_i,x_i$ label values and set names. To execute it, we should start by adding the set equation to the WDB with the new set name $res$,

\begin{align*}
res = \Rec\; p. \setof{l\!:\!x \in t \mid \varphi(x,l,p) }.
\end{align*}

\noindent
The set name $res$ denoting the result of the recursion query should represent a subset of $t$ where 
only some of $l_i\!:\!x_i$ will participate. It is computed iteratively as an increasing sequence 
$p_k$ of subsets of $t$:

\begin{align*}
p_0 = \; & \setof{} \;\; \mbox{(empty set)} \\
p_1 = \; & p_0 \cup \setof{l\!:\!x \in t \mid \varphi(x,l,p_0) }\rhd p_1=P_1 \\
p_2 = \; & p_1 \cup \setof{l\!:\!x \in t \mid \varphi(x,l,p_1) }\rhd p_2=P_2 \\
         & \ldots
\end{align*}

\noindent
This sequence of equations with new set names $p_k$ (in fact, intermediate results) should be generated iteratively, with each new set equation generated after the previous one. Each of these complicated equations is reduced essentially by using the above process of reduction for the ordinary separation construct giving rise to a subset $P_k$ of $t$. As these subsets are inflating, and $t$ is finite, this process should be halted when $P_k = P_{k+1}$ (stabilisation). Note that checking equality between these sets does not require the computation of bisimulation as each iterative set $p_k$ is an ``explicit'' subsets of $t$ (elements of the bracket expression $P_k$ are exactly, i.e.\ not up to bisimulation, some of $l_i:x_i$ from the right-hand side of the equation for $t$). Now, simplify the initial equation $res = \Rec\; p.\setof{...}$ by replacing it with  $res = P_k$:

\begin{align*}
res = \Rec\; p. \setof{l\!:\!x \in t \mid \varphi(x,l,p) }\rhd res = P_k.
\end{align*}

\noindent
Note that the subprocesses of the above process

\[
res_{ik}=\varphi(x_i,l_i,p_k)\rhd\ldots
\]

\noindent
(where $\varphi$ can be quite complicated formula involving complicated subterms) may introduce new set names with their corresponding set equations. Of course, they should also be considered as the part of the result of this computation (as soon as they are contained in the transitive closure of $res$). Thus, it has been demonstrated how to resolve the $\Delta$-term recursive separation.

\subsection{Decoration}\label{sec:impl_dec}

Although the decoration operator can be explained sufficiently easily on the intuitive level (see \cite{A88} and Section~\ref{sec:dec_operation}), its implementation should be done particularly carefully and precisely. To resolve the query
\begin{align*}
\Dec(g,v)
\end{align*}

\noindent
over a WDB with $g$ and $v$ arbitrary set names, i.e.\ to simplify the equation
\begin{align*}
res=\Dec(g,v)\rhd res=\setof{\ldots},
\end{align*}

\noindent
let us firstly consider some auxiliary queries which deserve to be included as library query declarations  and, most importantly, add an intermediate conceptual level of abstraction in the description of the operational semantics for the decoration operator.

\subsubsection{Auxiliary (library) queries useful for computing decoration}\label{sec:impl_dec_aux}

Let us now define several auxiliary queries dealing with representation of graphs as sets of ordered pairs.

\paragraph{\texttt{Nodes}:}
Now, consider a set name \verb+g+ with the flat%
\footnote{Recall that the query system considers WDB as a flat system of set equations, and all set equations it eventually produces are also flat.  (Only at the very last step of outputting the query result will the system produce set equations with reasonably nested right-hand sides.)
}
WDB-equation

\begin{small}
\begin{verbatim}
     g = { ..., l:p, ...}
\end{verbatim}
\end{small}

\noindent
with \verb+l:p+ any labelled set name appearing in the right-hand side (which can be a name of an ordered pair or just of an arbitrary set). The (abstract) set values \verb+First(p)+ and \verb+Second(p)+ are called \emph{$g$-nodes}%
\footnote{
Recall that \texttt{First(p)} and \texttt{Second(p)} are library queries defined in Section~\ref{par:first_second} and Appendix~\ref{app:predefined}.
}
so that

\[
\verb+First(p)+\stackrel{l}{\longrightarrow}\verb+Second(p)+
\]

\noindent
serves as an \emph{$g$-edge}, and therefore the (absolutely arbitrary) set $g$ plays the role of a \emph{graph}. Alternatively, we could ignore those \verb+p+ in \verb+g+ which are not ordered pairs -- the approach adopted below. Note that different set names may denote the same set, in particular, the same $g$-node, so that we will need to choose canonical \verb+g+-node names in the algorithm considered below.

\medskip

\noindent
The set of \verb+g+-nodes can be formally defined in $\Delta$ as library query declaration
\begin{small}
\begin{verbatim}
    set query Nodes (set g) =
        union separate { m : p in g  | call isPair ( p ) }
\end{verbatim}
\end{small}

\noindent
The set \verb+Nodes(g)+ (the union of two element sets \verb+p+ in \verb+g+) contains exactly all \verb+g+-nodes, but, strictly speaking, each \verb+g+-node in this set (being an element of some $p$ in $g$) has a label \verb+fst+ or \verb+snd+ and possibly appears twice, under both of these labels. However, this feature (which could be corrected by replacing these labels by the neutral ``empty'' label \verb+null+) will play no role in the following considerations. On the other hand, preserving this information on the nodes in \verb+Nodes(g)+ might be useful in other examples of using this query declaration.

\paragraph{\texttt{Children}:}
We also need the concept of \emph{$g$-children} of a node $x$ in a graph $g$ (as a set of ordered pairs), which is essentially the set of all outgoing edges from $x$ in $g$. This can be defined set theoretically by the following library query declaration (with three occurrences of the \verb+call+ keyword omitted to simplify reading):
\begin{small}
\begin{verbatim}
    set query Children(set x,set g)=
        collect {l:Second(p)
            where l:p in g
                  and ( isPair(p) and First(p)=x )
        }
\end{verbatim}
\end{small}

\noindent
Evidently, if the set \verb+x+ is not the value of \verb+First(p)+ for some pair \verb+p+ as required in this declaration then \verb+Children(x,g)={}+ (the empty set).

\paragraph{\texttt{Regroup}:}
Let us now define the set valued library operation \verb+Regroup(g)+ that can reorganise (without losing any essential information) any graph $g$ into something closely similar to the system of set equations represented by this graph. (For simplicity we again omit all \verb+call+ keywords.)  Pay attention to the use of the label \verb+null+ which can be considered here as the ``empty'' label (some label is formally necessary according to the BNF of the language):

\begin{small}
\begin{verbatim}
    set query Regroup(set g)=
              collect {'null':Pair(x, Children(x,g))
              where m:x in Nodes(g)
              }
\end{verbatim}
\end{small}

\noindent
Informally, each pair \verb+Pair(x,Children(x,g))+ collected in \verb+Regroup(g)+ is considered as \emph{abstractly} representing a set equation, where:
\begin{itemize}
\item{
first element \verb+x+ of the pair (understood as the abstract set denoted by \verb+x+) plays the role of a node of \verb+g+ or of an abstract set name -- the left-hand side of the intended equation, 
and 
}
\item{
second element, set \verb+Children(x,g)+, plays the role of the right-hand side of this equation -- the evident bracket expression enumerating the labelled elements (\verb+g+-nodes) of this set.
}
\end{itemize}

\noindent
It is crucial here that the set of ordered pairs \verb+Regroup(g)+ is \emph{functional} in the sense that for each (abstract set) \verb+x+ there exist at most one (abstract) pair \verb+Pair(x,c)+ in \verb+Regroup(g)+ with the first element \verb+x+ (and with \verb+c+ uniquely defined by \verb+x+ as \verb+c=Children(x,g)+). In fact, \verb+Regroup(g)+ defines abstractly the correct system of set equations where each abstract set name (a set in \verb+Nodes(g)+) has exactly one (abstract) equation with this name as the left-hand side. The operation \verb+Regroup(g)+ will make it easier extracting from \verb+g+ the required system of set equations, described in the main algorithm for computing decoration operation below.

\paragraph{An assumption.}
\emph{Now, let us assume that the fragment of the $\Delta$-language without decoration operation has already been implemented}. Then we can make calls to the above library queries applied to appropriate set name arguments in a given WDB, such as the set name \verb+g+ (representing a set of ordered pairs) in the call \verb+Regroup(g)+. The latter call will be used in the implementation of decoration operator in the next section.

\medskip

As usually, when executed by the query system, these library operations generate new set names and set equations and add them to the WDB. In particular, considering set names generated by the query system, the result of \verb+Regroup(g)+ is, informally, a set of ordered pairs of the form \verb+{'fst':x,'snd':Children_x}+ where \verb+x+ and \verb+Children_x+ 
(denoted as \verb+c+ in the algorithm below) are now set names%
\footnote{In further detail, when executing the query \texttt{Regroup(g)}, a new set name~\texttt{r} and set equation \texttt{r=Regroup(g)} are generated. Then, the implemented reduction process ($\rhd$) executing this query will give rise to a flat equation \texttt{r=\{..., 'null':e, ...\}} with each set name \texttt{e} in the right-hand side having the equation \texttt{e=\{'fst':x,'snd':Children\_$\!$\_$\!$\_x\}}.
}%
. Moreover, according to the natural implementation of the declaration for the query \verb+Children(x,g)+, the right-hand side of the equation for each set name \verb+Children_x+,
\begin{small}
\begin{verbatim}
     Children_x = { ..., l:y, ... },
\end{verbatim}
\end{small}

\noindent
contains labelled set names (in fact, \verb+g+-node names) \verb+l:y+ for all (labelled) $g$-children of the $g$-node named by~\verb+x+. Note that the algorithm described in the next section operates with these $g$-node names.

\subsubsection{Algorithm for computing decoration}
\label{sec:decoration-algorithm}

We will show how the decoration operation \verb+decorate(g,v)+ can be implemented over a given WDB (with \verb+g+ and \verb+v+ any set names from the WDB) exploiting the above library query declarations. This can be done as follows:


\begin{enumerate}
\item[]{{\bf START with the current version of WDB and the term $\Dec(g,v)$ for a given set names $g$ and $v$.}
}
\item{
{\bf Extend current version of WDB} by the equation $res=\Dec(g,v)$ where $res$ is a new set name. 
}
\item {\bf Regroup \verb+g+ and canonise \verb+g+-node names}.
	\begin{enumerate}
	\item\label{item:regroup}{
	Call the query \verb+Regroup(g)+. 	
	This amounts to simplifying the extended system of set equations 
	WDB + (\verb+r=Regroup(g)+) for \verb+r+ a new set name, which results in some new 
	(auxiliary) set names and flat set equations, including the flattened version 
	\verb+r={..., 'null':e, ...}+ of \verb+r=Regroup(g)+, and, for each \verb+e+ in~\verb+r+, 
	
	\verb+e={'fst':x,'snd':c}+, 
	\verb+c={..., l:y, m:z, ...}+.
	}
	\item Canonise \verb+g+-node names:
		\begin{enumerate}
		\item \label{item:extract} 
		Extract \verb+g+-\emph{node names} (all \verb+x+, \verb+y+, \verb+z+, \verb+...+) from the result  
		in (\ref{item:regroup}), 
		\item Compare which of them, considered as sets, are equal between themselves 
		(bisimilar as set names, represent the same abstract \verb+g+-node).
		\item For each \verb+g+-node name \verb+u+ find its canonical representative 
		\verb+Can_u+ as the first in the lexicographical order \verb+g+-node name bisimilar to \verb+u+. 
		(Thus, \verb+u+ is bisimilar to \verb+Can_u+. Note that \verb+Can_u+ is not a new set name 
		--- just one of those extracted in the step~\ref{item:extract}.)
			\item \label{item:x->Can_x} In the resulting set equations in (\ref{item:regroup}) 
			
				\verb+e={'fst':x,'snd':c}+, 
				\verb+c={..., l:y, m:z, ...}+
			
			(for each \verb+e+ in~\verb+r+) 
			replace \verb+g+-node names \verb+x+ and \verb+y+,$\ldots$, respectively, 
			by \verb+Can_x+ and \verb+Can_y+,$\ldots$, thereby transforming these equations to 
			
				\verb+e={'fst':Can_x,'snd':c}+, 
				\verb+c={.., l:Can_y, m:Can_z,..}+,
				\linebreak
				$\ldots$.
			
			(The original versions of these equations should be deleted.)
			
			\item\label{item:x->Can_x-Clean1} If for another pair of such equations (for \verb+e'+ 
			in \verb+r+),

				\verb+e'={'fst':Can_x','snd':c'}+,
				\verb+c'={..., l':Can_y', ...}+, 
				
			set names \verb+Can_x+ and \verb+Can_x'+ in \verb+e+ and \verb+e'+, respectively, coincide 
			then omit one of these pairs (does not matter which), and repeat this until no such coincidence 
			of canonical node names will exist.

			\item\label{item:x->Can_x-Clean2}
			Eliminate possible repetitions of labelled canonical node names \verb+l:Can_y+ in each \verb+c+ 
			(which can arise, e.g. due to replacements in (\ref{item:x->Can_x}) as \verb+l:Can_y+ 
			can literally coincide with some \verb+m:Can_z+ in \verb+c+ 
			for different \verb+g+-node names \verb+y+ and~\verb+z+). 

		\end{enumerate}
	
	From now on, these \verb+Can_u+ serve as \emph{canonical} \verb+g+-\emph{node names}. 
	Only these node names will be used below as uniquely representing \verb+g+-nodes. 
\end{enumerate}
\item{{\bf Does a canonical \texttt{g}-node name bisimilar to \texttt{v} exist?} 
Find a canonical \verb+g+-node name~\verb+w+ bisimilar to set name~\verb+v+ (or just coinciding with \verb+v+ if \verb+v+ is itself a canonical \verb+g+-node name). Two answers are possible:

	{\bf No - } 
	The required canonical \verb+g+-node name \verb+w+ bisimilar to \verb+v+
	does not exist 
	(and thus \verb+v+ can be treated as naming an isolated \verb+g+-node):
  \begin{enumerate}
  \item{Simplify the equation \verb+res = decorate(g,v)+ to \verb+res = {}+ (empty set).
  \linebreak
  Then move to {\bf END} of the algorithm}.
  
  \end{enumerate}
}

	{\bf Yes - } The required canonical \verb+g+-node name \verb+w+ does exist 
	(and thus \verb+v+ can be treated as naming a proper \verb+g+-node):
  \begin{enumerate}
  \item Generate new set equations for duplicated canonical \verb+g+-node names:
  	\begin{enumerate}
  	\item For each set name \verb+s+ which is a canonical \verb+g+-node name create a new duplicate set 
        name \verb+Dupl_s+ (in particular, \verb+Dupl_w+, \verb+Dupl_Can_x+, etc.).
		\item For the equations 
				
				\verb+e={'fst':Can_x,'snd':c}+, 
				\verb+c={...,l:Can_y,m:Can_z,...}+,	
				
				obtained in 
		(\ref{item:x->Can_x}, \ref{item:x->Can_x-Clean1}, \ref{item:x->Can_x-Clean2}) for each 
		\verb+e+ in \verb+r+,  
				extend further the current extension of WDB by new set equations: 
				
				\verb+Dupl_Can_x = {..., l:Dupl_Can_y, m:Dupl_Can_z, ...}+, 
				
				thereby constructing a system of set equations for duplicate 
				names whose graph is isomorphic to the abstract graph \verb+g+.
		\end{enumerate}
				In particular, this will add to the WDB the equation for \verb+Dupl_w+:
\begin{small}
\begin{verbatim}
        Dupl_w = W
\end{verbatim}        
\end{small}
         with the right-hand side a bracket expression \verb+W+ defined as described above 
         (and involving only duplicated canonical \verb+g+-node names). 
				
   \item{
         Simplify the equation, \texttt{res=decorate(g,v)} by replacing it with the (flat) equation
\begin{small}
\begin{verbatim}
        res = W.
\end{verbatim}        
\end{small}        
         (End of algorithm.)
        }
  
  \end{enumerate}%

\item[]{{\bf END with the (simplified) set equation $res = \setof{l_1\!:\!x_1, ..., l_n\!:\!x_n}$ (and the associated equations for set names in \verb+W+, etc.).}}

\end{enumerate}

\noindent
In the case of the query, $res = \Dec(G,V)$ where $G$ and $V$ are $\Delta$-terms and not just set names (as above), the ``splitting'' rule should be invoked first, which will result in three equations $g=G$, $v=V$ and $res=\Dec(g,v)$ for the new set names $g$ and $v$. Then these equation should be simplified, in particular, by using the above algorithm for the decoration.

\subsection{Transitive closure}\label{sec:impl_tc}

Let us now consider implementation of the transitive closure operation $\TC(a)$, where $a$ is considered as a set name with the flat equation $a = \setof{l_1\!:\!x_1, ..., l_m\!:\!x_m}$ for $l_i, x_i$ label values and set names, as the following (recursive) algorithm:

\begin{enumerate}
\item[]{
{\bf START with the current version of WDB and the transitive closure term $\TC(a)$ where $a$ is set name, and WDB contains flat set equation $a = \setof{l_1\!:\!x_1, ..., l_m\!:\!x_m}$.}
}
\item{
{\bf Extend current version of WDB} by the equation $res=\TC(a)$ where $res$ is a new set name. 
}
\item{\label{step:tc_create_new_set_equation}
{\bf Replace the original set equation $res=\TC(a)$ by the new (temporary) set equation $res = \setof{'null'\!:\!a}$} (singleton set) for the same set name $res$. 
(This will be further populated below.)
}
\item{\label{step:tc_recursive_jump}
{\bf Find the first labelled element $m\!:\!z$} of $res = \setof{\ldots,m\!:\!z,\ldots}$ such that $z\not\subseteq res$. (Elements for which $z\subseteq res$ should be marked and put at the end of the current bracket expression for $res$ so that they will not be considered again and again. For efficiency, the bracket expression for $res$ can be organised as a directed ``loop'' structure with some point of entrance. Each time when $z\subseteq res$ holds at the entrance point then this point in the loop will be marked and the entrance point shifted to the next one to repeat the inclusion test.)

{\bf If} it does not exist (the currently observed element and hence all $m\!:\!z$ are marked), go to the {\bf END}. 

{\bf Else} replace the current equation $res = \setof{\ldots,m\!:\!z,\ldots}$ with the $m\!:\!z$ found 
(at the current entrance point) by 
\[
res=\setof{\ldots,m\!:\!z,\ldots}\cup (z\setminus res) 
\]
(inserting elements of $z\setminus res$ in the loop immediately after $m\!:\!z$, then marking $m\!:\!z$ 
as now $z\subseteq res$ for the extended $res$ 
and shifting the entrance point from $m\!:\!z$ 
to the next point of so extended loop --- the first element in $z\setminus res$).

(Computing $z\setminus res$ can evidently also use the loop structure of $res$ with marking ignored.)

{\bf Repeat~\ref{step:tc_recursive_jump}}. 

}
\item[]{
{\bf END with the set equation for $res$. }
}
\end{enumerate}

\noindent
Note that in fact $\TC(a)=\bigcup \setof { \setof{a},a, \bigcup a, \bigcup\bigcup a, ... }$.  

\section{Representation of query output}\label{sec:impl_output}

Recall that the implemented query system works internally with (WDB represented as) a flat system of set equations, and produces query results in this flat form. The resulting set equations also use internally generated (local) set names having no mnemonics. It appears that some nesting in the outputted equations might be desirable which would simultaneously eliminate some internal set names by substituting them with bracket expressions. This substitution can be repeated giving rise to possibly deeply nested results. Consider, for example the result of the \emph{restructuring query} from Section~\ref{sec:restructuring-query} obtained after some such automatic substitutions:

\begin{small}
\begin{verbatim}
    Query is well-formed, well-typed and executable

    Result = {
      'publication':res2,
      'publication':res0,
      'publication':res1,
      'publication':{
        'type':"Book",
        'refers-to':res1,
        'refers-to':res2
      }
    }

    res0 = {
      'type':"Paper",
      'author':"Smith",
      'title':"Databases",
      'refers-to':res1
    }

    res1 = {
      'type':"Paper",
      'type':"Book",
      'author':"Jones",
      'title':"Databases"
    }

    res2 = {
      'type':"Paper",
      'refers-to':res0
    }

    Finished in: 1866 ms (query execution is 1864 ms, and 
    postprocessing time is 2 ms)

    Comment(s):
    Double quotation denotes atomic values like "atom" representing 
    singleton sets "atom" = {'atom':{}}, etc.
\end{verbatim}
\end{small}

\noindent
Note that, in this example further substitutions could be made to eliminate even those few local names \verb+res0+, \verb+res1+, \verb+res2+, so that there would be just one deeply nested equation \verb+result={...}+. However, this would be a rather inconvenient form as set names to be substituted occur several times, and identical subexpressions could be repeated many times making the query result difficult to grasp. Thus, the system makes such suitable nesting to avoid multiple substitutions in the whole system of equations. Additionally, nested bracket expressions like \verb+{Paper:{}}+ which imitate atomic values in our approach are replaced, quite naturally, by \verb+"Paper"+. Note that in the later case there may be multiple substitutions and replacements of the same expression. Similarly, set names for the empty set are always replaced by \verb+{}+. In this way query results become sufficiently readable. Lastly, in the case of cycles substitutions could be infinitely repeated. To avoid this, the system should only substitute those set names $res_i$ with the corresponding bracket expression if $res_i \not\in \TC(res_i)$ holds (in addition to the other rules for substitutions above). Also, the computation of transitive closure should be restricted to those new set names resulting from the execution of the query, thus, in principle, this can be done quickly on only local set names.

\medskip

However, any such postprocessing of the query result can sometimes lead to unnatural looking output, for example in the above query result there is some undesirable extra nesting for one of the publications. In other cases (such as showing a graph as a set of ordered pairs) such nesting appears more reasonable. Also atomic values and explicitly shown empty sets\verb+{}+ are very natural. Of course it would be better if the user could choose the preferred form, or the result could be optionally visualised as a graph.

%
%


\chapter{$\Delta$ Query Syntax}\label{chap:syntax}

\section{Parsing (well-formed queries)}\label{sec:implementation_delta_syntax}

\nocite{L97}\nocite{WM95}\nocite{W89}

\subsection{Implemented $\Delta$-language grammar}\label{sec:G+D}

The syntax of the implemented language was discussed in Chapter~\ref{chap:theoretical_language}, with the full syntax appearing in  Appendix~\ref{app:BNF}. The implemented language is described as \emph{Extended Backus-Naur form} (EBNF or, shortened, BNF), defined as a set of production rules, with each production describing one syntactical category represented as a non-terminal. For example, the production rule 

\begin{small}
\begin{verbatim}
    <query> ::=
        "boolean query" <delta-formula> | "set query" <delta-term>
\end{verbatim}
\end{small}

\noindent
defines the \verb+<query>+ syntactical category (also called \emph{non-terminal}) by stipulating in general that a terminal can be substituted by a sequence of \emph{terminals} such as \verb+"boolean query"+ and other non-terminals such as \verb+<delta-formula>+. Here the symbol \verb+|+ allows to describe alternative productions. (There are also other ways in the BNF to describe more complicated alternations in production rules.) Continuing such substitutions by using production rules for \verb+<delta-formula>+, etc., a sequence consisting only of terminals can be obtained. Further, as terminals are strings of symbols, the final concatenation is also a string of symbols which, properly speaking, is called \emph{well-formed query}, provided it was generated starting from the non-terminal \verb+<query>+. (Quite similarly we can consider well-formed \emph{delta formulas}, \emph{delta terms}, etc.) Thus, the BNF defines how to construct any query in $\Delta$. In fact, each $\Delta$-query, if well-formed, generates a parse tree (by using BNF-forks discussed below) which should be subsequently checked for well-typedness (see Section~\ref{sec:well-typed_delta_queries}).

\subsection{BNF forking}\label{sec:bnf_forking}

Firstly a general note on the BNF grammar. Each production rule from the BNF (except some auxiliary ones which can be eliminated as we will see below) can be represented as one, several, or even infinitely many alternative \emph{forks} \verb+F1,F2,...+ each having the same label (syntactical category or non-terminal) on the root of the fork. For example, the rule 

\begin{small}
\begin{verbatim}
    <A> ::=	<B><C> | <B><D><E>
\end{verbatim}
\end{small}

\noindent
splits into two rules
\begin{small}
\begin{verbatim}
    <A> ::=	<B><C>
    <A> ::=	<B><D><E>,
\end{verbatim}
\end{small}

\noindent
evidently corresponding to two forks with the branching degree two and three, whose roots are labelled by \verb+<A>+ and leafs labelled, respectively, as \verb+<B>+, \verb+<C>+ and \verb+<B>+, \verb+<D>+, \verb+<E>+. Let us analogously consider the production rule

\begin{small}
\begin{verbatim}
    <set constant declaration> ::= "set constant" <set constant>
                                   ("be"|"=") <delta term>
\end{verbatim}
\end{small}

\noindent
which generates two unique forks depending on whether \verb+"be"+ or \verb+"="+ is used -- each fork has a branching degree of four.

\medskip

Thus whole BNF grammar can then be represented as a set of all such forks. In fact, the parse tree of a query is constructed of such forks. However, not all BNF production rules are so simple and literally split into forks as will be discussed below.

\subsubsection{Recursion by Kleene operators}\label{sec:bnf_recursion}

Recursive BNF rules using repetition by the Kleene star and plus (\verb+*+ and \verb@+@) operators generates an infinite set of forks; \verb+*+ represents zero or more repetitions, and \verb%+% represents one or more repetitions. For example the following rule represents a sequence of declarations:

\begin{small}
\begin{verbatim}
    <declarations> ::= <declaration> ( "," <declaration> )*
\end{verbatim}
\end{small}

\noindent
Each fork has a root labelled by \verb+<declarations>+ and any number of leaves labelled by \verb+<declaration>+, separated by the terminal leaves labelled by \verb+","+. Evidently, the branching of these forks have an arbitrary odd degree because of the separator \verb+","+ considered formally as a leaf. Analogously the following syntactic categories are also considered:

\begin{small}
\begin{verbatim}
    <variables>, <parameters>, <multiple union>, <conjunction>
    <disjunction>, <quasi-implication>, <labelled terms>
\end{verbatim}
\end{small}

\subsubsection{Identifier forks}\label{sec:indentifier_forks}

There is further simplification to the BNF forks and to parse trees by eliminating the ``intermediate'' \verb+<identifier>+ category playing rather an auxiliary role. Thus, we will replace corresponding production rules by those generating infinitely many simple (one child) forks:

\begin{small}
\begin{verbatim}
    <boolean query name> ::=	( (A-Z) | (a-z) | (0-9) | "_" | "-" )+
    <set query name> ::=		( (A-Z) | (a-z) | (0-9) | "_" | "-" )+

    <label variable> ::=		( (A-Z) | (a-z) | (0-9) | "_" | "-" )+
    <label constant> ::=		( (A-Z) | (a-z) | (0-9) | "_" | "-" )+

    <set variable> ::=		( (A-Z) | (a-z) | (0-9) | "_" | "-" )+
    <set constant> ::=		( (A-Z) | (a-z) | (0-9) | "_" | "-" )+
\end{verbatim}
\end{small}

\noindent
There are infinitely many of such identifier forks because there are infinitely many sequences of \emph{alphanumeric characters} (just those characters participating in the identifier forks) which can serve as a leaf label of a fork for each of the above syntactical categories.

\medskip

Root nodes of these forks of the corresponding nodes in a parse tree are called \emph{Identifier Nodes}
(IN). In general, every occurrence of \verb+<identifier>+ in the right-hand sides of production rules in BNF is replaced by:

\begin{small}
\begin{verbatim}
    ( (A-Z) | (a-z) | (0-9) | "_" | "-" )+
\end{verbatim}
\end{small}

\noindent
There is, however, restrictions on these alphanumeric strings: they should not coincide with keywords of $\Delta$ language.

\subsubsection{Set name forks}

Let us recall the production rules related with \emph{full set names} represented by the syntactical category \verb+<set name>+. This important category, including some additional auxiliary productions, appears as follows:

\begin{small}
\begin{verbatim}
    <set name> ::=     <URI> "#" <simple set name>
    
    <URI> ::=          ( <web prefix> | <local prefix> ) <file path>
    
    <web prefix> ::=   "http://" <host> "/" [ "~" <identifier> "/" ]
    <local prefix> ::= "file://" ( (A-Z) | (a-z) ) ":/"
    
    <host> ::=         <identifier> [ "." <host> ]
    <file path> ::=    <identifier> ( "/" <file path> | <extension> )
    <extension> ::=    ".xml"
    
    <simple set name> ::=	<identifier>
    <identifier> ::=	( (A-Z) | (a-z) | (0-9) | "_" | "-" )+
\end{verbatim}
\end{small}

\noindent
Here all the syntactical categories, besides \verb+<set name>+, play an auxiliary role. Therefore, by composing them, all these production rules will produce two kind of one child forks for set names

\begin{footnotesize}
\begin{verbatim}
    <set name> ::= "http://... " "#" ( (A-Z) | (a-z) | (0-9) | "_" | "-" )+
\end{verbatim}
\end{footnotesize}

\noindent
or

\begin{footnotesize}
\begin{verbatim}
    <set name> ::= "file://... " "#" ( (A-Z) | (a-z) | (0-9) | "_" | "-" )+
\end{verbatim}
\end{footnotesize}

\noindent
Here \verb+"http://..."+ and \verb+"file://..."+ represent any string of symbols allowed by the \verb+<URI>+ production rule. Therefore, the production rule \verb+<set name>+ generates an infinite number of (one child) forks with the root \verb+<set name>+ and the leaf a string of characters as defined in the above productions.

\medskip

We will not consider other cases of defining BNF forks relying on the readers' intuition which should be  based on the above examples. Assertions~1-3 from the next section should summarise and give more understanding on the way which BNF forks are defined.

\subsubsection{Assertions on BNF forks}\label{sec:ass_bnf_forks}

After defining the set of forks of the BNF, we can make the following assertions.

\begin{ass:bnf}
Only Identifier Nodes (IN) can have just one child leaf labelled by a sequence of alphanumeric characters.
\end{ass:bnf}

\begin{proof}
Inspection of the whole BNF (and the definitions above) show that only IN can have just one child leaf labelled by a sequence of alphanumeric characters.
\end{proof}

Note that \verb+<set name>+ forks, although one child, have leafs containing non-alphanumeric characters \texttt{":"}, \texttt{"/"} and \texttt{"\#"}. 

\medskip

\begin{ass:bnf}
In fact, parsing of any given query generates a corresponding query parse tree constructed from these forks connected in the evident way. Here it is assumed that all keywords like "forall", "let", etc are included in the parse tree as terminals (except they are not allowed to be leafs of identifier forks).
\end{ass:bnf}

\medskip

\begin{ass:bnf}[Uniqueness of forks\footnote{This assertion will be used in the syntactical category renaming algorithm in Section~\ref{sec:SCR_algorithm}}]\label{ass:unique_forks}
Two different forks can have coinciding leaf labels (in the natural order) only if each of them is an identifier fork (see above). That is, if one of the two forks F1 and F2 is not an identifier fork and both forks have the same leaves then (their roots coincide and) F1 = F2. Or equivalently, the syntactic category of any fork, except for identifier forks, can be determined according to the syntactic categories of its children.
\end{ass:bnf}

\begin{proof}
We should check all possible cases. Assuming that two forks F1 and F2 have the same leaves and one of them has the root labelled not as identifier fork, show that F1 = F2.
\\\\
\noindent
\emph{Example:} If F1 or F2 has the root \verb+<quantified formula>+ then both have the same first leaf e.g. \verb+<forall>+ (or \verb+<exists>+). Then, according to the BNF, another fork must also have the root \verb+<quantified formula>+ and therefore F1 = F2, as required.
\\\\
\noindent
\emph{Example:} If F1 or F2 has the root \verb+<forall>+ then both have the same first leaf \verb+"forall"+ and the leaf \verb+"in"+ (or \verb+"<-"+). Inspection of all BNF forks shows that any fork containing both these leafs must have the root \verb+<forall>+. Therefore F1 = F2.
\\\\
\noindent
\emph{Example:} If F1 or F2 has the root \verb+<union>+ then both have the same first leaf \verb+"union"+ (or \verb+"U"+) and second leaf \verb+<delta-term>+. Inspection of all BNF forks shows that any fork containing both these leafs must have the root \verb+<union>+. Therefore F1 = F2.
\\\\
\noindent
All other cases follow as above.
\end{proof}

\begin{note}\label{note:ambiguity}\em
Despite this Assertion which means a kind of unambiguity of parsing (actually only a conditional and partial unambiguity) we will see in Section~\ref{sec:ambiguous_grammar} that parsing according to the BNF of $\Delta$ is actually quite ambiguous. This means that the same query can have parse trees of the same form, but with different labelling of nodes by syntactical categories. Later we will consider contextual analysis algorithm dealing with typing which will resolve this kind of ambiguity. 
\end{note}
}

\subsection{Query parsing}

The parser for the BNF syntax of the language Delta can easily be implemented which can transform any query $q$ into parse tree. The process of parsing $q$ involves matching of BNF production rules (represented rather in the form of forks defined above) starting at the root production rule for \verb+<top level command>+ until all possibilities are exhausted. The output of parsing the query $q$ is the query parse tree $qt$.

\medskip

During the process of parsing, successful matching of production rules creates new nodes in the parse tree connected by fork edges from the previous node, except for the root production rule which itself has no parent node. Successful matching of terminals creates new nodes labelled by the sequence of matched characters.

\subsubsection{Example query parse tree}\label{sec:Example query parse tree}

\noindent
Let us consider the simple example of query
\begin{small}
\begin{verbatim}
    boolean query 
      let label constant l='Robert'
      in l='Rob*'
    endlet;
\end{verbatim}
\end{small}

\noindent
and the corresponding query parse tree,


\begin{figure}[!ht]
\hspace{-2.5em} 
\centering
\includegraphics[scale=1.0]{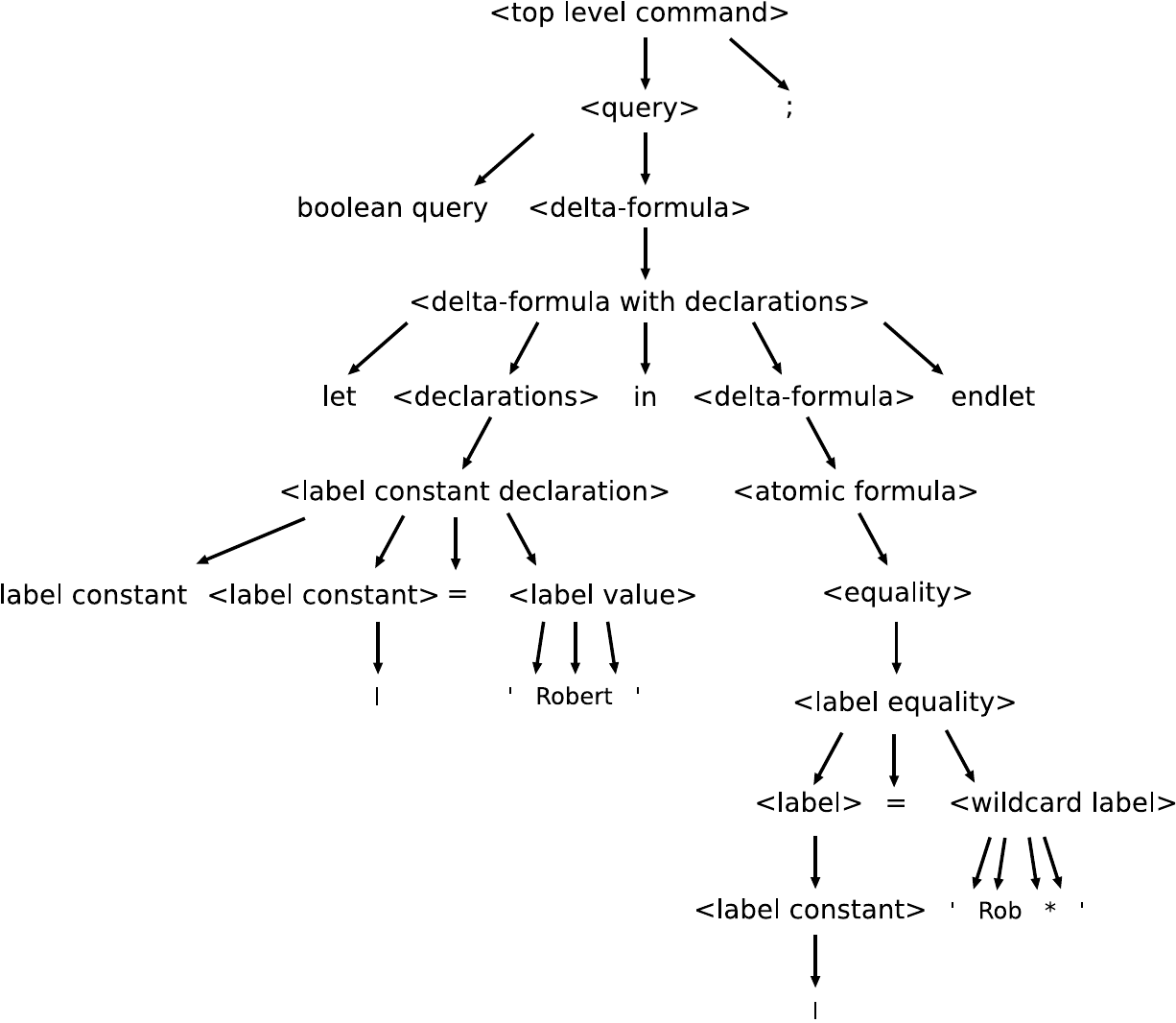}
\caption{Example parse tree}\label{fig:parse-tree}
\end{figure}

\noindent
Strictly speaking, some parts of this parse tree are omitted for brevity. Say, according to Section~\ref{sec:bnf_recursion}, between \texttt{<declarations>} and \texttt{<label constant declaration>} we should have a tree node \texttt{<declaration>}.

\subsubsection{Aims of query parsing}

Well-formedness of any query is determined according to the rules of the BNF grammar. However, when all possibilities for matching productions are 
unsuccessfully exhausted in any attempt to construct a parse tree then the query is considered as 
non-well-formed with appropriate error messages outputted.

\medskip

Moreover, to further aid contextual analysis (see Section~\ref{sec:well-typed_delta_queries}) the parser should output, in addition to the parse tree of the query, the list of all Identifier Nodes (see Section~\ref{sec:indentifier_forks}) in the parse tree labelled by:

\begin{small}
\begin{verbatim}
    <boolean query name>, <set query name>, <label variable>,
    <label constant>, <set variable>, <set constant>.
\end{verbatim}
\end{small}

\subsection{Parsing ambiguities}\label{sec:ambiguous_grammar}

The syntax of the implemented $\Delta$-language (expressed as BNF) is intended for any user to understand the constructs of $\Delta$, and how to write valid $\Delta$-queries -- well-formed and well-typed. However, the implemented parser alone cannot guarantee well-typedness of queries. Note that, well-typedness is checked by the contextual analysis algorithms described later in Section~\ref{sec:well-typed_delta_queries}.

\medskip

The problem is that the grammar of our implemented $\Delta$-language is ambiguous concerning types as we briefly commented this in Note~\ref{note:ambiguity} above. Thus, the typing of identifiers, say as label constant or variable, or set constant or variable, etc., is actually decided from the context. For example, let us consider the equality query: 

\begin{small}
\begin{verbatim}
    boolean query a=b;
\end{verbatim}
\end{small}

\noindent
Parsing of this query could realise two unique parse trees, where the statement \verb+a=b+ represents either \verb+<label equality>+ or \verb+<set equality>+. Thus, the syntactical category of this statement depends wholly on the interpretation of the identifiers \verb+a+ and \verb+b+ as either, label constants or variables, or set constants or variables, respectively. The parse tree presented above in Figure~\ref{fig:parse-tree} is also not unique one because the syntactic category \verb+<label constant>+ under \verb+<label>+ could be formally replaced according to syntax by \verb+<label variable>+, however, intuitively contradicting the label constant declaration \texttt{let label constant l = ...}.

\medskip

\noindent
Furthermore, let us even strengthen the above example,
\begin{small}
\begin{verbatim}
    boolean query let 
        label constant l='Robert', 
        label constant m='John'
      in 
        l=m endlet;
\end{verbatim}
\end{small}

\noindent
where the statement \verb+l=m+ intuitively represents the syntactic category \verb+<label equality>+ because according to the context the identifiers \verb+l+ and \verb+m+ are label constants. However, the BNF formally allows that \verb+<label equality>+ could be replaced with \verb+<set equality>+ and \verb+l+ and \verb+m+ are are taken as \verb+<delta-term>+s, independently of the declarations that \verb+l+ and \verb+m+ are both label constants. Even the following query can be formally parsed, i.e.\ is well-formed,
\begin{small}
\begin{verbatim}
    boolean query let 
        label constant l='Robert', 
        set constant m={}
      in 
        l=m endlet;
\end{verbatim}
\end{small}

\noindent
despite being evidently non-well-typed by equating label with set.

\medskip

Therefore, the syntax (expressed as BNF) alone is insufficient and requires guessing which rule to apply to make the parse tree (and to guarantee that the parsed query is)
\linebreak
\mbox{well-typed}. Therefore, such guesses by the parser should be subsequently checked, to ensure no contradictions with the actual typing of identifiers. Moreover, the syntactic categories of all nodes, not just IN, should be checked and possibly renamed (according to the grammar) without changing the structure of the parse tree. Such renaming is done by the \emph{contextual analysis algorithm}, detailed in
Section~\ref{sec:well-typed_delta_queries}, whose role is to ensure query well-typedness and eliminate potential ambiguities, as above.

\subsection{Grammar classification}
Note that the syntax of $\Delta$-query language, fully presented as BNF in Appendix~\ref{app:BNF}, can be classified as \emph{context-free grammar} according to Chomsky's definitions of formal languages. Taking the definition from the textbook about parsing \cite{W89}, all production rules of a context free grammar have the form:

\[
A \longrightarrow \gamma
\]

\noindent
where $A$ represents a unique non-terminal, and $\gamma$ represents an ordered list of terminals and/or non-terminals (possibly empty). Context free grammars are those where each non-terminal $A$ can be transformed by a production rule into corresponding $\gamma$ without any additional criteria of context. Our grammar satisfies this property and therefore cannot grasp contexts which are necessary for correct typing of queries. Thus, an additional contextual analysis algorithm working jointly with the parser is required which we discuss in the following section.

\section{Contextual analysis (well-typed queries)}\label{sec:well-typed_delta_queries}

\subsection{Aim of contextual analysis}\label{sec:aim_contextual_analysis}

The aim of contextual analysis is to determine whether every \emph{identifier occurrence} in a query $q$ is declared%
\footnote{\label{foot:free/closed}
An identifier occurrence in some expression $e$ (not necessary a full-fledged query; $e$ can be a fragment of a query $q$) which is non-declared inside $e$ can also be called \emph{free} in $e$, whereas those correctly declared inside $e$ identifier occurrences are called \emph{closed}.
Therefore the terms ``declared'' and ``closed'', and ``non-declared'' and ``free'', have the same meaning. (This agreement on terminology is, however, non-traditional in the particular case of (set or label) constants for which it is more habitual to use the terms ``declared'' or ``non-declared'' instead of ``closed'' or ``free''.) We assume that each full-fledged query $q$ must be closed in this sense (all its identifiers must be declared inside $q$).
}%
, thereby having type, and whether the whole query is well-typed (all types are coherent). Each identifier occurrence should be appropriately typed as either: \emph{set constant} or \emph{variable}, \emph{label constant} or \emph{variable} or \emph{query name} of some type%
\footnote{
To simplify terminology, we consider \emph{variable} or \emph{constant} or \emph{query name} as typing information of some identifier, alongside the proper types \emph{set} or \emph{label} or \emph{boolean} or the complex type (\ref{eq:query-type}).
}%
. Note that query names can have more complicated types than variables or constants,

\begin{equation}\label{eq:query-type}
(type_1, type_2, ... , type_n \longrightarrow type)
\end{equation}

\noindent
where each participating $type_i$ is either \emph{set} or \emph{label}, and $type$ after the arrow is either \emph{set} or \emph{boolean}%
\footnote{
Note that, we formally have no queries or query names in $\Delta$ of the type \emph{label}. However, label values can be represented in the same way as atomic values, i.e.\ as singleton sets of the form $\setof{l:\emptyset}$.
}%
. Each $type_i$ is the expected type of $i$-th parameter of the query name $q$, and $n$ is the required number of parameters -- according to the declaration of this query name. From this type it should be already clear that the identifier $q$ is a (set or boolean) query name, how many arguments it has, and the typing of each argument.

\medskip

Furthermore, an identifier occurrence is considered declared if it is contained within the scope of an appropriate identifier declaration, and well-typed if both the identifier occurrence and identifier declaration have the same types. Moreover, for query to be well-typed, coherence of typing (for equalities, as in the examples above, membership statements and query calls) should be additionally required.

\subsubsection{Strategies for computing contextual analysis}

In principle there are two possible algorithms for performing  contextual analysis of any query $q$, both algorithms are named after the way in which they walk the parse tree of $q$:

\begin{itemize}
\item{
{\bf Top-down contextual analysis} -- The parse tree is walked in breadth first manner starting at the root node, creating a list of the identifier declarations (called the context) which is used to check that all other identifier occurrences are closed and well-typed according to these declarations.
}
\item{
{\bf Bottom-up contextual analysis} -- Walking of the parse tree starts from any identifier occurrence leaf $i$%
\footnote{
For example, the second leaf labelled by the identifier $l$ in Fig.~\ref{fig:parse-tree} above
}
ascending up the corresponding branch of the parse tree, searching for an identifier declaration which declares $i$%
\footnote{
In Fig.~\ref{fig:parse-tree} above the corresponding node would be \texttt{<delta-formula with declarations>} having the declaration of the label constant $l$ under it. Note that quantifiers and other quantifier-like constructs, called binders (see Section~\ref{sec:context_definitions}), are also considered as identifier declarations.
}%
. The existence of a corresponding identifier declaration indicates that the identifier occurrence is declared. Moreover, the real types of all such $i$ can be extracted from the corresponding declarations and compared with syntactical categories of these nodes $i$ in the parse tree. In the case of coherence, the parse tree and hence the query is considered \emph{well-typed}. Otherwise, syntactical categories of the parse tree nodes could be possibly corrected by (another bottom-up procedure of) renaming syntactical categories of some non-leaf nodes by the iterative algorithm described below in Section~\ref{sec:contextual_in_detail}. If such a renaming is successful -- giving rise to a correct parse tree according to both the BNF and the typing, then the resulting version of tree and the original query are also considered \emph{well-typed}, otherwise \emph{non-well-typed}.
}
\end{itemize}

\subsection{Some useful definitions}\label{sec:context_definitions}

\begin{defn}[Identifier Node]\label{def:identifier_node}
\emph{Identifier Nodes} (IN) were introduced in Section~\ref{sec:indentifier_forks}, as those nodes in the parse tree labelled by one of the following syntactic categories:
\begin{small}
\begin{verbatim}
    <boolean query name>, <set query name>, <label variable>,
    <label constant>, <set variable>, <set constant>.
\end{verbatim}
\end{small}

\noindent
Additionally, let us define \emph{Identifier Node Name} (INN) as string of symbols labelling the unique child (in fact, a leaf called above as $i$) of the corresponding IN fork in the parse tree.
\end{defn}

\begin{defn}[Binder Node]\label{def:binder_node}
\emph{Binder (or binding) Nodes} (BN) are those nodes in the parse tree labelled by one of the following syntactic categories:
\begin{small}
\begin{verbatim}
   <delta-term with declarations>, <delta-formula with declarations>,
   <collect>, <separate>, <recursion>, <quantified formula>.
\end{verbatim}
\end{small}

\noindent
Binder nodes can have appropriate declarations like \verb+"let..."+, \verb+"forall..."+, 
\linebreak
\verb+"exists..."+, etc., as described in Definition~\ref{def:identifier_declaration_node}, and thereby can \emph{bind} identifier occurrences (or IN).
\end{defn}

\begin{figure}[!ht]
\hspace{-1.5em} 
\centering
\includegraphics[scale=0.90]{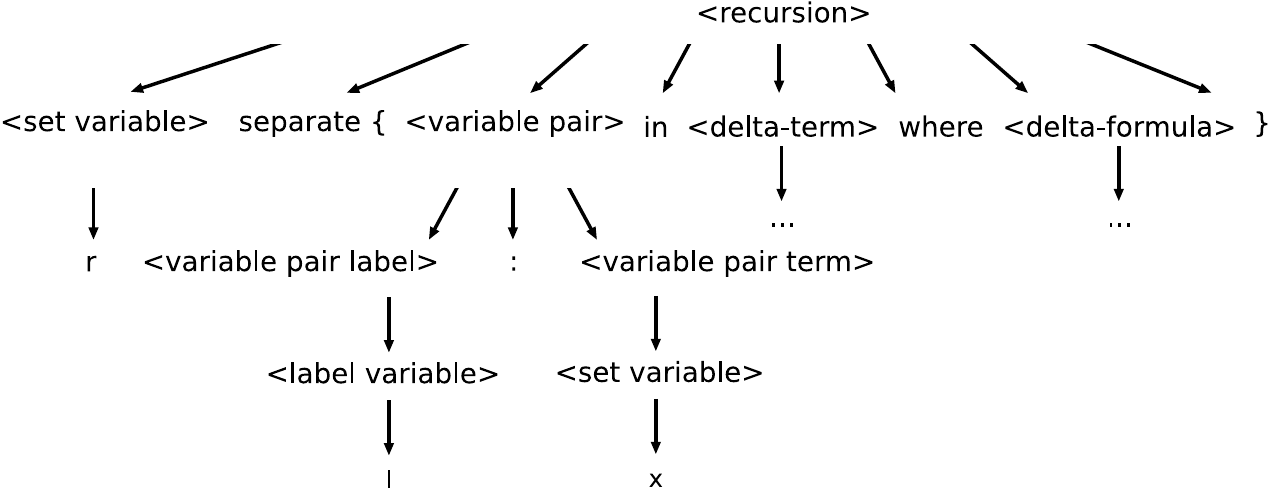}
\caption{Fragment of a query parse tree}\label{fig:fragment_parse_tree}
\end{figure}

\begin{defn}[Identifier Declaration Node]\label{def:identifier_declaration_node}
Following from Definition~\ref{def:binder_node} those declarations belonging to BN are called \emph{identifier declarations nodes (IDN) of a BN} and defined as follows.
\begin{itemize}
\item{
For BNs \verb+<delta-formula with declarations>+ with \verb+"let"+ declaration(s), and \verb+<delta-term with declarations>+ with \verb+"let"+ declaration(s) the IDNs are:
        \begin{itemize}
        \item{\verb+<set constant declaration>+ grandchild of \verb+<declarations>+,}
        \item{\verb+<label constant declaration>+ grandchild of \verb+<declarations>+,}
        \item{\verb+<set query declaration>+ grandchild of \verb+<declarations>+, and}
        \item{\verb+<boolean query declaration>+ grandchild of \verb+<declarations>+.}
        \end{itemize}
}
\item{
For BNs \verb+<separate>+ and \verb+<collect>+ the IDNs are:
        \begin{itemize}
        \item{\verb+<label variable>+ grandchild of \verb+<variable pair>+, and}
        \item{\verb+<set variable>+ grandchild of \verb+<variable pair>+.}
        \end{itemize}
}
\item{
For BN \verb+<recursion>+ the IDNs are:
        \begin{itemize}
        \item{\verb+<set variable>+ child of \verb+<recursion>+,}
        \item{\verb+<label variable>+ grandchild of \verb+<variable pair>+, and}
        \item{\verb+<set variable>+ grandchild of \verb+<variable pair>+.}
        \end{itemize}
}
\item{
For BN \verb+<quantified formula>+ the IDNs are:
        \begin{itemize}
        \item{\verb+<label variable>+ grandchild of \verb+<variable pair>+, and}
        \item{\verb+<set variable>+ grandchild of \verb+<variable pair>+.}
        \end{itemize}
}
\end{itemize}

\noindent
For example, Figure~\ref{fig:fragment_parse_tree} depicts a fragment of a query parse tree, where the root node
\linebreak
\verb+<separate>+ is a BN and the corresponding IDN nodes (described above) can 
be found by walking the paths from the \verb+<separate>+ node,
\begin{itemize}
\item[]{
\verb+<variable pair>+ $\rightarrow$ \verb+<variable pair label>+ $\rightarrow$ \verb+<label variable>+
}
\item[]{
\verb+<variable pair>+ $\rightarrow$ \verb+<variable pair set>+ $\rightarrow$ \verb+<set variable>+
}
\end{itemize}
\noindent
All other cases follow as the above. Note that there may be many IDNs of a given BN. Any IDN declares one or more identifiers (IN) each of which has its name as a string of symbols (the leaf under~IN).

\end{defn}

\begin{defn}[Bounding Term or Formula or Label Value Node ]\label{def:binding_term_node}
\hfill
\begin{itemize}
\item[(a)]
Following from Definition~\ref{def:binder_node}, the \emph{bounding} term or formula or label value nodes 
\linebreak
(\mbox{BTFLVN}) of a BN
\begin{verbatim}
     <collect>
     <separate>
     <recursion>
     <quantified formula>
     <delta-term with declarations> 
     <delta-formula with declarations>
\end{verbatim}
\noindent
is defined, respectively, as 
\begin{itemize}
\item{
a unique \verb+<delta-term>+ child of:
	\begin{itemize}
	\item{\verb+<collect>+ or \verb+<separate>+ or \verb+<recursion>+ or}
	\item{\verb+<forall>+ child of \verb+<quantified formula>+ or}
	\item{\verb+<exists>+ child of \verb+<quantified formula>+ or}
	\item{any \verb+<set constant declaration>+  grandchild of 
	\\
	\verb+<delta-term with declarations>+ or 
	\\
	\verb+<delta-formula with declarations>+ or
	}
	\item{any \verb+<set query declaration>+ grandchild of 
	\\
	\verb+<delta-term with declarations>+ or 
	\\
	\verb+<delta-formula with declarations>+, or
	}
	\end{itemize}
}
\item{
a unique \verb+<label value>+ child of:
	\begin{itemize}
	\item{any \verb+<label constant declaration>+ grandchild of
	\\
	\verb+<delta-term with declarations>+ or 
	\\
	\verb+<delta-formula with declarations>+ or
	}
	\end{itemize}
}
\item{
a unique \verb+<delta-formula>+ child of:
	\begin{itemize}
	\item{any \verb+<boolean query declaration>+ child of 
	\\
	\texttt{<delta-term with declarations>} or 
	\\
	\texttt{<delta-formula with declarations>}.
	}
	\end{itemize}
}
\end{itemize}

\item[(b)]
Each BTFLVN of a BN restricts the range of the value of some INs (variables, constants or query names) which BN binds%
\footnote{Which was briefly hinted in the Definition~\ref{def:binder_node}
}
and which we also call \emph{bounded or restricted IN(s) by the BTFLVN}%
\footnote{
Moreover, the IN bounded by BTFLVN should not be free in the BTFLVN (i.e., if present in the BTFLVN, it should be declared inside this BTFLVN)
as we will discuss later as one of the conditions to be checked by contextual analysis algorithm. This is the reason why we need Definition~\ref{def:binding_term_node}.
}%
. These INs are defined as follows:
\begin{itemize}
\item{
In the case of BNs \texttt{<collect>}, \texttt{<separate>}, \texttt{<recursion>} and 
\linebreak
\texttt{<quantified formula>}, the bounded INs are respectively 
\linebreak
\texttt{<label variable>} and \texttt{<set variable>} grandchildren of \texttt{<variable pair>}.
}
\item{
Additionally, in the case of BN \texttt{<recursion>} one more bounded IN is its immediate \texttt{<set variable>} child.
}
\item{
In the case of BNs \texttt{<delta-formula with declarations>} or
\linebreak
\texttt{<delta-term with declarations>}, the bounded IN is either the declared \texttt{<set constant>} or \texttt{<label constant>}, or \texttt{<set query name>}, or 
\linebreak
\texttt{<boolean query name>}.
}
\end{itemize}
\end{itemize}
\end{defn}

\medskip

\noindent
For example, Figure~\ref{fig:fragment_parse_tree} depicts the query parse tree for an expression $e$ (fragment of a query $q$), where the root node \verb+<recursion>+ is a BN and the corresponding BTFLVN and the bounded INs can be found by walking the paths,
\begin{itemize}
\item[]{\verb+<recursion>+ $\rightarrow$ \verb+<delta-term>+  (BTFLVN)}
\item[]{\verb+<recursion>+ $\rightarrow$ \verb+<set variable>+ (IN)}
\item[]{
\verb+<recursion>+ $\rightarrow$ \verb+<variable pair>+  $\rightarrow$ 
\\
\verb+     <variable pair label>+ $\rightarrow$ \verb+<label variable>+ (IN)
}
\item[]{
\verb+<recursion>+ $\rightarrow$ \verb+<variable pair>+ $\rightarrow$ 
\\
\verb+     <variable pair term>+ $\rightarrow$ \verb+<set variable>+ (IN)
}
\end{itemize}

\noindent
whereas \texttt{<label variable>} ($l$) and \texttt{<set variable>} ($x$) are INs bounded by this \linebreak \verb+<delta-term>+ (BTFLVN). Additional (recursion) \texttt{<set variable>} ($r$) is IN also
\linebreak
bounded by \verb+<delta-term>+ (BTFLVN).

\subsection{Bottom-up contextual analysis in detail}\label{sec:contextual_in_detail}

As stated in the brief description in Section~\ref{sec:aim_contextual_analysis}, contextual analysis should check that the given well-formed query (according to the parser) is also well-typed. To this end, the bottom-up contextual analysis algorithm, first of all, iteratively searches for the nearest identifier declaration for each identifier occurrence, i.e. each IN in the parse tree. We assume that before starting contextual analysis the parser generates a list of all INs (not those INs of the declarations in IDNs) along with their currently chosen typing (immediately seen from syntactical categories of these INs, say, \verb+<set variable>+, etc.) during the parsing process. The parser outputs this list if the query is well-formed.

\subsubsection{Identifier declaration search (IDS) algorithm}\label{sec:IDS_algorithm}

Single iteration of the search for the nearest identifier declaration of an IN is determined by the \emph{Identifier Declaration Search} (IDS) algorithm. The inputs to this algorithm is any $qt$ (query parse tree) and some IN in $qt$. The output of the IDS algorithm is the ordered triple $<BN, IDN, IN>$ (if the required one exists at all) consisting of: BN (Binding Node), IDN (Identifier Declaration Node) and the given IN.

\medskip

Note that, IDN contains typing information of the declared identifier (including the information whether it is a constant or variable, or a query name -- also a kind of typing information). In fact, the IDN is recoverable from BN and IN in the parse tree, however, it is convenient to have IDN included in the triple obtained during this process.


\noindent
\paragraph*{\bf Identifier Search Algorithm $IDS(qt,IN)$:}

\begin{enumerate}
\item[]{
{\bf START with a given IN belonging to $qt$.}
}
\item{
{\bf Make this node (IN) the \emph{current node}.}
}
\item{
{\bf Ascend from the \emph{current node} traversing up $qt$} to its unique parent node, making this node the \emph{current node}.
}
\item{
{\bf Is the \emph{current node} a BN?}
\\
\\{\bf No} -- Move to step~\ref{IDS_SEARCH_ROOT}.
\\
\\{\bf Yes} -- Iterate from right to left through IDNs of the BN, searching for the first%
\footnote{
Formally, it is not forbidden that the same identifier name could be multiply declared even in the same binder, but only the right most one is that which binds the IN considered and which assigns a type to IN.
}
suitable candidate identifier declaration whose declared identifier has the same name (INN) as the given IN. If a suitable candidate IDN exists then construct the ordered triple $<BN, IDN, IN>$ (end of algorithm), \emph{otherwise} move to step~\ref{IDS_SEARCH_ROOT}.
}
\item\label{IDS_SEARCH_ROOT}{{\bf Is the \emph{current node} the root node of $qt$?}
\\
\\{\bf Yes} -- No suitable candidate identifier declaration could be found, and therefore, the IN is non-declared. Output ordered triple $<NULL, NULL, IN>$ (end of algorithm).
\\
\\{\bf No} -- Continue searching for a suitable identifier declaration by moving to step~2.
}
\item[]{
{\bf END with the ordered triple $<BN, IDN, IN>$ if a suitable identifier declaration exists, otherwise with $<NULL, NULL, IN>$.}
}

\end{enumerate}

\noindent
The IDS algorithm should iteratively generate the triples as above for all INs (actually, for those identifier occurrences not in a declaration) of the given parse tree $qt$. If all these are non-null triples then the query $q$ is considered as \emph{closed} (yet possibly not well-typed). Thus, any closed query $q$ has all INs declared with preliminary typing according to the declarations (IDN)  from the corresponding triples. For non-closed query an error message should be generated by the implementation saying that the query has non-declared identifiers. Moreover, any closed query $q$ and its parse tree $qt$ are considered also well-typed if all identifiers have coherent typing both in respect to their corresponding declarations and syntactical categories of 
the parse tree~$qt$. More precisely, this means that:


\begin{enumerate}
\item{
Syntactical categories of IN (e.g.\ \texttt{<set variable>} or \texttt{<boolean query name>}, etc.) should be the same as declared in IDN (in  corresponding triple), and
}
\item{
Types of participating parameters in query calls should agree with types discovered from IDNs declaring corresponding query names.
}
\end{enumerate}

\noindent
If these two clauses do hold then in other nodes the BNF itself supports correct typing and/or syntactical categories (such as \texttt{<set equality>} vs.\ \texttt{<label equality>}, etc.). Otherwise, an appropriate renaming of syntactical categories of the nodes in $qt$ should be tried (as detailed in the next section), based on the initial partial correcting only the discrepancies in the clauses~(1)~and~(2), with the aim to recover well-typed version of $qt$ and conclude that the query $q$ is \emph{well-typed}. If such a renaming is impossible, then $q$ is considered as \emph{non-well-typed}.

\subsubsection{Syntactic category renaming (SCR) algorithm}\label{sec:SCR_algorithm}

It is required that renaming should lead to a correct parse tree. This means that the 
\emph{syntactic category renaming} (SCR) algorithm,
\begin{itemize}
\item{
takes a parse tree with some already correctly renamed nodes (such as INs, by removing the discrepancies mentioned above, and may be some other nodes as we will see below) and formally marked as ``correct'', and
}
\item{
if necessary, attempts to rename other nodes ensuring that the parse tree remains faithful to the $\Delta$-language BNF syntax (well-formed).
}
\end{itemize}

\noindent
Thus, the \emph{input} is a given parse tree $qt$ with some (non-leaf) labels \emph{already relabelled}$\,$%
\footnote{
Note that, INs are formally non-leaf nodes, although neighboring to leafs. As we will see below in Section~\ref{sec:main_cont_anal_algo}, not only INs should be initially relabelled in the input parse tree. These may be also query call \texttt{<parameters>} which, unlike INs, may be far away from leaves in the parse tree.
} 
and additionally marked as ``correct'', with the output being either: (i) parse tree with all other nodes successfully relabelled ($q$ is well-formed), or (ii) an error state ($qt$ is inconsistent with the $\Delta$-language syntax, even after further relabelling).

\medskip

The procedure of relabelling starts from the leafs of the parse tree, and, while going bottom-up along the tree relabels according to the $\Delta$-language BNF syntax (if necessary) those nodes which have not already been relabelled. Newly relabelled nodes are additionally marked as ``correct'', and visited nodes are marked also as ``seen'' as described formally below. At each stage of the computation some nodes are already marked by this procedure as ``correct'', and only a node $N$ can be relabelled and then also marked as ``correct'' and ``seen'' which, (i) has not yet marked as ``seen'' (although probably marked as ``correct'' by the input marking), and (ii) all its children, $Children(N)$, have already marked as both ``seen'' and ``correct''.


\paragraph*{Syntactical renaming algorithm $SCR(qt)$:}

\begin{enumerate}
\item[]{{\bf START with parse tree $qt$.}}
\item{
{\bf Initially mark some nodes as ``seen'' and ``correct''.} Mark all leaf nodes,  INs, IDNs and \verb+<set name>+ nodes both as ``seen'' and ``correct''%
\footnote{
In fact, as we discussed above, INs and query call \texttt{<parameters>} are already marked as correct in the input parse tree $qt$.
}%
.
\\
\\{\bf Note:} Syntactic categories of ``correct'' nodes will not be renamed by this algorithm. Furthermore, \verb+<set name>+ nodes should not be renamed (and thus, these are initial marked as ``correct'') as they evidently have unambiguous type \emph{set} and definitely require no renaming.
}
\item\label{SCR_SUITABLE_NODES}{
{\bf Find any node suitable for correcting.} Find node $N$, which is not marked as ``seen'', and whose all children are marked both as ``correct'' and ``seen'' (giving rise to a fork $N \longrightarrow Children(N)$ in $qt$). Does the required $N$ exist in $qt$?
\\
\\{\bf No} -- Therefore, by induction, all nodes in the tree are already marked as ``correct'', (end of algorithm).
\\
\\{\bf Yes} - Check and (if necessary, and possible) rename according to BNF the syntactical category of $N$:
	\begin{enumerate}
	\item\label{SCR_SUITABLE_F}{
	{\bf Find a suitable fork $F$ in the BNF that matches the children of $N$.}
	Find a fork $F$ from the BNF whose leaves match with $Children(N)$.
	As N is not an identifier node, it follows from Assertion~\ref{ass:unique_forks} 
	from Section~\ref{sec:bnf_forking} that there 
	can exists only one such fork $F$, if any.
	
	\smallskip
	
	{\bf If} the required fork $F$ does not exist in the BNF, output error message ``query is not well-typed'' 
	indicating the statement in the query $q$ corresponding to the node $N$ which 
	``cannot be properly typed'', and halt (end of algorithm).
	
	\smallskip
	
	{\bf Otherwise}, if $F$ exists,  move to step~\ref{SCR_NOT_CORRECT}~or~\ref{SCR_CORRECT} depending on 
	whether $N$ is already marked as ``correct'' or not.
	
	\smallskip
	
	{\bf Note:} The term `matching' means that the branching degree should be the same and the matching 
	children nodes (in the natural order) have the same labels.
	The labels of $N$ and the root of $F$ are not required to coincide for matching to be successful.
	}
	\item\label{SCR_NOT_CORRECT}{
	{\bf \emph{N} is not marked as ``correct''} - relabel syntactical category of $N$ exactly as the root of $F$,
	mark $N$ as ``correct'' and ``seen'', and move to step~\ref{SCR_SUITABLE_NODES}.
	}
	\item\label{SCR_CORRECT}{
	{\bf \emph{N} is marked as ``correct''} - if the label of the root of $F$ \emph{coincides} with
	the label on $N$ then mark $N$ also as ``seen'' and move to step~\ref{SCR_SUITABLE_NODES}.
	
	However, if the label of the root of $F$ \emph{differs} from the label on $N$, generate the error 
	message ``query is not well-typed; conflicts with the expected syntax'' and indicate which syntactic category name 
	(and corresponding place in the query) requires renaming. (End of algorithm.)
    }
  \end{enumerate}
}
\item[]{
{\bf END with either correctly relabelled parse tree, or an appropriate error state.}
}
\end{enumerate}

\noindent
The successful result of this algorithm would give us a full guarantee that the resulting relabelled tree is still the correct parse tree of the given query which is therefore well-formed. Most importantly%
\footnote{
also, taking into account appropriate renaming of syntactical categories of query call \texttt{<parameters>} considered below
}%
, it will also guarantee that the query is well-typed: parse tree labelling is fully coherent, both with the typing and all other details in declarations of identifiers (such as to be a constant or variable or query name).

\subsubsection{Contextual analysis algorithm}\label{sec:main_cont_anal_algo}

The complete algorithm for bottom-up contextual analysis consists of the following (macro) steps. The input is any query parse tree $qt$ and the list of INs (both obtained from the parser). The output being either: (i) correctly relabelled query parse tree ($q$ is well-typed), or (ii) an error message ($q$ is non-well-typed).

\medskip

\noindent
{\bf Contextual analysis algorithm $CA(qt,\mathrm{the\ list\ of\ INs})$:}

\begin{enumerate}
\item[]{
{\bf START with the list of INs of the query parse tree $qt$.}
}
\item\label{CA_IDS_SEARCHES}{
{\bf Find suitable candidate declaration (BN and IDN) for each identifier occurrence (each IN).} That is, iterate over the given list of INs calling IDS algorithm for each IN (see Section~\ref{sec:IDS_algorithm}). The result of these identifier declaration searches is the list of declaration triples for all INs.

For those INs for which the algorithm IDS outputs $< NULL, NULL, IN >$ the corresponding error messages ``identifier non-declared'' should be outputted concerning all such identifier occurrences in the query $q$ and additionally that the ``query is not well typed''. 

If IDS outputted NULL triple for some IN then end of algorithm; otherwise move to step~\ref{CA_RELABEL}.
}
\item\label{CA_RELABEL}{{\bf Relabel syntactical categories of some parse tree nodes according to step~\ref{CA_IDS_SEARCHES}.}
	\begin{enumerate}
	\item{
	{\bf Relabel syntactical categories of identifier occurrences.}
	Labels of nodes (i.e.\ syntactical categories) generated by the parser contain the preliminary 
	information on the typing (assigned by the parser and possible contradicting the actual type). 
	The real typing of any IN and, in fact, the real syntactical categories (the node labels) of the INs
	can be correctly determined using the IDN from the declaration triple of IN.
	The parse tree labelling for these INs should be updated accordingly 
	(may be vacuously if the given IN, in fact, does not need updating 
	according to the IDN) with marking these nodes as ``correct''.
	This can be done straightforwardly for all INs (in particular for query names to be discussed below).
	Thus after relabelling, all INs will be actually marked as ``correct''.
	}
	\item{{\bf Relabel syntactical categories of query call parameters}%
\footnote{
In some cases similar to query parameters the parser already assumes some typing. For example, in the membership statement $l:a \in b$ the syntactical categories of $l,a,b$ must be, respectively, \texttt{<label>}, \texttt{<delta-term>} and \texttt{<delta-term>}, according to the BNF. In the case of equality $a=b$, the expressions $a$ and $b$ must be of the same type according to BNF, although the choice of type is ambiguous as shown by those examples in Section~\ref{sec:ambiguous_grammar}. But, the case of query call parameters requires our special attention in the currently described algorithm.
}{\bf.}
	In the case of INs which are query names in query calls some additional renaming of some (possibly)
	non-IN nodes (query parameters) is required as described below.
	\\
	\\If we have a query call $q(t_1, ..., t_n)$ with the query name $q$ of the type
\[
(type_1, type_2, ..., type_m \longrightarrow type)
\]
	obtained from the appropriate IDN by the algorithm IDS (where all participating $type_i$ are
	\emph{set} or \emph{label}, and the type after arrow is \emph{set} or \emph{boolean})
	then we should:
		\begin{enumerate}
		\item{
		Check whether $m=n$; if not, the query is not well-typed, and the algorithm should halt with an appropriate error message.
		}
		\item{
		If $m=n$, rename (possibly vacuously) syntactical categories of parameter nodes $t_i$ 
		(\verb+<delta-term>+ or \verb+<label>+) according to the types $type_i$ (\emph{set} or 
		\emph{label}), and mark them as ``correct''.
		}
		\end{enumerate}
	}
	\end{enumerate}
}
\item{
{\bf Relabel syntactical categories of all other parse tree nodes.} Apply SCR algorithm (Section~\ref{sec:SCR_algorithm}) to the resulting partially relabelled parse tree. Thereby other nodes of the parse tree will also be potentially renamed.
	\begin{enumerate}
	\item{{\bf Were all other nodes successfully renamed?}
	
	{\bf Yes} - If the SCR algorithm renamed and marked all nodes as ``correct'', 
	then move to Step~\ref{CA_ADDITIONAL_REQUIREMENTS}
	to check for additional 
	requirement (that query is properly ``bounded'').
	
	{\bf No} - Parsing agreeing with typing is impossible, and appropriate error messages from 
	SCR algorithm should be outputted. End of algorithm.
	}
	\end{enumerate}
}
\item\label{CA_ADDITIONAL_REQUIREMENTS}{
{\bf Additional requirements on bounding terms or formulas (BTFLVNs)}
	\begin{enumerate}
	\item{
	{\bf Check that (the names of) bounded identifiers (INs) of: \texttt{<separate>},
	\linebreak
	\texttt{<recursion>}, \texttt{<collect>}, \texttt{<delta-formula with declarations>},
	\linebreak
	\texttt{<delta-term with declarations>}, and 
	\texttt{<quantified formula>}
	\linebreak
	have no non-declared occurrences inside the bounding term or formula \\ (BTFLVN)}.
	\\ 
	\\
	For convenient implementation of this clause we assume additionally that the parser also generates
	for each bounding term or formula (BTFLVN) the sub-list of INs (from the list of all INs generated by 
	the parser) lying \emph{under} BTFLVN in $qt$%
\footnote{
If BTFLVN is LVN -- a label value node -- then this list is, of course, empty.
}.
	In other words, these are some of the identifiers occurring in the query~$q$.
	This can be represented as lists (for each BTFLVN) of the form:
\[
< BTFLVN, IN_1,\ldots, IN_k >.
\]
	Using the list of these IN$_i$ under the given BTFLVN and the declaration triples of the form 
	$<BN,IDN,IN>$ generated by the IDS algorithm, it should be checked that each IN$_i$ from the above 
	list whose name coincides with the name of some bounded IN by the given BTFLVN 
	(see Definition~\ref{def:binding_term_node}~(b)) is declared in this BTFLVN.
	The latter means that such an IN has its own binding node BN (from the appropriate unique triple),
	and this BN lies under or coincides with the given BTFLVN.
	This should hold for each BTFLVN in $qt$.
	Otherwise contextual analysis should be aborted with corresponding error message.
	
	In particular, in the case or recursion, we should check that the recursion binding set variable,
	as well as variables from the binding variable pair, do not occur free in the bounding term.
	Also, each query name should not occur free in the defining term or formula, and set constant
	should not occur free (non-declared) in the defining term, etc.
	However, in the case of set constants and query names we need to add the following additional
	requirements.
	}
	\item{
	Check that for each \texttt{<set constant declaration>} the defining
	\linebreak
	\texttt{<delta-term>} 
	has all of its 
	set or label \emph{variables} declared within this term. That is, intuitively,
	\texttt{<delta-term>} defining a set constant should have a constant value. However, constants 
	and query names  inside this \texttt{<delta-term>} may be declared in the query outside this term.
	
	To do this, use the list of INs of \emph{variables} lying under the node 
	\texttt{<delta-term>} of \texttt{<set constant declaration>} 
	and the identifier declaration triples of the form $<BN,IDN,IN>$ generated by the above 
	IDS algorithm, and check that each BN of such a variable IN lies in the \texttt{<delta-term>} 
	node subtree. Otherwise, such a variable IN of the \texttt{<delta-term>} 
	is considered as free, and the contextual analysis should be aborted 
	with the corresponding error message.
	}
	\item{
	Check that for each \texttt{<set query declaration>} the defining 
  \linebreak 
	\texttt{<delta-term>} 
	has all its set or label 
	\emph{variables} declared (quantified, etc.) either inside this term or in the given 
	\texttt{<set query declaration>} as
	\linebreak
	\texttt{<variables>} parameters of the declared query. Constants, and 
	query names inside this \texttt{<delta-term>} may be declared in the query outside this term.
	Quite similarly check for each \texttt{<boolean query declaration>} and corresponding \texttt{<delta-formula>}.
	}
	\item{
	The remaining check that \verb+<label constant declaration>+ uses closed \verb+<label value>+ 
	is evidently vacuous, as actually there is nothing to check.
	}
	\end{enumerate}
}
\item[]{
{\bf END with a correctly relabelled and well-typed and properly bounded parse tree (``query is well-formed and well-typed''), or a partially relabelled parse tree plus additional error messages (``query is well-formed but not well-typed'', etc.).}
}
\end{enumerate}

\subsection{Extension of contextual analysis to support libraries}

That the library declarations are well-formed and well-typed can be checked by reducing these declarations to the ordinary queries, as it was shown in Section~\ref{sec:library-implementation}, and applying parsing and contextual analysis algorithm described above to the resulting query.


\chapter{XML Representation of Web-like Databases (XML-WDB Format)}\label{chap:xml-wdb}

\nocite{ME06}
\nocite{SF05}

\section{Represention of WDB by graph or set equations}\label{sec:hypersets_approach_to_wdb}

As we discussed in Chapter~\ref{chap:ssd_wdb} the (hyper)set theoretic approach \cite{LS01,LS97,LS99,S87,S93,S06} to WDB is based on the concept of hereditary finite sets or, more generally, hyperset theory \cite{A88,BM96}. Such semi-structured data is represented as abstract sets (of sets of sets, etc.) with the possibility for membership relation to form cycles. 

\begin{figure}[ht]
\centering
\includegraphics{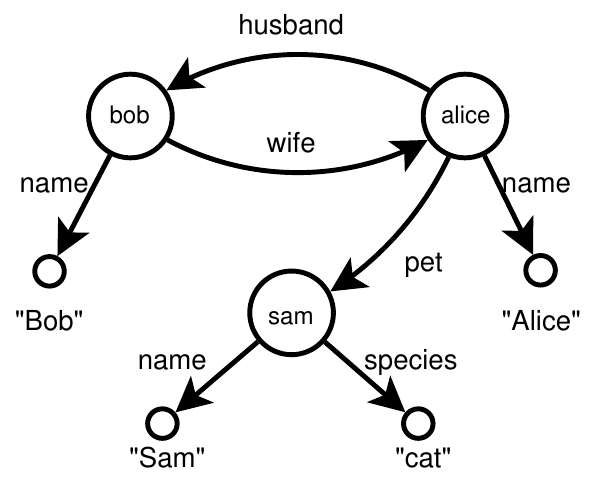}
\caption{Example WDB representing a fictitious family}\label{fig:WDB_family}
\end{figure}

\noindent
For visualisation purposes hyperset databases are represented as \emph{graphs} (see Figure~\ref{fig:WDB_family}) where nodes correspond to set  names and labelled edges to membership relation. When considering implementation (and also intuitively from the set theoretic view) it is far more appropriate to represent WDB as \emph{system of set equations}. Each set equation consists of a \emph{set name} equated to a \emph{bracket expression}; \emph{labelled elements} of such sets may be either atomic values, nested bracket expressions, or set names (described in some other equations). For example, system of \emph{flat} set equations corresponding to the WDB graph in Figure~\ref{fig:WDB_family} looks as follows:

\begin{verbatim}
    bob   = { name:"Bob", wife:alice }
    alice = { name:"Alice", husband:bob, pet:sam }
    sam   = { name:"Sam", species:"cat" }
\end{verbatim}

\noindent
or, equivalently, with the \emph{nesting} allowed: 
\begin{verbatim}
    bob   = { name:"Bob", wife:alice }
    alice = { name:"Alice", husband:bob,
              pet:{name:"Sam", species:"cat"} }
\end{verbatim}

\noindent
In particular, this demonstrates that the specific form of set names (e.g. \verb+bob+, \verb+alice+,  \verb+sam+) however helpful intuitively are formally not important. They can always be renamed (say by numbered ``object identities'' e.g. \verb+&23+, etc.) or substituted as above. In general, the role of set names in any system of set equations depends on its position. Those set names occurrences on the left-hand side of set equation (simple set names) are also called \emph{defined} set names, whereas, all other set name occurrences are called \emph{referenced} set names. Each referenced set name should be defined somewhere in the system, and only once.

\medskip

The implemented query system considers WDB as systems of flat set equations (without any nesting). As described below, WDB is represented practically as a system of XML files each containing a fragment of the whole system of set equations of the WDB, which proves convenient. From the perspective of any database designer, the informational content of WDB is carried by:
\begin{itemize}
\item{Labels on WDB-graph edges e.g. \verb+name+, \verb+wife+, \verb+husband+, etc.}
\item{Atomic data (see Note~\ref{note:atomic-data}) on leaves e.g. \verb+"Bob"+, \verb+"Alice"+, etc.}
\item{Graph structure or, respectively, set-element nesting.}
\end{itemize}

\smallskip

\begin{note}[Atomic data]\label{note:atomic-data}\em
Atomic data is, in fact, treated as singleton sets consisting of a labelled empty set or, equivalently, as labels on additional leaf edges in the WDB graph. For example, the atomic value \verb+"Bob"+ from the above example is formally represented as

\begin{verbatim}
    {Bob:{}}
\end{verbatim}

\noindent
or, respectively, as the labelled edge with the target node being a leaf,

\begin{figure}[!ht]
\hspace{0.9cm}
\includegraphics{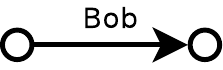}
\end{figure}

\end{note}

\noindent
For example, taking into account the above description, the corresponding system of (almost) flat set equations (with atomic values simulated as labelled empty sets) representing the WDB graph depicted in Figure~\ref{fig:WDB_family} should actually be:

\begin{small}
\begin{verbatim}
    bob        = { name:bob_name, wife:alice }
    bob_name   = { Bob:{} }
    alice      = { name:alice_name, husband:bob, pet:sam }
    alice_name = { Alice:{} }
    sam        = { name:sam_name, species:cat_name }
    sam_name   = { Sam:{} }
    cat_name   = { cat:{} }
\end{verbatim}
\end{small}

\noindent
To completely flatten this system we need to further replace all nested occurrences of \verb+{}+, say, by the set name \verb+empty+ and add one more equation \verb+empty = {}+. Of course, nesting is a reasonable notion, and atomic values are more user friendly from the external point of view. Thus, these concepts are included in the XML representation of WDB considered below, although the query system internally uses only completely flat set equations%
\footnote{
Note that WDB may (briefly) involve complicated equations, such as $res = q$ where $q$ is an arbitrarily complicated term or formula, during the execution of queries $q$ or after invoking the ``splitting'' rule during reduction. But, this extended system is, in fact, reduced to the flat form, and it is technically more convenient to work with other given WDB equations if they are presented in the flat form.
}%
.

\section{Practical representation of WDB as XML}

Although set equations represent WDB in the most natural and intuitive way, directly suggesting that such data are hypersets, it makes sense to relate this approach to the popular XML representation of semi-structured data and use appropriate existing techniques. Thus, numerous and independently existing XML data can be treated by our approach, making its application considerably wider.

\medskip

Extensible Markup Language (XML) is popular model for ordered (typically) tree-like semi-structured data. The portability, scaleability and tree (but extendable to graph) structure of XML has given rise to its wide spread useage. As such, systems of set equations, possibly allowing deep nesting, although very intuitively appealing could be represented practically as XML documents also based on the idea of representation of nesting data. However, the primary goal of our approach is not the implementation of XML querying, as much research and practical work has already been devoted to the latter:
\emph{CDuce} \cite{BCF03},
\emph{Lore} \cite{GMW99}, \nocite{AQMWW97lorel,MAGQW97lore}
\emph{Quilt} \cite{CRF01}
\emph{XML-GL} \cite{CCDFPT99},
and \emph{XML-QL} \cite{DFFLS99}; 
as well as the W3C standards
\emph{XSLT}  \cite{XSLT},
\emph{XPath} \cite{XPath},
and \emph{XQuery} \cite{XQuery} (based on Quilt).

\medskip

The main idea of the proposed XML-WDB format is to represent WDB systems of set equations as XML documents of a special form, and the most essential step consists in recursively replacing any labelled bracket expression

\begin{verbatim}
    label : {...}
\end{verbatim}

\noindent
by the XML element:

\begin{verbatim}
    <label>...</label>
\end{verbatim}

\noindent
Additionally, XML-WDB documents require: {\bf (i)} the special root element \verb+<set:eqns>+ which denotes system of set equations, and {\bf (ii)} the nested elements \verb+<set:eqn>+ denoting particular set equations. Defined set names participate as values of the \verb+set:id+ attribute of \verb+<set:eqn>+ tags, and referenced set names as values of the \verb+set:ref+ attribute (and also \verb+set:href+ attribute discussed later) of any other tags. Note that, as stated above, XML represents \emph{ordered} tree-like semi-structured data, however, our set-theoretic approach to WDB ignores order. Thus, such XML documents are treated by our approach ignoring the order (and possible repetition) of elements.

\medskip

Let us consider the system of set equations (with nesting allowed) in Section~\ref{sec:hypersets_approach_to_wdb} (depicted visually in Figure~\ref{fig:WDB_family}) and its representation as an XML document in XML-WDB file~\ref{WDB_XML_family}. The names of the special elements (\verb+set:eqns+ and \verb+set:eqn+) and special attributes (\verb+set:id+, \verb+set:ref+ and \verb+set:href+) should appeal to the readers' intuition that the XML-WDB document below corresponds to the above system of set equations.

\begin{xml-file}[!ht]
\begin{small}
\begin{verbatim}

    <?xml version="1.0"?>
    <set:eqns xmlns:set="http://www.csc.liv.ac.uk/~molyneux/XML-WDB">

      <set:eqn set:id="bob">
        <name>Bob</name>
        <wife set:ref="alice" />
      </set:eqn>

      <set:eqn set:id="alice">
        <name>Alice</name>
        <husband set:ref="bob" />
        <pet>
          <name>Sam</name><species>cat</species>
        </pet>
      </set:eqn>

    </set:eqns>
\end{verbatim}
\end{small}
\caption{Family database (cf. Figure~\ref{fig:WDB_family}).}
\label{WDB_XML_family}
\end{xml-file}


Recall that atomic data such as \verb+name:"Bob"+ is interpreted as \verb+name:{Bob:{}}+, and should therefore be translated into \verb+<name><Bob></Bob></name>+ or, equivalently, into \verb+<name><Bob/></name>+. This might seem to contradict 
XML-WDB file~\ref{WDB_XML_family} where rather \verb+<name>Bob</name>+ is used, but the inverse translation in Section~\ref{sec:rewriting_xml_to_sse} (Rule~2) shows that the empty element \verb+<Bob></Bob>+ or \verb+<Bob/>+ is treated equivalently as text data \verb+Bob+. Here it appears as text data for the readers' convenience.

\subsection{XML-WDB document format}

In general, an arbitrary XML-WDB document is defined as follows.

\begin{defn}[XML-WDB; see also Section~\ref{sec:xml-wdb-schema} for the corresponding XML schema]\label{def:XML-WDB}
A well-formed and valid XML-WDB	file is an XML document with the root element
\linebreak
\verb+<set:eqns>+ containing possibly several \verb+<set:eqn>+ sub-elements. The \verb+<set:eqns>+ element should contain no attributes, whereas, the element \verb+<set:eqn>+ should contain the required \verb+set:id+ attribute only.
%
%
The value of the attribute \verb+set:id+ should have a unique value (across the whole document) called the \emph{defined set name} and can only be be a string of symbols which is any \emph{simple set name} (according to the syntactical category \verb+<simple set name>+ in the BNF). The elements \verb+<set:eqns>+, \verb+<set:eqn>+, and the attribute  \verb+set:id+ are not allowed to appear anywhere else in the document. The element \verb+<set:eqn>+ can contain possibly several arbitrary XML sub-elements. The attributes \verb+set:ref+ and \verb+set:href+ can appear (at any depth) in those arbitrary elements under \verb+<set:eqn>+.
The values of the attributes \verb+set:ref+ and \verb+set:href+ are called \emph{referenced set names}, and must correspond to some existing \verb+set:id+ value in the same XML-WDB document in the case of \verb+set:ref+, or \verb+set:id+ value in some other XML-WDB document in the case of \verb+set:href+. To this end, the value of the attribute \verb+set:href+ should be \emph{full set name} (as discussed in Section~\ref{sec:distributed_xml-wdb}; cf. the syntactical category \verb+<set name>+ in the BNF) consisting of an (XML-WDB file) URL and simple set name defined in that file (delimited by \#).

\bigskip

\noindent
Everything else allowed by XML standard, what is not forbidden by the above restrictions, 
is permitted in the XML-WDB format.
\end{defn}

\begin{note}\em
The important feature of this definition is that XML-WDB documents can contain quite arbitrary XML elements under \verb+<set:eqn>+, thus allowing to include arbitrary XML data with any nesting, any text data and any attributes%
\footnote{
In general, arbitrary attributes are treated by the Rule~1 in Section~\ref{sec:rewriting_xml_to_sse} below.
}
(except \verb+set:id+, and with restrictions on values of \verb+set:ref+ and \verb+set:href+, as described above) into our hyperset approach to WDB. However, the order and repetitions of data will be irrelevant for our approach, and the usual XML attributes (except the attributes \verb+set:ref+ and \verb+set:href+ which have a special role, as described above) will be treated rather as tags which permit no further nesting.
\end{note}

\subsection{Distributed WDB}\label{sec:distributed_xml-wdb}

Any WDB system of set equations may be divided into several subsystems (as XML-WDB files) with the possibility for the set names $s$ participating in one subsystem (XML-WDB file) to be defined by set equations $s=\setof{\ldots}$ either in the same or in some other subsystems (XML-WDB files). Thus, strictly speaking, we should always consider the corresponding full versions of set names defined in set equations of distributed WDB, even when a simple set name is used for simplicity. That is, each simple set name occurring as a value of \texttt{set:id} or \texttt{set:ref} attributes within an WDB-XML file should be understood as full set name obtained from the URL of this file by concatenating it with the simple name using \verb+#+ to delimite these parts. Moreover, this technique allows to avoid unintended simple set name clashes without cooperation or collaboration between the authors of distributed WDB-XML files. (Unfortunately, unintended clashes for using the same label for different intuitive meanings is still possible, however, this is not formal contradiction in our approach. Here the well-known idea of namespaces in XML could be used.)

\medskip
\begin{figure}[ht]
\centering
\includegraphics{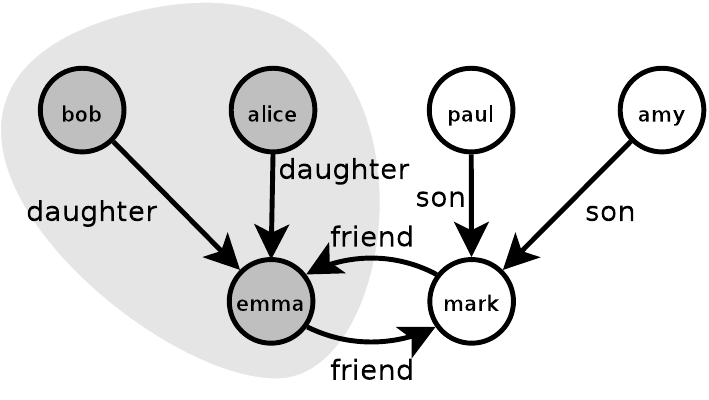}
\caption{Example distributed WDB representing two fictitious families, divided into two fragments represented as white and grey nodes}\label{fig:WDB_families}
\end{figure}

\noindent
Defined set names appearing in some XML-WDB file can participate as referenced set names in the same or other XML-WDB files. Those set names defined in the same XML-WDB file are referenced as simple set name values of the attribute \verb+set:ref+, whereas, set names defined in some other XML-WDB file are referenced as full set name values of the attribute \verb+set:href+. It is required that each full set name should refer to an existing XML-WDB file and the set equation within that file for the simple set name part (after the \verb+#+ symbol).

\medskip

Let us now consider an example of distributed WDB, representing two families (visualised in Figure~\ref{fig:WDB_families}) and the corresponding XML-WDB files \verb+family1.xml+ and \verb+family2.xml+ (XML files~\ref{WDB_XML_family1_distributed} and~\ref{WDB_XML_family2_distributed}) appearing below. Both simple and full set names participate as referenced set names in this example distributed WDB. For example, take the labelled element \verb+daughter:emma+ represented in XML-WDB file \verb+family1.xml+ as
\begin{small}
\begin{verbatim}
    <daughter set:ref="emma" />
\end{verbatim}
\end{small}

\noindent
where the attribute \verb+set:ref+ refers to simple set name \verb+emma+ defined within the same file. As an illustration of distribution, consider the labelled element \verb+friend:mark+ represented as
\begin{small}
\begin{verbatim}
    <friend set:href="...family2.xml#mark" />
\end{verbatim}
\end{small}

\noindent
where the attribute \verb+set:href+ refers to set name \verb+mark+ defined in the file \verb+family2.xml+. Note that, the URL in this example has shorted for the sake of simplicity.

\begin{xml-file}[!ht]
\begin{small}
\begin{verbatim}

    <?xml version="1.0"?>
    <set:eqns xmlns:set="http://www.csc.liv.ac.uk/~molyneux/XML-WDB">

      <set:eqn set:id="bob">
        <daughter set:ref="emma" />
      </set:eqn>

      <set:eqn set:id="alice">
        <daughter set:ref="emma" />
      </set:eqn>

      <set:eqn set:id="emma">
        <friend set:href="...family2.xml#mark" />
      </set:eqn>

    </set:eqns>
\end{verbatim}
\end{small}
\caption{Family database fragment (cf. grey nodes Figure~\ref{fig:WDB_families}): family1.xml}
\label{WDB_XML_family1_distributed}
\end{xml-file}

\smallskip
\begin{xml-file}[!h]
\begin{small}
\begin{verbatim}

    <?xml version="1.0"?>
    <set:eqns xmlns:set="http://www.csc.liv.ac.uk/~molyneux/XML-WDB">

      <set:eqn set:id="paul">
        <son set:ref="mark" />
      </set:eqn>

      <set:eqn set:id="amy">
        <son set:ref="mark" />
      </set:eqn>

      <set:eqn set:id="mark">
        <friend set:href="...family1.xml#emma" />
      </set:eqn>

    </set:eqns>
\end{verbatim}
\end{small}
\caption{Family database fragment (cf. white nodes Figure~\ref{fig:WDB_families}): family2.xml}
\label{WDB_XML_family2_distributed}
\end{xml-file}

\pagebreak

\noindent
The analogy of WDB with the WWW and, in particular possible distributed character of WDB does not imply it is necessarily so huge and unorganised as the WWW. It could be distributed between several sites, and supported by specialised WDB servers of some departments of an organisation owning this WDB and maintaining some specific structure of this WDB.

\medskip

Thus, WDB might, in fact, be much more structured than the WWW, however, the general approach imposes no restrictions. Therefore, the concept of WDB \emph{schema} or \emph{typing} relation between hypersets or graphs (much more flexible than for the relational databases and based on the notion of bisimulation or ``one-way'' simulation) relativised to some typing relation on labels/atomic values can be considered for such databases \cite{BDFS97,LS97,S93, D02}. Here we will not go into details of this important topic as our main concern is the straightforward implementation of querying WDB which does not take into account any such WDB schemas.

\subsection{Transformation rules from XML to systems of set equations}\label{sec:rewriting_xml_to_sse}

Let us show how any XML-WDB document, as described above, can be treated as a system of set equations by using the following simple transformations (applicable, in fact, to arbitrary XML documents, but giving the desired system of set equations only for the
\linebreak
\mbox{XML-WDB} documents). There are however currently some restrictions on XML-WDB in these transformation rules which can easily be relaxed, for example attributes having many values \texttt{attr="value1 value2 ..."} are not taken into account.

\subsubsection{Elimination of attributes and text data}

The first two transformation rules, applied recursively, will eliminate attributes and atomic (text) data from arbitrary XML element by treating them as tags.

\medskip

\noindent
{\bf Rule 1} (Attribute elimination, except attributes \verb+set:id+, \verb+set:ref+ and \verb+set:href+).

%
%
%

\medskip

\noindent
XML tags which have attributes,
\begin{small}
\begin{verbatim}
    <tag attr="value" other-attributes> 
      some-content
    </tag>
\end{verbatim}
\end{small}

\noindent
transform to

\begin{small}
\begin{verbatim}
    <tag other-attributes>
      <attr>value</attr>
      some-content
    </tag>
\end{verbatim}
\end{small}

\noindent
where \verb+attr+ is restricted to be any attribute name except the distinguished attributes \verb+set:id+, \verb+set:ref+ and \verb+set:href+ belonging to the \verb+set+ namespace which will be considered later. Additionally, \verb+some-content+ means arbitrary XML content of an XML element.

\medskip

\noindent
In the case of empty element with attributes,

\begin{small}
\begin{verbatim}
    <tag attr="value" other-attributes />
\end{verbatim}
\end{small}

\noindent
transformation quite analogously gives the similar result,

\begin{small}
\begin{verbatim}
    <tag other-attributes>
      <attr>value</attr>
    </tag>
\end{verbatim}
\end{small}

\noindent
This rule is applied until all attributes, except those attributes beglonging to the \verb+set+ namespace (\verb+set:id+, \verb+set:ref+ and \verb+set:href+), are eliminated. This way attributes are actually treated as tags.

\medskip

\noindent
{\bf Rule 2} (Atomic data elimination).

\medskip

\noindent
Text data with no white spaces

\begin{small}
\begin{verbatim}
    any-text-data 
\end{verbatim}
\end{small}

\noindent
transforms to the empty XML element

\begin{small}
\begin{verbatim}
    <any-text-data/>
\end{verbatim} 
\end{small}

\noindent
In the case of text data containing white characters (spaces, carriage-returns, tabs),

\begin{small}
\begin{verbatim}
    any text    data 
\end{verbatim}
\end{small}

\noindent
all white characters are ignored, and the result is the corresponding sequence of the empty elements,

\begin{small}
\begin{verbatim}
    <any/><text/><data/>
\end{verbatim} 
\end{small}

\noindent
As our set theoretic approach ignores order and repetitions (in contrast with the ordinary XML approach) this, in fact, means that a sentence (any text data) is considered rather as an unordered set of words. This way text data are actually treated as tags. (An another alternative would be to replace all white characters by the underscore symbol, thus giving rise to \verb+<any_text_data/>+, like above.)

\bigskip

\noindent
Iterated application of rules 1 and 2 eliminates all atomic (text) data and attributes except those attributes belonging to the \verb+set+ namespace (\verb+set:id+, \verb+set:ref+ and \verb+set:href+).

\subsubsection{Elimination of tags}

The remaining rules below allow transformation of XML elements with (simple) attributes and text data eliminated by the above rules into bracket expressions (possibly involving set names), and into set equations if there are tags \texttt{set:eqns} and \texttt{set:eqn} occurring as described in Definition~\ref{def:XML-WDB}. In the intermediate steps, the expression transformed will be in the mixed language.

\medskip
\noindent
{\bf Rule 3} (Tag elimination, except the tags \texttt{set:eqns} and \texttt{set:eqn}).

\medskip

\noindent
For arbitrary XML tags, except \texttt{set:eqns} and \texttt{set:eqn}, which have no attributes,
\begin{small}
\begin{verbatim}
    <tag>
      some-content
    </tag>
\end{verbatim}
\end{small}

\noindent
transforms into
\begin{small}
\begin{verbatim}
    tag:{some-content}.
\end{verbatim}
\end{small}

\noindent
Those possibly remaining tags in sub-elements of \verb+some-content+ will be eliminated recursively by application of transformation rules 3 and 4. Quite analogously for the case of the empty element,
\begin{small}
\begin{verbatim}
    <tag/>
\end{verbatim}
\end{small}

\noindent
transforms to
\begin{small}
\begin{verbatim}
    tag:{}
\end{verbatim}
\end{small}

\medskip

\noindent
{\bf Rule 4} (Elimination of tags with \verb+set:ref+ and \verb+set:href+ attributes).

\begin{small}
\begin{verbatim}
    <tag set:ref="set-name" />
\end{verbatim}
\end{small}

\noindent
transforms to the sequence
\begin{small}
\begin{verbatim}
    tag:set-name
\end{verbatim}
\end{small}

\noindent
Recall that other attributes were already eliminated by Rule~1. Furthermore, according to the definition of well-formed XML document an attribute name must only appear once in any tag, however, \verb+set:ref+ and \verb+set:href+ may participate together in any tag. The above elimination is considered as typical if only the attribute \verb+set:ref+ or \verb+set:href+ occurs.

\medskip

\noindent
Additionally, we must consider the following more general, however unlikely case when some content is present:
\begin{small}
\begin{verbatim}
    <tag set:ref="set-name1" set:href="set-name2">
      some-content
    </tag> 
\end{verbatim}
\end{small}

\noindent
transforms to
\begin{small}
\begin{verbatim}
    tag:set-name1,
    tag:set-name2,
    tag:{some-content}.
\end{verbatim}
\end{small}

\noindent
However, to be consistent with the first version of Rule~4, if \verb+some-content+ is empty, then (as an exception) the result should not contain the labelled element, \verb+tag:{}+.

\bigskip

The above rules hold also for the case of the attribute \verb+set:href+, or when both \verb+set:ref+ and \verb+set:href+ are present within a tag. Note that after applying Rule 4, the difference between these two attributes is not taken into account in generating the result.
Recall that \verb+set:ref+ refers to a simple set name, whereas, \verb+set:href+ refers to a full set name which is actually an URL together with simple set name (see Section~\ref{sec:distributed_xml-wdb}). Such syntax explicitly differentiating between simple and full set names is convenient for implementation. After applying this rule this feature will disappear, but the difference between the shapes of simple and full set names will remain, so that nothing essential will be lost.

\medskip

\noindent
{\bf Rule 5} (Elimination of tags \verb+set:eqn+ and \verb+set:eqns+).

\begin{small}
\begin{verbatim}
    <set:eqn set:id="simple-set-name">some-content</set:eqn>
\end{verbatim}
\end{small}

\noindent
is replaced by the equation,

\begin{small}
\begin{verbatim}
    simple-set-name = {some-content}
\end{verbatim}
\end{small}

\noindent
and,

\begin{small}
\begin{verbatim}
    <?xml ... >
    <set:eqns>some-content</set:eqns>
\end{verbatim}
\end{small}

\noindent
is replaced by

\begin{small}
\begin{verbatim}
    some-content
\end{verbatim}
\end{small}

\noindent
that is, by system of set equations (in the case of a well-formed XML-WDB document; cf.\ Definition~\ref{def:XML-WDB} above).

\bigskip

\noindent
Note that, all the above rules can be applied in arbitrary order, leading to a unique system of set equations.

\subsection{XML schema for XML-WDB format}\label{sec:xml-wdb-schema}

A well-formed and valid XML-WDB document must conform to Definition~\ref{def:XML-WDB}. As our general goal is implementation, let us also present the XML schema%
\footnote{
also available at \url{http://www.csc.liv.ac.uk/~molyneux/XML-WDB/schema/xml-wdb.xsd}
}
(at the end of this section) which corresponds to this definition almost completely (as XML schemes are, in fact, insufficiently expressible).

\medskip

First of all, the schema requires that all the declared elements \verb+eqns+ and \verb+eqn+, and attributes \verb+id+, \verb+ref+ and \verb+href+
are qualified under the namespace \url{http://www.csc.liv.ac.uk/~molyneux/XML-WDB}. In practice the author of any XML-WDB document can declare this namespace as the mnemonic \verb+set+%
\footnote{
In fact, the namespace \url{http://www.csc.liv.ac.uk/~molyneux/XML-WDB} could be declared 
\linebreak
by any chosen mnemonic, let us say \texttt{s}.
}
and use \texttt{set:eqns} instead of just \texttt{eqns}, etc. to emphasise these special elements/attributes are subject to the rules of this schema.

\medskip

\noindent
The root element \verb+eqns+ of an XML-WDB document is declared in the schema as having the complex type \verb+system_of_set_equations+, as follows,
\begin{small}
\begin{verbatim}
    <xsd:element name="eqns" type="system_of_set_equations"/>.
\end{verbatim}
\end{small}

\noindent
The complex type \verb+system_of_set_equations+ is defined as
\begin{small}
\begin{verbatim}
    <xsd:complexType name="system_of_set_equations">
       <xsd:sequence minOccurs="0" maxOccurs="unbounded">
         <xsd:element name="eqn" type="set_equation"/>
       </xsd:sequence>
    </xsd:complexType>
\end{verbatim}
\end{small}

\noindent
where an arbitrary number ($\ge 0$) of set equations can participate in any XML represented system of set equations. Note that, by definition only, \verb+eqn+ subelements can participate under an \verb+eqns+ element. Here, \verb+eqn+ elements represent set equations by the given complex type \verb+set_equation+, which is defined by two elements:
\begin{small}
\begin{verbatim}
    <xsd:sequence minOccurs="0" maxOccurs="unbounded">
      <xsd:any namespace="##any" processContents="lax"/>
    </xsd:sequence>

    <xsd:attribute form="qualified" 
                   name="id" 
                   type="xsd:ID" 
                   use="required"/>
\end{verbatim}
\end{small}
Thus, any \verb+eqn+ element must contain the required attribute \verb+id+, and may contain arbitrary XML sub-elements. Note that, by definition, only one attribute, \verb+id+, must appear in \verb+eqn+ elements. The corresponding value of the \verb+id+ attribute must be unique over the entire
\linebreak
\mbox{XML-WDB} document according the type \verb+xsd:ID+. However, the schema only ensures the well-formedness with \verb+lax+ processing of arbitrary XML sub-elements, and therefore does not check that such elements are XML-WDB valid according to Definition~\ref{def:XML-WDB}. In particular this schema says nothing about \verb+ref+ and \verb+href+ attributes and how they can be used. Thus, our implementation additionally ensures the following:
\begin{itemize}
\item{
The elements \verb+eqns+ and \verb+eqn+ and attribute \verb+id+ qualified under the \url{http://www.csc.liv.ac.uk/~molyneux/XML-WDB/} namespace can not participate in arbitrary
\linebreak
XML sub-elements.
}
\item{
The attribute \verb+ref+ must have simple set name value, defined by the \verb+id+ attribute in the same XML-WDB file. Furthermore, the attribute \verb+href+ must have full set name value whose simple set name part is defined in some other well-formed and valid XML-WDB file.
}
\end{itemize}

\noindent
Thus, any well-formed XML document is considered as valid XML-WDB document if it can be successfully validated against the above schema and conforms to these additional rules. However, our $\Delta$ language query implementation deals directly with systems of set equations, therefore it is necessary to rewrite from valid XML-WDB files into systems of set equations, by treating them with the rules from Section~\ref{sec:rewriting_xml_to_sse}. The inverse transformation from systems of set equations to XML-WDB format is also implemented.

\clearpage

\begin{xml-schema}[ht]
\begin{small}
\begin{verbatim}


  <?xml version="1.0" encoding="UTF-8"?>

  <xsd:schema  xmlns:xsd="http://www.w3.org/2001/XMLSchema"
    targetNamespace="http://www.csc.liv.ac.uk/~molyneux/XML-WDB"
    xmlns="http://www.csc.liv.ac.uk/~molyneux/XML-WDB"
    elementFormDefault="qualified"
    attributeFormDefault="unqualified">

    <xsd:complexType name="system_of_set_equations">

       <xsd:sequence minOccurs="0" maxOccurs="unbounded">
         <xsd:element name="eqn" type="set_equation"/>
       </xsd:sequence>

    </xsd:complexType>

    <xsd:complexType name="set_equation">

       <xsd:sequence minOccurs="0" maxOccurs="unbounded">
         <xsd:any namespace="##any" processContents="lax"/>
       </xsd:sequence>

       <xsd:attribute form="qualified" name="id"
         type="xsd:ID" use="required"/>

    </xsd:complexType>

    <xsd:element name="eqns" type="system_of_set_equations"/>

  </xsd:schema>
\end{verbatim}
\end{small}
\caption{XML-WDB file schema: xml-wdb.xsd}
\label{wdb_schema}
\end{xml-schema}

\clearpage


%
%


\part{Evaluation}\label{part:evaluation}


\chapter{Comparative analysis}\label{chap:comp_anal}

\section{Preliminary comparison}

There have been many proposed approaches for modelling and querying semi-structured data. Many of these approaches are based on the graph model, which has become the prevalent model for representation of semi-structured data. For example, the graphical Object Exchange Model (OEM)  \cite{PGMW95} was used in the integration of heterogeneous information sources in Tsimmis \cite{GMPQRSUVW97} and the semi-structured query language Lorel \cite{AQMWW97lorel,MAGQW97lore}. Moreover, there has been some trend toward the XML document model, which is essentially the graph model restricted to ordered trees, but arbitrary graphs can be imitated by using the attributes \verb+id+ and \verb+ref+ to define links between tree branches. In fact, Lore (implementation of the Lorel language) was later migrated to XML~\cite{GMW99}.

\medskip

The most natural and intuitive way of querying graphs employed in most approaches is path navigation by using path expressions. However, path expressions are evidently sufficiently complicated syntactical means to achieve expressive power in queries. This is practically very reasonable and means path expressions are a strong technical tool. But, on a logical level (in the wide sense of this word) such complicated things are always considered as definable in terms of some other more fundamental concepts. Thus, in foundation of mathematics such fundamental concepts are set, membership relation, logical quantifiers, etc.\ allowing to express all other concepts, constructions and proofs in mathematics and (theoretical) computer science. In a sense, the graph approach to semi-structured databases lacks natural logically fundamental concepts, and in these circumstances path expressions are included as the main tool for achieving expressive power. On the other hand, the set theoretic approach to semi-structured databases presented in this thesis does not require path expressions%
\footnote{besides the related classical operation of transitive closure of a set and a general recursion operator --- classical inductive definitions
} 
to achieve high expressive power which in fact captures exactly all ``generic'' polynomial time computable operations over hypersets \cite{LS97, LS99,S87,S93}. Therefore, the language can be considered theoretically as having in this sense no ``gaps''. But, from the point of view of practical usability and efficiency of implementation, path expressions should be eventually included in our implementation of the $\Delta$-language although not increasing its expressive power (see \cite{S06}).

\medskip

From the traditional theoretical point of view polynomial time computability of queries in $\Delta$ (which is usually theoretically considered as ``feasible computability'') allows to consider $\Delta$ as computationally viable. However, in a practical sense, we cannot insist on this usage of the term ``feasible'' because polynomials can be of high degree and with huge coefficients. Also, this makes less sense in the context of those most expensive computational steps assuming downloading numerous files from the World-Wide Web. Thus, we rather consider this characteristic not as a witness of efficiency of $\Delta$ but as a good witness of expressive power of the language. Anyway, when comparing this approach with others, it can be considered as top-down from theory to practice. In particular, this explains again our attitude to not include path expressions in the main conceptual version of the $\Delta$-language, being a definable concept, and considering them only as technical ``conservative'' extension, although very important practically.


\medskip

Recall that  hypersets representing WDB can be visualised as graphs, and thus, in principle, our approach can treat graph structured data from other approaches, but assuming that the order and repetition of such data does not matter. As the latter is not always the case, the precise comparison with other approaches is not so straightforward. Similarly, our implementation can query arbitrary XML elements, rewriting from XML-WDB to systems of set equations and ignoring order. Although the aim of the project was not XML querying, this accomplishment extends possible applicability of our implementation.

\medskip

Now, after these preliminary general comments, let us consider several known approaches 
to semi-structured databases and to set theoretic programming.

\section{SETL}

An important practical predecessor of our work is the set theoretic programming language SETL \cite{SDDS86, S70, S73} which deals with hereditarily-finite well-founded sets (without cycles) and tuples. (Note that tuples or, more generally, records $[a_1:x_1,\ldots,a_n:x_n]$ can be trivially treated in our approach as sets $\setof{a_1:x_1,\ldots,a_n:x_n}$ in which all labels $a_i$ are different.) This general purpose programming language exploits the notion of set as fundamental data structure with its set theoretic style of constructs like collection in $\Delta$. It is, however, an imperative language using such traditional operators as the assignment operator, loops, etc. For example, let us consider the SETL program:

\pagebreak

\begin{small}
\begin{verbatim}
    A = {1,2,3,4,5};
    B = { x: x in A | x >= 3 };
    print(B);
\end{verbatim}
\end{small}

\noindent
where the statement on the second line reminds us of the $\Delta$-term collect. In fact, the result of executing this SETL program is the output set of \verb+B+, which is, in fact, defined as those numbers \verb+x+ belonging to the set \verb+A+ such that the number \verb+x+ is greater than or equal to three, as follows:
\begin{small}
\begin{verbatim}
    {3,4,5}.
\end{verbatim}
\end{small}

\noindent
Furthermore, in SETL, equality between sets is understood as ``deep'' set equality implemented as the following (recursive) procedure taken from \cite{SDDS86}:
\begin{small}
\begin{verbatim}
    proc equal(S1,S2);
        if # S2 /= # S1 
        then return false;
        else
            (forall x in S1) 
                if x notin S2 then return false; 
                end if;
            end forall loop;
            return true; -- S1 and S2 are equal
        end if;
    end proc;
\end{verbatim}
\end{small}

\noindent
That is, the two sets \verb+S1+ and \verb+S2+ are equal if they have the same cardinality and each element \verb+x+ of the set \verb+S1+ participates as a member in the set \verb+S2+. In fact, this equality procedure will be called recursively for each membership test \verb+notin+ (where, like in our case, $x\in y\iff\exists x'\in y\,.\,\texttt{Equal}(x,x')$). Hence, \verb+S1+ and \verb+S2+ are equal if their elements are equal and their elements are also equal, and so on. This is similar to bisimulation equivalence which is an important concept in our hyperset theoretic approach. The use of cardinality operator $\#$ either witnesses that hereditarily-finite sets are  represented in SETL implementation in strongly extensional form and, anyway, assumes further recursive call of equality. In contrast to SETL, the implemented $\Delta$ language is actually a declarative query language to semi-structured or Web-like databases and, as such, is not intended to be a universal language. The degree of universality of $\Delta$ is characterised by its expressive power equivalent to polynomial time. Also, SETL does not have any construct similar to the decoration operator within the $\Delta$-language which allows for restructuring, but its universal character should allow to define decoration for acyclic graphs. In contrast to SETL, the main characteristic feature of $\Delta$ is the extension of the ideas of descriptive complexity theory \cite{I82,I99,S80,V82} (usually considered in connection with the relational approach to databases) from finite relational structures to hereditarily-finite (hyper)sets and, thereby, to semistructured databases.

\medskip

The most recent development on the SETL language was the implementation described in \cite{B00}, which introduced Internet programming using sockets into the SETL language. In fact, these latest considerations further support that SETL is actually a general purpose programming language, and in this sense differs from $\Delta$ which is a query language.

\section{UnQL}\label{sec:comparisons_unql}

The UnQL query language \cite{BDHS96,BFS00} is closest to our approach as it is based on bisimulation, with its operators also being bisimulation invariant as in our case. However, despite considering bisimulation, UnQL is based on the graph model, and the op.\ cit.\ do not even mention hyperset theory. UnQL can also be characterised as a bottom-up approach from graphs to something reminding us of hypersets. Moreover, there is no  operator for testing equality between graph vertices (neither literal nor based on bisimulation) in the UnQL language. However, bisimulation should be used in defining the semantics of path expressions (patterns in their terminology) in the UnQL language, as shown in \cite{S06} and in our example in Section~\ref{sec:imitating_path_expressions}, ensuring that its operations really are bisimulation invariant. Much of the UnQL approach is devoted to the rather complicated way in which they deal with graphs, which appears more technical compared to the intuitive  denotational and operational semantics of the hyperset approach. In a sense, UnQL has defined only operational semantics over graphs, which is bisimulation invariant. No abstract concept like hyperset and corresponding (hyper)set theoretical style of thought is explicitly described. Moreover, operational semantics of the structural recursion operator is rather complicated by working with multiple ``input'' and ``output'' vertices considered as essential part of graphs to be queried by UnQL. Therefore, semi-structured data represented in UnQL does not exactly correspond to hypersets, although it can be imitated by hypersets as shown in \cite{S06}. Also, the UnQL language and related language UnCal were shown in \cite{S06} to be embeddable within $\Delta$, but, as reasonably conjectured, not vice versa. This embedding, although done in purely set theoretic terms, is based on the interpretation of arbitrary graphs as sets of ordered pairs. The bisimulation invariant operations on graphs of UnQL are defined set theoretically but as operations on graphs rather than as operations on abstract entities denoted by these graphs (with multiple ``inputs'' and ``outputs'') considered up to simulation. In particular, the main structural recursion construct of UnQL is definable in $\Delta$ by manipulating graphs using recursive separation and concluded by applying decoration operation to get a hyperset imitating the result (with multiple ``inputs'' and ``outputs''). In fact, many of the operations in UnQL are based on various ways of appending such kind of graphs (via ``inputs'' and ``outputs''), including structural recursion, all of which may be considered as a special versions of the decoration operator. However, the full version of the powerful decoration operator (which is much simpler and logically more fundamental than its particular versions mentioned) is neither considered nor definable in UnQL (according to the conjecture in \cite[page 813]{S06}).

\section{Lore}

Lore (Lightweight Object REpository) \cite{MAGQW97lore} is the implementation of the Lorel query language \cite{AQMWW97lorel} based on the OEM graph model \cite{PGMW95}. Lorel is an extention of the Object Query Language (OQL) \cite{C93} and, in fact, statements written in the Lorel are translated to OQL. Moreover, additional features of Lorel (such as path expressions, and type coercion) are syntactical sugaring of OQL. The OEM model is similar to the data model used in UnQL, but unlike UnQL and also our approach, does not consider graphs up to bisimulation. Therefore, bisimulation invariance is not pursued in this approach, hence, in this way it is crucially different from UnQL and $\Delta$. In the OEM model equality is between graph nodes (OIDs) rather than value equality using bisimulation. Lorel also uses ordinary equality between sets of OIDs, which, however, is not the ``deep'' set equality assumed by bisimulation. Therefore, Lorel would treat some of our examples differently, and thus, only very informal and superficial comparison is possible, unlike the comparison with UnQL. However, the \verb+select+ operator of Lorel is very similar to our \verb+collect+ construct, as illustrated in the following example Lorel query:
\begin{small}
\begin{verbatim}
    SELECT pub
    FROM pub in BibDB
    WHERE pub.author = "Smith"
\end{verbatim}
\end{small}

\noindent
and the (strikingly similar) corresponding $\Delta$-query,
\begin{small}
\begin{verbatim}
    set query collect {
        'null':pub
        where pub-type:pub in BibDB
        and author:"Smith" in pub
    }
\end{verbatim}
\end{small}

\noindent
Note that only OIDs are \verb+select+ed in Lorel, whereas in $\Delta$ (OIDs or) set names denote (hyper)sets which are, in fact (on the level of abstract semantics) \verb+collect+ed. Note that, OIDs in Lorel denote just themselves and nothing more. Lorel can not express restructuring queries, unlike $\Delta$ which can perform restructuring queries with the decoration operation (at the final stage). Thus, informally (as formal comparison is impossible due to the above differences in data models -- graphs vs.\ hypersets represented by graphs) Lorel (and also UnQL) can be said to be also strictly embeddable in $\Delta$%
\footnote{ignoring so called path variables which may potentially lead to exponential complexity and, for simplicity, some less essential aspects like typing and coercion
}%
. Finally, there is also no recursion operator (except for Kleenes star in path expressions) and nothing similar to decoration operator (important for deep restructuring).

\section{Strudel}

Strudel is a Web site management system \cite{FFKLS98} for creating Web pages from heterogeneous data sources via the StruQL query language \cite{FFLS97} (see also \cite{ABS00}). In particular, the \verb+link+ clause in StruQL is able to do simple restructuring. In fact, Strudel allows to generate real Web sites in a declarative way from a site graph (a graphical ``plan'' of a site) that encodes the Web site's structure. The latter feature resembles the decoration construct although outside of hyperset approach. In Studel data is integrated from heterogeneous sources by mediators which rewrite from various data sources (such as XML files, bibtex files, etc.) to Strudel data graphs. StruQL queries over these data graphs, in fact, define the Web site structure creating Web pages and hyperlinks between Web pages.

\section{G-Log}

G-Log \cite{CDQT02} is another query language for semi-structured data represented as arbitrary labelled graphs. However, unlike the other approaches consider so far (Lorel, UnQL, $\Delta$) any query, as well as data, in G-log is represented graphically as a set of schematical red/green coloured ``rule'' graphs. Querying in G-log (in general, updating) is based on matching the query rule graph with the ``concrete'' black coloured data graph. This matching assumes one of three possible kinds of bisimulation (in particular, isomorphic embedding) of the red part of the rule with a subgraph of the black concrete data graph, and using the green part for updating the concrete data graph. This procedure is essentially non-deterministic and, in fact, can be executed in non-deterministic polynomial time (rather than polynomial time in the case of $\Delta$). The expressive power of G-log in its present form, or its potential extensions, is unclear, as well as precise comparison with $\Delta$. Granted, both are based on bisimulation but in a somewhat different way. The rule graphs of G-log can be described in some logical form, but it is unclear how to systematically relate this with the syntax of $\Delta$ to have a better comparison. In principle, extending $\Delta$ by quantification over the subset of a set, \mbox{$\forall x\subseteq t, \exists x\subseteq t$}, together with definability in $\Delta$ the necessary versions of bisimulation over graphs could make it possible to imitate matching of a rule graph with a subgraph of the data graph. But, it seems unclear whether there exists a natural unifying conceptual framework for both approaches. Furthermore, G-log is an open ended language with some ideas of its extension discussed in \cite{CDQT02}. In any case, we can conclude that UnQL and even Lorel%
\footnote{
ignoring that Lorel does not consider bisimulation
}
are syntactically, as well as in terms of operational semantics, much closer to $\Delta$ than G-log. However, matching with a subgraph is somewhat similar to the idea of path expressions which appear in both UnQL and Lorel, the latter being imitated in $\Delta$ as illustrated in Section~\ref{sec:imitating_path_expressions}.

\section{Tree (XML) model approaches}

The XML data model is based on ordered trees, whereas the other approaches to querying semi-structured databases discussed so far deal with  arbitrary graphs. (However, as we already mentioned, using attributes \verb+id+ and \verb+ref+ in XML allows imitate arbitrary graphs.) It might seem that querying XML data is formally outside of the (hyper)set theoretic view as the XML document model assumes a fixed order on the children of any node. Despite this our approach is able to query restricted XML documents (XML-WDB files which, however, can involve arbitrary nested XML elements) interpreted as systems of set equations. 

\medskip

The following comparisons focus on three contemporary XML data model approaches, XSLT, XQuery and XPath, all of which were developed by W3C working groups. In fact, these languages are the successors to many other XML model approaches, for example, XQuery is based on the Quilt query language \cite{CRF01}. However, for brevity no comparisons will be made with these predecessors.

\subsection*{XSLT}

XSLT (eXtensible Stylesheet Language transformations) \cite{XSLT} is a rule based language for transforming the structure of an XML document, that is, XSLT rewrites an XML document to another XML document with different structure. Thus, XSLT does allow convenient manipulation of XML documents. XSLT rules are composed of template rules which \emph{match} attributes/elements using XPath-like expressions (discussed below) and create new XML elements/attributes or apply other template rules. This style of language and its operational semantics is rather different from the $\Delta$-query language. In particular XSLT is typically used to visualise XML documents by transforming them into HTML Web pages.

\subsection*{XQuery}

XQuery \cite{XQuery} is declarative query language for XML documents, and was derived from Quilt \cite{CRF01}, Lorel \cite{AQMWW97lorel} (described above) and XML-QL \cite{DFFLS99}. XQuery is, in fact, Turing complete and thus can be considered as more than just a query language but also, in a sense, as a general purpose programming language.

\subsection*{Path expressions (XPath)}

XQuery and XSLT include XPath path expressions in its syntax. XPath is a language especially created to express paths navigating over XML document trees, and, in fact, XPath itself can serve as a query language.

\medskip

Currently path expressions are not included in the implemented $\Delta$-query language, however, they were shown to be definable in the original language \cite{S06}, and a simple example demonstrating how $\Delta$ could be extended syntactically to have path expressions and how it can define their meaning was shown in Section~\ref{sec:imitating_path_expressions}. Thus, our language is rich enough by fundamental operators over sets so that, at least theoretically, path expressions are unnecessary. Of course, practically they are very desirable and must be included in $\Delta$ to make it more practically convenient and user friendly. Moreover, path expressions, if implemented well, would make execution time of queries better than queries imitating path expressions in the current version of $\Delta$.

\bigskip

In general, comparison of $\Delta$ with query languages for XML can be done only on a rather superficial level. In fact, they do not share a common data model and the levels of abstraction are so different that more detailed comparison in general terms is difficult. We can only repeat that the closest approach to ours is UnQL where comparisons can be done in quite precise mathematical formulations~\cite{S06}.


\chapter{Conclusion and future outlook}\label{chapter:conclusion}

In this thesis we explored the experimental implementation of the hyperset approach to
\linebreak
semi-structured or Web-like databases and the query language $\Delta$ originally known only on a pure theoretical level. The primary goal was to demonstrate working practically with the $\Delta$-query language, and secondly, some considerations towards one crucial aspect of efficiency of such querying in the case of distributed WDB. The latter involves some theoretical considerations in Chapter~\ref{chap:local} and empirical testing in Section~\ref{sec:testing_real_oracle}.

\medskip

This chapter begins by reviewing the hyperset approach to semi-structured databases in the context of this thesis. 
In Section~\ref{sec:novel_contributions} we summarise the main results of our work which, in brief, consist in (i)~the implementation of the query language $\Delta$ and (ii)~development the concept of local/global bisimulation and running experiments demonstrating 
its fruitfulness in making query execution more efficient when equality (bisimulation) is involved. Some further simple optimisations used in our implementation are also discussed. Then we recapitulate briefly in Section~\ref{sec:concluding-comparisons} comparisons of $\Delta$ with other 
most close query languages. Finally, we conclude in Section~\ref{sec:further_work} with some closing discussion towards possible future extensions and optimisations.

\section{Hyperset approach to semi-structured databases}

First of all, the hyperset approach to semi-structured or Web-like databases and their querying was described in this thesis on the base of the earlier theoretical work done in \cite{LS97,S93,S06}. This approach considers hypersets as the abstract data model for WDB where the concrete representation of hypersets is given by systems of set equations which can be saved either as plain text files or as XML-WDB files. Likewise in relational databases where the abstract data model is relations, our approach focuses on abstract hypersets and strongly distinguishes them from their concrete representations by set equations (or corresponding XML-WDB form). Set theory is known to play an extraordinary foundational role in mathematics, and here we wanted to demonstrate in a practical context that very general set theoretic approach towards
\linebreak
semi-structured or Web-like databases is also quite reasonable.

\medskip

Systems of set equations can also be trivially represented as graphs where the latter, if considered literally, lead to the more traditional approach to semi-structured databases. To visualise our considerations we also use graphs, but they play only an auxiliary role. Abstractly, graph nodes as well as corresponding set names in set equations, denote hypersets. In fact, it is assumed that any user of our query system should mainly rely on pure set theoretic style of thought which is (mostly) simple and intuitive.%
\footnote{
The most subtle concept in our approach is the decoration operation.
}
Otherwise it would not be so widely accepted both in the foundation of mathematics, and in everyday mathematical practice. As graphs or corresponding systems of set equations can involve cycles, their nodes or set names denote, in general, hypersets. They differ from the ordinary concept of sets in the fact that hypersets are not necessary well-founded. Based on well-developed and understood hyperset theory \cite{A88,BM96}, such sets pose no conceptual difficulty in our approach. This approach demonstrates on a practical level that hypersets are no more difficult than the usual concept of sets, and are quite useful by allowing arbitrary semi-structured data to be represented in a completely set theoretic manner.

\medskip

An additional feature of our data model is its distributed character, that is any system of set equations representing a WDB is allowed to be distributed, with set names used in one \mbox{(XML-WDB)} file possibly described by set equations in the others files. This leads to distinctions between simple set names described in the same file, and full set names involving also the URL of the file where this set name is described. This does not change the hyperset approach but extends its possible applicability. On the other hand, this distributed character of a WDB poses an additional challenge on how to check practically whether two set names (possibly described in remote files) denote the same abstract hyperset, i.e.\ whether two given set names or graph nodes are bisimilar. However, the problem of computing bisimulation in the distributed case was shown here to be, in principle, resolvable practically, as remarked later in Section~\ref{sec:conclusion-local-global}.

\medskip

Respectively, the $\Delta$-query language considered here is set theoretic with the denotation $\Delta$ bearing from logic and set theory and traditionally emphasising its bounded character. The latter guarantees that all queries in $\Delta$ are computable in finite, in fact, polynomial time with respect to the the size of the input WDB. Moreover, it is known to have expressive power exactly corresponding to polynomial time (see \cite{LS99,S93} and particularly \cite{LS97,S93} for precise formulations of the labelled case considered here).

\section{Novel contributions}\label{sec:novel_contributions}

The main results of this work are the implementation of the hyperset approach to
\linebreak
semi-structured databases and the query language $\Delta$, and, secondly, the local/global approach towards efficient computation of bisimulation in the case of distributed WDB.

\subsection{Implementation of the hyperset approach to semi-structured databases}

The implemented version of the language $\Delta$ is quite complex and even somewhat comparable with practical programming languages. In fact, there was not enough time to create the most optimal implementation. The general problem of efficiency is so difficult and involving so many various aspects (see e.g.\ \cite{GMUW08}) that it is mostly outside the scope of this thesis (with one exception which is most essential to our hyperset approach; see Section~\ref{sec:conclusion-local-global}). Taking this into account, the main criteria were correctness of the implementation and its user friendliness so that the language could be demonstrated to a more practically oriented, rather than just a mathematically inclined, audience. As far as we see, the implementation satisfies these criteria based on our testing and also writing and running the worked examples in  Sections~\ref{sec:example_queries}--\ref{sec:lin-ord}. This query system was also used by my supervisor, Vladimir Sazonov, as demonstration tool for undergraduate students. This initial practical goal of the project lead to the successful development of:

\begin{itemize}
\item{
{\bf Implementation of the $\Delta$-query language} as a declarative language, based on those theoretical constructs in the original $\Delta$-language. Furthermore, for the convenience of writing queries some important features were included in the implemented language, such as \emph{library declarations} and \emph{query declarations} which, although very useful as the reader can see from the example queries, do not extend the theoretical expressive power of the language.
}
\item{
{\bf Algorithms for checking the validity of queries} to ensure both well-formedness and well-typedness. These algorithms add important low-level details for our implementation serving also as a sufficiently strong guarantee that the implementation was done correctly. The aim of the \emph{parsing} algorithm is to ensure well-formedness, according to the BNF grammar in Appendix~\ref{app:BNF}; whereas the aim of the \emph{contextual analysis} algorithm is to ensure well-typedness (which required considerable efforts to develop).
}
\end{itemize}

\noindent
The above syntactical considerations were highly important for implementation, and much time was dedication to ensuring these algorithms were  described and implemented correctly. In fact, the following developments strongly rely on these algorithms:

\begin{itemize}
\item{
{\bf Implementation of operational semantics of $\Delta$ language} according to reduction rules in \cite{S06} with some additional low-level 
descriptions for the operators \texttt{recursion}, \texttt{decoration} and \texttt{TC} also given here to aid implementation.
}
\item{
{\bf XML representation of WDB} by developing the XML-WDB format for systems of set equations and implementing algorithms rewriting from XML-WDB documents into systems of set equations, and vice versa. Currently we accepted this XML-WDB format as the standard way of representing WDB. These files can be saved on various sites and hyperlinked via full set names as we discussed above, and thus, WDB can be distributed (and queried) over the Internet. In fact, the XML-WDB format allows our approach to treat arbitrary nested XML elements within a WDB. The aim of this practical representation of WDB as XML is the ability, in principle, to query any existing XML data in our hyperset approach (assuming order and repetition in these data play no essential role).
}
\end{itemize}

\subsection{Local/global approach towards efficient implementation of bisimulation}\label{sec:conclusion-local-global}

Bisimulation between WDB graph nodes or set names (i.e.\ whether they denote the same hypersets) is a crucial concept for the whole hyperset approach to WDB. The equality symbol (\verb+=+) in our language means, abstractly, the identity between hypersets. But, from the point of view of implementation which deals with set names, rather than with abstract hypersets, the equality operator (\verb+=+) means bisimulation which assumes sufficiently complicated computation. Thus, if we want to remain faithful to this approach and really value this set theoretic style then we should not only implement bisimulation, as it is described in Chapter~\ref{chap:bisimulation}, but also work towards optimising this expensive operation. It can be particularly expensive in the case of distributed WDB when computing bisimulation would assume potentially downloading lots of (possibly) remote WDB files, and we pay special attention to this challenge.

\medskip

The main idea of the local/global approach consists in computing the (global) bisimulation relation ($\bis$) on the whole distributed WDB from many couples of local approximation relations ($\bis^L_+$ and $\bis^L_-$) for each WDB site (or even for each WDB file), and that the latter relations are easily derivable locally. This way the global task is distributed between the main agent (Bisimulation Engine) and local agents (servers of WDB sites). Furthermore, empirical testing suggested that the exploitation of local approximations in the computation of global bisimulation relation $\bis$ can considerably improve performance. Also, the idea that the Bisimulation Engine is working in background time (similarly to Google) to compute the global bisimulation relation from local approximations was crucial in this performance improving strategy. Experiments described in Section~\ref{sec:testing_real_oracle} suggested that bisimulation, although a very challenging problem, especially in distributed case, is not so hopeless practically as it might seem. In particular, taking such optimisations into account the hyperset approach to WDB seems also potentially feasible practically.

\subsection{Further optimisation}

The work done on local/global bisimulation was the main focus of our attempts to optimise our implementation of the hyperset approach in the case of distributed WDB. Also, some additional consideration was given on writing more efficient queries in the current implemented version of $\Delta$, such as the removal of redundancies by using the so called canonisation query \verb+Can(x)+. In fact, this query does not change its input (\verb+Can(x)=x+ as abstract hypersets) but transforms its representation into an equivalent  strongly extensional (non-redundant) form. The effect of using \verb+Can+ in one particular example (in the query which linear orders any hyperset, Section~\ref{sec:lin-ord}) is quite impressive. Another general optimisation related with the recursion operator (and also crucially improving execution time of the linear ordering query mentioned above) is based on the possibility of replacing bisimulation to compare the iteration steps by simple comparison of participating set names only. Of course, further work on optimising the implementation of $\Delta$ (in comparison with writing optimal queries, for example exploiting \verb+Can+ above) remains to be done (see Section~\ref{sec:further_work} below).

\section{Comparisons with other approaches}\label{sec:concluding-comparisons}

After considering various approaches in Chapter~\ref{chap:comp_anal} we have found that the UnQL and Lorel query languages are closest to our approach. However conceptually, i.e., in fact, from the point of view of the hyperset approach, UnQL is the most close to $\Delta$. The implemented $\Delta$-language does not include yet path expressions typical for other approaches. But, this language is already a very expressive, and, in a sense, subsumes both the UnQL and (the main features of) Lorel languages.

\section{Further work}\label{sec:further_work}

In short, the primary goal of implementation and attempts towards optimisation described in this thesis can be considered as successful. However, development of the implementation and the experiments was very time consuming, and there was insufficient time to implement all potential ideas. Many useful features have yet to be implemented, such as:

\begin{itemize}
\item{
{\bf Extending the implemented $\Delta$-query language to make it more user friendly} with quantification over multiple variables. Also, similarly for the case of collection, separation and recursion constructs. 
}
\item{
{\bf Improving the library function}, in particular to allow multiple or user defined libraries.
}
\item{
{\bf Extending the implemented $\Delta$-query language to include path expressions} which are typically included in other approaches towards semi-structured databases and, additionally, are very useful practically. In principle, path expressions could be implemented by rewriting them into $\Delta$-queries according to definitions in \cite{S06}. But, straightforward implementation should be more efficient.
}
\item{
{\bf Extending the implemented $\Delta$-query language by update queries.}
}
\item{
{\bf More user friendly interface} for inputting queries and WDB, as well as for outputting query results. In particular, the graphical visualisation of WDB and query results (developing a special WDB browser, as well as an editor for WDB files).
}
\end{itemize}

\noindent
Additionally, suitable techniques should be developed for creating WDB, taking into account its hyperset theoretic character:

\begin{itemize}
\item{
{\bf Using WDB schemas} in the context of hyperset approach to impose restriction on the structure of WDB, just like in the relational approach but not necessarily so rigid. In fact, enforcing structure makes queries easier to write, and, additionally, can serve to eliminate possible unintended redundancies in set equations which could arise otherwise due to poor WDB design.
}
\end{itemize}

\noindent
Furthermore, although some suggestions towards efficiency were made here, there remains much work towards development of a practically efficient implementation:

\begin{itemize}
\item{
{\bf Adapting known and developing new optimisation techniques} such as indexing, hashing and other data structures helping to implement efficient searching as described in \cite{U88} to the case of semi-structured data. Redundancies in set equations arising during computation should be regularly eliminated, thus allowing writing queries without explicit using the canonisation query. In this case equality between sets trivially becomes the identity relation rather than the bisimulation relation. Also, identical query calls should be executed only once.} 
\item{
{\bf Dealing with redundancies} in various circumstances by developing various techniques and methodology e.g.\ related with redundancies (bisimilarities) arising due to local updates in a WDB file (answering questions such as: are redundancies possibly arising in such local way easy to eliminate? under which conditions? etc.), or due to mirroring WDB sites, etc.
}
\item{
{\bf Further improvements on the bisimulation engine} transforming it from imitational to a more realistic version (Web service) assuming several levels (granularity) of locality (WDB-files, WDB-sites, the whole WDB) and extending the range of experiments with this engine.
}
\item{
{\bf Adopting known \cite{DPP04, F89} and developing new techniques for optimisation of bisimulation} which, for example, may take advantage of WDB scheme (see above).
}
\end{itemize}

\noindent
There is great scope for further theoretical and practical work. In summary, this could mean developing a full-fledged WDB management system and also WDB design techniques, and other methodologies based on the hypeset approach. Of course, the hyperset approach could be further evolved, e.g.\ it can be extended to also involve standard datatypes like integers, reals, strings as atomic data or label values with arithmetical and other operations over them (completely lacking in the current version of $\Delta$), etc. Also, multi-hypersets \cite{LS07}, records, lists, etc. could be allowed. Another version of the $\Delta$ language capturing LogSpace \cite{LS01, LS97logspace} (currently for well-founded sets only) could be either implemented in its present form or, firstly, theoretically extended to the case of hypersets. Anyway, working on the theoretical level in various directions and simultaneously developing more practically oriented implementations, like in this thesis, seems a fruitful style of research.


%
%

\cleardoublepage

\fancyhead[LE]{\thepage\hspace{2em}\footnotesize{\nouppercase{\appendixname}}}
\pagestyle{fancy}




%
%

\appendix
\chapter{Appendix}

\section{Implemented BNF grammar of $\Delta$-query language}\label{app:BNF}

The grammar of the implemented $\Delta$-language is represented by the metasyntax notation Extended \mbox{Backus-Naur} Form (EBNF) which allows for example to define the repetition of syntactical categories using \verb+*+ or \verb=+= (unlike regular BNF which does not have these features). For example, the EBNF production rule of \verb+<declarations>+ in Section~\ref{app:BNF_declarations} defines an infinite number of possible forks, with any number of leaves labelled by \verb+<declaration>+ each separated by the terminal leaf labelled by \verb+","+.

\medskip

\noindent
The EBNF notation (used here to express the $\Delta$-language grammar) defines production rules as sequence of terminals (symbols) or non-terminals,

\medskip

\begin{tabular}{ll}
\verb+"xxx"+	& - Terminal\\
\verb+<yyy>+	& - Non-terminal\\
\end{tabular}	

\medskip

\noindent
where production rules are constructed (from those terminals or non-terminals) according to the following rules,

\medskip

\begin{tabular}{ll}
Parentheses, \verb+()+	& - Grouping \\
Vertical bar, \verb+|+	& - Alternation \\
Square brackets, \verb+[]+	& - Optional \\
& \\
Kleene star, \verb+*+	& - Repeat 0 or more times \\
Kleene plus, \verb=+=	& - Repeat 1 or more times \\
& \\
\end{tabular}

\subsection*{Top level commands}

\begin{small}
\begin{verbatim}
<top level command> ::=
    ( "library" <library command> | <query> | "exit" ) ";"

<query> ::=
    "boolean query" <delta-formula> | "set query" <delta-term>
\end{verbatim}
\end{small}

\subsection*{Library commands}
\begin{small}
\begin{verbatim}
<library command> ::=
    "add" <declarations> |
    "list" [ "verbose" ]
\end{verbatim}
\end{small}

\subsection*{Declarations}\label{app:BNF_declarations}

\begin{small}
\begin{verbatim}
<declarations> ::= <declaration> ( "," <declaration> )*

<declaration> ::=
    <set constant declaration> | <label constant declaration> |
    <set query declaration> | <boolean query declaration>

<set constant declaration> ::=
    "set constant" <set constant> ("be"|"=") <delta-term>

<label constant declaration> ::=
    "label constant" <label constant> ("be"|"=") <label value>

<set query declaration> ::=
    "set query" <set query name> "(" <variables> ")" ("be"|"=")
    <delta-term>

<boolean query declaration> ::=
    "boolean query" <boolean query name> "(" <variables> ")"
    ("be"|"=") <delta-formula>

<variables> ::= <variable> ( "," <variable> )*
<variable> ::=  ( "set" <set variable> | "label" <label variable> )

<parameters> ::= <parameter> ( "," <parameter> )*
<parameter> ::=  ( <delta-term> | <label> )

<boolean query name> ::= <identifier>
<set query name> ::=     <identifier>
\end{verbatim}
\end{small}

\newpage
\subsection*{$\Delta$-terms}

\begin{small}
\begin{verbatim}
<delta-term> ::= <set variable> |
                 <set constant> |
                 <set name> |
                 <atomic value> |
                 <enumerate> |
                 <union> |
                 "(" <multiple union> ")" |
                 <collect> |
                 <separate> |
                 <transitive closure> |
                 <recursion> |
                 <decoration> |
                 <if-else term> |
                 <set query call> |
                 <delta-term with declarations>

<set name> ::= <URI> "#" <simple set name>

<atomic value> ::= """ <identifier> """

<enumerate> ::= "{" <labelled terms> "}"

<union> ::=     (  "U" | "union" ) <delta-term>

<multiple union> ::=
    <delta-term> ( ( "U" | "union" ) <delta-term> )*

<collect> ::=
    "collect" "{" <labelled term> ( "where" | "|" ) <variable pair>
    ("in"|"<-") <delta-term> [ "and" <delta-formula> ] "}"

<separate> ::=
    "separate" "{" <variable pair> ("in"|"<-") <delta-term>
    ( "where" | "|" ) <delta-formula> "}"

<transitive closure> ::=
    ( "tc" | "TC" | "transitiveclosure" ) <delta-term>

<recursion> ::=
    "recursion " <set variable> " {" <variable pair> (" in "| "<-")
    <delta-term> ( "where" | "|" ) <delta-formula> "}"

<decoration> ::=   "decorate" "(" <delta-term> ", " <delta-term> ")"

<if-else term> ::= "if" <delta-formula> "then" <delta-term>
                   "else" <delta-term> "fi"

<set query call> ::= "call" <set query name> "(" <parameters> ")"

<delta-term with declarations> ::=
    "let " <declarations> "in" <delta-term> " endlet"

<URI> ::=          ( <web prefix> | <local prefix> ) <file path>
<web prefix> ::=   "http://" <host> "/" [ "~" <identifier> "/" ]
<local prefix> ::= "file://" ( (A-Z) | (a-z) ) ":/"
<host> ::=         <identifier> [ "." <host> ]
<file path> ::=    <identifier> ( "/" <file path> | <extension> )
<extension> ::=    ".xml"
<simple set name> ::= <identifier>
\end{verbatim}
\end{small}

\subsection*{$\Delta$-formulas}

\begin{small}
\begin{verbatim}
<delta-formula> ::=	<atomic formula> |
                    "(" <conjunction> ")" |
                    "(" <disjunction> ")" |
                    "(" <quasi-implication> ")" |
                    <quantified formula> |
                    <negated formula> |
                    <if-else formula> |
                    <delta-formula with declarations>

<atomic formula> ::=
    <equality> | <label relationship> | <membership> |
    <boolean query call> | "true" | "false"

<equality> ::=     <set equality> | <label equality>

<set equality> ::= <delta-term> "=" <delta-term>

<label equality> ::=
    <label> "=" <wildcard label> | <wildcard label>  "=" <label>
\end{verbatim}
\end{small}

\pagebreak

\begin{small}
\begin{verbatim}
<wildcard label> ::=
    ["*"] ( <label variable> | <label constant> ) ["*"] |
    "'" ["*"] <identifier> ["*"] "'"

<label relationship> ::= <label> "<" <label>
                         <label> ">" <label>
                         <label> "<=" <label>
                         <label> ">=" <label>

<membership> ::=        <labelled term> ("in"|"<-") <delta-term>

<boolean query call> ::= "call" <boolean query name>
    "(" <parameters> ")"

<if-else formula> ::= "if" <delta-formula> "then" <delta-formula>
                      "else" <delta-formula> "fi"

<delta-formula with declarations> ::=
    "let" <declarations> "in" <delta-formula> "endlet"

<conjunction> ::=       <delta-formula> ( "and" <delta-formula> )*

<disjunction> ::=       <delta-formula> ( "or" <delta-formula> )*

<quasi-implication> ::= <delta-formula>
    ( <quasi-implication connective> <delta-formula> )*

<quasi-implication connective> ::=
    "<=" | "=>" | "implies" | "iff" | "<=>"

<quantified formula> ::= <forall> <delta-formula> |
                         <exists> <delta-formula> |

<forall> ::=
    "forall" <variable pair> ("in"|"<-") <delta-term> [ "." ]

<exists> ::=
    "exists" <variable pair> ("in"|"<-") <delta-term> [ "." ]

<negated formula> ::= "not" <delta-formula>
\end{verbatim}
\end{small}

\subsection*{Variables, constants, literals etc.}

\begin{small}
\begin{verbatim}
<label> ::= <label variable> | <label value> | <label constant>
<label variable> ::= <identifier>
<label constant> ::= <identifier>
<label value> ::=    "'" <identifier> "'"

<set variable> ::=   <identifier>
<set constant> ::=   <identifier>

<labelled terms> ::= <labelled term> ( "," <labelled term> )*
<labelled term> ::=  <label> ":" <delta-term>

<variable pair> ::=  <variable pair label> ":" <variable pair term>
<variable pair label> ::= <label variable> | <label value>
<variable pair term> ::= <set variable>

<identifier> ::=     ( (A-Z) | (a-z) | (0-9) | "_" | "-" )+
\end{verbatim}
\end{small}

\clearpage

\section{Example XML-WDB files}\label{app:XML-WDB_files}


\begin{xml-file}[!h]
\begin{small}
\begin{verbatim}


    <?xml version="1.0"?>

    <set:eqns 
     xmlns:xsi="http://www.w3.org/2001/XMLSchema-instance" 
     xsi:noNamespaceSchemaLocation=
     "http://www.csc.liv.ac.uk/~molyneux/XML-WDB/schema/xml-wdb.xsd"
     xmlns:set="http://www.csc.liv.ac.uk/~molyneux/XML-WDB">

      <set:eqn set:id="BibDB">
        <paper set:href=
         "http://www.csc.liv.ac.uk/~molyneux/t/BibDB-f2.xml#p1"/>
        <paper set:href=
         "http://www.csc.liv.ac.uk/~molyneux/t/BibDB-f2.xml#p2"/>
        <paper set:href=
         "http://www.csc.liv.ac.uk/~molyneux/t/BibDB-f2.xml#p3"/>
        <book set:ref="b1"/>
        <book set:ref="b2"/>
      </set:eqn>

      <set:eqn set:id="b1">
        <refers-to set:ref="b2"/>
        <refers-to set:href=
         "http://www.csc.liv.ac.uk/~molyneux/t/BibDB-f2.xml#p1"/>
      </set:eqn>

      <set:eqn set:id="b2">
        <author>Jones</author>
        <title>Databases</title>
      </set:eqn>

    </set:eqns>
\end{verbatim}
\end{small}
\caption{XML-WDB file http://www.csc.liv.ac.uk/\~{}molyneux/t/BibDB-f1.xml (cf. Section~\ref{sec:example_queries}).}
\label{WDB_XML_BibDB_URL1}
\end{xml-file}

%
%

\clearpage

\begin{xml-file}[th]
\begin{small}
\begin{verbatim}


    <?xml version="1.0"?>

    <set:eqns 
     xmlns:xsi="http://www.w3.org/2001/XMLSchema-instance" 
     xsi:noNamespaceSchemaLocation=
     "http://www.csc.liv.ac.uk/~molyneux/XML-WDB/schema/xml-wdb.xsd"
     xmlns:set="http://www.csc.liv.ac.uk/~molyneux/XML-WDB">

      <set:eqn set:id="p1">
        <refers-to set:ref="p2"/>
      </set:eqn>

      <set:eqn set:id="p2">
        <author>Smith</author>
        <title>Databases</title>
        <refers-to set:ref="p3"/>
      </set:eqn>

      <set:eqn set:id="p3">
        <author>Jones</author>
        <title>Databases</title>
      </set:eqn>

    </set:eqns>
\end{verbatim}
\end{small}
\caption{XML-WDB file {http://www.csc.liv.ac.uk/\~{}molyneux/t/BibDB-f2.xml} (cf. Section~\ref{sec:example_queries}).}
\label{WDB_XML_BibDB_URL2}
\end{xml-file}

\clearpage

\section{Predefined library queries}\label{app:predefined}\label{app:lin-ord-declarations}

\begin{small}
\begin{verbatim}
    set query Pair (set x,set y) be
        { 'fst':x, 'snd':y },

    boolean query isPair (set p) be (
            exists l: x in p . (
                l='fst'
                and
                forall m:z in p . ( m='fst' => z=x )
            )
            and
            exists l:y in p . (
                l='snd'
                and
                forall m:z in p .( m='snd' => z=y )
            )
        ),

    set query First (set p) be
        union separate { l:x in p where l='fst' },

    set query Second (set p) be
        union separate { l:x in p where l='snd' },

    set query CartProduct (set x,set y) be
        union collect {
            'null':collect {
                'null':call Pair ( xx, yy )
                where l:yy in y
            }
            where m : xx in x
        },

    set query Square (set z) be
        call CartProduct ( z, z ),

    set query LabelledPairs (set v) be
        collect { l:{ 'fst':v, 'snd':u } where l:u in v },

    set query Nodes (set g) be
        union separate { m:p in g where call isPair ( p ) },
\end{verbatim}
\end{small}

\begin{small}
\begin{verbatim}
    set query Children (set x,set g) be
        collect {
            l:call Second ( p )
            where l:p in g
            and (
                call isPair ( p )
                and
                call First ( p ) = x
            )
        },

    set query Regroup (set g) be
        collect {
            'null':call Pair ( x, call Children ( x , g ) )
            where m : x in call Nodes ( g )
        },

    set query CanGraph (set x) be
        union collect {
            'null':call LabelledPairs ( v )
             where m:v in TC ( x )
        },

    set query Can (set x) be
        decorate ( call CanGraph ( x ), x ),

    set query TCPure(set x) be
        collect{ 'null':v where l:v in TC ( x ) },

    set query HorizontalTC (set g) be
        recursion p {
            'null':pair in call Square ( call Nodes ( g ) )
            where (
                call First ( pair ) = call Second ( pair )
                or
                exists m:z in call Nodes ( g ) . (
                   'null':call Pair ( call First ( pair ), z ) in p
                   and
                   'null':call Pair ( z, call Second ( pair ) ) in g
                )
            )
        },

    set query TC_along_label (label l,set z) be 
        recursion p {
            k:x in TC ( z )
            where (
                ( x=z and k = 'null' )
                or
                ( k=l and exists m:y in p . l:x in y )
            )
        },

    set query SuccessorPairs (set L) be 
        separate {
            l:pair in L
            and not exists l:x in call Nodes(L) . (
                'null':call Pair ( call First ( pair ),x ) in L
                and 
                'null':call Pair ( x, call Second ( pair ) ) in L
            )
        },

    boolean query Precedes5(set R,label l,set x,label m,set y) be (
            l < m
            or (
                l=m
                and
                exists 'null':p in R . (
                    'fst':x in p and 'snd':y in p
                )
            )
        ),

\end{verbatim}
\end{small}

\newpage

\begin{small}
\begin{verbatim}
    set query StrictLinOrder_on_TC (set z) be
        recursion R {
            'null':p_xy in call Square( call Can(call TCPure(z)) )
            where (
                (
                  not 'null':p_xy in R
                  and
                  not exists 'fst':xx in p_xy .
                        exists 'snd':yy in p_xy .
                          exists 'null':inv_p in R . (
                            'fst':yy in inv_p
                            and
                            'snd':xx in inv_p
                          )
                )
                and
                exists 'snd':yyy in p_xy .
                  exists lu:u in yyy . (
                    exists 'fst':xxx in p_xy .
                      forall lv:v in xxx . (
                        call Precedes5(R,lu,u, lv,v)
                        or
                        call Precedes5(R,lv,v, lu,u)
                      )
                    and
                    forall fs:xy in p_xy .
                      forall lw:w in xy . (
                        call Precedes5(R,lu,u, lw,w) =>
                        exists 'fst':xxxx in p_xy .
                          exists lp:p in xxxx .
                            exists 'snd':yyyy in p_xy .
                              exists lq:q in yyyy . (
                                not call Precedes5(R,lp,p, lw,w) and
                                not call Precedes5(R,lw,w, lp,p) and
                                not call Precedes5(R,lq,q, lw,w) and
                                not call Precedes5(R,lw,w, lq,q)
                              )
                      )
                  )
            )
        }
\end{verbatim}
\end{small}


%
%
\clearpage

\backmatter{}

\fancyhead[LE]{\thepage\hspace{2em}\footnotesize{\nouppercase{\leftmark}}}

\addcontentsline{toc}{chapter}{Bibliography}
\bibliographystyle{plain}
\bibliography{thesis_references}


\end{document}